\documentclass[a4paper,fleqn]{cas-sc}
\usepackage{amsmath}
\usepackage{amssymb}
\usepackage{graphicx}
\usepackage{mathrsfs}
\usepackage{amsfonts}
\usepackage{amsthm}
\usepackage{color}
\usepackage{titletoc}
\usepackage{hyperref}
\usepackage{physics}
\usepackage{ulem}

\titlecontents{section}
              [1.1cm]
              { \normalsize}%
              {\contentslabel{2em}}%
              {}%
              {\titlerule*[0.3pc]{$\cdot$}\normalsize\contentspage\hspace*{0cm}}%
\titlecontents{subsection}
              [2cm]
              {\normalsize}%
              {\contentslabel{2em}}%
              {}%
              {\titlerule*[0.3pc]{$\cdot$}\normalsize\contentspage\hspace*{0cm}}%
\titlecontents{subsubsection}
              [2.9cm]
              {\normalsize}%
              {\contentslabel{2.5em}}%
              {}%
              {\titlerule*[0.3pc]{$\cdot$}\normalsize\contentspage\hspace*{0cm}}%

\usepackage{lipsum}
\usepackage{gensymb}
\usepackage[numbers,sort&compress]{natbib}
\usepackage[numbers]{natbib} 

\def\tsc#1{\csdef{#1}{\textsc{\lowercase{#1}}\xspace}}
\tsc{WGM}
\tsc{QE}
\tsc{EP}
\tsc{PMS}
\tsc{BEC}
\tsc{DE}

\begin{document}
\let\WriteBookmarks\relax
\def\floatpagepagefraction{1}
\def\textpagefraction{.001}
\shorttitle{Tao Yu, Xi-Han Zhou, Gerrit E. W. Bauer, and I. V. Bobkova}
\shortauthors{Tao Yu et~al.}

\title [mode = title]{Electromagnetic Proximity Effect: Superconducting Magnonics and Beyond}

\author[a]{Tao Yu}
[orcid=https://orcid.org/0000-0001-7020-2204]
\cormark[1]
\ead{taoyuphy@hust.edu.cn}

\address[a]{School of Physics, Huazhong University of Science and Technology, Wuhan 430074, China}

\author[a]{Xi-Han Zhou}

\author[b,c]{Gerrit E. W. Bauer}[orcid=https://orcid.org/0000-0002-3615-8673]
\address[b] {Kavli Institute for Theoretical Sciences, University of  the Chinese Academy of Sciences, Beijing 100190, China}
\address[c]{WPI-AIMR and Institute for Materials Research and CSRN, Tohoku University, Sendai 980-8577, Japan}

\author[d,e]{I. V. Bobkova}
[orcid=https://orcid.org/0000-0003-1469-186]
\address[d]{Moscow Institute of Physics and Technology, Dolgoprudny, 141700 Moscow region, Russia}
\address[e]{National Research University Higher School of Economics, 101000 Moscow, Russia}

\cortext[cor1]{Corresponding author.}

\begin{abstract}
The exchange interaction at interfaces between superconductors (SCs) and ferromagnets (FMs) has been a central topic in condensed matter physics for many decades, starting with the prediction of exotic phases such as the Fulde–Ferrell–Larkin–Ovchinnikov states and leading to the discovery of triplet superconductivity. This review focuses on new phenomena in SC|FM heterostructures caused by the \textit{non-contact dipolar interaction} between magnons, i.e., the quanta of spin wave excitations in the ferromagnet, and the superconducting order. A universal non-relativistic spin-orbit coupling locks the polarization and momentum of their evanescent stray magnetic fields and leads to chiral screening by proximate superconductors. The interaction-induced hybrid quasiparticles are magnon-Meissner collective modes, magnon-cooparon, Josephson plasmonic modes, and nodal magnon-photon polaritons. Superconducting and normal metallic gates modulate and control the magnetodipolar interaction and thereby magnetization and energy transport at interfaces and in thin films.\\
Here we review our understanding of dipole-coupled SC|FM systems obtained in the past five years, including the chiral gating of magnons in magnetic films by superconducting or metallic gates, proximity effects such as enhanced magnon transport, novel magnonic crystals, superconducting magnon transistors, ultrastrong magnon-photon coupling, and composite quasiparticles of magnons screened by triplet Cooper pairs. We emphasize the need for a self-consistent treatment of the coupled magnon-Cooper pair dynamics in solving the coupled Landau-Lifshitz-Gilbert equation and Maxwell equations. By replacing London’s penetration depth with the microwave skin depth, we capture the difference between the screening by superconducting and highly conducting metals. The normal metal|ferromagnet bilayer system dynamics can then be represented by a non-Hermitian chiral Hamiltonian that leads to the prediction of a switchable magnon Peltier effect, a chiral magnon non-Hermitian skin effect, nodal magnon-photon polaritons, anomalous scattering of spin waves at semi-transparent potential barriers, and dipolar spin pumping. Finally, we point out the new functionalities offered by SC|FM devices for both classical and quantum information technologies and present perspectives for future research.
	
\end{abstract}


\maketitle

\tableofcontents

\section{Introduction}

The spontaneous emergence of order in an interacting many-body system at a critical temperature is arguably one of the most fascinating phenomena in physics~\cite{Anderson_book}. The symmetry breaking associated with these phase transitions induces dramatic changes in the physical properties of the systems. Textbooks discuss, for example, the crystal formation out of liquid or gaseous phases, the superfluidity of atomic condensates, the superconductivity of metals, and the macroscopic magnetic/electric fields generated by ferromagnetism/ferroelectricity in solids, respectively.

\begin{figure}[htp!]
    \centering
    \includegraphics[width=0.9\linewidth]{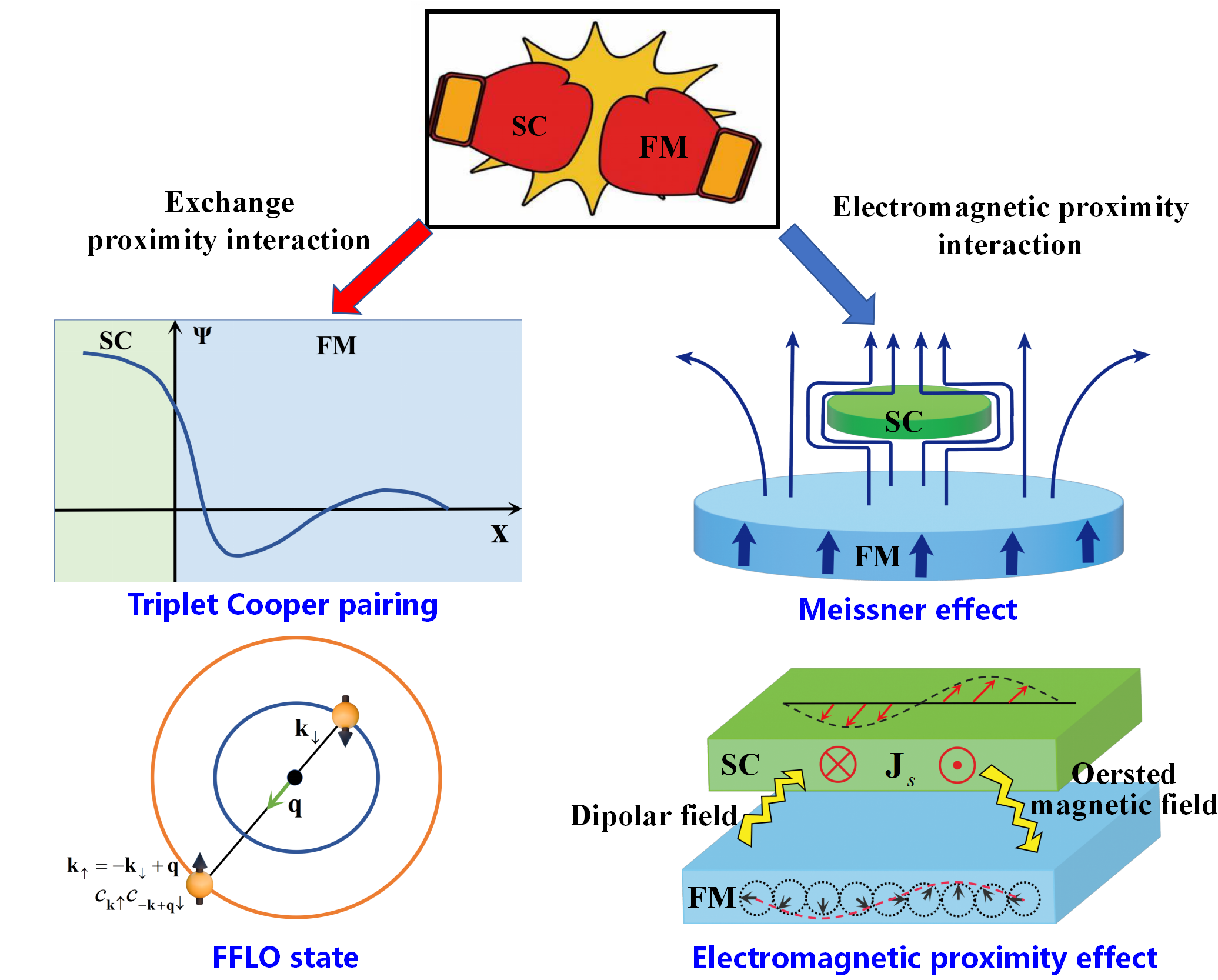}
    \caption{Overview of the main coupling mechanisms between a superconductor and a ferromagnet, viz. the $s$-$d$ exchange interaction and the dipolar interaction. The exchange interaction either in the magnet or at the interface can lead to triplet Cooper pairing and the Fulde-Ferrell-Larkin-Ovchinnikov (FFLO) state. The dipolar proximity interaction gives rise to the Meissner and electromagnetic proximity effects.}
    \label{overview_introduction}
\end{figure}

Different orders may coexist in a friendly manner within a given material system, such as ferroelectricity and magnetism in multiferroic materials. On the other hand, magnetism is, in general, a foe to superconductivity since it suppresses Cooper pair formation, as illustrated in Fig.~\ref{overview_introduction}. Superconducting ferromagnets only exist in exotic states such as the spatially modulated Fulde–Ferrell–Larkin–Ovchinnikov (FFLO) phase~\cite{Larkin3,Fulde,RevModPhys.76.263},  in the form of triplet Cooper pairs with finite orbital angular momentum intrinsically in materials such as uranium-based heavy-fermion superconductors~\cite{doi:10.1073/pnas.2403067121}, in some other unconventional superconductors, including borocarbides \cite{Muller2001}, ternary borides, chalcogenides~\cite{Buzdin1984}, and iron-based superconductors~\cite{Zapf2017,Kim2022}, or in the form of a long-range proximity effect~\cite{Bergeret_2005,Klapwijk}.

Advanced thin film material growth and microfabrication techniques allow the study of heterostructures such as bi- and multilayers of differently ordered materials. In the case of ferromagnets (FMs) on superconductors (SCs), the contact exchange interaction dominates only at atomic-scale distances from the interfaces. The effects can still be drastic when the respective films have a thickness smaller than the superconducting coherence length. Ferromagnet(FM)|superconductor(SC) bi- and trilayers allow the controlled study of the LOFF oscillations~\cite{Klapwijk}. Numerous reviews address the theory and experiments of equilibrium and transport properties of coupled FM$|$SC heterostructures~\cite{Bergeret_2005,RevModPhys.77.935,Linder_045005,Linder2015,Eschrig_2015,Melnikov:2022,RevModPhys.96.021003}. Novel multiferroic composites can be synthesized in the form of multilayers between ferroelectric and magnetic materials~\cite{RevModPhys.77.1083,10.1063/1.4921545,Spaldin2017,10.1093/nsr/nwz023,RevModPhys.96.015005}.

Despite the enormous bulk of published research results, we perceive the need to review the developments in the study of SC|FM heterostructures that address new insights into the effects of non-local electromagnetic interactions. Diamagnetism gives rise to phenomena such as the Meissner effect and the levitation of superconducting disks on top of ferromagnets, as shown in Fig.~\ref{overview_introduction}. On a microscopic scale, electromagnetic interactions are not considered important since conduction electrons efficiently screen the stray fields in equilibrium and low-frequency phenomena. However, the elementary excitations of the magnetic order, i.e., spin waves with quanta referred to as magnons~\cite{lenk,Bauer2012,Chumak,Grundler,Demidov,RevModPhys.91.035004,Yu_chirality}, are typically oscillating in the gigahertz (GHz) regime. Their stray fields can penetrate metals and superconductors over length scales that can easily exceed the thickness of ultrathin films. Since the response of the conductors in turn affects the magnetization, the dynamics of the entire system pose a self-consistency problem, also sketched in Fig.~\ref{overview_introduction}. Over the past few years, solutions for various configurations and devices have revealed rich ramifications that are measurable and potentially relevant to applications, implying that the time is ripe to compile the evidence in a comprehensive review.

The present review focuses on the basic as well as novel physics of dynamical properties of multilayers that incorporate magnets, normal metals, and superconductors, with some discussion of ferroelectrics as well. It is ordered as follows. In Sec.~\ref{Magnetization_dynamics_and_Cooper_pairs}, we introduce the energy and elementary excitations of magnetic orders~\cite{lenk,Chumak,Grundler,Demidov,Brataas,Barman,Yu_chirality,Cornelissen,Zou,Wang,Flebus,ani1,ani2,ani3,DM_1,DM_2,mag_book,quenching,kittel_book,spin_wave_book,Jackson,magnon_photon,Holstein,LL_eq,normalized1,normalized2} as well as the fundamental equations describing the dynamics of superconductors~\cite{Leggett_book,Cooper_pairs_1956,BCS_1957,HS_pairing_symmetry,FS_order_parameter,Sigrist,Abrikosov,Gorkov,Bergeret,Tokatly,de:Gennes,Eilenberger,Larkin,Larkin1,Larkin2,Larkin3,Usadel,Eliashberg,Shelankov,Shelankov1,Eckern,Rammer, Kuprianov,Kopnin,Chandrasekhar,Takane,Bauer,Konschelle,Espedal,Ginzburg,Zhao,Yamashita,M.:J.:Stephen,Wolfle,Einzel,Wolfle2,Aronov,Combescot,Hara,Andreev,Klam,Manske,Buzdin,Tokatly_1,Schmid,Bobkova2021,Linder_045005,Linder2015,Eschrig_2015,Melnikov:2022}, 
 at low frequencies by the two-fluid model~\cite{33_cavity_S_FI_coupling_Jacobsen,two_fluid1,two_fluid2,two_fluid3,two_fluid4,two_fluid5,two_fluid6,two_fluid7,two_fluid8,two_fluid9,two_fluid10,two_fluid11,two_fluid12,two_fluid13,two_fluid14,two_fluid15,two_fluid16,London_eq,two_fluid_superconductor,two_fluid_Janssonn,two_fluid_Tao}. In Sec.~\ref{Dynamic_dipolar_interaction_between_a_ferromagnet_and_a_superconductor}, we present the general theory of the electromagnetic interaction between a ferromagnet and a superconductor by addressing the dipolar field emitted by the magnon that drives the Meissner supercurrent in the superconductor. The Oersted magnetic fields in turn modulate the magnons~\cite{46_Efficient_Gating,21_Fraerman,2_FS_for_Magnonic_application_Ustinov,48_Giant_enhancement_of_magnon_transport,24_SF_shift_van_der_sar}. The reaction of the superconductor to the magnon depends on the characteristics of the dipolar magnetic and electric fields, which is governed by a chirality that locks the propagation, spin, and surface normal by a right-hand rule~\cite{PhysRevA.84.025602,khanikaev2013photonic,Tao_chiral_pumping,Tao_chiral_excitation,DEmode,DE2,Yu_chirality,Tao_Magnon_Accumulation,Yu_Chiral_Coupling2021,BLIOKH20151,spin_density,PhysRevApplied.22.034042,SOC1,SOC2,SOC3,SOC4,SOC5,spin_density1,spin_density2,spin_density3} and leads to the chiral gating of magnons~\cite{46_Efficient_Gating,Golovchanskiy_gating2,Golovchanskiy_gating4,21_Fraerman,55_Burmistrov,42_Centala,Golovchanskiy_gating1_exp,Golovchanskiy_gating3_exp,8_Pt_S_F_vortex_Blamire,13_Passamani,54_L.Tao} and chiral enhancement of magnon transport  by superconductors~\cite{48_Giant_enhancement_of_magnon_transport}. 
We also review magnonic crystals realized by proximate superconductors owing to the static Meissner effect~\cite{5_magnon_fluxon_Nat_Chumak,19_magnon_fluxonics_Chumak,34_Berakdar,39_Cherenkov_radiation_Buzdin,37_klos,Golovchanskiy_gating3_exp}. In addition, we address the chiral electric field emitted by ferrons~\cite{ferron,bauer_ferron_trans,ferroelectric,Capacitors,Point_Contacts,ferron_transport_exp,Coherent_exp,Diffuse_exp,fe_polarition,bulkmode_ferron2,surface_ferron,bulkmode_ferron,ferron_landau,ferron_landau1,LiNbO3,LiNbO3_g,LKT1,LKT2,LKT3,LKT4}, i.e., the collective mode excitation of ferroelectric dipoles, which can analogously interact strongly with the superconductor.
Section~\ref{Ultrastrong_coupling_in_SC_FM_SC_Josephson_junctions} focuses on the superconductor$|$ferromagnet$|$superconductor Josephson junctions in which the dipolar field emitted by the spin waves is reflected back and forth by the adjacent superconducting gates. The associated magnon confinement shifts the ferromagnetic resonance frequency~\cite{4_sun,7_SFS_shift_Blamire,10_SFS_shift_Ustinov,25_SFS_shift_Stolyarov,bai_shift,20_shift_Silaev,47_Gating_ferromagnetic_resonance,16_Giant_demagnetization_effects_Buzdin,shift_yangguang} and enhances the magnon-photon coupling to become ultrastrong~\cite{18_photon-to-magnon_coupling_Ustinov,52_Persistent_nodal_magnon-photon_polariton_in_ferromagnetic_heterostructures,23_magnon_photon_coupling_Silaev}. 
The supercurrent in Josephson junctions may also couple to magnons, which generates a Josephson plasmonic mode~\cite{41_Eremin,1_Efetov,36_Bobkov,Josephson_book}. 
In Sec.~\ref{Exchange_interaction_between_magnons_and_Cooper_pairs}, for the completeness of the physical picture, we discuss the dynamical effect of the exchange interaction, which results in the appearance of magnon-cooparons, i.e., the quasiparticles caused by the dynamical screening cloud of triplet Cooper pairs induced by the spin-wave exchange magnetic field~\cite{PhysRevLett.127.207001,Bobkova2022,Bobkov2023,Bergeret2018,Buzdin,Bergeret,Nembach2015,Demokritov2001,Cornelissen2015,Goennenwein2015,Kamra2018,Sadovnikov2015,Wang2018}, which exhibit increased effective mass, screened spin, and spatial non-locality. In Sec.~\ref{Superconducting2normal_metals}, we expand the applicability
of previous results for superconductors to normal metals in which the chiral stray field drives a normal rather than a supercurrent in the contacts. We show that we only have to replace London's penetration depth $\lambda_L$ of the superconductor by a complex skin depth $\delta$ of the normal metal. We show that the normal metal contacts damp the magnetization dynamics in a chiral fashion~\cite{LIU2024172556,50_Chiral-damping-enhanced_magnon_transmission,45_Imaging_spin_wave_damping_underneath_metals,Chiral_Damping_of_Magnons}, enabling a non-Hermitian skin effect~\cite{50_Chiral-damping-enhanced_magnon_transmission}, magnon transistors~\cite{49_Chirality_enables_thermal_magnon_transistors,53_van_Wees,15_Liu}, and nodal magnon-photon polaritons~\cite{52_Persistent_nodal_magnon-photon_polariton_in_ferromagnetic_heterostructures}. We also review the experimental evidence for spin transfer mediated by the dipolar interaction and present theoretical predictions of the chiral locking of an injected magnon flow and associated spin accumulation directions~\cite{31_Reversal_enabling_magnon_memory_Grundler,38_Magnetization_Reversal_Grundler,inject_exp,Chiral_Injection_of_Magnons}.  We summarize this review article and give an outlook for future studies in Sec.~\ref{summary}.

\section{Magnetization dynamics and Cooper pairs}
\label{Magnetization_dynamics_and_Cooper_pairs}

\subsection{Magnetic order}

The magnetic moment ${\pmb \mu}$ is a basic physical quantity in magnetism. Classically, it is the leading term of a multipole expansion of the stray magnetic fields of an ensemble of magnetic charges and electric currents.
The simplest magnetic moment  ${\pmb \mu}_l$ is generated by an electron that moves in a simple loop. It is proportional to the orbital angular momentum ${\bf l}$ via ${\pmb \mu}_l=-|g_L \mu_B| {\bf l}$, in which $g_L$ is the $g$-factor of the orbital magnetic moment and $\mu_B$ is the Bohr magneton.
According to relativistic quantum mechanics, the intrinsic angular momentum of the electrons or spin $\bf s$,  causes a spin magnetic moment ${\pmb \mu}_s=- |g_S\mu_B| {\bf s}$, where $g_S$ is the $g$-factor of the spin magnetic moment. In the non-relativistic limit,  $g_S\sim 2$ and $g_L= 1$, and the total magnetic moment is the sum of the orbital and spin contributions~\cite{mag_book}
\begin{align}
    {\pmb \mu}={\pmb \mu}_l+{\pmb \mu}_s \approx -{\mu_B} ({\bf l}+2{\bf s}).
\end{align}
In magnetic materials, the electron spins order spontaneously to generate persistent magnetization distributions.   
Inside a ferromagnet, the magnetization distribution ${\bf M}({\bf r})=\sum_i {\pmb \mu}_i/\delta V$ is an average over the $i$-th  magnetic moment inside a mesoscopic volume $\delta V$. When a crystal field hinders the 
formation of current loops and suppresses the formation of an orbital magnetic moment~\cite{quenching}, the magnetization ${\bf M}({\bf r})=-\gamma\hbar{\bf s}({\bf r})$ is proportional to the spin density ${\bf s}({\bf r})$, where $-\gamma=-|g_S\mu_B|/\hbar$ is the gyromagnetic ratio of an electron.

Magnetic moments interact with external and internal effective magnetic fields by a Zeeman interaction \(H_Z\).  When ${\bf B}_0=\mu_0 {\bf H}_0$ is the applied magnetic flux density in the magnetic medium, and disregarding internal fields for the moment
\begin{align}
    H_Z=-\mu_0 \int  {\bf M}({\bf r})\cdot {\bf H}_{ 0}d{\bf r}=\mu_0\gamma \hbar \int  {\bf s}({\bf r})\cdot {\bf H}_{ 0} d{\bf r},
\end{align}
where $\mu_0$ is the permeability of the vacuum and \({\bf H}_0\) is the magnetic field strength, which is minimized by
${\bf M} \parallel{\bf H}_0$.

The main cause for magnetism is the exchange interaction, a modulation of the Coulomb interaction by the Pauli exclusion principle. In ions with open 3$d$ or 4$f$ shells, the exchange interaction favors finite local moments of spins that remain stable inside the condensed matter. But neighboring spins in a crystal also interact by an exchange coupling. According to the Heisenberg model, the free energy of an ensemble of spins reads 
\begin{align}
    H_{\rm ex}=-\sum_{i,j}J_{ij} {\bf s}_i\cdot {\bf s}_j,
\end{align}
where ${\bf s}_j$ represents the electron spin at site ${j}$ and $J_{ij}=J_{ji}$ is the direct exchange coupling constant between spins at sites $i$ and $j$. Since the atomic wave functions of local atomic moments are exponentially localized, the interaction is often significant only between nearest neighbors. In a simple crystal, $J_{ij}=J$ is the same with all neighbors such that
\begin{align}
    H_{\rm ex}=-J \sum_{i}\sum_{j}^{n.n.}  {\bf s}_i\cdot {\bf s}_j,
    \label{ex_sum_k}
\end{align}
where \textit{j} is summed only over the nearest neighbors of $i$. In reciprocal space
\begin{align}
    {\bf s}_i=\sum_{\bf k}e^{i {\bf k}\cdot {\bf r}_i}{\bf s}_{\bf k},
    \label{fourier_s}
\end{align}
and $H_{\rm ex}=-J \sum_{i}\sum_{j}^{n.n.}  \sum_{\bf k, {\bf k}'} e^{i {\bf k}\cdot {\bf r}_i} e^{i {\bf k}'\cdot {\bf r}_j} {\bf s}_{\bf k}\cdot {\bf s}_{\bf k'}$.
For the simple cubic lattice with six nearest neighbors 
\begin{align}
    \sum_j^{n.n.}e^{i {\bf k}'\cdot {\bf r}_j}=
    e^{i {\bf k}'\cdot {\bf r}_i} (\cos{k_x' a}+\cos{k_y'a}+\cos{k_z' a}),
    \label{sum_prime}
\end{align}
where $a$ is the distance between nearest neighbors, and 
\begin{align}
    H_{\rm ex}=-2NJ\sum_{\bf k} (\cos{k_x a}+\cos{k_ya}+\cos{k_z a}) {\bf s}_{-\bf k}\cdot {\bf s}_{{\bf k}},
    \label{Hex_in_k}
\end{align}
where $N$ is the number of spins. 
When the magnetization modulation changes slowly on the scale of \(a\), only small wave numbers $|{\bf k}|\ll 1/a$ contribute to the energy. We may then expand the cosine term $\cos{k_x a}\approx 1-(k_xa)^2/2$ and
\begin{align}
    H_{\rm ex}=-NJ\sum_{\bf k} (6-k^2a^2){\bf s}_{-{\bf k}}\cdot{\bf s}_{\bf k}.
    \label{exchange_k}
\end{align}
More generally, for simple cubic, face-centered cubic, and body-centered cubic lattices with $Z$ nearest neighbors 
\begin{align}
    H_{\rm ex}=-NJ\sum_{\bf k} \left(Z-\dfrac{Zk^2a^2}{6}\right){\bf s}_{-{\bf k}}\cdot{\bf s}_{\bf k}.
\end{align}
In the long wavelength limit ${\bf s}({\bf r})$ is effectively a continuous function of ${\bf r}$ such that
\begin{align}
    {\bf s}_{\bf k}=\dfrac{1}{N a^3}\int d{\bf r} e^{-i {\bf k}\cdot {\bf r}} {\bf s}({\bf r}),
\end{align}
where $a^3$ is the volume per spin. Moreover, we may replace $\sum_{\bf k}\rightarrow V/(2\pi )^3 \int d{\bf k}=Na^3/(2\pi )^3 \int d{\bf k}$.
Hence
\begin{align}
 -NJZ\sum_{\bf k}{\bf s}_{-{\bf k}}\cdot{\bf s}_{\bf k}=&-\dfrac{JZ}{a^3(2\pi)^3}\int d{\bf k}d{\bf r}d{\bf r}' e^{i{\bf k}\cdot({\bf r}-{\bf r}')} {\bf s}({\bf r})\cdot {\bf s}({\bf r}')=-NJZs^2,
\end{align}
where $s^2={\bf s}({\bf r})\cdot {\bf s}({\bf r})$ is the constant squared modulus of the atomic spin, and
\begin{align}
    \sum_{\bf k}\dfrac{NJZa^2k^2}{6} {\bf s}_{-{\bf k}}\cdot{\bf s}_{\bf k}=& -\dfrac{JZ }{6 a (2\pi)^3}\int d{\bf k}d{\bf r}d{\bf r}'   {\bf s}({\bf r})\cdot {\bf s}({\bf r}')(-k^2)e^{i{\bf k}\cdot({\bf r}-{\bf r}')}\nonumber\\
    =&-\dfrac{JZ }{6 a } \int d{\bf r} {\bf s}({\bf r})\cdot \nabla^2{\bf s}({\bf r}).
\end{align}
The total exchange energy in the continuum limit reads
\begin{align}
    H_{\rm ex}=&E_0+ \left(\dfrac{-JZ}{6 a}\right)\int d{\bf r} {\bf s}({\bf r})\cdot\nabla^2 {\bf s}({\bf r}),
    \label{h_ex2}
\end{align}
where $E_0=-JZNs^2$. 
The spin ${\bf s}({\bf r})$ is related to the magnetization ${\bf M}({\bf r})$ via ${\bf s}({\bf r})=s{\bf M}({\bf r})/M_s$, where \(M_s=\gamma\hbar s\) is the saturated magnetization. Equation~({\ref{h_ex2}}) can be interpreted as the interaction of the magnetization with an effective magnetic field ${\bf H}_{\rm ex}({\bf r})$, i.e.,
\begin{align}
    H_{\rm ex}=E_0 -\dfrac{\mu_0}{2}\int d{\bf r} {\bf M}({\bf r})\cdot {\bf H}_{\rm ex}({\bf r}),
\end{align}
where ${\bf H}_{\rm ex}({\bf r})=\alpha_{\rm ex}\nabla^2 {\bf M}({\bf r})$ and $\alpha_{\rm ex}=JZs^2/(3\mu_0a M_s^2)$ is the exchange stiffness parameter.
The exchange coupling constant is proportional to the Curie temperature  $T_c$~\cite{kittel_book} 
\begin{align}
    J=\dfrac{3 k_B T_c}{2 Z s(s+1)},
\end{align}
where $k_B$ is the Boltzmann constant, so the  exchange stiffness
\begin{align}
    \alpha_{\rm ex}=\dfrac{k_B T_c s^2 }{2 s(s+1)\mu_0 M_s^2 a}.
    \label{lapex}
\end{align}
For example, the ferrimagnet yttrium iron garnet (YIG) has a cubic crystal structure with the lattice constant $A=1.2\times 10^{-9}$~m. 40 Fe$^{3+}$ ions in a unit cell occupy 16 octahedral and 24 tetrahedral sites~\cite{YIG_structure} with average volume $a^3\approx A^3/40$. 
Fe$^{3+}$ has a half-filled 3$d$ shell and spin $s=5/2$. According to Eq.~({\ref{lapex}}) with $T_c=559~$K and $\mu_0M_s=0.177$~T, $\alpha_{\rm ex}\approx 3.15\times 10^{-16}$ m$^{-2}$, which is close to  $\alpha_{\rm ex}=3\times 10^{-16} $~m$^{-2}$ reported in the literature~\cite{spin_wave_book}.

Not only do the neighboring spins interact with each other through the exchange coupling, but the long-range dipolar interaction couples the magnetization at a distance.
Indeed, the magnetization emits a dipolar magnetic field that spreads over a long range, which is governed by Maxwell's equations~\cite{Jackson} 
\begin{subequations}
    \begin{align}
\nabla\times {\bf E}({\bf r},t)&=-\dfrac{\partial{\bf  B}({\bf r},t)}{\partial t},&& \nabla\cdot \mu_0[{\bf H}({\bf r},t)+{\bf M}({\bf r},t)]=0,   \label{maxwell_equations_1} \\ 
\nabla\cdot {\bf E}({\bf r},t)&=\dfrac{\rho_f}{\varepsilon_r}  ,&&\nabla\times {\bf H}({\bf r},t)={\bf J}({\bf r},t)+\varepsilon_r \dfrac{\partial{\bf  E}({\bf r},t)}{\partial t},
    \label{maxwell_equations_2}
\end{align}
\end{subequations}
where ${\bf J}({\bf r},t)$ is the electric current of electrons, $\varepsilon_r$ is the dielectric constant that depends on the materials, and $\rho_f$ is the electron charge density. 
For the ferromagnetic insulators, ${\bf J}({\bf r}, t)=0$ and $\rho_f=0$. Taking the curl of the second equation of Eq.~({\ref{maxwell_equations_2}}) and substituting it into Eqs.~(\ref{maxwell_equations_1}), we find the stray magnetic field ${\bf H}({\bf r},t)$ governed by 
\begin{align}
     \nabla^2 {\bf H}=-\nabla(\nabla\cdot {\bf M})+\mu_0 \varepsilon_r \partial_t^2({\bf H}+{\bf M}),
     \label{m2h}
\end{align}
in which the second term on the right-hand side is responsible for the radiation of electromagnetic field~\cite{Jackson}.
For the magnetization fluctuation of long wavelength in the ferromagnets, the magnetization and its emitted stray magnetic field vary in the frequency of around tens of gigahertz and wavelength $\lambda_s\sim 1~\mu {\rm m}$, leading to the wave vector $k_s\sim 6\times 10^6$~m$^{-1}$. According to Eq.~(\ref{m2h}), the spatial fluctuation $\nabla^2 {\bf H}\sim k_s^2 {\bf H}\sim 4\times 10^{13}{\bf H}$ is much larger than the time derivative $\mu_0 \varepsilon_r \partial_t^2({\bf H}+{\bf M})\sim \mu_0 \varepsilon_r \omega^2({\bf H}+{\bf M}) \sim 10^5 ({\bf H}+{\bf M})$ for $\omega\sim 100$~GHz, assuming ${\bf H}$ and ${\bf M}$ in the same order. Therefore, it is safe to disregard the radiative time-derivative term in Eq.~(\ref{m2h}) such that the quasistatic approximation applies.
We notice that when the wavelength of the spin wave is comparable to that of light, \textit{i.e.,} $k_s^2\approx (\omega/c)^2\sim 2.6\times 10^{4}$~m$^{-2}$, $\nabla^2 {\bf H}\sim \mu_0\varepsilon_r \partial_t^2({\bf H}+{\bf M})$, in which case the radiation of the electromagnetic field plays a vital role in the magnetization dynamics~\cite{magnon_photon}. Accordingly, in the short-wavelength regime, the exchange interaction governs the dynamics of the magnetization, while in the long-wavelength regime, the dipolar interaction dominates.

By the quasistatic approximation, the magnetization fluctuation generates the stray magnetic field via
\begin{align}
     \nabla^2 {\bf H}=-\nabla(\nabla\cdot {\bf M}),
     \label{quansi_M_H}
\end{align}
which is generated by the magnetic charge $\rho_M=\nabla\cdot {\bf M}$.  
It is then convenient to express the stray magnetic field in terms of the magnetic potential ${\bf H}=-\nabla\phi$, which leads to the Poisson equation
\begin{align}
    \nabla^2\phi=\nabla\cdot{\bf M}.
\end{align}
With the magnetic scalar potential 
\begin{align}
    \phi=-\dfrac{1}{4\pi}\int d {\bf r}' \dfrac{\nabla'\cdot {\bf M}({\bf r}')}{|{\bf r}-{\bf r}'|},
    \label{phi}
\end{align}
the dipolar field
\begin{align}
  {\bf H}_{\rm dip}({\bf r})=-\nabla\phi=\dfrac{1}{4\pi }\nabla\int d {\bf r}'\dfrac{\nabla'\cdot {\bf M}({\bf r}')}{|{\bf r-r}'|}.
  \label{dipoar_field_H}
\end{align}
The dipolar interaction contributes to the Hamiltonian
\begin{align}
    H_{\rm dip}=\dfrac{\mu_0}{2}\int {\bf M}({\bf r})\cdot {\bf H}_{\rm dip}({\bf r})d{\bf r}.
\end{align}

We note that in the presence of SOC, there are other vital interactions in magnetism, \textit{i.e.,}  $H_{\rm DM}$ caused by the Dzyaloshinskii-Moriya interaction (DMI)~\cite{DM_1,DM_2} and $H_{\rm ani}$ caused by the magnetocrystalline anisotropy~\cite{ani1,ani2,ani3}.

\subsection{Magnons}
\label{Magnons}

Around the equilibrium magnetic configuration, the collective excitations in the ordered magnetic materials are spin waves, or their quanta, magnons~\cite{lenk,Chumak,Grundler,Demidov,Brataas,Barman,Yu_chirality,Cornelissen,Zou,Wang,Flebus}. Their typical frequencies range from gigahertz to terahertz scales. 
Supposing a uniform magnetic field $H_0 \hat{\bf z}$ applied along the $\hat{\bf z}$-direction that biases the magnetization along the same direction, the transverse fluctuation $M_{x,y}({\bf r})\ll M_z({\bf r})\approx M_s$, where $M_s$ is the saturation magnetization of the ferromagnets.
The magnetization ${\bf M}({\bf r})$ is associated with the spin operator $\hat{\bf s}({\bf r})$ via ${\bf M}({\bf r})=-|\gamma|\hbar\hat{\bf s}({\bf r})$. The spin is opposite to the magnetization with the $z$-component along the $-\hat{\bf z}$-direction.

To quantize the spin waves in the continuum model, we first use the Holstein-Primakoff transformation~\cite{Holstein} with the bosonic operator $\hat{b}({\bf r})$ to express the spin operators, which in the linear response regime read 
\begin{align}
\hat{s}_+({\bf r})&=\hat{s}_x({\bf r})+i\hat{s}_y({\bf r})	=\hat{b}^\dagger({\bf r})\sqrt{2S-\hat{b}^\dagger({\bf r}) \hat{b}({\bf r})}\approx\sqrt{2S}\hat{b}^\dagger({\bf r}),\nonumber\\
\hat{s}_-({\bf r})&=\hat{s}_x({\bf r})-i\hat{s}_y({\bf r})=\sqrt{2S-\hat{b}^\dagger ({\bf r})\hat{b}({\bf r})}\hat{b}({\bf r})\approx\sqrt{2S}\hat{b}({\bf r}),\nonumber\\
\hat{s}_z({\bf r})&=-S+\hat{b}^\dagger ({\bf r})\hat{b}({\bf r})\approx -S,
\label{H_P_transformation}
\end{align}
in which $S=M_s/(\gamma\hbar)$. 
$\hat{b}({\bf r})$ is expanded by the magnon modes labeled by ``$p$'' via mixing the magnon annihilation $\hat{m}_{p}$ and creation $\hat{m}_{p}^\dagger$  operators, i.e., 
\begin{align}
\hat{b}({\bf r})&=\sum_{p}\left({\cal A}_{p}({\bf r})\hat{m}_{p}+{\cal B}_{p}({\bf r})\hat{m}_{p}^\dagger\right),\nonumber\\
\hat{b}^\dagger({\bf r})&=\sum_{p} \left({\cal A}_{p}^*({\bf r})\hat{m}_{p}^\dagger+{\cal B}_{p}^*({\bf r})\hat{m}_{p}\right),
    \label{magnon}
\end{align}
where ${\cal A}_{p}({\bf r})$ and ${\cal B}_{p}({\bf r})$ are the amplitudes of the annihilation $\hat{m}_{p}$ and creation $\hat{m}_{p}^\dagger$ operators. 
According to Eq.~(\ref{H_P_transformation}), the spin operators become 
\begin{align}
\hat{s}_x({\bf r})&=\dfrac{\sqrt{2S}}{2}\sum_{p} \left(\left({\cal A}_{p}({\bf r})+{\cal B}_{p}^*({\bf r})\right)\hat{m}_{p}+{\rm H.c.}\right),\nonumber\\
\hat{s}_y({\bf r})&=\dfrac{\sqrt{2S}}{2}\sum_{p} \left(i\left({\cal A}_{p}({\bf r})-{\cal B}_{p}^*({\bf r})\right)\hat{m}_{p}+{\rm H.c.}\right).
\label{Spin-magnon}
\end{align}
Such a mixing of annihilation and creation magnon operators render the orthonormal condition for ${\cal A}_{p}({\bf r})$ and ${\cal B}_{p}({\bf r})$ different from that of photons and phonons.
The bosonic characteristic of $\hat{b}({\bf r})$ leads to two relations  
\begin{align}
    &[\hat{b}({\bf r}),\hat{b}^\dagger({\bf r'})]=\delta({\bf r-r'})\rightarrow\sum_{p} \left[{\cal A}_{p}({\bf r}){\cal A}_{p}^*({\bf r'})-{\cal B}_{p}({\bf r}){\cal B}_{p}^*({\bf r'})\right]=\delta({\bf r-r'}),\nonumber\\
    &[\hat{b}({\bf r}),\hat{b}({\bf r'})]=0\rightarrow\sum_{p}\left[{\cal A}_{p}({\bf r}){\cal B}_{p}({\bf r'})-{\cal B}_{p}({\bf r}){\cal A}_{p}({\bf r'})\right]=0.
    \label{normalized_k}
\end{align}

We consider the typical spin Hamiltonian composed of the Zeeman coupling to the external magnetic field $H_0\hat{\bf z}$, the exchange interaction, and the dipolar interaction: 
\begin{align}
 \hat{H}=&-\mu_0\int d{\bf r} \hat{M}_z({\bf r}) H_0-\dfrac{\mu_0}{2}\int d{\bf r}\hat{\bf M}({\bf r})\cdot \left( {\bf H}_{\rm ex}(\hat{\bf M},{\bf r})+ {\bf H}_{\rm dip}(\hat{\bf M},{\bf r})\right)\nonumber\\
=&\mu_0 \gamma \hbar \int d{\bf r}\hat{s}_z({\bf r}) H_0  -\dfrac{\mu_0 (\gamma \hbar)^2 }{2} \int d{\bf r} \hat{\bf s}({\bf r})\cdot \left( {\bf H}_{\rm ex}(\hat{\bf s},{\bf r})+ {\bf H}_{\rm dip}(\hat{\bf s},{\bf r})\right),
\label{Hamiltonian_example}
\end{align}
where the exchange magnetic field ${\bf H}_{\rm ex}(\hat{\bf M},{\bf r})=\alpha_{\rm ex }\nabla^2 \hat{\bf M}({\bf r})$, and the dipolar magnetic field [Eq.~(\ref{dipoar_field_H})]
\begin{align}
  {\bf H}_{\rm dip}(\hat{\bf M}, {\bf r})=\dfrac{1}{4\pi }\nabla\int d {\bf r}'\dfrac{\nabla'\cdot \hat{\bf M}({\bf r}')}{|{\bf r-r}'|}.
\end{align}
In terms of the equations of motion of the magnetization, we diagonalize the Hamiltonian and demonstrate the normalization condition of the mode ``wavefunction" ${\cal A}_{p}({\bf r})$ and ${\cal B}_{p}({\bf r})$.

 We focus on the magnetic samples in which the saturation magnetization is uniform, such as the magnetic film, sphere, ellipsoid, and cylinder~\cite{Depolarization_Tensor,Demagnetizing_Factors,Demagnetizing_Factors2}, which with Hamiltonian \eqref{Hamiltonian_example} implies that the static dipolar field inside the sample is aligned with that of the saturation magnetization,  since, otherwise, the saturation magnetization precesses around the static dipolar field. In this situation, the equilibrium spin distribution
 \begin{align}
 s_e({\bf r})=\left\{	
 \begin{array}{cc}
		S, & {\rm {\bf r}\in \ the \ sample}, \\
		0, & {\rm {\bf r}\notin \ the \ sample},
	\end{array}\right..
 \end{align}
 According to the Heisenberg equation, the equation of motion of $\hat{s}_x({\bf r})$ and $\hat{s}_y({\bf r})$ from Hamiltonian \eqref{Hamiltonian_example} reads 
\begin{align}
  \partial_t \hat{s}_x({\bf r})=&-\mu_0 \gamma H_0 \hat{s}_y ({\bf r})+ {\mu_0 (\gamma^2  \hbar)}\left[\hat{s}_y({\bf r})\left({H}_{{\rm ex},z}(\hat{\bf s},{\bf r})+{H}_{{\rm dip},z}(\hat{\bf s},{\bf r})\right)-\hat{s}_z({\bf r})\left({H}_{{\rm ex},y}(\hat{\bf s},{\bf r})+{H}_{{\rm dip},y}(\hat{\bf s},{\bf r})\right)\right],\nonumber \\
  \partial_t \hat{s}_y({\bf r})
    =&\mu_0 \gamma H_0 \hat{s}_x ({\bf r})+ {\mu_0 (\gamma^2  \hbar)}\left[-\hat{s}_x({\bf r})\left({H}_{{\rm ex},z}(\hat{\bf s},{\bf r})+{H}_{{\rm dip},z}(\hat{\bf s},{\bf r})\right)+\hat{s}_z({\bf r})\left({H}_{{\rm ex},x}(\hat{\bf s},{\bf r})+{H}_{{\rm dip},x}(\hat{\bf s},{\bf r})\right)\right].
    \label{eom_S}
 \end{align} 
In the linear-response regime, ${H}_{{\rm ex},z}(\hat{\bf s},{\bf r})\approx {H}_{{\rm ex},z}(-S\hat{\bf z},{\bf r})=-{H}_{{\rm ex},z}(S\hat{\bf z},{\bf r})$ and ${H}_{{\rm dip},z}(\hat{\bf s},{\bf r})\approx -{H}_{{\rm dip},z}(S\hat{\bf z},{\bf r})$ are static  in Eq.~\eqref{eom_S}, such that  the fluctuated magnetic field is governed by 
\begin{align}
  \hspace{-0.6cm}{H}_{{\rm ex},y}(\hat{s}_y\hat{\bf y},{\bf r})+{H}_{{\rm dip},y}(\hat{\bf s}_\perp,{\bf r})=&H_{{\rm dip},y}(S\hat{\bf z},{\bf r})+\dfrac{\partial_t \hat{ s}_x({\bf r})+\mu_0 \gamma H_0 \hat{s}_y({\bf r})}{\mu_0 \gamma^2\hbar S}+\dfrac{\hat{s}_y({\bf r})\left({H}_{{\rm dip},z}(S\hat{\bf z},{\bf r})+{H}_{{\rm ex},z}(S\hat{\bf z},{\bf r})\right)}{S}, \nonumber\\
 \hspace{-0.6cm} {H}_{{\rm ex},x}(\hat{s}_x\hat{\bf x},{\bf r})+{H}_{{\rm dip},x}(\hat{\bf s}_\perp,{\bf r})=&{H}_{{\rm dip},x}({ S}\hat{\bf z},{\bf r})+\dfrac{-\partial_t \hat{ s}_y({\bf r})+\mu_0 \gamma H_0 \hat{s}_x({\bf r})}{\mu_0 \gamma^2\hbar S}+\dfrac{\hat{s}_x({\bf r})\left({H}_{{\rm dip},z}(S\hat{\bf z},{\bf r})+{H}_{{\rm ex},z}(S\hat{\bf z},{\bf r})\right) }{S},
  \label{eom_linear_s}
\end{align}
where the transverse spin fluctuation $\hat{\bf s}_\perp=\hat{s}_x\hat{\bf x}+\hat{s}_y\hat{\bf y}$. Also, in the linear response regime, 
\[\hat{s}_z({\bf r})=-\sqrt{s^2_e({\bf r})-(\hat{s}_x^2({\bf r})+\hat{s}_y^2({\bf r}))}\approx -s_e({\bf r})+(\hat{s}_x^2({\bf r})+\hat{s}_y^2({\bf r}))/(2S)=-s_e({\bf r})+\tilde{s}(\bf r),
\]
where $\tilde{s}({\bf r})=(\hat{s}_x^2({\bf r})+\hat{s}_y^2({\bf r}))/(2S)$ relates to the magnitude of spin fluctuation. 
The Hamiltonian~\eqref{Hamiltonian_example} in the linear response regime becomes
\begin{align}
\hat{H}=&E_0+\dfrac{\hbar}{2S}\int d{\bf r} \left[{\hat{s}_x({\bf r})\partial_t \hat{s}_y({\bf r})-\hat{s}_y({\bf r}) \partial_t \hat{s}_x({\bf r}) }\right]\nonumber\\
-&\dfrac{\mu_0 (\gamma \hbar)^2 }{2} \int d{\bf r} \left[
\tilde{s}({\bf r})H_{{\rm dip},z}(S\hat{\bf z},{\bf r})-s_e({\bf r})H_{{\rm dip},z}(\tilde{s}\hat{\bf z},{\bf r})
 \right] +\left[ \tilde{s}({\bf r})H_{{\rm ex},z}(S\hat{\bf z},{\bf r})-s_e({\bf r})H_{{\rm ex},z}(\tilde{s}\hat{\bf z},{\bf r}) \right] \nonumber\\
 -&\dfrac{\mu_0 (\gamma \hbar)^2 }{2} \int d{\bf r} s_e({\bf r})H_{{\rm dip},z}({\bf \hat{s}}_{\perp},{\bf r}),
\label{H_mi}
\end{align}
where the constant
$E_0=-\mu_0 \gamma \hbar\int d{\bf r} H_0 s_e({\bf r})-(\mu_0 \gamma^2 \hbar^2 /2)\int d{\bf r} 
\left[s_e({\bf r})H_{{\rm ex},z}(S\hat{\bf z},{\bf r})+s_e({\bf r})H_{{\rm dip},z}(S\hat{\bf z},{\bf r})\right]$.
The second integral in Eq.~\eqref{H_mi} relates to the fluctuation of the magnetization or the effective magnetic field along the $\hat{\bf z}$-direction, which we demonstrate vanishes:
\begin{align}
    \int d{\bf r}\tilde{s}({\bf r})H_{{\rm dip},z}(S\hat{\bf z},{\bf r})=&\dfrac{1}{4\pi}\int d{\bf r}d{\bf r}' \tilde{s}({\bf r})\nabla_z\dfrac{\nabla_{z'}{s}_e({\bf r}')}{|{\bf r}-{\bf r}'|}=\int d{\bf r} s_e({\bf r})H_{{\rm dip},z}({\tilde{s}\hat{\bf z}},{\bf r}),
\end{align}
such that $\int d{\bf r} \left[
\tilde{s}({\bf r})H_{{\rm dip},z}(S\hat{\bf z},{\bf r})-s_e({\bf r})H_{{\rm dip},z}(\tilde{s}\hat{\bf z},{\bf r})
 \right]=0$.
Similarly, 
\begin{align}
    \int d{\bf r} \tilde{s}({\bf r})H_{{\rm ex},z}(S\hat{\bf z},{\bf r})=&\alpha_{\rm ex}\int d{\bf r} \tilde{s}({\bf r})\nabla^2s_e({\bf r})=\int d{\bf r} s_e({\bf r})H_{{\rm ex},z}(\tilde{s}\hat{\bf z},{\bf r}),
\end{align}
such that $\int d{\bf r} \left[ \tilde{s}({\bf r})H_{{\rm ex},z}(S\hat{\bf z},{\bf r})-s_e({\bf r})H_{{\rm ex},z}(\tilde{s}\hat{\bf z},{\bf r}) \right]=0$.
The third integral in Eq.~\eqref{H_mi} 
 \begin{align}
\int d{\bf r} s_e({\bf r})H_{{\rm dip},z}({\bf \hat{s}}_{\perp},{\bf r})&=\dfrac{1}{4\pi}\int d{\bf r}d{\bf r}' \left[s_e({\bf r})\nabla_z\dfrac{\nabla_{x'}\hat{s}_x({\bf r}')}{|{\bf r}-{\bf r}'|}+s_e({\bf r})\nabla_z\dfrac{\nabla_{y'}\hat{s}_y({\bf r}')}{|{\bf r}-{\bf r}'|}\right]\nonumber\\
	&=\int d{\bf r} \left[s_x({\bf r}) H_{{\rm dip},x}(S\hat{\bf z},{\bf r})+s_y({\bf r}) H_{{\rm dip},y}(S\hat{\bf z},{\bf r})\right]
\end{align}
 vanishes as well since for the uniform saturation magnetization the static transverse dipolar field $H_{{\rm dip},y}(S\hat{\bf z},{\bf r})=0$ and $H_{{\rm dip},x}(S\hat{\bf z},{\bf r})=0$. Accordingly, the Hamiltonian \eqref{H_mi} becomes 
\begin{align}
    \hat{H}=E_0+\dfrac{\hbar}{2S}\int d{\bf r} \left({\hat{s}_x({\bf r})\partial_t \hat{s}_y({\bf r})-\hat{s}_y({\bf r}) \partial_t \hat{s}_x({\bf r}) }\right).
    \label{H_pt}
\end{align}

Since $\hat{m}_{p}$ are the magnon operators, $\hat{m}_{p}\propto e^{-i\omega_{p} t}$  and $\hat{m}^\dagger_{p}\propto e^{i\omega_{p} t}$, where $\omega_{p}$ is the eigenfrequency of mode ``$p$''. According to Eq.~(\ref{Spin-magnon}),
\begin{align}
\partial_t\hat{s}_x({\bf r})=&-{i\sqrt{S/2}}\sum_p\omega_{p} \left[\left({\cal A}_{p}({\bf r})+{\cal B}_{p}^*({\bf r})\right)\hat{m}_{p}-\left({\cal A}^*_{p}({\bf r})+{\cal B}_{p}({\bf r})\right)\hat{m}_{p}^\dagger\right],\nonumber\\
\partial_t\hat{s}_y({\bf r})=&-{\sqrt{S/2}}\sum_p\omega_{p}\left[ \left({\cal B}_{p}^*({\bf r})-{\cal A}_{p}({\bf r})\right)\hat{m}_{p}-\left({\cal A}^*_{p}({\bf r})-{\cal B}_{p}({\bf r})\right)\hat{m}_{p}^\dagger\right].
\label{pt_s}
\end{align}
Combining Eqs.~(\ref{pt_s}), (\ref{Spin-magnon}) and (\ref{normalization}), we diagonalize the  Hamiltonian (\ref{H_pt}) according to 
\begin{align}
    \hat{H}=&E_0+\hbar \int d {\bf r} \sum_{ pp'} \omega_{p'} \left[\left( {\cal A}_{p}({\bf r}){\cal A}_{p'}^*({\bf r})-{\cal B}_{p}^*({\bf r}){\cal B}_{p'}({\bf r})\right)\hat{m}_{p}\hat{m}_{p'}^\dagger+\left( {\cal A}_{p}^*({\bf r}){\cal A}_{p'}({\bf r})-{\cal B}_{p}({\bf r}){\cal B}^*_{p'}({\bf r})\right)\hat{m}_{p}^\dagger\hat{m}_{p'}\right.\nonumber\\
    +&\left.\left({\cal B}_{p}^*({\bf r}){\cal A}_{p'}({\bf r})-{\cal A}_{{p}}({\bf r}){\cal B}_{p'}^*({\bf r})\right)\hat{m}_{p}\hat{m}_{p'}+\left({\cal B}_{p}({\bf r}){\cal A}_{p'}^*({\bf r})-{\cal A}_{{p}}^*({\bf r}){\cal B}_{p'}({\bf r})\right)\hat{m}^\dagger_{p}\hat{m}_{p '}^\dagger\right]\nonumber\\
    =&E_0+\frac{1}{2}\sum_{p} \hbar \omega_{p}(\hat{m}_{p}\hat{m}_{p}^\dagger+\hat{m}_{p}^\dagger\hat{m}_{p}),
\end{align}
by the assumption that 
\begin{subequations}
\begin{align}
    &\int d{\bf r}[{\cal A}_{p}({\bf r}){\cal A}_{p'}^*({\bf r})-{\cal B}_{p}^*({\bf r}){\cal B}_{p'}({\bf r})]=\delta_{pp'},\\
    &\int d{\bf r}[{\cal B}_{p}({\bf r}){\cal A}_{p'}^*({\bf r})-{\cal B}_{p'}({\bf r}){\cal A}_{p}^*({\bf r})]=0.
\end{align}
\end{subequations}
This implies that the inverse transformation of the operator $\hat{b}({\bf r})$ exists: when we express
\begin{align}
     \hat{m}_{p}&= \int d{\bf r} \left[{\cal A}_{{p}}^*({\bf r}) \hat{ b}({\bf r})-{\cal B}_{{p}}({\bf r})\hat{b}^\dagger({\bf r})\right],\nonumber\\
 \hat{m}_{p}^\dagger&=\int d{\bf r} \left[{\cal A}_{p}({\bf r}) \hat{b}^\dagger({\bf r})-{\cal B}_{p}^*({\bf r})\hat{b}({\bf r})\right],
        \label{inverse_mk}
    \end{align}
the bosonic characteristic of the operator $\hat{m}_{p}$ leads to the other two relations 
\begin{align}
    &[\hat{m}_{p},\hat{m}^\dagger_{{p}'}]=\delta_{pp'}\rightarrow\int d{\bf r}[{\cal A}_{p}({\bf r}){\cal A}_{p'}^*({\bf r})-{\cal B}_{p}^*({\bf r}){\cal B}_{p'}({\bf r})]=\delta_{pp'},\nonumber\\
    &[\hat{m}_{p},\hat{m}_{p'}]=0\rightarrow\int d{\bf r}[{\cal B}_{p}({\bf r}){\cal A}_{p'}^*({\bf r})-{\cal B}_{p'}({\bf r}){\cal A}_{p}^*({\bf r})]=0.
    \label{normalization}
\end{align}
Equations.~({\ref{normalized_k}}) and (\ref{normalization}) can be regarded as completeness relations that ensure $\hat{m}_{p}$ forms a complete set.

${\cal A}_{p}({\bf r})$ and ${\cal B}_{p}({\bf r})$ are related to the magnetization amplitudes ${\cal M}_{x,y}^{p}({\bf r})$. 
Linearization of the Landau-Lifshitz (LL) equation  $\partial_t{\bf M}^p=-\mu_0 \gamma {\bf M}^p\times(H_0 \hat{\bf z}+{\bf H}_{\rm ex}+{\bf H}_{\rm dip})$~\cite{LL_eq} with ${\bf M}^p({\bf r},t)=\{{\cal M}_{x}^{p}({\bf r})e^{-i\omega_p t},{\cal M}_{y}^{p}({\bf r})e^{-i\omega_p t},M_s\}$ leads to
\begin{align}
&	\mu_0 \gamma M_s \left(H_{{\rm ex},y}\left[ {\cal M}_{y}^{p}({\bf r})\hat{\bf y}\right] 
	+{H}_{{\rm dip},y}\left[ {\cal M}_{x}^{p}({\bf r})\hat{\bf x}+{\cal M}_{y}^{p}({\bf r})\hat{\bf y}\right] 
	\right)\nonumber\\
&=-i\omega_{p}{\cal M}_{x}^{p}({\bf r})+\mu_0 \gamma{\cal M}_{y}^{p}({\bf r})\left( H_0+ {H}_{{\rm dip},z}(M_s\hat{\bf z},{\bf r})+{H}_{{\rm ex},z}(M_s\hat{\bf z},{\bf r})\right),\nonumber \\
    &	\mu_0 \gamma M_s \left(	H_{{\rm ex},x}\left[ {\cal M}_{x}^{p}({\bf r})\hat{\bf x}\right] 
	+{H}_{{\rm dip},x}\left[ {\cal M}_{x}^{p}({\bf r})\hat{\bf x}+{\cal M}_{y}^{p}({\bf r})\hat{\bf y}\right] 
	\right)\nonumber\\
	&=i\omega_{p}{\cal M}_{y}^{p}({\bf r})-\mu_0 \gamma{\cal M}_{x}^{p}({\bf r})\left( H_0+ {H}_{{\rm dip},z}(M_s\hat{\bf z},{\bf r})-{H}_{{\rm ex},z}(M_s\hat{\bf z},{\bf r})\right). 
    \label{eom_m}
\end{align}
To obtain the equation of ${\cal A}_{p}({\bf r})$ and ${\cal B}_{p}({\bf r})$, based on Eq.~\eqref{eom_linear_s} we take the commutator 
\begin{align}
  &\left[{H}_{{\rm ex},y}(\hat{s}_y\hat{\bf y},{\bf r})+{H}_{{\rm dip},y}(\hat{\bf s}_\perp,{\bf r}),\hat{m}_p^\dagger\right]\nonumber\\
 & =\left[H_{{\rm dip},y}(S\hat{\bf z},{\bf r})+\dfrac{\partial_t \hat{ s}_x({\bf r})+\mu_0 \gamma H_0 \hat{s}_y({\bf r})}{\mu_0 \gamma^2\hbar S}+\dfrac{\hat{s}_y({\bf r})\left({H}_{{\rm dip},z}(S\hat{\bf z},{\bf r})+{H}_{{\rm ex},z}(S\hat{\bf z},{\bf r})\right)}{S},\hat{m}_p^\dagger\right], \nonumber\\
&\left[ {H}_{{\rm ex},x}(\hat{s}_x\hat{\bf x},{\bf r})+{H}_{{\rm dip},x}(\hat{\bf s}_\perp,{\bf r}),\hat{m}_p^\dagger\right]\nonumber\\
&=\left[{H}_{{\rm dip},x}({ S}\hat{\bf z},{\bf r})+\dfrac{-\partial_t \hat{ s}_y({\bf r})+\mu_0 \gamma H_0 \hat{s}_x({\bf r})}{\mu_0 \gamma^2\hbar S}+\dfrac{\hat{s}_x({\bf r})\left({H}_{{\rm dip},z}(S\hat{\bf z},{\bf r})+{H}_{{\rm ex},z}(S\hat{\bf z},{\bf r})\right) }{S},\hat{m}_p^\dagger\right],
  \label{commutation_eom_linear_s}
\end{align}
leading to 
\begin{align}
&\mu_0 \gamma M_s \left(	H_{{\rm ex},y}\left[ i({\cal A}_{p}({\bf r})-{\cal B}_{p}^*({\bf r}))/2\hat{\bf y}\right] +{H}_{{\rm dip},y}\left[ ({\cal A}_{p}({\bf r})+{\cal B}_{p}^*({\bf r}))/2\hat{\bf x}+i({\cal A}_{p}({\bf r})-{\cal B}_{p}^*({\bf r}))/2\hat{\bf y} \right] 
\right) \nonumber\\
&=-i\omega_{p}\left({\cal A}_{p}({\bf r})+{\cal B}_{p}^*({\bf r})\right)/2+i\left({\cal A}_{p}({\bf r})-{\cal B}_{p}^*({\bf r})\right)/2\left[ H_0+ {H}_{{\rm dip},z}(M_s\hat{\bf z},{\bf r})+{H}_{{\rm ex},z}(M_s\hat{\bf z},{\bf r})\right], \nonumber\\
&\mu_0\gamma M_s \left(	H_{{\rm ex},x}\left[({\cal A}_{p}({\bf r})+{\cal B}_{p}^*({\bf r}))/2\hat{\bf y}\right] 
	+{H}_{{\rm dip},x}\left[ ({\cal A}_{p}({\bf r})+{\cal B}_{p}^*({\bf r}))/2\hat{\bf x}+i({\cal A}_{p}({\bf r})-{\cal B}_{p}^*({\bf r}))/2\hat{\bf y} \right] \right)\nonumber\\
    &=i\omega_p i\dfrac{{\cal A}_p({\bf r})-{\cal B}_p^*({\bf r})}{2}+\mu_0 \gamma \dfrac{{\cal A}_p({\bf r})+{\cal B}_p^*({\bf r})}{2}\left[ H_0+ {H}_{{\rm dip},z}(M_s\hat{\bf z},{\bf r})+{H}_{{\rm ex},z}(M_s\hat{\bf z},{\bf r})\right],
\label{eom_A_B}
\end{align}
in which $M_s= \gamma \hbar S$.
By comparing Eqs.~\eqref{eom_m} and \eqref{eom_A_B}, we find ${\cal A}_p({\bf r})$ and ${\cal B}_p({\bf r})$ are related to the mode amplitudes ${\cal M}_{x,y}^p({\bf r})$ via
\begin{align} 
    {\cal M}_x^{p}({\bf r})=&({1}/{2})\left({\cal A}_{p}({\bf r})+{\cal B}_{p}^*({\bf r})\right),\nonumber\\
    {\cal M}_y^{p}({\bf r})=&({i}/{2})\left({\cal A}_{p}({\bf r})-{\cal B}_{p}^*({\bf r})\right).
    \label{M_A_B}
\end{align}
The spin operators can then be expanded by the magnon operators $\{\hat{m}_p, \hat{m}_p^\dagger\}$ in terms of the magnetization amplitudes ${\cal M}_{x,y}^p({\bf r})$ as
\begin{align}
\hat{s}_x({\bf r})&={\sqrt{S/2}}\sum_{p} \left[\left({\cal A}_{p}({\bf r})+{\cal B}_{p}^*({\bf r})\right)\hat{m}_{p}+{\rm H.c.}\right]=\sqrt{2S}\sum_{p} \left({\cal M}^{p}_{x}({\bf r})\hat{m}_{p}+{\rm H.c.}\right),
\nonumber
\\
\hat{s}_y({\bf r})&={\sqrt{S/2}}\sum_{p} \left[i\left({\cal A}_{p}({\bf r})-{\cal B}_{p}^*({\bf r})\right)\hat{m}_{p}+{\rm H.c.}\right]=\sqrt{2S}\sum_{p} \left({\cal M}^{p}_{y}({\bf r})\hat{m}_{p}+{\rm H.c.}\right),
\label{magnon_quant}
\end{align}
where the amplitudes is normalized according to Eq.~(\ref{normalization}), i.e.,~\cite{normalized1,normalized2}
\begin{subequations}
\begin{align}
  \int d{\bf r} \left({\cal M}_x^{p}({\bf r}){\cal M}_y^{{p}'*}({\bf r})-{\cal M}_x^{{p}'*}({\bf r}){\cal M}_y^{p}({\bf r})\right) &=\int d{\bf r} \dfrac{i}{2}\left( -{\cal A}_{p}({\bf r}){\cal A}_{p'}^*({\bf r})+{\cal B}_{p}^*({\bf r}){\cal B}_{p'}({\bf r})\right)=-\dfrac{i}{2}\delta_{pp'},
  \label{renormalization_relation}\\
    \int d{\bf r} \left({\cal M}_y^{p}({\bf r}){\cal M}_x^{p'}({\bf r})-{\cal M}_x^{p}({\bf r}){\cal M}_y^{p'}({\bf r})\right)&=\int d{\bf r} \dfrac{i}{2} \left({\cal A}_{p}({\bf r}){\cal B}_{p'}^*({\bf r})-{\cal B}_{p}^*({\bf r}){\cal A}_{p'}({\bf r})\right)=0.
    \label{normalization_22}
    \end{align}
\end{subequations}

\subsection{Pairing symmetry of Cooper pairs}

Since the discovery of superconductivity in Hg when the temperature is cooled below the liquid nitrogen temperature~\cite{Onnes,Onnes1,Onnes2}, the investigation of superconductors and their heterostructures has lasted for more than one hundred years. Since 1956, Bardeen, Cooper, and Schrieffer (BCS) proposed the microscopic mechanism for the conventional superconductivity~\cite{Cooper_pairs_1956,BCS_1957}, in which the key idea is the formation of Cooper pairs bonded by an even weak attractive potential between two electrons to lower the free energy of the system. 
Such potential makes the gapless Fermi sphere unstable, thus favoring a state holding an energy gap at the Fermi energy. 
In conventional superconductivity, the attractive potential comes from the electron-phonon interaction~\cite{BCS_1957}. Afterward, the emergence of the heavy-Fermion superconductivity ~\cite{Heavy_fermion_first,HF_RMP_1984,HF_RMP_2009,HF_science_2004}, the copper-based superconductivity ~\cite{LBCO_Muller,YBCO_Wu,HS_Scalapino,Anderson_RVB,HS_superexchange,HS_pairing_symmetry,HS_doping_Mott,HS_t_J,HS_pseudogap}, the iron-based superconductivity ~\cite{FS_Norman,FS_order_parameter,FS_Ding_Hong,FS_transport,FS_RMP_2011} challenge the BCS theory as well as the electron-phonon mechanism ~\cite{HF_RMP_1984,HF_RMP_2009,Anderson_RVB,HS_superexchange,HS_pairing_symmetry,HS_doping_Mott,HS_t_J,HS_pseudogap,FS_RMP_2011}. Nevertheless, many experimental phenomena can be still well understood via the analysis of the pairing symmetry of Cooper pairs~\cite{Sigrist,HS_pairing_symmetry,FS_order_parameter}.

Mathematically, the Cooper pairing is described by the anomalous correlation between two electrons~\cite{Sigrist,HS_pairing_symmetry,Leggett_book} 
\begin{equation}
F_{\sigma\sigma'}({\bf r}_1-{\bf r}_2,t_1-t_2)=\langle\hat{\Psi}_{\sigma}({\bf r}_1,t_1)\hat{\Psi}_{\sigma'}({\bf r}_2,t_2)\rangle,
\label{general_anomalous_correlation}
\end{equation}
where $\hat{\Psi}_{\sigma}({\bf r},t)$ represents the field operator of the electron with spin $\sigma$ and the space-time coordinate $({\bf r},t)$. 
In terms of the Fourier components in the frequency and momentum space, the anomalous correlation  
\begin{equation}
F_{\sigma\sigma'}({\bf r}_1-{\bf r}_2,t_1-t_2)=f({\bf r},t)S_{\sigma\sigma'}
=\Big[\int\frac{d\omega}{2\pi} \frac{d{\bf k}}{(2\pi)^d}e^{i{\bf
      k}\cdot({\bf r}_1-{\bf r}_2)-i\omega(t_1-t_2)}f(\omega,{\bf k})\Big]S_{\sigma\sigma'},
\label{anomalous_correlation_further}
\end{equation}
where $f({\bf r},t)$ and $S_{\sigma\sigma'}$ are the ``wavefunction" of the Cooper pairs in the real and spin spaces~\cite{Sigrist,Leggett_book}, $d$ is the dimension, $f(\omega,{\bf k})$ is the Fourier components $(\omega,{\bf k})$ of the wavefunction.
Since any Fermion obeys the Fermi-Dirac statistics, exchanging the space, time, and spin indexes of the two field operators, the anomalous correlation function \eqref{general_anomalous_correlation} changes sign. Such an operation of exchanging the space and time is equivalent to replacing $-{\bf k}$ and $-\omega$ in Eq.~\eqref{anomalous_correlation_further}. Thus, depending on the spin singlet and triplet in the spin wavefunction $S_{\sigma\sigma'}$, we can classify the Cooper pairing in terms of the parity dependence on the frequency and momentum in $f(\omega,{\bf k})$ ~\cite{Sigrist,HS_pairing_symmetry,FS_order_parameter,pairing_table,Eschrig_2015}.

When the spin wavefunction $S_{\sigma\sigma'}$ is in its spin singlet state, exchanging the spin index, the spin wavefunction changes sign. So $f(\omega,{\bf k})=f(-\omega,-{\bf k})$ remains unchanged in the real-space wavefunction by the Fermi-Dirac statistics. 
Accordingly, $f(\omega,{\bf k})$ is either odd in the frequency and momentum or even in the frequency and momentum. Furthermore, when the Cooper pairing function $f(\omega,{\bf k})$ is even in the momentum, the associated superconductivity is in the $\{s,d,...\}$ waves; when $f(\omega,{\bf k})$ is odd in the momentum, it lies in the $\{p,f,...\}$ waves.
On the other hand, when the spin wavefunction $S_{\sigma\sigma'}$ is in the spin-triplet state, exchanging the spin index, the spin wavefunction remains unchanged. According to the Fermi-Dirac statistics, 
$f(\omega,{\bf k})=-f(-\omega,-{\bf k})$. Thus, $f(\omega,{\bf k})$ is either even in the frequency and odd in the momentum or is odd in the frequency and even in the momentum. Table~\ref{waves} summarizes the various Cooper pairing~\cite{pairing_table}.

\begin{table}[h]
\caption{Classification of Cooper pairings $f(\omega,{\bf k})$ in the frequency and momentum spaces according to the spin singlet and triplet pairings.}
\begin{tabular}{ccccc}
\hline
\begin{minipage}[m]{.15\textwidth}
	\centering\vspace*{10pt}
Spin
\vspace*{10pt}
\end{minipage} & \begin{minipage}[m]{.15\textwidth}\centering
Frequency
\end{minipage} &  \begin{minipage}[m]{.15\textwidth}\centering
Momentum
\end{minipage} & \begin{minipage}[m]{.15\textwidth}
~
\end{minipage} & \begin{minipage}[m]{.15\textwidth}
~
\end{minipage} \\ \hline
		\multirow{2}{*}{\begin{tabular}[c]{@{}c@{}}\begin{minipage}[m]{.15\textwidth}
		\centering	\vspace*{15pt}		
		Singlet~(odd)\\~\\~\\ $\uparrow\downarrow-\downarrow\uparrow$		
		\end{minipage}\end{tabular}}      & \begin{minipage}[m]{.15\textwidth}
	\centering
Even
\end{minipage}      & Even     & \begin{minipage}[m]{.15\textwidth}
\centering\vspace*{4pt}
\includegraphics[width=2cm]{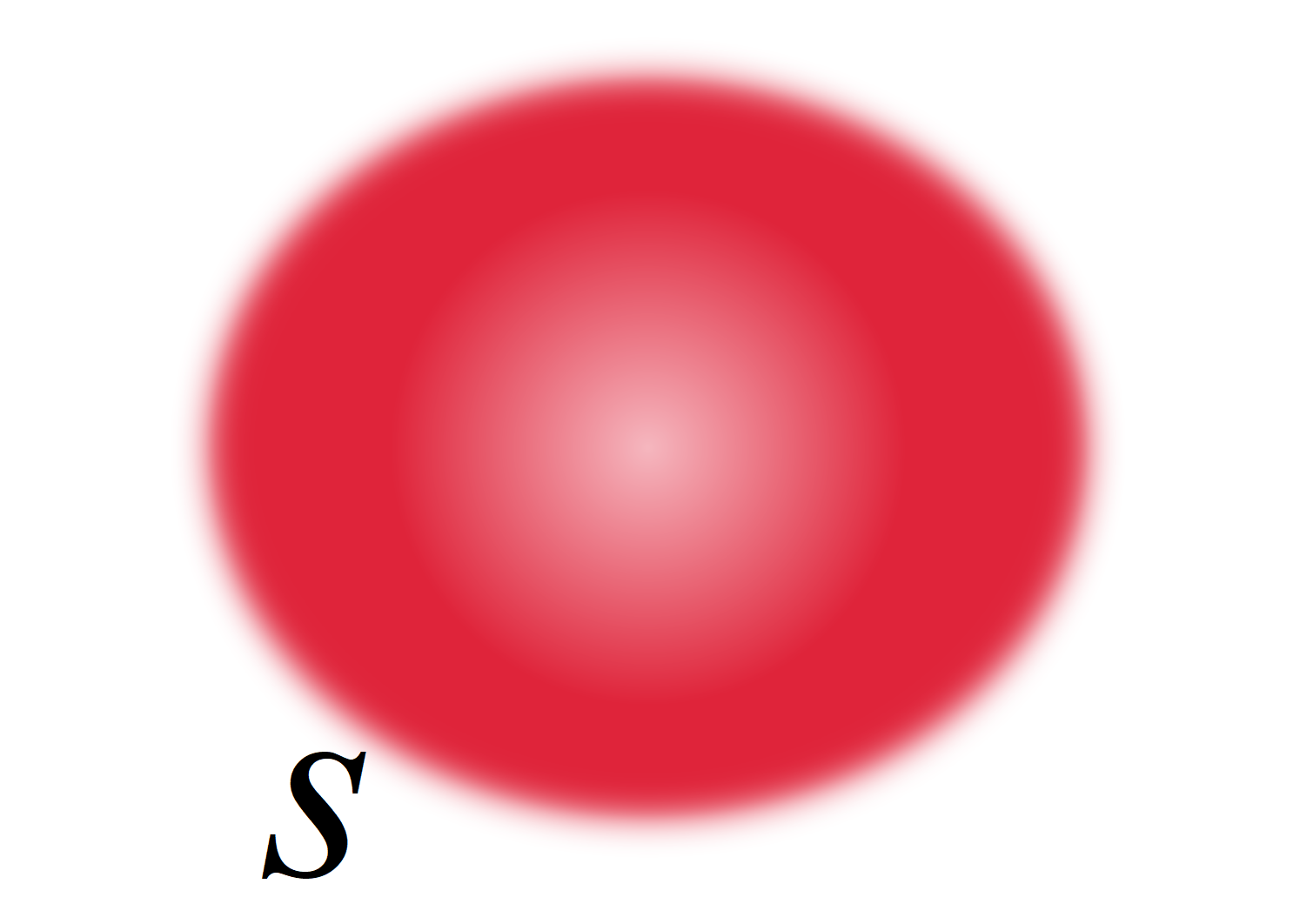}
\vspace*{2pt}
\end{minipage}   & \begin{minipage}[m]{.15\textwidth}
\centering\vspace*{4pt}
\includegraphics[width=2.1cm]{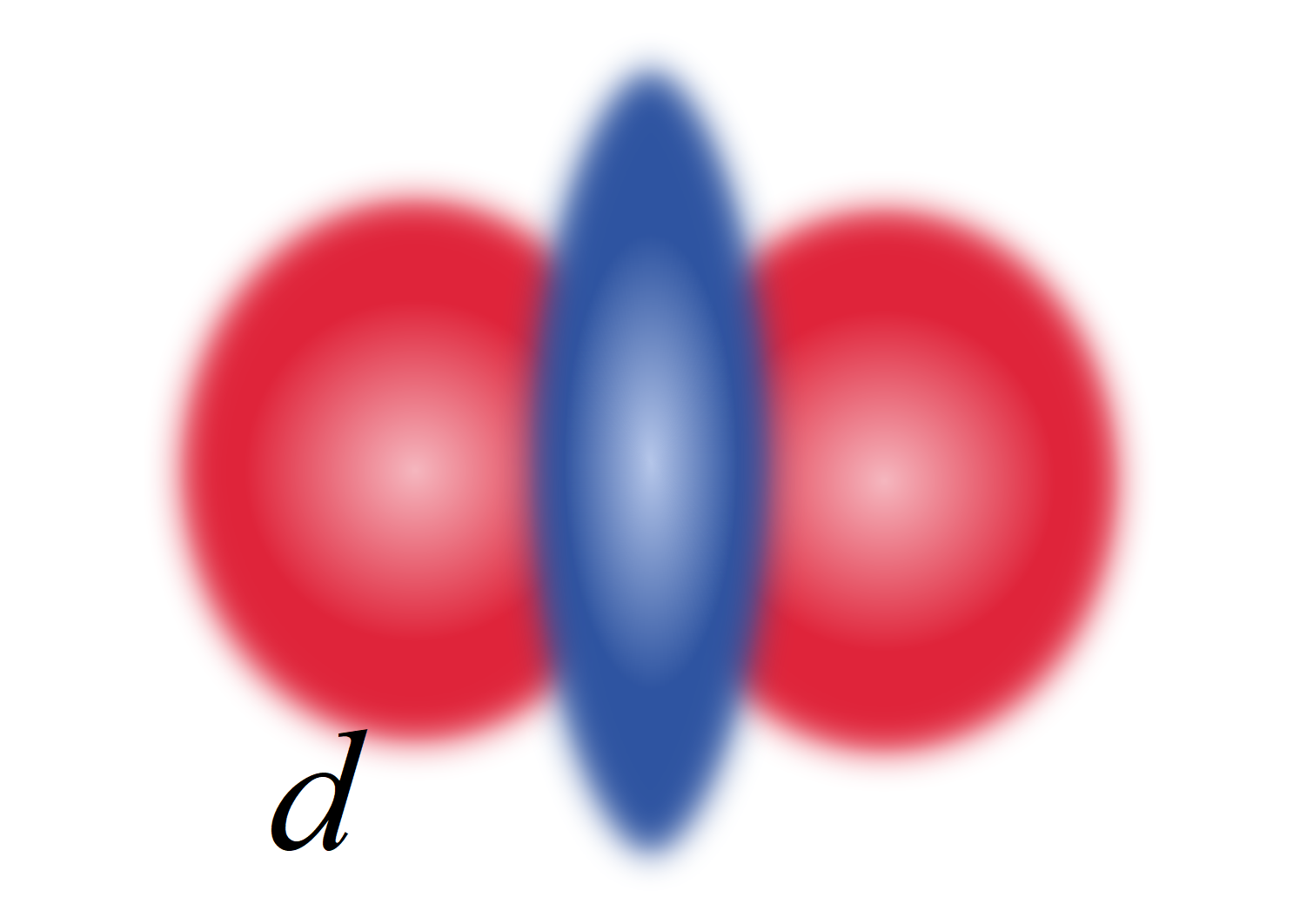}
\vspace*{2pt}
\end{minipage}  \\ \cline{2-5} 
& Odd       & Odd      
& \begin{minipage}[m]{.15\textwidth}
\centering\vspace*{4pt}
\includegraphics[width=2.1cm]{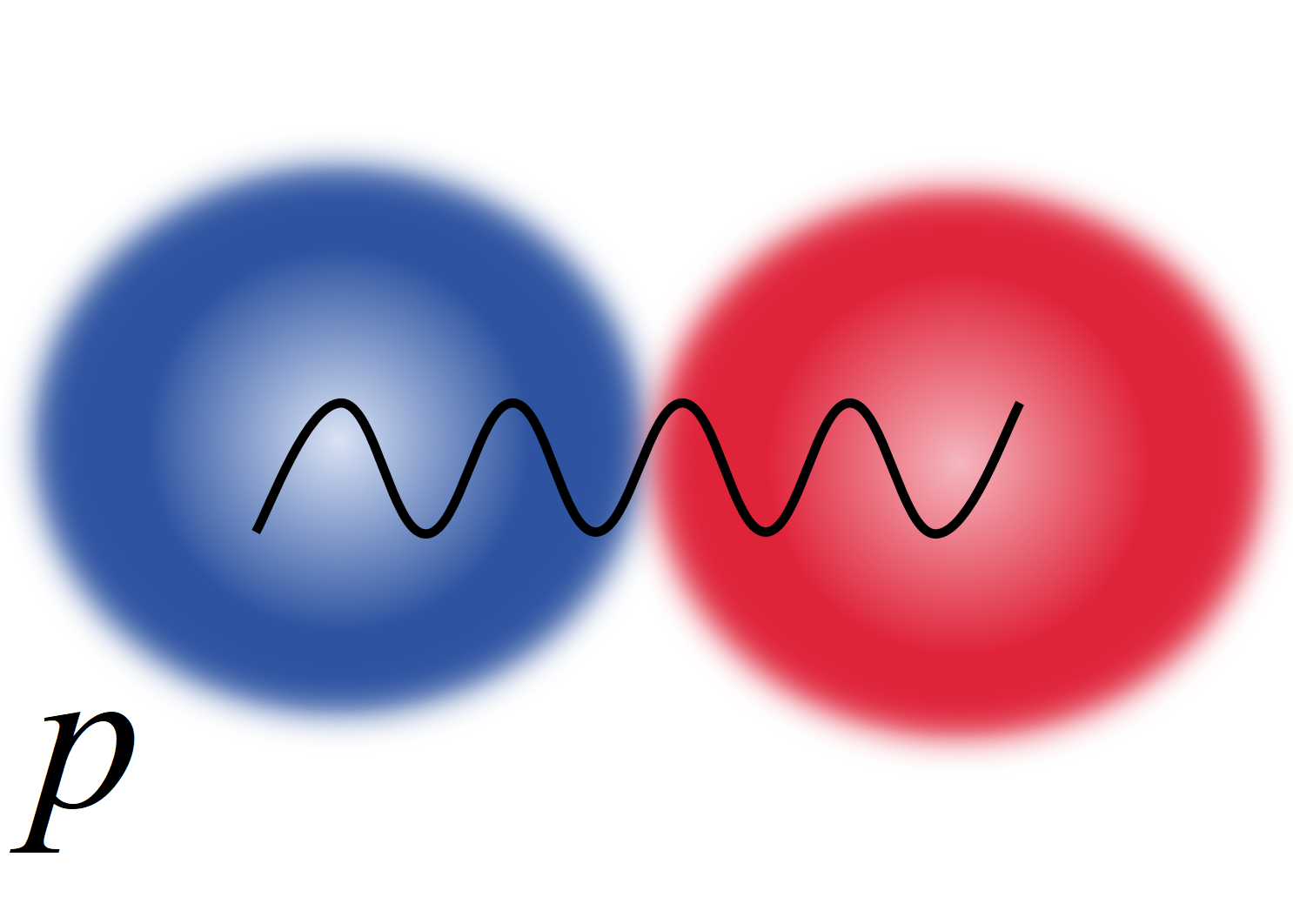}
\vspace*{2pt}
\end{minipage} & 
\begin{minipage}[m]{.15\textwidth}
\centering\vspace*{4pt}
\includegraphics[width=2cm]{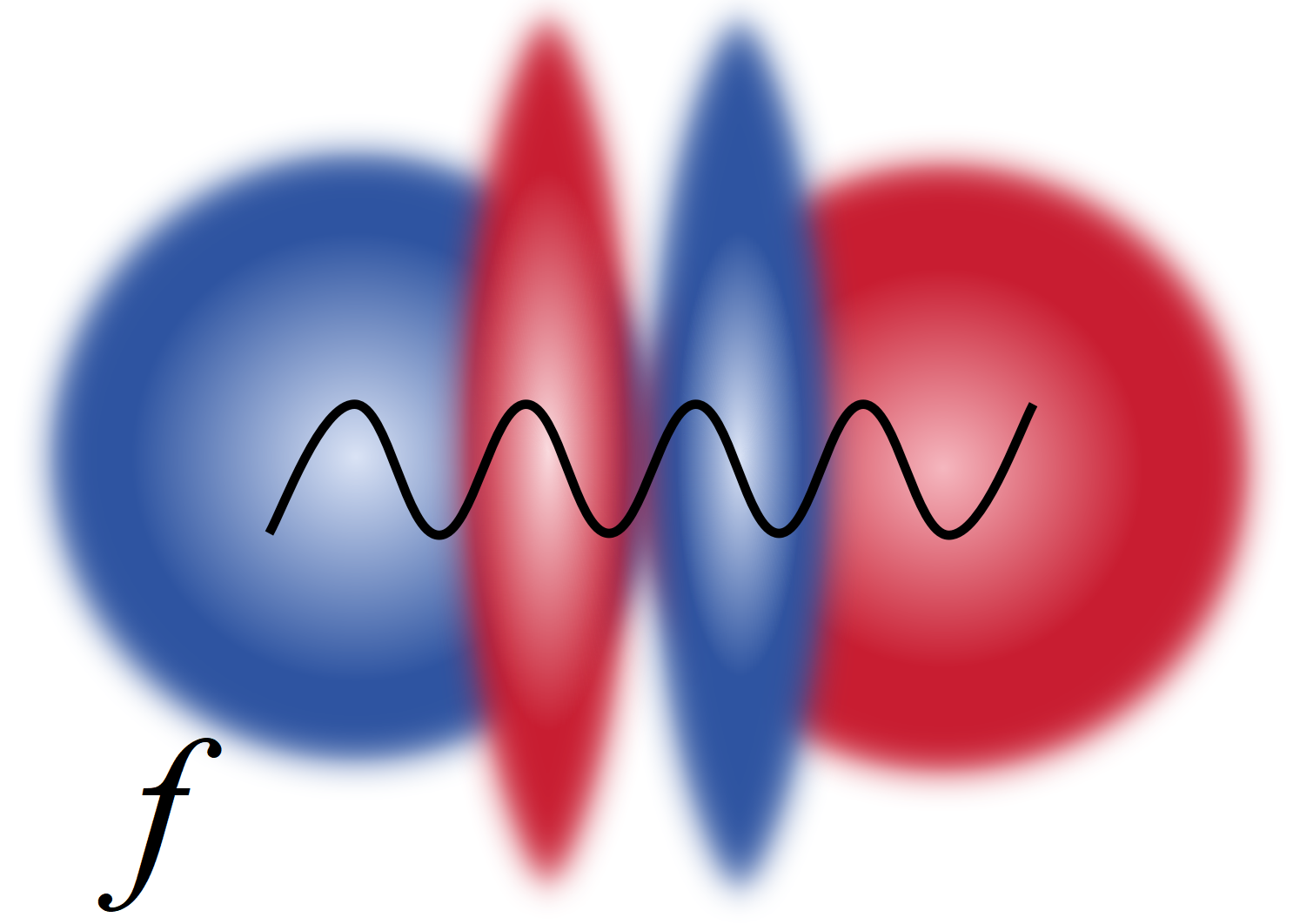}
\vspace*{2pt}
\end{minipage} \\ \hline
\multirow{2}{*}{\begin{tabular}[c]{@{}c@{}}\begin{minipage}[m]{.15\textwidth}
\centering	\vspace*{6pt}			$\uparrow\uparrow~~~~~~\downarrow\downarrow$	\\~\\
Triplet (even)\\~\\ $\uparrow\downarrow+\downarrow\uparrow$		
\end{minipage}\end{tabular}} & Even      & Odd      & \begin{minipage}[m]{.15\textwidth}
\centering\vspace*{4pt}
\includegraphics[width=2.2cm]{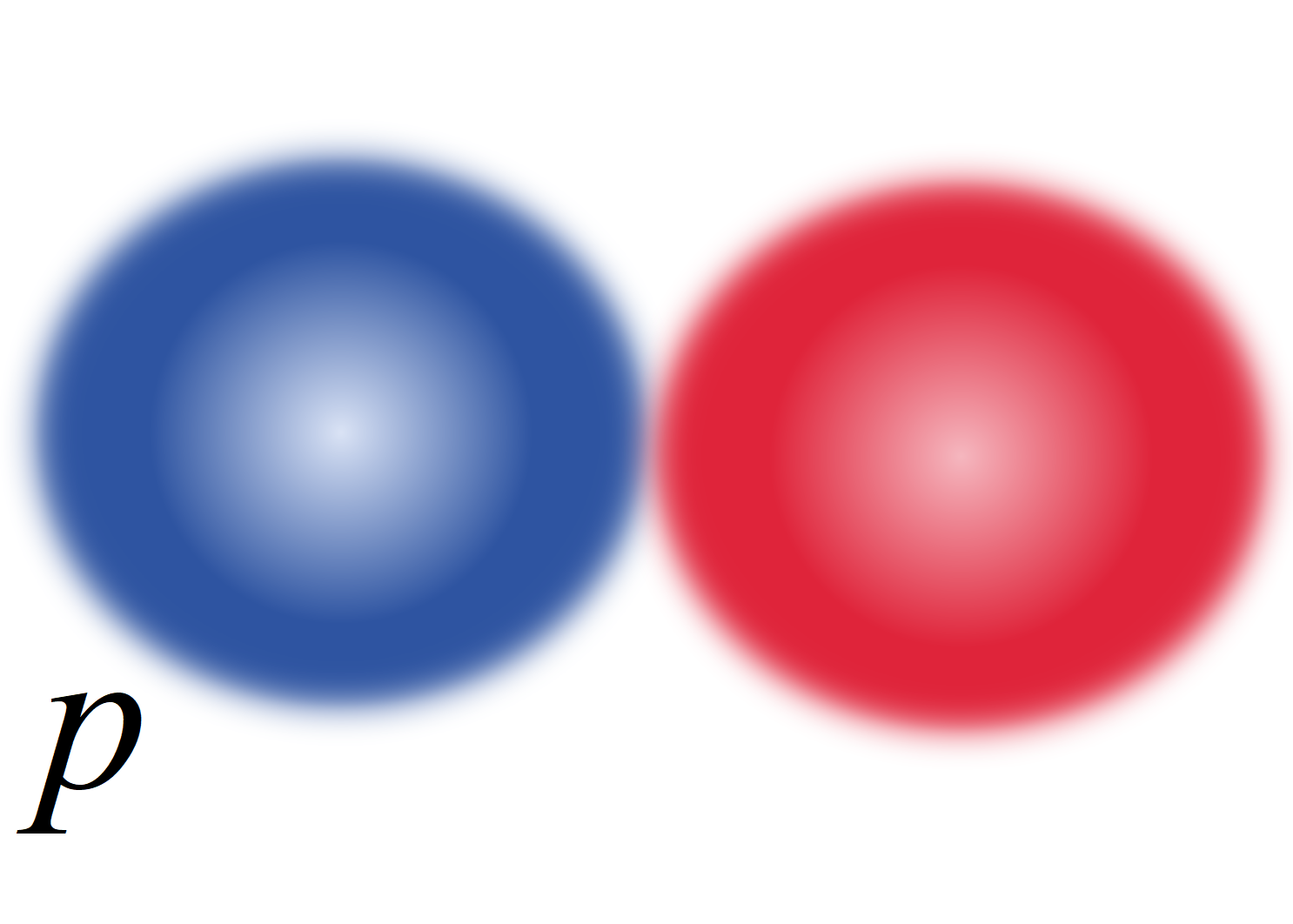}
\vspace*{2pt}
\end{minipage} & \begin{minipage}[m]{.15\textwidth}
\centering\vspace*{4pt}
\includegraphics[width=2cm]{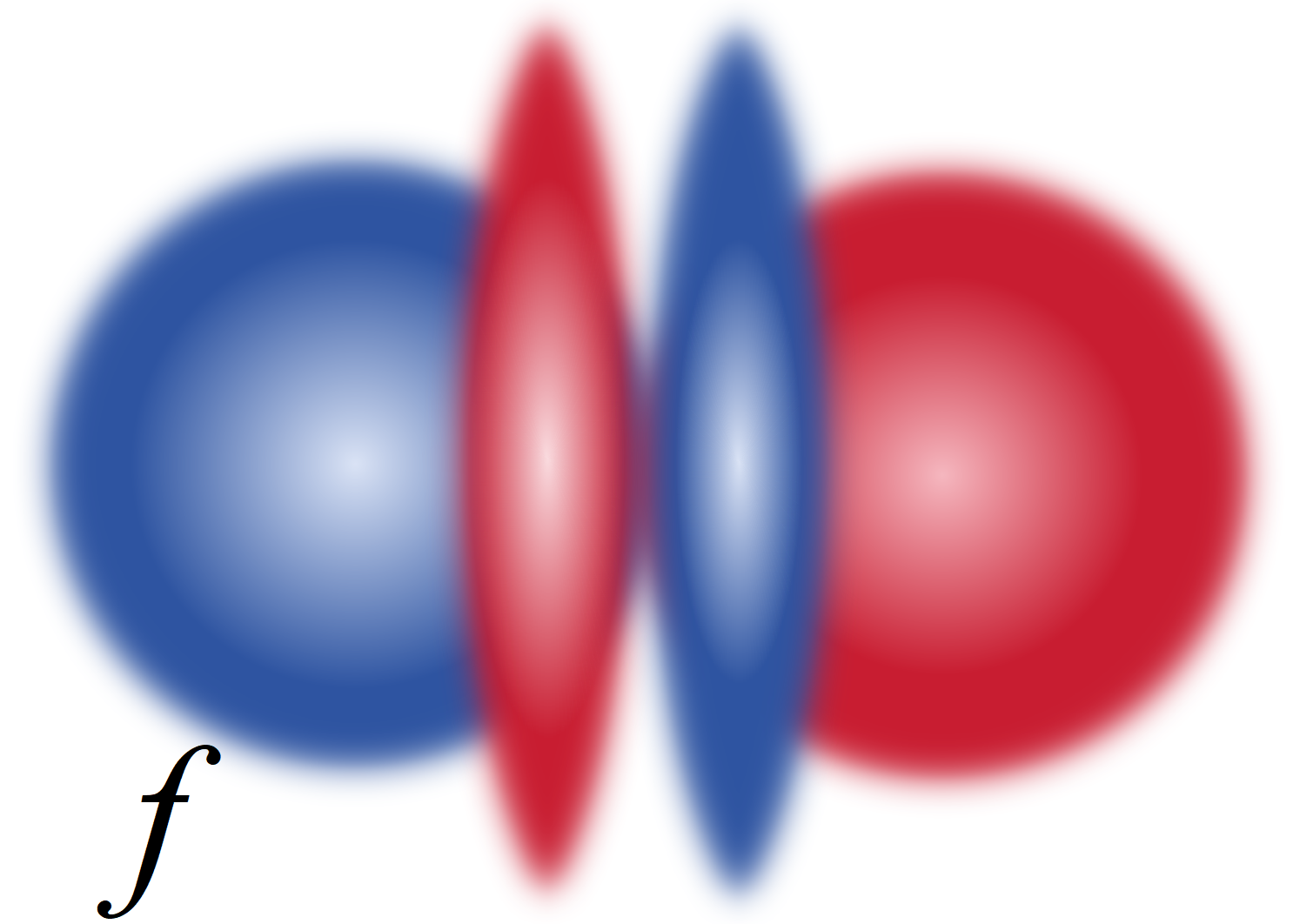}
\vspace*{2pt}
\end{minipage} \\ \cline{2-5} 
		& Odd       & Even     & \begin{minipage}[m]{.15\textwidth}
\centering\vspace*{4pt}
\includegraphics[width=1.8cm]{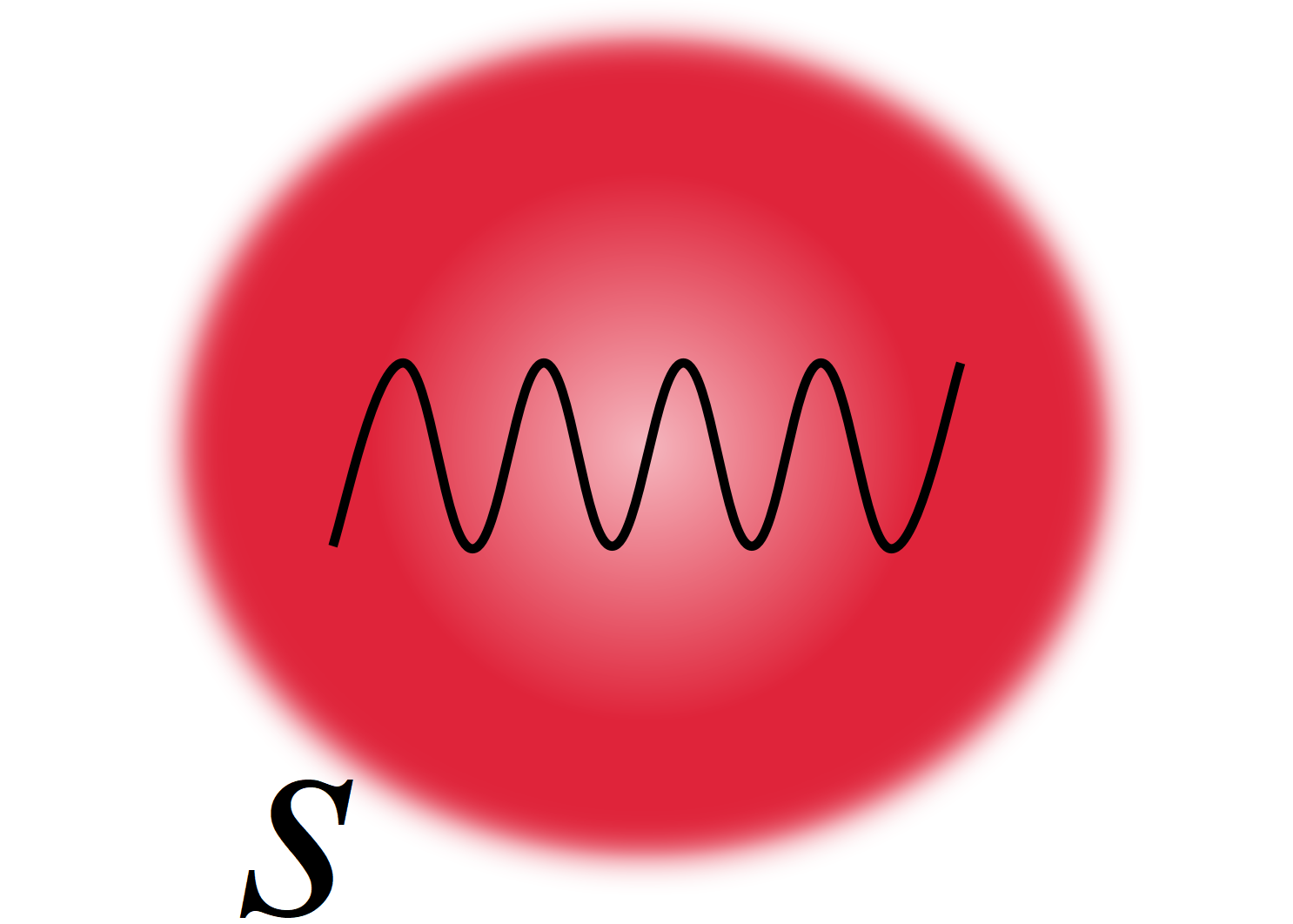}
\vspace*{2pt}
\end{minipage} & \begin{minipage}[m]{.15\textwidth}
\centering\vspace*{4pt}
\includegraphics[width=2cm]{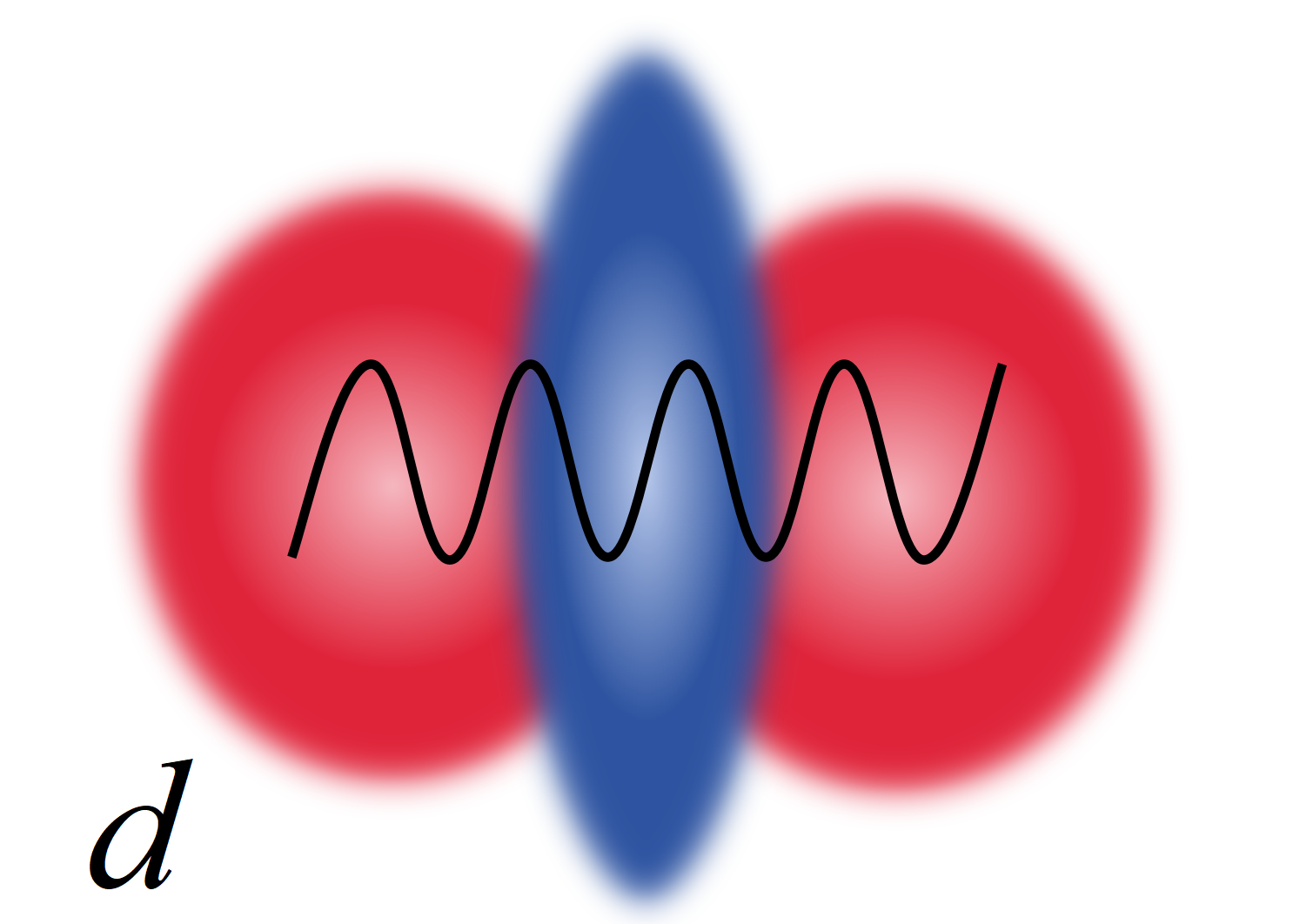}
\vspace*{2pt}
\end{minipage} \\ \hline
\end{tabular} \label{waves} 
\end{table}

When the Cooper pairing \eqref{general_anomalous_correlation} is known, in the framework of the weak-coupling theory, the Landau order parameter is calculated according to the gap equation~\cite{Sigrist,HS_pairing_symmetry,Leggett_book}
\begin{equation}
\Delta_{\sigma\sigma'}({\bf k})=\sum_{{\bf k}',\omega, \sigma_1, \sigma_2}V_{\sigma \sigma',\sigma_1 \sigma_2}({\bf k},{\bf
  k}')F_{\sigma_1 \sigma_2}({\bf k}',\omega),
\label{gap_equation_intro}
\end{equation}
 where $V_{\sigma \sigma',\sigma_1 \sigma_2}({\bf k},{\bf k}')$ is the pairing potential of the superconductivity. In the absence of the SOC it can be factorized as $V_{\sigma \sigma',\sigma_1 \sigma_2}({\bf k},{\bf k}') = V({\bf k},{\bf k}')\Gamma_{\sigma \sigma',\sigma_1 \sigma_2}$.  In conventional singlet $s$-wave superconductivity $V({\bf k},{\bf k}')$ is taken as a constant $V_0$, $\Gamma_{\sigma \sigma',\sigma_1 \sigma_2}=(1/2)(i \sigma_y)_{\sigma \sigma}(-i\sigma_y)_{\sigma_1 \sigma_2}$  and the summation over $\omega$ is restricted by the Debye frequency $\Omega_D$.
We note that the symmetry of the superconducting order parameter does not always follow from that of the superconducting pairing since the symmetry of the pairing potential is also essential according to the gap equation (\ref{gap_equation_intro})~\cite{Sigrist}. 
For example, when the triplet, even-frequency, $p$-wave pairing in the superconductor exists, the Cooper pairing is odd in momentum according to Table.~\ref{waves}. 
For the singlet $s$-channel with constant pairing potential $V_0$ and $\Gamma_{\sigma \sigma',\sigma_1 \sigma_2}=(1/2)(i \sigma_y)_{\sigma \sigma}(-i\sigma_y)_{\sigma_1 \sigma_2}$, according to the gap equation \eqref{gap_equation_intro}, there exists no triplet order parameter. From this point of view, one concludes that the realization of the triplet order parameter requires an exceptional pairing potential~\cite{Sigrist}.

For example, we provide a detailed explanation of triplet superconductivity. Theoretically, there are three levels to describe the triplet superconductivity from the micro to macroscopic point of view~\cite{Leggett_book,Sigrist}. Firstly, at the most microscopic level, the Cooper pairing can characterize the triplet superconductivity, which can be represented as a $2\times 2$ matrix $[{\bf f}(\omega,{\bf k}')\cdot{\pmb \sigma}]i\sigma_y$ in the spin space, where the ${\bf f}$ vector depends on both the momentum ${\bf k}'$ and the frequency $\omega$. 
Secondly, when the dependence on the frequency in the Cooper pairing is integrated out in the integral \eqref{anomalous_correlation_further}, the anomalous correlation is described by the ${\bf l}$ vector: $[{\bf l}({\bf k}')\cdot{\pmb \sigma}]i\sigma_y$. Finally, at the most macroscopic level, when the dependence on the momentum in the anomalous correlation is further integrated out with the momentum dependence of the pairing potential in the triplet channel $V_{\sigma \sigma',\sigma_1 \sigma_2} = (1/2)V({\bf k}, {\bf k}') (i \pmb \sigma \sigma_y)_{\sigma \sigma'}(-i \sigma_y \pmb \sigma)_{\sigma_1 \sigma_2}$ in the gap equation \eqref{gap_equation_intro}, we obtain the Landau superconducting order parameter, which is described by the ${\bf d}$ vector: $\hat{\Delta}=[{\bf d}({\bf k})\cdot{\pmb \sigma}]i\sigma_y$.

The triplet superconductivity can usually be divided into the unitary and non-unitary classes using the ${\bf d}$ vector of the order parameter~\cite{Leggett_book,Sigrist}. In this point of view, the order-parameter matrix $\hat{\Delta}$ in the spin space is treated as a wave function of the Cooper pairs. The expectation of the spin polarization of the triplet Cooper pairs is then calculated by~\cite{Leggett_book,Sigrist}
\begin{equation}
{\rm tr}[\hat{\Delta}^{\dagger}\hat{\pmb \sigma}\hat{\Delta}]\propto i{\bf d}\times {\bf d}^*. 
\end{equation}
Accordingly, when $i{\bf d}\times {\bf d}^*=0$, the associated triplet Cooper pairs have no net spin polarization, which satisfies ${\bf d}^{*}={\bf d}$. Such triplet superconductivity is called unitary triplet superconductivity; otherwise, it is called non-unitary triplet superconductivity~\cite{Leggett_book,Sigrist}. 
Several possible systems are candidates of triplet superconductivity, including the superfluid $^3$He ~\cite{Leggett_book,Osheroff,Fay,Anderson_PRL,Anderson_PRA,Leggett} and unconventional superconductor Sr$_2$RuO$_4$~\cite{Fay,Anderson_PRL,Anderson_PRA,Leggett_book,Rice,Mackenzie,Ishida,Eremin,Maeno,Read,Ivanov,Mackenzie2}. The triplet Cooper pairing may also appear in  conventional superconductors~\cite{Gorkov2,Yang,X.:Liu,Reeg,Triola,Bobkova_184518} with interface SOC, non-centrosymmetric superconductor~\cite{Sigrist,Frigeri,Bauer}, and the conventional superconductor-ferromagnet interface~\cite{Buzdin,Bergeret,Tokatly,Tokatly_1,Alidoust,Bernardo,Kalcheim,Linder2015,Eschrig_2015}.

\subsection{Dynamics of superconductors}
\label{quasi-classical_equations}

According to the BCS theory, the conventional superconductivity may be well described by the  Bogoliubov de-Gennes (BdG) Hamiltonian under the mean-field approximation~\cite{Abrikosov,Gorkov}
\begin{equation}
\hat{H}=\hat{H}_0+\hat{H}_{\rm int}+\hat{H}_{\rm im}.
\label{Hamiltonian_quasi-classical}
\end{equation}
With the field operator $\hat{\Psi}=(\hat{\psi}_{\uparrow},\hat{\psi}_{\downarrow}^{\dagger})^T$ under the Nambu representation, the kinetic energy 
\begin{equation}
\hat{H}_0=\int d{\bf
  r}\hat{\Psi}^{\dagger}({\bf r})\Big(\frac{\displaystyle {\bf p}^2}{\displaystyle
  2m}-\mu\Big)\tau_3\hat{\Psi}({\bf r}),
\end{equation}
where $m$ and $\mu$ are, respectively, the mass of the electrons and the chemical potential of the system; $\tau_3={\rm diag}(1,-1)$ is the metric in the Nambu representation. The superconducting order parameter $\Delta({\bf
r})=\lambda\langle\hat{\psi}_{\downarrow}({\bf r})\hat{\psi}_{\uparrow}({\bf r})\rangle$ under the mean-field approximation~\cite{Abrikosov,Gorkov} is responsible for the superconductivity phenomena, 
where $\lambda<0$ is the electron-electron effective attractive potential, with which the pairing Hamiltonian
\begin{equation}
 \hat{H}_{\rm int}=\int d{\bf
  r}\hat{\Psi}^{\dagger}({\bf r})\tilde{\Delta}({\bf r})\tau_3\hat{\Psi}({\bf
  r}),
\label{pair}  
\end{equation}
in which $\tilde{\Delta}({\bf r})=\left(\begin{array}{ccc}
0&-\Delta({\bf r})\\
\Delta^*({\bf r})&0
\end{array}\right)$.
The electron-impurity interaction Hamiltonian
\begin{equation}
\hat{H}_{\rm im}=\int d{\bf r}\hat{\Psi}^{\dagger}({\bf r})U({\bf
  r})\tau_3\hat{\Psi}({\bf r}),
\end{equation}
where $U({\bf r})$ is the electron-impurity interaction potential. Starting from the Heisenberg equation of motion, we can derive the BdG equation for the field operator $\hat{\Psi}$ as
\begin{equation}
i\partial_t\hat{\Psi}=\hat{H}\hat{\Psi}.
\label{BdG_operator}
\end{equation}

The dynamics of conventional superconducting metals is described by the dynamic equation based on the quasi-classical approximation~\cite{Bergeret,Buzdin,Tokatly,Tokatly_1,de:Gennes,Eilenberger,Larkin,Larkin1,Larkin2,Larkin3,Usadel,Eliashberg, Schmid,Shelankov,Shelankov1,Eckern,Rammer, Kuprianov,Kopnin,Chandrasekhar,Takane,Bauer,Konschelle,Espedal}. The quasi-classical approximation may be understood in terms of the two-particle Green's function $G({\bf r}_1,{\bf r}_2;t_1,t_2)$. Making the Fourier transformation ${\bf r}_1 - {\bf r}_2 \to {\bf p}$ and $t_1 - t_2 \to E$, it is convenient to work further with the Green's function $G({\bf p}, {\bf R}; E,T)$, where ${\bf R}=({\bf r}_1+{\bf r}_2)/2$ is the center-of-mass coordinate and $T=(t_1+t_2)/2$ is the center-of-mass time. 
The dependence on the momentum {\bf k} can be further decomposed into the dependence on the modulus value of the momentum or the normal-state energy $\xi_{{\bf p}}$ and the direction of the momentum $\hat {\bf p}= {\bf p}/|{\bf p}|$.
The quasi-classical approximation is generally applied to the case that the system has a large Fermi surface, such that $\Delta/E_F \ll 1$, where $E_F$ is the corresponding Fermi energy. 
In this case, the main superconductivity-related physics occurs near the Fermi surface~\cite{Bergeret,Buzdin,Tokatly,Tokatly_1,de:Gennes,Eilenberger,Larkin,Larkin1,Larkin2,Larkin3,Usadel,Eliashberg, Schmid,Shelankov,Shelankov1,Eckern,Rammer, Kuprianov,Kopnin,Chandrasekhar,Takane,Bauer,Konschelle,Espedal}. 
In this situation, the dependence of the Green's function on $\xi_{{\bf k}}$ may be integrated out, while its dependence on $\hat {\bf p}$, ${\bf R}$, $E$ and $T$ should be dealt with explicitly. 
It is worth noting that for systems with small Fermi surfaces, the quasi-classical approximation may not be applicable, as all electrons participate in the dynamics. In this situation, we may alternatively choose the quasiparticle approximation~\cite{Yu,Yu2}. 
In the quasiparticle approximation,  the information about the frequency $E$ in the Green's function is integrated out, while the Green's function explicitly depends on the momentum, the center-of-mass coordinates, and the center-of-mass time~\cite{Zhao,Yamashita,M.:J.:Stephen,Wolfle,Einzel,Wolfle2,Aronov,Combescot,Hara,Andreev,Klam,Manske}. In general, the quasiparticle approximation can only be applied when the perturbation to the system's superconducting order parameter is small, and then the quasiparticle energy spectrum is well defined.

Using the quasi-classical approximation~\cite{Bergeret,Buzdin,Tokatly,Tokatly_1,de:Gennes,Eilenberger,Larkin,Larkin1,Larkin2,Larkin3,Usadel,Eliashberg, Schmid,Shelankov,Shelankov1,Eckern,Rammer, Kuprianov,Kopnin,Chandrasekhar,Takane,Bauer,Konschelle,Espedal}, we can set up the Eilenberger equation~\cite{Eilenberger,Eliashberg}, the Usadel equation~\cite{Usadel}, and the macroscopic Ginzburg-Landau equation~\cite{Ginzburg} that are commonly used for the superconductor dynamics from the fully microscopic Gor'kov equation~\cite{Abrikosov,Gorkov}. In general, from the most microscopic level, the Gor'kov equation can be used as a starting point for studying the dynamics of superconductors. However, the Gor'kov equation involves calculating all the degrees of freedom of the Green's function in the dynamic process, which is a challenging calculation.
When the Fermi energy $E_F$ of the metal is much larger than the superconducting order parameter $\Delta$, the quasi-classical approximation can be applied to simplify the Gor'kov equation. Such a simplified equation is called the Eilenberger equation~\cite{Eilenberger,Eliashberg}. Furthermore, the Eilenberger equation can be reduced to the Usadel equation in the dirty sample limit~\cite{Usadel}.
When the temperature $T$ is close to the superconducting transition temperature $T_c$, the superconducting order parameter of the system becomes small. In this situation, the Gor'kov and Eilenberger equations can be further reduced to the macroscopic Ginzburg-Landau equation about the superconducting order parameter~\cite{Ginzburg}. In this part, we elaborate on the derivation of the Gor'kov~\cite{Abrikosov,Gorkov}, Eilenberger~\cite{Eilenberger,Eliashberg}, Usadel~\cite{Usadel}, and Ginzburg-Landau equations~\cite{Ginzburg} in conventional superconducting metals based on the Keldysh Green's function method~\cite{Rammer,Kopnin, Chandrasekhar}.

\subsubsection{Gor'kov equation}

The Gor'kov equation~\cite{Abrikosov,Gorkov} is governed by the nonequilibrium contour Green's function~\cite{Rammer,Kopnin, Chandrasekhar} under the Hamiltonian \eqref{Hamiltonian_quasi-classical}. 
Under the Nambu representation, the contour Green's function of the system is defined as~\cite{Rammer}
\begin{align} 
\underline{\tilde{G}}_{12}=-i\tau_3\langle
T_c\hat{\Psi}_1\hat{\Psi}_2^{\dagger}\rangle
=\left(\begin{array}{ccc}
\tilde{G}_{12}^{--}&\tilde{G}_{12}^{-+}\\
\tilde{G}_{12}^{+-}&\tilde{G}_{12}^{++}
\end{array}\right),
\label{Keldysh_pre}
\end{align}
which is a $4\times 4$ matrix. 
In Eq.~\eqref{Keldysh_pre}, ``1" and ``2" indicate the spatiotemporal coordinate points $(t_1,{\bf r}_1)$ and $(t_2,{\bf r}_2)$. The components $\tilde{G}_{12}^{--}=-i\tau_3\langle
T\hat{\Psi}_1\hat{\Psi}_2^{\dagger}\rangle$, $\tilde{G}_{12}^{-+}=i\tau_3\langle \hat{\Psi}_2^{\dagger}\hat{\Psi}_1\rangle$,
$\tilde{G}_{12}^{+-}=-i\tau_3\langle 
\hat{\Psi}_1\hat{\Psi}_2^{\dagger}\rangle$, and $\tilde{G}_{12}^{++}=-i\tau_3\langle
\tilde{T}\hat{\Psi}_1\hat{\Psi}_2^{\dagger}\rangle$, where $T$ and $\tilde{T}$ indicate the chronological and inverse chronological products of operators. 
In the Keldysh-space~\cite{Rammer}, the non-equilibrium Green's function \eqref{Keldysh_pre} is transformed to  
\begin{equation} 
\underline{\tilde{G}}_{12}=\left(\begin{array}{ccc}
\tilde{G}_{12}^{R}&\tilde{G}_{12}^{K}\\
0&\tilde{G}_{12}^{A}
\end{array}\right),
\label{Keldysh}
\end{equation}
in which
\begin{subequations}
\begin{align}
\tilde{G}_{12}^{R}=G_{12}^{--}-G_{12}^{-+}=G_{12}^{+-}-G_{12}^{++},\\
\tilde{G}_{12}^{A}=G_{12}^{--}-G_{12}^{+-}=G_{12}^{-+}-G_{12}^{++},\\
\tilde{G}_{12}^{K}=G_{12}^{++}+G_{12}^{--}=G_{12}^{+-}+G_{12}^{-+}.
\label{definition}
\end{align}
\end{subequations}
Furthermore, following the BdG equation \eqref{BdG_operator},  the Green's function \eqref{Keldysh} obeys the Gor'kov equation~\cite{Abrikosov,Gorkov}
\begin{subequations}
\begin{align}
\label{Gorkov_0}
&\int dx_3 \underline{\stackrel{\rightarrow}{\mathcal{G}}}_{13}\underline{\tilde{G}}_{32}=\delta(x_1-x_2)I_{4\times
4}+\int dx_3 \underline{\tilde{\Sigma}}_{13}\underline{\tilde{G}}_{32},\\
&\int dx_3\underline{\tilde{G}}_{13}\underline{\stackrel{\leftarrow}{\mathcal{G}}}_{32}
=\delta(x_1-x_2)I_{4\times
4}+\int dx_3 \underline{\tilde{G}}_{13}\underline{\tilde{\Sigma}}_{32},
\label{anti_Gorkov}
\end{align}
\end{subequations}
in which  
\begin{subequations}
\begin{align} 
&\underline{\stackrel{\rightarrow}{\mathcal{G}}}_{13}=\left(\begin{array}{ccc}
i\frac{\displaystyle \partial}{\displaystyle \partial
  t_1}\tau_3+\frac{\displaystyle \partial_1^2}{\displaystyle 2m}+\mu-\tilde{\Delta}_1&0\\
0&i\frac{\displaystyle \partial}{\displaystyle \partial
  t_1}\tau_3+\frac{\displaystyle \partial_1^2}{\displaystyle 2m}+\mu-\tilde{\Delta}_1
\end{array}\right)\delta(x_1-x_3),\\
&\underline{\stackrel{\leftarrow}{\mathcal{G}}}_{32}=\left(\begin{array}{ccc}
-i\frac{\displaystyle \stackrel{\leftarrow}{\partial}}{\displaystyle \partial t_2}\tau_3
+\frac{\displaystyle \stackrel{\leftarrow}{\partial_2^2}}{\displaystyle 2m}+\mu-\tilde{\Delta}_2&0\\
0&-i\frac{\displaystyle \stackrel{\leftarrow}{\partial}}{\displaystyle \partial t_2}\tau_3
+\frac{\displaystyle
  \stackrel{\leftarrow}{\displaystyle \partial_2^2}}{\displaystyle 2m}+\mu-\tilde{\Delta}_2
\end{array}\right)\delta(x_3-x_2),
\end{align}
\end{subequations}
and
\begin{equation}
{\underline{\tilde{\Sigma}}}_{13}=U_1 \delta(x_1-x_3) \underline{\tilde{G}}_{13}
\end{equation}
is the self-energy contributed by the electron-impurity interaction~\cite{Mahan}.

The Gor'kov equation \eqref{Gorkov_0} and \eqref{anti_Gorkov} contain all the microscopic information of the system, which is then challenging to solve universally.
Fortunately, starting from the Gor'kov equation, the kinetic equations obeyed by the quasiparticles can be obtained under the quasiparticle approximation~\cite{Zhao,Yamashita,M.:J.:Stephen,Wolfle,Einzel,Wolfle2,Aronov,Combescot,Hara,Andreev,Klam,Manske}. On the other hand, under the quasi-classical approximation~\cite{Rammer,Kopnin,Chandrasekhar},  the Eilenberger equation for the dynamics of superconducting order parameter can be derived~\cite{Eilenberger,Eliashberg}.

\subsubsection{Eilenberger equation}

\label{Eilenberger_conventional}

In this part, we set up the Eilenberger equation~\cite{Eilenberger,Eliashberg} from the Gor'kov equation [\eqref{Gorkov_0} and \eqref{anti_Gorkov}]. We first simplify the notation, i.e., $\int dx_3 A_{13}B_{32}=(A\otimes B)(1,2)$. With such notation, the Gor'kov equation \eqref{Gorkov_0} and \eqref{anti_Gorkov} are expressed as~\cite{Rammer,Kopnin,Chandrasekhar}
\begin{subequations}
\begin{align}
&(\underline{\stackrel{\rightarrow}{\mathcal{G}}}\otimes
\underline{\tilde{G}})(1,2)=\delta(x_1-x_2)
+(\underline{\tilde{\Sigma}}\otimes \underline{\tilde{G}})(1,2),
\label{Gorkov1}\\
&(\underline{\tilde{G}}\otimes\underline{\stackrel{\leftarrow}{\mathcal{G}}})(1,2)
=\delta(x_1-x_2)+(\underline{\tilde{G}}\otimes\underline{\tilde{\Sigma}})(1,2).
\label{anti_Gorkov1}
\end{align}
\end{subequations}
Taking the difference of Eqs.~\eqref{Gorkov1} and \eqref{anti_Gorkov1} leads to  
\begin{equation}
(\underline{\stackrel{\rightarrow}{\mathcal{G}}}\otimes
\underline{\tilde{G}})(1,2)-(\underline{\tilde{G}}\otimes\underline{\stackrel{\leftarrow}{\mathcal{G}}})(1,2)=
[\underline{\tilde{\Sigma}}\otimes \underline{\tilde{G}}]_{-}(1,2),
\label{difference_0}
\end{equation} 
where $[\underline{\tilde{\Sigma}}\otimes \underline{\tilde{G}}]_{-}(1,2)\equiv (\underline{\tilde{\Sigma}}\otimes 
\underline{\tilde{G}})(1,2)-(\underline{\tilde{G}}\otimes\underline{\tilde{\Sigma}})(1,2)$.
With the center-of-mass coordinates $T=(t_1+t_2)/2$ and ${\bf
  R}=({\bf R}_1+{\bf R}_2)/2$ and the relative coordinates $t=t_1-t_2$ and ${\bf r}={\bf r}_1-{\bf r}_2$, Eq.~\eqref{difference_0} is alternatively expressed as 
\begin{equation}
i\Big(\frac{1}{2}\frac{\partial}{\partial T}+\frac{\partial}{\partial
  t}\Big)
\hat{\tau}_3\underline{\tilde{G}}_{12}+i\Big(\frac{1}{2}\frac{\partial}{\partial T}-\frac{\partial}{\partial
t}\Big)\underline{\tilde{G}}_{12}\hat{\tau}_3+\frac{1}{m}\frac{\partial}{\partial {\bf
    R}}\cdot\frac{\partial}{\partial {\bf
    r}}\underline{\tilde{G}}_{12}+[\hat{\Delta}\otimes\underline{\tilde{G}}]_{-}(1,2)=[\underline{\tilde{\Sigma}}\otimes
\underline{\tilde{G}}]_{-}(1,2),
\label{center_relative}
\end{equation}
where $\hat{\tau}_3={\rm diag}(\tau_3,\tau_3)$ and $\hat{\Delta}={\rm diag}(-\tilde{\Delta},-\tilde{\Delta})$. 
In the Fourier space, the Green's function
\begin{equation}
\underline{\tilde{G}}({\bf R},T;{\bf p},E)=\int dtd{\bf r}e^{iEt-i{\bf
    p}\cdot{\bf r}}\underline{\tilde{G}}({\bf R},T;{\bf
  r},t).
\end{equation}
Accordingly, Eq.~\eqref{center_relative} is written in the Fourier space as
\begin{align}
\nonumber
&\frac{i}{2}\frac{\partial}{\partial T}\Big\{\hat{\tau}_3,
\underline{\tilde{G}}({\bf R},T;{\bf p},E)\Big\}
+\Big[E\hat{\tau}_3,\underline{\tilde{G}}({\bf R},T;{\bf p},E)\Big]+\frac{i{\bf
    p}}{m}\cdot\frac{\partial}{\partial {\bf R}}\underline{\tilde{G}}({\bf
  R},T;{\bf p},E)+[\hat{\Delta}\otimes_{\mathrm{full}}\underline{\tilde{G}}]_{-}(X,p)\\
&
=[\underline{\tilde{\Sigma}}\otimes_{\mathrm{full}}
\underline{\tilde{G}}]_{-}(X,p).
\label{to_be_further}
\end{align}
In Eq.~\eqref{to_be_further}, $\{\ ,\ \}$ and $[\ ,\ ]$ indicate the anticommutator and commutator, respectively, and 
\begin{equation}
(A\otimes_{\mathrm{full}} B)(X,p)=\int dx e^{-ipx}\int
dx_3A(1,3)B(3,2) = \exp[i(\partial_X^A \partial_p^B - \partial_p^A \partial_X^B)/2]A(X,p)B(X,p),
\end{equation}
where $X\equiv (T,{\bf R})$, $p\equiv (E,{\bf p})$, $px\equiv-Et+{\bf p}\cdot{\bf r}$, $\partial_X \equiv (-\partial_T, \nabla_{\mathbf R})$, $\partial_p \equiv (-\partial_E, \nabla_{\mathbf p})$, and $\partial_X^A \partial_p^B \equiv -\partial_T^A \partial_E^B + \nabla_{\mathbf R}^A \cdot \nabla_{\mathbf p}^B$.

In Eq.~\eqref{to_be_further}, the last term on the left-hand side and the term on the right-hand side are still complicated. They can be simplified by exploiting the gradient expansion over the center-of-mass coordinate $\mathbf R$ and the momentum $\mathbf p$~\cite{Rammer,Kopnin,Chandrasekhar}, i.e.,
\begin{align}
&(A\otimes_{\mathrm{full}} B)(X,p) \nonumber \\
&= A(X,p) \otimes B(X,p)
+\frac{i}{2}\left(\partial_{\mathbf R}^A\partial_{\mathbf p}^B-\partial_{\mathbf p}^A\partial_{\mathbf R}^B\right)A(X,p) \otimes B(X,p) + ...   \nonumber \\
& \approx A(X,p) \otimes B(X,p),
\label{gradient}
\end{align}
where 
\begin{align}
A(X,p) \otimes B(X,p) = \exp[-i(\partial_T^A \partial_E^B - \partial_E^A \partial_T^B)/2]A(X,p)B(X,p).
\label{gradient_2}
\end{align}
All gradient terms, which are disregarded in Eq.~(\ref{gradient}), contain small factor $1/(\xi_S p_F) \sim \Delta/E_F$ in different powers \cite{Rammer,Serene1983} and, consequently, are small as compared to the leading order term $A(X,p) \otimes B(X,p)$. Here $\xi_S$ is the superconducting coherence length. At the same time, in general non-stationary situation the gradient terms containing derivatives with respect to the center-of-mass time and energy cannot be disregarded, see for example, Refs.~\cite{Bobkova2021,Bobkova2022,Bobkov2023}.

Retaining the lowest order in the spatial and momentum gradient expansion \eqref{gradient} yields 
\begin{subequations}
\begin{align}
&[\hat{\Delta}\otimes_{\mathrm{full}}\underline{\tilde{G}}]_{-}(X,p)\approx\big[\hat{\Delta}({\bf
  R},T),\underline{\tilde{G}}({\bf R},T;{\bf p},E)\big]_\otimes,\label{Delta}\\
&[\underline{\tilde{\Sigma}}\otimes_{\mathrm{full}}
\underline{\tilde{G}}]_{-}(X,p)\approx\big[\underline{\tilde{\Sigma}}({\bf R},T;{\bf p},E),
\underline{\tilde{G}}({\bf R},T;{\bf p},E)\big]_\otimes, 
\label{scattering}
\end{align}
\end{subequations}
where $[A,B]_\otimes = A \otimes B - B \otimes A$. In Eq.~\eqref{scattering}, the self-energy 
$\underline{\tilde{\Sigma}}({\bf R},T;{\bf p},E)$ contributed by the electron-impurity interaction can be calculated by the Feynman diagram in the momentum space. After averaging the position of the impurity in the sample~\cite{Rammer,Kopnin,Chandrasekhar,Mahan}, we obtain 
\begin{equation}
\underline{\tilde{\Sigma}}({\bf R},T;{\bf
  p},E)=\frac{1}{2\pi\tau}\int\frac{d\Omega_{{\bf p}'}}{4\pi}\int d\xi_{{\bf
    p}'}\underline{\tilde{G}}({\bf R},T;{\bf p}',E),
\label{self-energy}
\end{equation}
where $\tau$ is the momentum relaxation time due to the electron-impurity scattering and $\Omega_{{\bf p}}$ is the solid angle. Substituting Eqs.~\eqref{Delta}, \eqref{scattering} and \eqref{self-energy} into \eqref{center_relative}, we arrive at 
\begin{align}
\nonumber
&\big[E\hat{\tau}_3+\hat{\Delta}({\bf R},T), \underline{\tilde{G}}({\bf R},T;{\bf
  p},E)\big]_\otimes +\frac{i{\bf p}}{m}\cdot \frac{\partial}{\partial {\bf R}}\underline{\tilde{G}}({\bf R},T;{\bf
  p},E)\\
&-\frac{1}{2\pi\tau}\int\frac{d\Omega_{{\bf p}'}}{4\pi}\int d\xi_{{\bf
    p}'}\big[\underline{\tilde{G}}({\bf R},T;{\bf p}',E), \underline{\tilde{G}}({\bf R},T;{\bf
  p},E)\big]_\otimes=0,
\label{start}
\end{align}
where it is taken into account that $(i/2)\partial_T\big\{\hat{\tau}_3, \underline{\tilde{G}}({\bf R},T;{\bf
  p},E)\big\}+\big[E\hat{\tau}_3, \underline{\tilde{G}}({\bf R},T;{\bf
  p},E)\big] = \big[E\hat{\tau}_3, \underline{\tilde{G}}({\bf R},T;{\bf
  p},E)\big]_\otimes$. It is worth noting that Eq.~\eqref{start} has been significantly simplified compared to the Gor'kov equation. Considering the quasi-classical approximation~\cite{Rammer,Kopnin,Chandrasekhar},  Eq.~\eqref{start} can be further simplified to obtain the Eilenberger equation~\cite{Eilenberger,Eliashberg}.

When the metallic system has a large Fermi energy, the main physics occurs near the Fermi surface.
In this situation, the physical quantities of the system may mainly depend on the direction of the momentum ${\bf p}$, while the modulus of the momentum can be simply set to the Fermi momentum $p_F$. 
Correspondingly, the dependence on the momentum modulus can be integrated out, with which the quasi-classical Green's function is defined as~\cite{Rammer,Kopnin,Chandrasekhar}
\begin{equation}
\underline{\tilde{g}}({\bf R},T;\hat{{\bf p}},E)=\frac{i}{\pi}\int
d\xi_{\bf p}\underline{\tilde{G}}({\bf R},T;{\bf
  p},E),
\label{quasi_classical}
\end{equation}
where $\hat{\bf p}\equiv {\bf p}/|{\bf p}|$ is the direction of the momentum. 
With Eq.~\eqref{start}, we obtain the Eilenberger equation satisfied by the quasi-classical Green's function~\cite{Eilenberger,Eliashberg}
\begin{align}
\big[E\hat{\tau}_3+\hat{\Delta}({\bf R}), \underline{\tilde{g}}({\bf R},T;\hat{{\bf
  p}},E)\big]_\otimes 
+i{\bf v}_F\cdot \frac{\partial}{\partial {\bf R}}\underline{\tilde{g}}({\bf R},T;\hat{{\bf
  p}},E)+\Big[\frac{i}{2\tau}\langle\underline{\tilde{g}}({\bf R},T;\hat{{\bf
    p}}',E)\rangle,
 \underline{\tilde{g}}({\bf R},T; \hat{{\bf
  p}},E)\Big]_\otimes=0,
\label{Eilenberger_general}
\end{align}
where $\langle\underline{\tilde{g}}({\bf R},T;\hat{{\bf p}}',E)\rangle=1/(4\pi)\int d\Omega_{{\bf p}'}\underline{\tilde{g}}({\bf R},T;\hat{{\bf p}}',E)$ and ${\bf v}_F\equiv {\bf p}_F/m$ is the Fermi velocity. 
In the steady state, the explicit dependence of all quantities on $T$ is absent and the $\otimes$-product is ~just the conventional product. Then Eq.~(\ref{Eilenberger_general}) is reduced to
\begin{equation}
\big[E\hat{\tau}_3+\hat{\Delta}({\bf R}), \underline{\tilde{g}}({\bf R};\hat{{\bf
  p}},E)\big]+i{\bf v}_F\cdot \frac{\partial}{\partial {\bf R}}\underline{\tilde{g}}({\bf R};\hat{{\bf
  p}},E)+\Big[\frac{i}{2\tau}\langle\underline{\tilde{g}}({\bf R};\hat{{\bf
    p}}',E)\rangle,
 \underline{\tilde{g}}({\bf R}; \hat{{\bf
  p}},E)\Big]=0.
\label{Eilenberger}
\end{equation}

In contrast to the original Gor'kov equation (\ref{Gorkov_0})-(\ref{anti_Gorkov}) the quasiclassical equation (\ref{Eilenberger}) is homogeneous, and
hence in some sense it leaves the ``norm" of $\underline{\tilde{g}}$ undetermined. Eilenberger~\cite{Eilenberger}
discovered the missing ``normalization condition" to be
\begin{equation} 
\underline{\tilde{g}}({\bf R},T;\hat{{\bf
    p}},E) \otimes \underline{\tilde{g}}({\bf R},T;\hat{{\bf
    p}},E)=I_{4\times 4}.
\label{condition}
\end{equation}
However, there are still redundant solutions in the system that need to be excluded with the help of asymptotic conditions appropriate to the given problem. Because of the need to solve the Eilenberger equation for a lot of quasiparticle trajectories $\hat{{\bf
    p}}$, the process is still relatively difficult. Usadel considered further simplifying the problem in the strong scattering limit, when the quasiclassical Green's function $\underline{\tilde{g}}$ tends to be isotropic in the momentum space, and derived the Usadel equation~\cite{Usadel}. Usadel equation~\cite{Usadel} is widely used to solve the problem of superconducting mesoscopics in stationary and non-stationary situations~\cite{Bergeret,Buzdin,Tokatly,Tokatly_1,Schmid,Shelankov,Shelankov1,Eckern,Rammer,Kuprianov,Kopnin,Chandrasekhar,Takane,Bauer,Konschelle,Espedal,Bobkova2021,Linder_045005,Linder2015,Eschrig_2015,Melnikov:2022}.

\subsubsection{Usadel equation}
\label{Usadel_conventional}

In the strong scattering limit, considering the Green's function tends to be isotropic in the momentum space, we are allowed to divide the quasi-classical Green's function \eqref{quasi_classical} into the isotropic and small anisotropic components in the momentum space. That is, we perform the angle expansion over the Green's function and retain the lowest orders with 
\begin{equation}
\underline{\tilde{g}}({\bf R},T;\hat{{\bf
    p}},E)=\underline{\tilde{g}}_s({\bf R},T;p_F,E)+\hat{\bf
  p}\cdot\underline{\tilde{\bf g}}_p({\bf R},T;p_F,E),
\label{separation}
\end{equation}
where $\underline{\tilde{g}}_s({\bf R},T;p_F,E)\equiv \int d{\Omega}_{\bf p}/(4\pi)\underline{\tilde{g}}({\bf R},T;\hat{{\bf
    p}},E)$ is the angle averaged quasi-classical Green's function, and the vector $\underline{\tilde{\bf g}}_p({\bf R},T;p_F,E)=\underline{\tilde{g}}_{p,x}({\bf
   R},T;p_F,E)\hat{\bf x}+\underline{\tilde{g}}_{p,y}({\bf R},T;p_F,E)\hat{\bf
   y}
+\underline{\tilde{g}}_{p,z}({\bf R},T;p_F,E)\hat{\bf z}$.
In particular, $\underline{\tilde{g}}_s({\bf R},T;p_F,E)$ and $\underline{\tilde{\bf g}}_p({\bf R},T;p_F,E)$
no longer depend on the direction of the momentum. We further assume that the system slightly deviates from the isotropic case, so 
$|\underline{\tilde{\bf g}}_p({\bf R},T;p_F,E)|\ll
 |\underline{\tilde{g}}_s({\bf R},T;p_F,E)|$.
 Substituting Eq.~\eqref{separation} into the Eilenberger equation \eqref{Eilenberger}, we find the even orders of $\hat{\bf p}$ obey 
\begin{equation}
\big[E\hat{\tau}_3+\hat{\Delta}({\bf R}), \underline{\tilde{g}}_s({\bf R},T;p_F,E)\big]_\otimes
+iv_F\hat{{\bf p}}\cdot\partial_{\bf R}\big[\hat{\bf
  p}\cdot\underline{\tilde{\bf g}}_p({\bf R},T;p_F,E)\big]=0;
\label{term_even}
\end{equation}
while the odd orders of $\hat{\bf p}$ obey 
\begin{align}
\big[E\hat{\tau}_3+\hat{\Delta}({\bf R}),
 \hat{\bf p}\cdot\underline{\tilde{\bf g}}_p({\bf R},T;p_F,E)\big]_\otimes+iv_F\hat{{\bf p}}\cdot
\partial_{\bf R}\underline{\tilde{g}}_s({\bf
R},T;p_F,E)+\frac{i}{2\tau}\big[\underline{\tilde{g}}_s({\bf R},T;p_F,E),
\hat{\bf p}\cdot\underline{\tilde{\bf g}}_p({\bf R},T;p_F,E)\big]_\otimes=0.
\label{term_odd}
\end{align}

With further considering the normalization condition \eqref{condition}, we obtain 
\begin{subequations}
\begin{align}
\label{one_norm}
&\underline{\tilde{g}}_s({\bf R},T;p_F,E) \otimes \underline{\tilde{g}}_s({\bf
  R},T;p_F,E)=1,\\
&\underline{\tilde{g}}_s({\bf
  R},T;p_F,E) \otimes \underline{\tilde{g}}_{p,\eta}({\bf R},T;p_F,E)
+\underline{\tilde{g}}_{p,\eta}({\bf R},T;p_F,E) \otimes \underline{\tilde{g}}_s({\bf
  R},T;p_F,E)=0,
\label{relation_0}
\end{align}
\end{subequations}
where $\eta=\{x,y,z\}$. Then with Eqs.~\eqref{one_norm} and \eqref{relation_0}, disregarding the first term in Eq.~\eqref{term_odd}, we find 
\begin{equation}
v_F\tau\frac{\partial}{\partial {R_{\eta}}}\underline{\tilde{g}}_s({\bf
  R},T;p_F,E)+\underline{\tilde{g}}_s({\bf
  R},T;p_F,E) \otimes \underline{\tilde{g}}_{p,\eta}({\bf R},T;p_F,E)=0.
\label{relation2}
\end{equation}
We multiply $\underline{\tilde{g}}_s({\bf R},T;p_F,E)$ on both sides of Eq.~\eqref{relation2}, substitute it into Eq.~\eqref{term_even}, and find 
\begin{equation}
\big[E\hat{\tau}_3+\hat{\Delta}({\bf R}), \underline{\tilde{g}}_s({\bf R},T;p_F,E)\big]_\otimes-iv_F^2\tau\hat{{\bf p}}\cdot
\frac{\partial}{\partial {\bf R}}\big[\underline{\tilde{g}}_s({\bf
  R},T;p_F,E) \otimes \hat{\bf
  p}\cdot
\frac{\partial}{\partial {\bf R}}\underline{\tilde{g}}_s({\bf
  R},T;p_F,E)\big]=0.
\label{term_final}
\end{equation}
Finally, by taking the angle average over Eq.~\eqref{term_final}, we arrive at the Usadel equation~\cite{Usadel}
\begin{equation}
\big[E\hat{\tau}_3+\hat{\Delta}({\bf R}), \underline{\tilde{g}}_s({\bf
  R},T;p_F,E)\big]_\otimes
-\frac{iv_F^2\tau}{3}\frac{\partial}{\partial {\bf R}}\cdot\Big[\underline{\tilde{g}}_s({\bf
  R},T;p_F,E)
\frac{\partial}{\partial {\bf R}}\underline{\tilde{g}}_s({\bf R},T;p_F,E)\Big]_\otimes=0.
\label{Usadel}
\end{equation}

\subsubsection{Ginzburg-Landau equation}

From a self-consistent point of view, when the temperature is close to the superconducting transition temperature and the superconducting order parameter is small, the Gor'kov equation~\cite{Abrikosov,Gorkov} and the Eilenberger equation~\cite{Eilenberger,Eliashberg} should reduce to the Ginzburg-Landau equation~\cite{Ginzburg}, which describes the dynamics of the macroscopic superconducting order parameter.

Before specifically deriving the Ginzburg-Landau equation, we consider the equations obeyed by the components of the quasi-classical Green's function
$\underline{\tilde{g}}=\left(\begin{array}{ccc}
\tilde{g}_{R}&\tilde{g}_{K}\\
0&\tilde{g}_{A}
\end{array}\right)$. Here, we only consider the clean limit. Starting from the Eilenberger equation \eqref{Eilenberger}, we obtain the equation satisfied by the components of the quasi-classical Green's function
\begin{equation}
\big[E\tau_3-\tilde{\Delta}, \tilde{g}_{R,A,K}\big]+i{\bf
  v}_F\cdot\partial_{\bf R}\tilde{g}_{R,A,K}=0.
\label{component_equation}
\end{equation}
In addition, the normalization condition \eqref{condition} provides another set of equations satisfied by the components of the quasi-classical Green's function:
\begin{align}
&\tilde{g}_{R,A}\tilde{g}_{R,A}=I_{2\times 2},\nonumber\\
&\tilde{g}_{R}\tilde{g}_{K}+\tilde{g}_{K}\tilde{g}_{A}=0.
\label{component_condition}
\end{align}

The components $\tilde{g}_{R,A,K}$ of the quasi-classical Green's function can be connected by the distribution function
$\tilde{h}$~\cite{Rammer,Kopnin,Chandrasekhar}, which is a $2\times 2$ matrix, i.e.,
\begin{equation}
\tilde{g}_{K}=\tilde{g}_{R}\tilde{h}-\tilde{h}\tilde{g}_{A}.
\label{arbitrary} 
\end{equation}
Equation~(\ref{arbitrary}) automatically satisfies the second of Eqs.~(\ref{component_condition}).  Such a distribution function $\tilde{h}$ can always be set to be diagonal~\cite{Rammer,Kopnin,Chandrasekhar}, as briefly demonstrated below. Making use of the first of Eq.~(\ref{component_condition}) it is easily seen that the transformation $\tilde{h}\rightarrow
\tilde{h}+\tilde{h}'=\tilde{h}+\tilde{g}_{R}\tilde{x}+\tilde{x}\tilde{g}_{A}$, where $\tilde{x}$ is an arbitrary matrix function, leaves the expression  \eqref{arbitrary} unchanged. 
In this way, we can always choose a special $\tilde{x}$ to set $\tilde{h}$ to be diagonal.
In particular, in the equilibrium state, $\tilde{h}$ is proportional to the identity matrix~\cite{Rammer,Kopnin,Chandrasekhar}
\begin{equation}
\tilde{h}_e=1-2f_0(\beta E)=\tanh(\beta E/2),
\label{imaginary} 
\end{equation}
where $f_0(x)\equiv 1/(e^x+1)$ is the Fermi-Dirac distribution function. 
Indeed, $\tilde{h}$  has the physical implication of a distribution function. In the non-equilibrium situation, $\tilde{h}$ 
can be solved by the kinetic equations.
In addition, by definition, the advanced and retarded quasi-classical Green's functions satisfy the relation
$\tilde{g}_{A}=-\tau_3\tilde{g}_{R}\tau_3$~\cite{Rammer,Kopnin,Chandrasekhar}.
Thereby, starting from $\tilde{g}_{R}$ and the distribution function $\tilde{h}$, we can find $\tilde{g}_{K}$ via \eqref{arbitrary}. With $\tilde{g}_{K}$, we can solve the superconducting order parameter~\cite{Rammer,Kopnin,Chandrasekhar}
\begin{equation}
\Delta({\bf R})=\frac{1}{8}|\lambda|
N_0\int\mbox{Tr}\big[(\tau_1-i\tau_2)\tilde{g}_{K}({\bf R};\hat{\bf
  p},E)\big]\frac{d\Omega_{\bf p}}{4\pi}\frac{dE}{2\pi},
\label{gap_equation}
\end{equation}
where $N_0$ is the density of states of the electrons at the Fermi surface in the normal state, and $\tau_1$ and $\tau_2$ are the $\hat{\bf x}$ and $\hat{\bf y}$ components of the Pauli matrix.

Near the superconducting transition temperature, the superconducting order parameter is small. The components of the quasi-classical Green's function can be expanded by the superconducting order parameter and its spatial gradients~\cite{Schmid,Shelankov,Shelankov1,Eckern,Kopnin} as
\begin{align}
\tilde{g}_{R,A,K}=\tilde{g}_{R,A,K}^{(0)}+\sum_{n=1}^{\infty}\tilde{g}_{R,A,K}^{(n)},
\end{align}
where $\tilde{g}_{R,A,K}^{(0)}$ is determined by its equilibrium and spatially homogeneous superconducting state and, when $n\ge
1$, $\tilde{g}_{R,A,K}^{(n)}\propto \Delta^n$ and contains spatial derivatives of the superconducting order parameter $\Delta$.

We first solve the quasi-classical Green's function $\tilde{g}_{R}^{(0)}$~\cite{Rammer,Kopnin,Chandrasekhar}. Neglecting spatial gradients of the superconducting order parameter from Eq.~\eqref{component_equation} and the normalization condition \eqref{condition}, $\tilde{g}^{(0)}_{R}$ obeys
\begin{subequations}
\begin{align}
&\big[E\tau_3-\tilde{\Delta}, \tilde{g}^{(0)}_{R}\big]=0,\\
&\tilde{g}^{(0)}_{R}\tilde{g}^{(0)}_{R}=I_{2\times 2},
\label{gR}
\end{align} 
\end{subequations}
from which we find $\tilde{g}_{R}^{(0)}=\frac{1}{\sqrt{E^2-|\Delta|^2}}\left(\begin{array}{ccc}
E&\Delta\\
-\Delta^*&-E \end{array}\right)$~\cite{Rammer,Kopnin,Chandrasekhar}. 
Accordingly, from Eqs.~\eqref{arbitrary} and \eqref{imaginary}, we obtain  
\begin{equation}
\tilde{g}_K^{(0)}=2\tanh\Big(\frac{\beta E}{2}\Big)\frac{1}{\sqrt{E^2-|\Delta|^2}}\left(\begin{array}{ccc}
E&\Delta\\
-\Delta^*&-E \end{array}\right).
\label{gK}
\end{equation}
Substituting Eq.~\eqref{gK} into the gap equation \eqref{gap_equation} yields  
\begin{equation}
1=N_0\frac{\lambda}{2}\int_{-\infty}^{\infty} dE\tanh\Big(\frac{\beta E}{2}\Big)\frac{1}{\sqrt{(E+i0^+)^2-\Delta^2}}=\frac{2\pi}{\beta}N_0|\lambda|\sum_{n=0}^{\infty}\frac{1}{\sqrt{\omega_n^2+\Delta^2}},
\label{positive_negative}
\end{equation}
where $\beta=1/(k_BT)$ and $\omega_n=(2n+1)\pi/\beta$ is the Matsubara frequency.
It is crucial to note that the summation here cannot be taken to infinity since a natural truncated energy exists, i.e., the Debye frequency $\Omega_{\rm D}$.
We then define a truncated integer $M_0$, governed by $2\pi(M_0+1/2)/\beta=k_B\Omega_{\rm D}$. Thus, $M_0(T)\approx
\Omega_D/(2\pi T)$ is dependent on the temperature. In this way, we can derive the critical temperature $T_c$ for superconductivity based on the gap equation \eqref{positive_negative}. At $T_c$, $\Delta=0$; thus by Eq.~\eqref{positive_negative} we obtain  
\begin{equation}
\frac{1}{N_0|\lambda|}=\sum_{n=0}^{M_0={\Omega_D}/({2\pi T_c})}\frac{2}{2n+1}\approx \gamma+\ln\big(\frac{\Omega_D}{2\pi T_c}\big)+2\ln2,
\label{summation}
\end{equation}
where $\gamma=\gamma_{k\rightarrow
\infty}\equiv 1+1/2+1/3+...+1/k-\ln k$ is the Euler constant~\cite{Math_gamma}.
According to Eq.~\eqref{summation}, the superconducting transition temperature $T_c=(2\Omega_D/\pi)e^{\gamma}e^{-1/(N_0|\lambda|)}$ ~\cite{Leggett_book}.

We proceed to derive $\tilde{g}_{R,A,K}^{(n\ge 1)}$, taking into account the diffusion of the fluctuation. 
We first derive the first-order correction $\tilde{g}_{R,A,K}^{(1)}$. 
Expressing the retarded Green's function  
$\tilde{g}_{R}=\tilde{g}_{R}^{(0)}+\tilde{g}_{R}^{(1)}$, from Eqs.~\eqref{component_equation}-\eqref{component_condition}, we obtain 
\begin{subequations}
\begin{align}
\label{how_tedious}
&\big[E\tau_3-\tilde{\Delta},\tilde{g}_{R}^{(1)}\big]=-i{\bf
  v}_F\cdot\partial_{\bf R}\tilde{g}_{R}^{(0)},\\
&\tilde{g}_{R}^{(0)}\tilde{g}_{R}^{(1)}+\tilde{g}_{R}^{(1)}\tilde{g}_{R}^{(0)}=0,\\
&\tilde{g}_{R}^{(0)}\tilde{g}_{R}^{(0)}=1.
\label{tedious}
\end{align}
\end{subequations}
The components in  $\tilde{g}_{R}^{(1)}=\left(\begin{array}{ccc}
g_{R,11}^{(1)}&g_{R,12}^{(1)}\\
g_{R,21}^{(1)}&g_{R,22}^{(1)}
\end{array}\right)$ can be solved from Eqs.~\eqref{how_tedious}-\eqref{tedious}. Specifically, 
\begin{equation}
g_{R,11}^{(1)}=\frac{\left(i{\bf v}_F\cdot\partial_{\bf
    R}g_{R,12}^{(0)}\right)g_{R,21}^{(0)}
-\left(i{\bf v}_F\cdot\partial_{\bf
  R}g_{R,21}^{(0)}\right)g_{R,11}^{(0)}}{4Eg_{R,21}^{(0)}+2\Delta
g_{R,12}^{(0)}-2\Delta^* g_{R,12}^{(0)}}.
\label{although_tedious}
\end{equation}
When $|\Delta|\ll |E|$, 
$g_{R,11}^{(1)}\approx i{\bf v}_F\cdot (\Delta^*\partial_{\bf
  R}\Delta-\Delta\partial_{\bf R}\Delta^*)/(4E^3)$. $g_{R,22}^{(1)}$
has a similar structure. So, $g_{R,11}^{(1)}$ and $g_{R,22}^{(1)}$ are proportional to the square of $\Delta$, which can be ignored in the linear expansion in terms of $\partial_{\bf
  R}\Delta$.
Thereby, solving $\tilde{g}_{R}^{(1)}$ only needs to retain the off-diagonal terms in the matrix. 
Accordingly, from Eqs.~\eqref{how_tedious}-\eqref{tedious}, we find 
\begin{subequations}
\begin{align}
&g_{R,12}^{(1)}\approx -i/(2E^2){\bf v}_F\cdot \partial_{\bf
  R}\Delta,\\
&g_{R,21}^{(1)}\approx -i/(2E^2){\bf v}_F\cdot \partial_{\bf
  R}\Delta^*.
\end{align}
\end{subequations}

The above analysis yields a straightforward iterative relation for the off-diagonal terms when $n\ge 2$, i.e.,
\begin{equation}
2E g_{R,12}^{(n)}=-i{\bf v}_F\cdot\partial_{\bf R}g_{R,12}^{(n-1)}.
\end{equation}
When $n=2$, 
\begin{equation}
g_{R,12}\approx
g_{R,12}^{(0)}+g_{R,12}^{(1)}+g_{R,12}^{(2)}=\frac{\Delta}{E}+\frac{|\Delta|^2\Delta}{2E^3}-\frac{i}{2E^2}{\bf
v}_F\cdot \frac{\partial\Delta}{\partial {\bf R}}-\frac{1}{4E^3}{\bf v}_F\cdot\frac{\partial}{\partial {\bf
R}}\Big({\bf v}_F\cdot\frac{\partial\Delta}{\partial {\bf R}}\Big).
\end{equation}
On the other hand, we take the equilibrium distribution of $\tilde{h}$. Thus, according to Eqs.~\eqref{arbitrary} and \eqref{imaginary}, we obtain  
\begin{equation}
g_{K,12}=2\tanh\Big(\frac{\beta E}{2}\Big)\Big[\frac{\Delta}{E}+\frac{|\Delta|^2\Delta}{2E^3}-\frac{i}{2E^2}{\bf
v}_F\cdot \frac{\partial \Delta}{\partial {\bf R}}-\frac{1}{4E^3}{\bf v}_F\cdot\frac{\partial}{\partial {\bf
R}}\Big({\bf v}_F\cdot\frac{\partial \Delta}{\partial {\bf R}}\Big)\Big].
\label{two_order}
\end{equation}

Substituting Eq.~\eqref{two_order} into the gap equation \eqref{gap_equation} yields
\begin{equation}
\Delta=\frac{2\pi
  N_0|\lambda|}{\beta}\sum\limits_{n=0}^{\infty}\Big(\frac{\Delta}{\omega_n}
-\frac{|\Delta|^2\Delta}{2\omega_n^3}+\frac{v_F^2}{12\omega_n^3}\frac{\partial^2 \Delta}{\partial {\bf
  R}^2}\Big).
\label{sum_sum}
\end{equation}
Here, two types of summation should be performed, noting that the truncation of the Debye frequency is needed in the summation over the Matsubara frequency [see Eq.~\eqref{summation}]. For the first summation 
\begin{equation}
\frac{2\pi
  N_0|\lambda|}{\beta}\sum\limits_{n=0}^{\infty}\frac{\Delta}{\omega_n}=\frac{2\pi
  N_0|\lambda|}{\beta}\sum\limits_{n=0}^{M_0(T_c)}\frac{\Delta}{\omega_n}+\frac{2\pi
  N_0|\lambda|}{\beta}\sum\limits_{M_0(T_c)}^{M_0(T)}\frac{\Delta}{\omega_n}=\Delta+\Delta N_0|\lambda| \ln\Big(\frac{T_c}{T}\Big);
\end{equation}
for the second and third summations,   
\begin{equation}
\sum\limits_{n=0}^{\infty}\frac{1}{\omega_n^3}=\frac{1}{8\pi^3T^3}\sum\limits_{n=0}^{\infty}\frac{1}{(n+1/2)^2}=\frac{7}{8\pi^3T^3}\zeta(3),
\end{equation}
in which the formula $\sum\limits_{n=0}^{\infty}{1}/{(n+1/2)^z}=(2^z-1)\zeta(z)$ is used~\cite{Stone}. Here $\zeta(z)$ is the Riemann function~\cite{Stone}. 
Finally, Eq.~\eqref{sum_sum} is written as 
\begin{equation}
\frac{\pi}{8T_c}\frac{7\zeta(3)v_F^2}{6\pi^3T_c}\frac{\partial^2 \Delta}{\partial {\bf R}^2}
+\ln\Big(\frac{T_c}{T}\Big)\Delta-\frac{7\zeta(3)}{8\pi^2T^2}|\Delta|^2\Delta=0,
\end{equation}
which is the Ginzburg-Landau equation for the conventional superconductivity~\cite{Ginzburg}.

\subsubsection{Electrodynamics of superconductors}
\label{two_fluid_model}

The general electrodynamics of superconductors relies on the frequency of the electromagnetic fields~\cite{33_cavity_S_FI_coupling_Jacobsen,two_fluid1,two_fluid2,two_fluid3,two_fluid4,two_fluid5,two_fluid6,two_fluid7,two_fluid8,two_fluid9,two_fluid10,two_fluid11,two_fluid12,two_fluid13,two_fluid14,two_fluid15,two_fluid16}, which goes beyond the scope of the present review article. Since we focus on the interaction between the magnons and superconductor electrons, the frequency of the associated electromagnetic field is sufficiently low $\lesssim 100$~GHz, such that we can exploit the two-fluid model~\cite{33_cavity_S_FI_coupling_Jacobsen,two_fluid1,two_fluid2,two_fluid3,two_fluid4,two_fluid5,two_fluid6,two_fluid7} to describe the electrodynamics of superconductors.

The electrodynamics of the superconductors based on the two-fluid model is developed by London \textit{et al.}~\cite{London_eq}. 
They considered that below the superconducting transition temperature $T_c$, there are two types of current carriers inside the superconductors, i.e., the normal current and the supercurrent.  
The normal fluid with density $\rho_n$ moves with the velocity ${\bf v}_n$ that can be scattered by impurities and phonons, while the superfluid with density $\rho_s$ moves with the velocity ${\bf v}_s$ is free of scattering. 
Here, $\rho_s+\rho_n$ equals the net electron density $\rho_e$.        
 For the electric field with the frequency $\omega$ oscillating in the material, it drives both the normal current and supercurrent, governed by
 \begin{align}
    &e{\bf E}-m_e\dfrac{{\bf v}_n}{\tau}= m_e\dfrac{\partial{\bf v}_n}{\partial
      t}=-i\omega m_e {\bf v}_n,\nonumber\\
    &e{\bf E}=m_e\dfrac{\partial{\bf v}_s}{\partial
      t}=-i\omega m_e {\bf v}_s,
      \label{accus_e}
 \end{align}
  where $m_e$ is the mass of the electrons and $\tau$ is the relaxation time for the momentum scattering.
   Thereby, the supercurrent and normal current read
 \begin{align}
     {\bf J}_s=\rho_s e{\bf v}_s=\dfrac{i\rho_s e^2}{m_e\omega}{\bf E},&&  {\bf J}_n=\rho_ne{\bf v}_n=\left[\dfrac{\rho_n e^2\tau}{m_e(1+\omega^2\tau^2)}+i\dfrac{\rho_n e^2 \tau}{m_e}\dfrac{\omega\tau}{1+\omega^2\tau^2}\right]{\bf E}.
     \label{jsjf}
 \end{align}
 The electric field generates both the normal current ${\bf J}_n$ and the supercurrent ${\bf J}_s$ with the net charge current ${\bf J}={\bf J}_n+{\bf J}_s= \tilde{\sigma}(\omega){\bf E}$ inside the superconductors with the conductivity~\cite{two_fluid_superconductor,two_fluid_Janssonn,two_fluid_Tao}
  \begin{align}
      \tilde{\sigma}(\omega)=\dfrac{\rho_ne^2\tau}{m_e}\dfrac{1}{1+\omega^2\tau^2}+i\left(\dfrac{\rho_ne^2\tau}{m_e}\dfrac{\omega\tau}{1+\omega^2\tau^2} +\dfrac{\rho_se^2}{m_e}\dfrac{1}{\omega}\right).
\end{align}   
The superfluid density $\rho_s$ and the normal fluid density $\rho_n$ governs the magnitude of both currents, which is strongly affected by the temperature $T$ via $\rho_s(T)=\rho_e[1-(T/T_c)^4]$ and $\rho_n(T)=\rho_e(T/T_c)^4$ when $T<T_c$. 
Since $\omega\tau\ll1$ for the microwave frequencies $\omega\lesssim 100$~GHz, we generally have   
\begin{align}
    \tilde{\sigma}(\omega,T)\approx\dfrac{\rho_n(T)e^2\tau}{m_e}+i\dfrac{\rho_s(T)e^2}{m_e}\dfrac{1}{\omega}=\sigma_n(T)+i\frac{1}{\omega\mu_0\lambda^2_L(T)},
    \label{conductivity}
\end{align}
where $\sigma_n(T)={\rho_n(T)e^2\tau}/{m_e}$ is the conductivity of normal metals and $\lambda_L(T)=\sqrt{m_e/(\mu_0\rho_s(T) e^2)}$ is London's penetration depth. 
Thereby, inside the superconductors, the total current ${\bf J}={\bf J}_s+{\bf J}_n =\tilde{\sigma} {\bf E}$ when $T<T_c$; otherwise, the superconductor recovers to the normal metal when $T>T_c$, such that $\tilde{\sigma}=\sigma_n$.

On the other hand, inside the superconductors, Maxwell's equation reads~\cite{Swihart_mode}
\begin{subequations}
    \begin{align}
\nabla\times {\bf E}({\bf r},t)&=-\mu_0\dfrac{\partial{\bf  H}({\bf r},t)}{\partial t},&& \nabla\cdot \mu_0{\bf H}({\bf r},t)=0,   \label{maxwell_equations_superconductor1} \\ 
\nabla\cdot {\bf E}({\bf r},t)&=0 ,&&\nabla\times {\bf H}({\bf r},t)={\bf J}({\bf r},t)+\varepsilon_0 \dfrac{\partial{\bf  E}({\bf r},t)}{\partial t},
    \label{maxwell_equations_superconductor2}
\end{align}
\end{subequations}
in which we assume the vacuum permeability $\mu_0$ for the superconductor. 
Taking the curl of the first equation in Eq.~(\ref{maxwell_equations_superconductor1}) and substituting Eq.~\eqref{maxwell_equations_superconductor2} into it, we obtain the equation of motion of the electric field ${\bf E}$ inside the superconductors
\begin{align}
    \nabla^2{\bf E}+k_S^2{\bf E}=0,
    \label{electric_field_SC}
\end{align}
 in which we use the relation ${\bf J}=\tilde{\sigma}{\bf 
 E}$. 
 The wave vector $k_S=\sqrt{\omega^2\mu_0\varepsilon_0+i\omega\mu_0\tilde{\sigma}}$ is not a real number. 
 At low temperatures $T\ll T_c$, $\tilde{\sigma}\approx i/(\omega\mu_0  \lambda_L^2)$, such that $k_S\approx\sqrt{\omega^2\mu_0\varepsilon_0-1/{\lambda_L^2}}$. For the conventional superconductors, $\lambda_L\sim 100$~nm;   $1/\lambda_L^2\approx 10^{14}\gg\omega^2\mu_0\varepsilon_0\approx1.1\times10^5$~m$^{-2}$ with the low frequency $\omega\sim 100$~GHz, such that $k_S\approx \pm i/\lambda_L$ is purely imaginary. In this situation, the electric field has the solution ${\bf E}\propto ^{-x/\lambda_L}$ inside the superconductors, which decays with the attenuation length $\sim \lambda_L$. Similarly, taking the curl of the second equation in Eq.~\eqref{maxwell_equations_superconductor2} and substituting Eq.~\eqref{maxwell_equations_superconductor1} into it yields 
     \begin{align}
    \nabla^2{\bf H}+k_S^2{\bf H}=0.
\end{align}
The magnetic field satisfies the same equation as the electric field Eq.~\eqref{electric_field_SC}, which only exists in the superconductors that are close to the surface with the characteristic penetration depth $\lambda_L$, showing the Meissner effect.

\section{Dynamic dipolar interaction between a ferromagnet and a superconductor}
\label{Dynamic_dipolar_interaction_between_a_ferromagnet_and_a_superconductor}

\subsection{General theory}
 \label{general_theory}

\begin{figure}[htp]
     \centering
     \includegraphics[width=\linewidth]{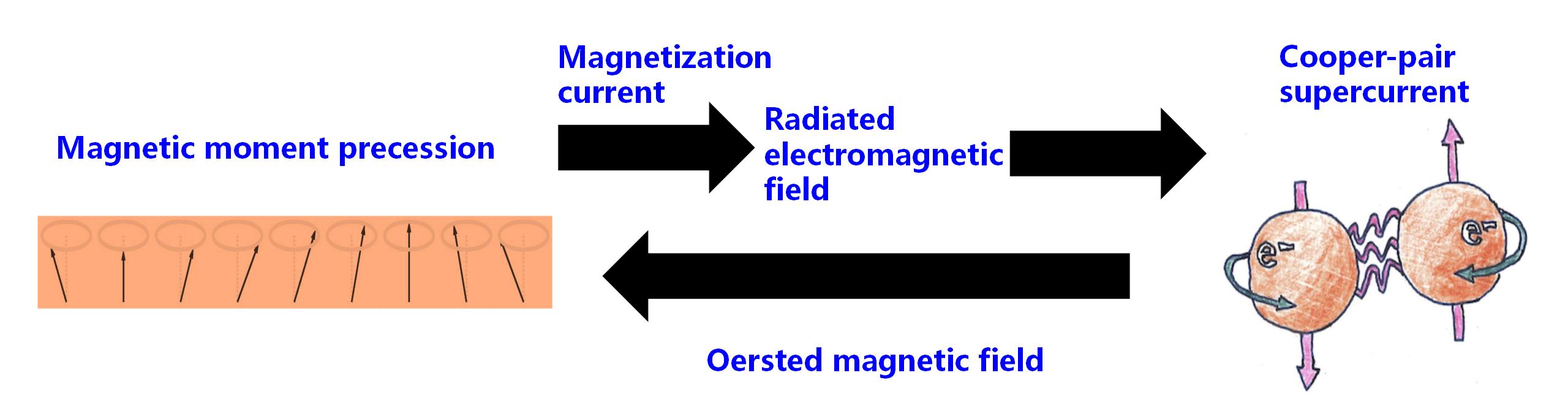}
     \caption{Coupling between spin waves and supercurrents via the electromagnetic proximity effect.}
     \label{electromagnetic_proximity_effect}
 \end{figure}

In superconductor$|$ferromagnet heterostructures, the dipolar field emitted by the magnetization ${\bf M}({\bf r},t)$ penetrates the superconductor and drives a diamagnetic Meissner supercurrent that in turn generates an Oersted magnetic field that couples to the magnetization dynamics. Here, we formulate a self-consistent treatment of the magnetization dynamics of a ferromagnet (FM) and the order parameter of a superconductor (SC) coupled by their mutual magneto-dipolar interaction, which we refer to as the electromagnetic proximity effect. 
 In the literature, the electromagnetic proximity effect is used to describe the screening response of the superconductor to a
 vector potential at (or near) the SC$|$FM interface~\cite{PhysRevB.99.104519,BESPALOV20221354032,10.1063/1.5037074,PhysRevB.100.020505}. We focus on the proximity effect mediated by electromagnetic radiation. 
Figure~\ref{electromagnetic_proximity_effect} illustrates the microscopic processes.  

\begin{figure}[htp!]
    \centering
    \includegraphics[width=0.6\linewidth]{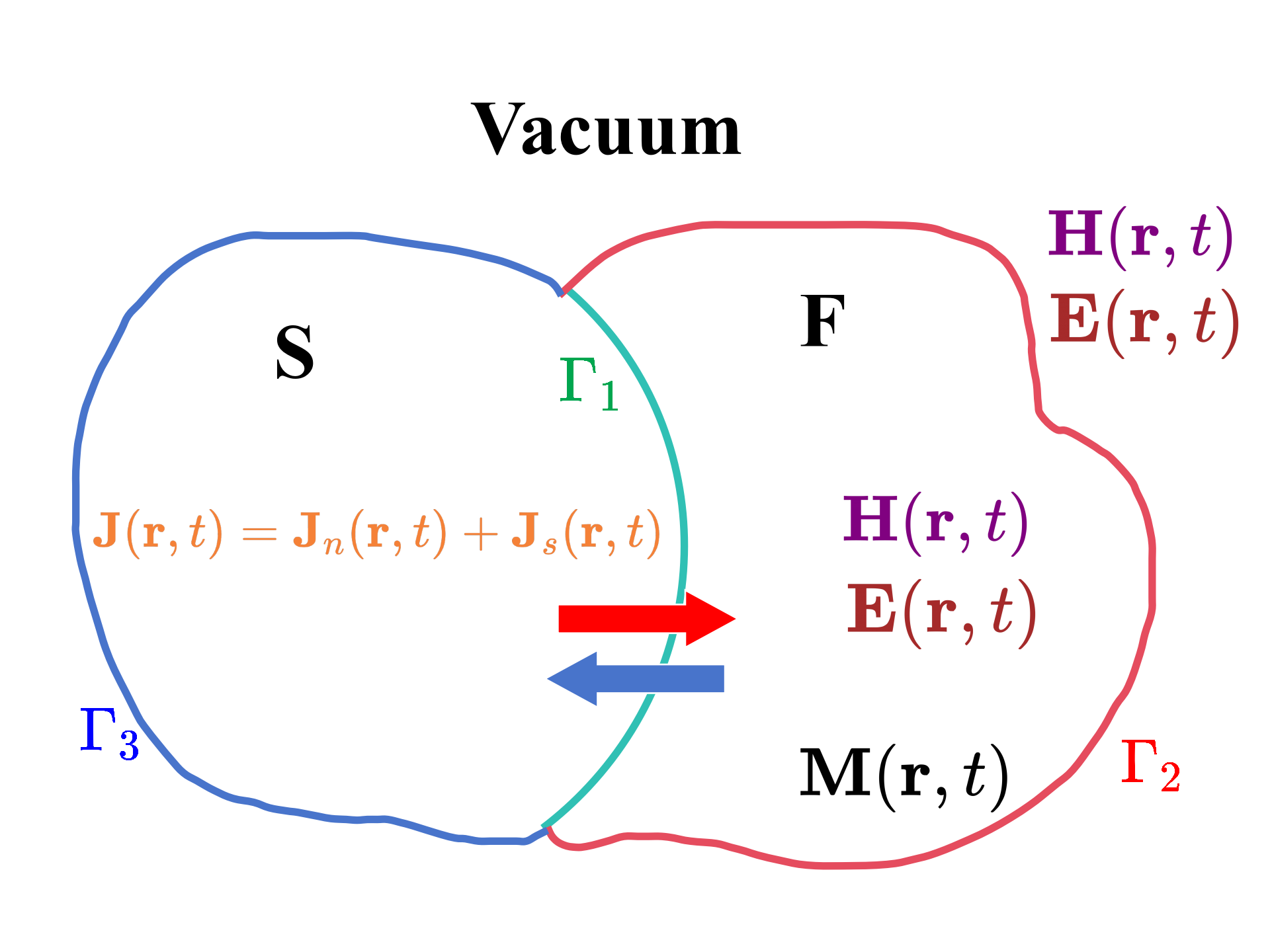}
    \caption{A superconductor$|$ferromagnet heterostructure contains three bulk regions (S, F, Vacuum) separated by three boundaries ($\Gamma_1, \Gamma_2, \Gamma_3$).  Inside the ferromagnet, the magnetization ${\bf M}\neq0$, and inside the superconductor, the total current ${\bf J}({\bf r},t)$ includes a normal ${\bf J}_n({\bf r},t)$ and a supercurrent ${\bf J}_s({\bf r},t)$. Maxwell's equations, with proper boundary conditions, govern the electromagnetic field throughout all space. }
    \label{SF_interaction}
\end{figure}

In Fig.~\ref{SF_interaction}, we divide the system into a vacuum, superconductor, and ferromagnet, and three boundaries, i.e., the superconductor$|$ferromagnet boundary $\Gamma_1$, the superconductor$|$vacuum boundary $\Gamma_2$, and the ferromagnet$|$vacuum boundary $\Gamma_3$. 
The dipolar interaction arises from the electromagnetic field that pervades all space and obeys Maxwell's equations. 
In the absence of free charges in the vacuum and electric insulators, the divergence of the electric field vanishes  $\nabla\cdot {\bf E}({\bf r},t)=0$.
In good metals and not too high frequencies, the free charge density $\rho_f$ obeys the charge conservation relation
\begin{align}
 \dfrac{\partial\rho_f}{\partial t}=-\nabla\cdot {\bf J},
\end{align}
while according to Ohm's law
\begin{align}
\nabla\cdot {\bf J}=\sigma\nabla\cdot {\bf E}=\dfrac{\sigma}{\varepsilon_r }\rho_f,
\end{align}
where \(\sigma\) is the conductivity and \(\varepsilon_r\) the dielectric constant. Hence $\rho_f(t)=\rho_0e^{-(\sigma/\varepsilon) t}$, where $\rho_0$ is the free charge density at $t=0$, so a time dependent perturbation $\rho_f(t)$ decays within the relaxation time $\tau=\varepsilon/\sigma$. 
In the classical conductor Cu $\tau\sim10^{-19}$~s, so fluctuations in the free charge vanish almost instantaneously. At microwave frequencies $\omega\lesssim 100$~GHz  $\omega\tau\ll1$ and $\nabla\cdot {\bf E}({\bf r},t)=0$ still holds except very close to boundaries or interfaces. Maxwell's equations in normal or magnetic metals then read
\begin{subequations}
    \begin{align}
\nabla\cdot {\bf E}({\bf r},t)&=0,&& \nabla\cdot \mu_0[{\bf H}({\bf r},t)+{\bf M}({\bf r},t)]=0,   \label{maxwell_equation_1} \\ 
\nabla\times {\bf E}({\bf r},t)&=-\dfrac{\partial{\bf  B}({\bf r},t)}{\partial t},&&\nabla\times {\bf H}({\bf r},t)={\bf J}({\bf r},t)+\varepsilon_r \dfrac{\partial{\bf  E}({\bf r},t)}{\partial t},
    \label{maxwell_equation_2}
\end{align}
\end{subequations}
where $\varepsilon_r=\varepsilon_{F}$ in a ferromagnet, while inside the superconductors or vacuum $\varepsilon_r=\varepsilon_0$ is the vacuum permittivity.

Inside the ferromagnet, the magnetization ${\bf M}$ and the magnetic field ${\bf H}$ contributes to the magnetic induction ${\bf B}=\mu_0 ({\bf H}+{\bf M})$.  Ohm's law reads ${\bf J}=\sigma_F {\bf E}$, with the conductivity $\sigma_F\ne 0$ for a metal and $\sigma_F=0$ a ferromagnetic insulators.
The induced currents generate electromagnetic fields that should be considered together with the driving field in a self-consistent manner. According to the two-fluid model (Sec.~\ref{two_fluid_model}), an electric field in a superconductor generates  ${\bf J}={\bf J}_n+{\bf J}_s$ with normal ${\bf J}_n$ and supercurrent ${\bf J}_s$ contributions.
Taking the curl of Eq.~(\ref{maxwell_equation_2}) leads to wave equations for the electric ${\bf E}$ and magnetic field ${\bf H}$ of frequency $\omega$, as summarized in Table~\ref{Equations_of_E_and_H}. The wave vector $k_0=\sqrt{\omega^2\mu_0\varepsilon_0}=\omega/c$ is governed by the light velocity $c=1/\sqrt{\mu_0\varepsilon_0}$, while those in ferromagnets and superconductors are affected by the material parameters. 

\begin{table}[htp!]
\centering
\renewcommand{\arraystretch}{1.5}
\caption{The wave equations of electric and magnetic fields in ferromagnets (F), superconductors (S), and the vacuum, where $k_F=\sqrt{\omega^2\mu_0\varepsilon_F+i\omega\mu_0 \sigma_F}$, $k_S=\sqrt{\omega^2\mu_0\varepsilon_0+i\omega\mu_0\tilde{\sigma}}$, and $k_0=\sqrt{\omega^2\mu_0\varepsilon_0}$.}
\begin{tabular*}{\textwidth}{@{\extracolsep{\fill}} |c|cc| }
	\hline
	& Electric field& magnetic field\\
	\hline
	in F & $\nabla^2 {\bf E}+k_F^2{\bf E}=-i\omega\mu_0 \nabla\times {\bf M}$ & $\nabla^2 {\bf H}+k_F^2{\bf H}=-\nabla(\nabla\cdot {\bf M})-k_F^2 {\bf M}$ \\
	\hline
	in S &$\nabla^2 {\bf E}+k_S^2{\bf E}=0$ & $\nabla^2 {\bf H}+k_S^2{\bf H}=0$\\
	\hline
	in Vacuum & $\nabla^2 {\bf E}+k_0^2{\bf E}=0$ & $\nabla^2 {\bf H}+k_0^2{\bf H}=0$ \\
	\hline
\end{tabular*}
\label{Equations_of_E_and_H}
\end{table}

According to Eq.~(\ref{maxwell_equation_2}), ${\bf E}_\parallel$ is continuous across planar interfaces. In the absence of surface currents, ${\bf H}_\parallel$ and since $\nabla\cdot {\bf B}=0$, ${\bf B}_\perp=\mu_0({\bf H_\perp+M_\perp})$ are also continuous at the boundaries.  Taking the divergence of the second equation in Eq.~(\ref{maxwell_equation_2}) $\nabla\cdot [{\bf J}({\bf r},t)+\varepsilon_r \partial_t{\bf E}({\bf r},t)]=0$ implies that the sum of the electric current ${\bf J}_\perp$ and the displacement current $\varepsilon_r \partial_t {\bf E}_\perp$ are continuous as well. 
The perpendicular component of the current then reads
\begin{align}
    &\text{in vacuum}: ~~~~~~~~~~~~~~~~~~~~~~~~~~~~~~\varepsilon_r=\varepsilon_0,&&{\bf J}_\perp=0,\nonumber\\
    &\text{in a metallic ferromagnet}: ~~~~~~~~~\varepsilon_r=\varepsilon_F,&&{\bf J}_\perp=\sigma_F {\bf E}_\perp,\nonumber\\
        &\text{in an insulating ferromagnet}: ~~~~~~~\varepsilon_r=\varepsilon_F,&&{\bf J}_\perp=0,\nonumber\\
    &\text{in a superconductor}:~~~~~~~~~~~~~~~~~~\varepsilon_r=\varepsilon_0,&& {\bf J}_\perp={\bf J}_{n\perp}+{\bf J}_{s\perp}=\tilde{\sigma} {\bf E}_\perp.
\end{align}
We argue above that the divergence of the electric field vanishes in the bulk of the vacuum and good metals, but not at boundaries that accumulate charges. The condition that ${\bf E}_\parallel $ and $\partial_\parallel {\bf E}_\parallel$ are continuous at interfaces combined with $\nabla\cdot {\bf E}=0$ implies that $\partial_\perp {\bf E}_{\perp}$ should be continuous at the interfaces as well, thereby completing the boundary conditions listed in Table.~\ref{boundary}.

\begin{table}[htp!]
\centering
\renewcommand{\arraystretch}{1.5}
\caption{The boundary conditions at different boundaries $\Gamma_1$, $\Gamma_2$ and $\Gamma_3$.}
	\begin{tabular}{|c|ccc|}
		\hline
		& $\Gamma_1$: Superconductor|Ferromagnet& $\Gamma_2$: Ferromagnet|Vacuum& $\Gamma_3$: Superconductor|Vacuum\\ \hline
	$\bf H_\parallel$  & ${\bf H}_\parallel|_S={\bf H}_\parallel|_F$   &${\bf H}_\parallel|_F={\bf H}_\parallel|_V$   &   ${\bf H}_\parallel|_S={\bf H}_\parallel|_V$  \\ \hline
	${\bf H}_\perp$  & ${\bf H}_\perp|_S=({\bf H}_\perp+{\bf M}_\perp)|_F$  & $({\bf H}_\perp+{\bf M}_\perp)|_F={\bf H}_\perp|_V$   &  ${\bf H}_\perp|_S={\bf H}_\perp|_V$  \\ \hline
	${\bf E}_\parallel$  & ${\bf E}_\parallel|_S={\bf E}_\parallel|_F$   & ${\bf E}_\parallel|_F={\bf E}_\parallel|_V$ &   ${\bf E}_\parallel|_S={\bf E}_\parallel|_V$  \\ \hline
	\multirow{2}{*}{$\bf E_\perp$} & $({\bf J}_\perp+\varepsilon_0  {\partial_t{\bf  E}_\perp})|_S=({\bf J}_\perp+\varepsilon_F  {\partial_t{\bf  E}_\perp})|_F$   &  $({\bf J}_\perp+\varepsilon_F  {\partial_t{\bf  E}_\perp})|_S=({\bf J}_\perp+\varepsilon_0  {\partial_t{\bf  E}_\perp})|_F$  &    $({\bf J}_\perp+\varepsilon_0  {\partial_t{\bf  E}_\perp})|_S=({\bf J}_\perp+\varepsilon_0  {\partial_t{\bf  E}_\perp})|_V$ \\ \cline{2-4} 
		& $\partial_\perp{\bf E}_\perp|_S=\partial_\perp{\bf E}_\perp|_F$   &     $\partial_\perp{\bf E}_\perp|_F=\partial_\perp{\bf E}_\perp|_V$ &    $\partial_\perp{\bf E}_\perp|_S=\partial_\perp{\bf E}_\perp|_V$ \\ \hline
	\end{tabular}
\label{boundary}
\end{table}

The electromagnetic proximity effect originates from the electric and magnetic stray fields emitted by the magnetization or electric-polarization dynamics. In the following, we address a typical feature of this coupling, viz., an induced chirality.

\subsection{Chiral electric and magnetic stray fields}
\label{chiral_fields}

\subsubsection{Photonic spin-orbit coupling of stray fields}

\label{photonic_SOC}

The coupling of the electron spin with its motion arises from the relativistic Dirac equation when expanded to leading order in the electron velocity relative to that of light. In condensed matter systems, this SOC may be interpreted in terms of a Zeeman interaction $-\boldsymbol{\mu}\cdot {\bf B}_{\mathrm{eff}}$ of the velocity-dependent magnetic field \({\bf B}_{\mathrm{eff}}\)  felt by an electron's magnetic moment \(\boldsymbol{\mu}\) moving in an applied or intrinsic crystal electric field. For free electrons,   \({\bf B}_{\mathrm{eff}}\) is proportional and normal to both the electric field and wave vector \textbf{k}  and the interaction is referred to as Rashba SOC~\cite{PhysRevA.84.025602}. The energy splitting of TE and TM modes in wave guides can be interpreted as a ``photonic SOC" since it is proportional to a photonic pseudo-spin and its wave vector~\cite{khanikaev2013photonic}.

Here, we explain that the photonic SOC in electromagnetic fields is a corollary of a more general principle. To this end, we point out a ubiquitous property of wave propagation, viz., a geometrical constraint that governs the propagation direction of all vector fields.

\begin{figure}[htp!]
    \centering
    \includegraphics[width=\linewidth]{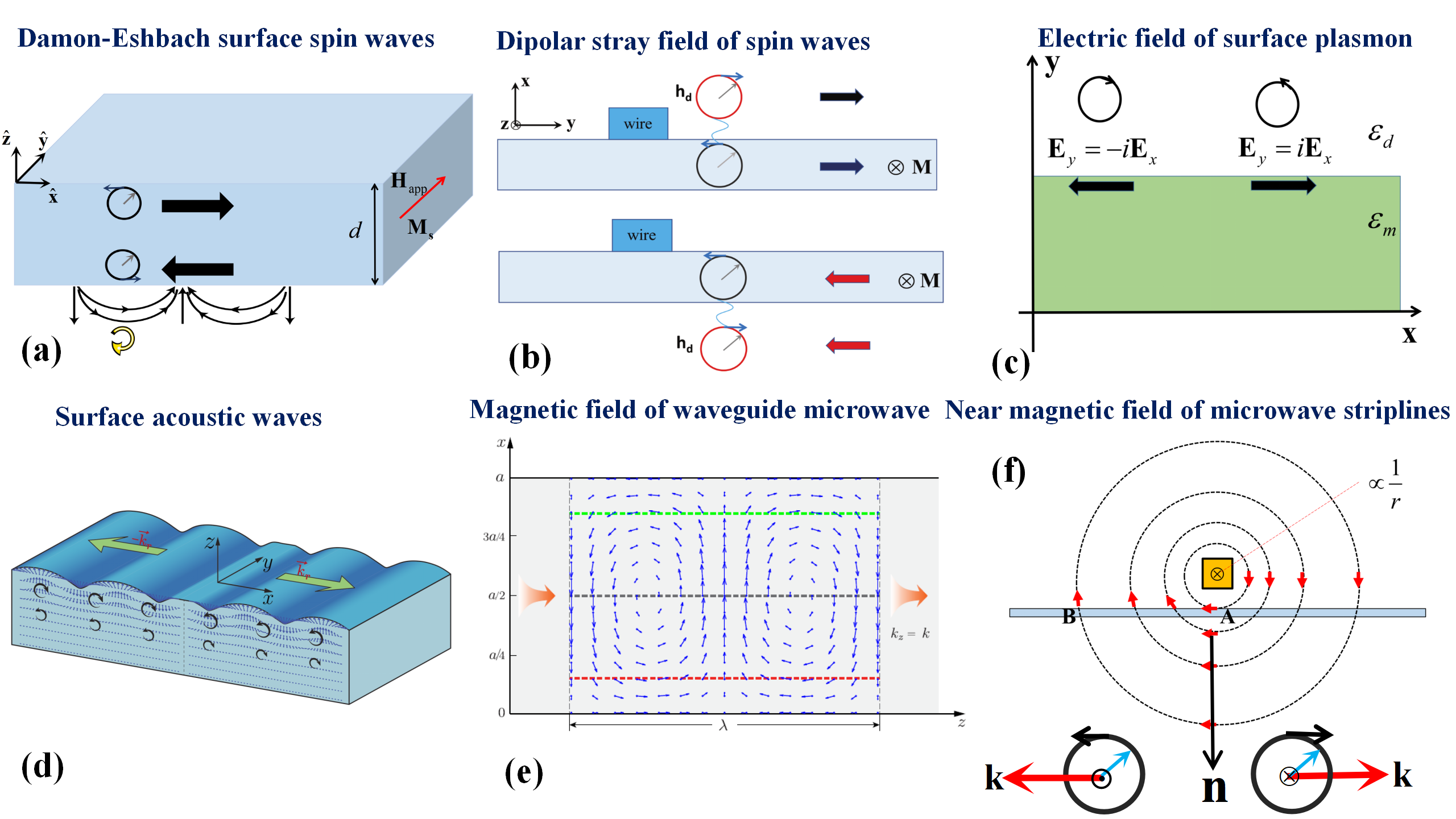}
    \caption{Chirality of classical waves of different nature and configurations.  (a) Damon-Eshbach surface spin waves; (b) the dipolar magnetic stray field of spin waves; (c) the electric field of surface plasmons; (d) surface acoustic waves; (e) the magnetic field of microwave guides;  (f)  near magnetic field of microwave striplines in a plane.\\
    \textit{Sources}: Fig.~(b) is taken from  Ref.~\cite{Tao_chiral_pumping}; Fig.~(c), (d), and (f) are taken from Ref.~\cite{Yu_chirality}; Fig.~(e) is taken from  Ref.~\cite{Tao_Magnon_Accumulation}.}
    \label{chiral_waves}
\end{figure}

For example,  the Damon-Eshbach spin waves at the surfaces of ferromagnets with an in-plane magnetization sketched in Fig.~\ref{chiral_waves}(a) propagate in opposite directions at the upper and lower surfaces of the magnetic film~\cite{DEmode,DE2},
governed by the cross product of the surface normal direction {\bf n} and in-plane magnetization direction {\bf M}. In thin magnetic films, the surface waves merge with a bulk wave to form perpendicular standing spin waves that are not chiral anymore.  However, its stray magnetic fields still are. Figure~\ref{chiral_waves}(b) illustrates that the stray magnetic field emitted by spin waves in thin magnetic films that propagate to the right (left) direction normal to the magnetization exist only above (below) the film~\cite{Tao_chiral_pumping,Tao_chiral_excitation}. 

The unidirectionality is associated with the locking of angular momentum vectors, such as the electron spin or the polarization of vector waves, with the propagation direction. Figure~\ref{chiral_waves}(c) shows the electric field of a surface plasmon polariton at the interface between a metal and a dielectric.  The polarization of waves propagating in opposite directions must be opposite~\cite{SPP1,SPP2,SPP3,SPP4,SPP5}.
The atoms in surface acoustic waves in dielectrics~\cite{surface_acoustic_wave,surface_magnetoelastic,surface_acoustic_wave2} follow a circular trajectory normal to the interface and propagation direction as in Fig.~\ref{chiral_waves}(d) with a direction that is locked to its wave vector. The magnetic field component of microwaves at the boundaries of metallic waveguides [Fig.~\ref{chiral_waves}(e)] and the magnetic field of microwave striplines sampled in a nearby plane [Fig.~\ref{chiral_waves}(f)] also subject to chirality.
The locking of spin and wave vectors becomes evident in Fourier space, in which the polarization or spin of a Fourier component can be expressed by a hand rule spanned by the surface normal, the wave vector, and the angular momentum~\cite{Yu_Chiral_Coupling2021}.

\begin{figure}[htp!]
    \centering
    \includegraphics[width=0.99\linewidth]{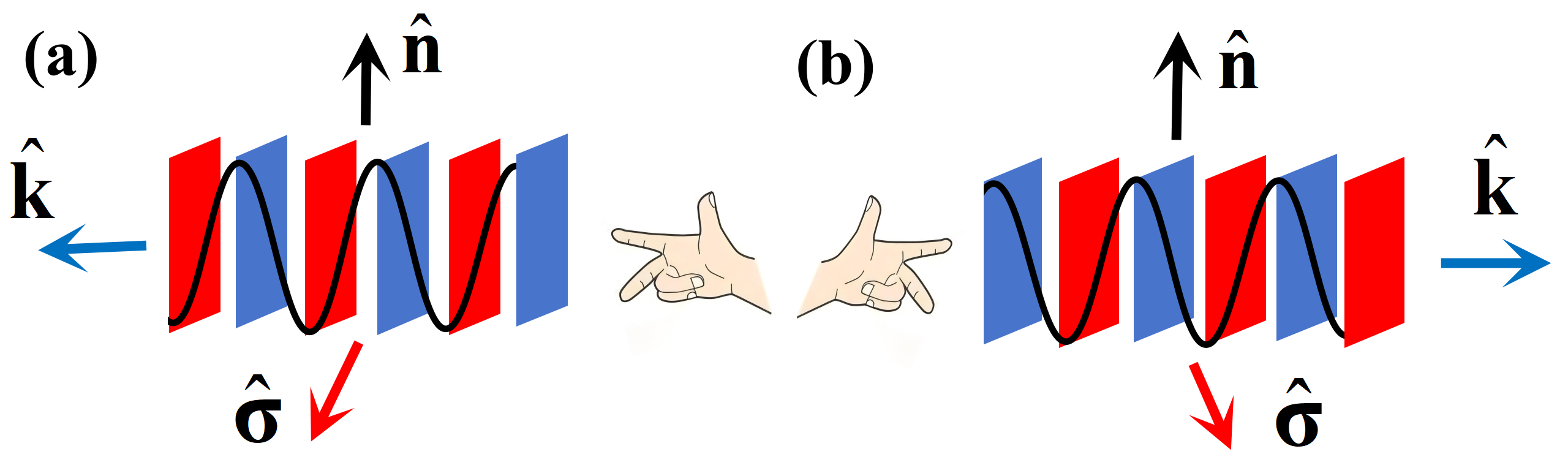}
    \caption{Right-handed classical waves [(a)] \textit{vs}. right-handed classical waves [(b)].}
    \label{chirality_index}
\end{figure}

In the examples above, chirality is perfect, but in general, we observe only a partial non-reciprocity with respect to reversing the propagation direction. It is therefore necessary to introduce a measure for a given situation that we introduce now, starting with the degree of chirality of plane classical waves that propagate in a plane, but are evanescent normal to it. 
We define three vectors shown in Fig.~\ref{chirality_index}. The wave vector {\bf k} is the propagation direction in the plane, {\bf n} is the surface normal such that \({\bf k} \cdot {\bf n} =0\). The third vector ${\pmb \sigma}$ characterize the ``spin", i.e., polarization or angular momentum, of the waves. For example, the spin density of an electromagnetic field with frequency $\Omega$  in vacuum reads~\cite{BLIOKH20151,Yu_chirality,spin_density}
\begin{align}
    {\pmb \sigma}=\frac{1}{4\Omega}{\rm Im}(\varepsilon_0{\bf E}^*\times{\bf E}+\mu_0{\bf H}^*\times{\bf H}),
\end{align}
where $\mu_0$ and $\varepsilon_0$ are, respectively, the vacuum permeability and permittivity. 
The chirality index 
 \begin{align}
     {\cal C}_k=\hat{\bf n}\cdot(\hat{\pmb \sigma}\times\hat{\bf k})
 \end{align}
measures the degree of locking between the propagation ${\bf k}$, the spin ${\pmb \sigma}$, and the surface-normal ${\bf n}$. The classical wave is ``right handed" when $0<{\cal C}_k\leq1$ and ``left handed" when $-1\leq {\cal C}_k<0$. 
Accordingly, a classical wave is perfectly chiral when the three vectors are perpendicular to each other and obey a right-hand or left-hand rule. A right-handed classical wave is locked as shown in Fig.~\ref{chirality_index}(a), {\bf n} is along the thumb direction, {\bf k} is along the index finger direction, and ${\pmb \sigma}$ is along the middle finger direction of the right hand, in which case, $C_k=1$. On the other hand, we use the left hand for a left-handed wave as in Fig.~\ref{chirality_index}(b). When fully chiral, $C_k=-1$. 

We now illustrate the surprising fact that an evanescent classical field \textbf{V} described by a vector with at least two components is always right-handed~\cite{PhysRevApplied.22.034042} at the hand of the generic form in the negative half-space ($x<0$) 
\begin{align}
    {\bf V}({\bf r},t)={\cal V}e^{i(k_yy+k_zz)+\sqrt{k_y^2+k_z^2}x}e^{-i\Omega t},
\end{align}
that propagates in the $y$-$z$ plane and is evanescent in the $\hat{\bf x}$-direction, where $k_y$ and $k_z$ are the respective wave vector components. We scale the field amplitudes as ${\cal\bf V}=(1,{\cal V}_ye^{i\phi_y},{\cal V}_ze^{i\phi_z})$, where ${\rm Im}{\cal V}_{y,z}=0$ and the phases $\{\phi_y,\phi_z\}\in[0,2\pi)$ are real.
Moreover, we assume that the vector field is source-free, which implies that its spin is normal to the wave vector. We show below that under these general conditions, the chirality index is larger than zero, which implies that \textbf{V} must be right-handed, irrespective of its physical nature.

In the absence of sources $\nabla\cdot {\bf V}({\bf r},t)=0$ and
\begin{align}
     &k_y{\cal V}_y\cos\phi_y+k_z{\cal V}_z\cos\phi_z=0,\notag\\
     &k_y{\cal V}_y\sin\phi_y+k_z{\cal V}_z\sin\phi_z=\sqrt{k_y^2+k_z^2}. 
     \label{nosource}
\end{align}
The components of the intrinsic angular momentum (or spin) density ${\bf S}({\bf r},t)\propto{\rm Im}[{\bf V}({\bf r},t)^*\times {\bf V}({\bf r},t)]$ read
\begin{align}
    &S_x(x,k_y,k_z)\propto-2{\cal V}_y{\cal V}_z\sin(\phi_y-\phi_z)e^{2\sqrt{k_y^2+k_z^2}x},\notag\\
    &S_y(x,k_y,k_z)\propto-2{\cal V}_z\sin\phi_ze^{2\sqrt{k_y^2+k_z^2}x},\notag\\
    &S_z(x,k_y,k_z)\propto2{\cal V}_y\sin\phi_ye^{2\sqrt{k_y^2+k_z^2}x}.
\end{align}
Since the field is transverse, spin and momentum are locked by
 \begin{align}
     {\bf k}\cdot{\bf S}\propto2(k_z{\cal V}_y\sin\phi_y-k_y{\cal V}_z\sin\phi_z)e^{2\sqrt{k_y^2+k_z^2}x}=0.\label{transfield}
 \end{align}
From Eqs.~(\ref{nosource}) and (\ref{transfield}), 
${\cal V}_ye^{i\phi_y}=k_ye^{i\phi_y}/(\sqrt{k_y^2+k_z^2}\sin\phi_y)$ and ${\cal V}_ze^{i\phi_z}=k_ze^{i\phi_z}/(\sqrt{k_y^2+k_z^2}\sin\phi_z)$.
The chirality index~\cite{PhysRevApplied.22.034042}
\begin{align}
    {\cal C}_k=-\hat{\bf x}\cdot(\hat{\bf S}\times\hat{\bf k})=\left[\frac{k_y^2k^2_z\sin^2(\phi_y-\phi_z)}{(k_y^2+k_z^2)^2\sin^2\phi_y\sin^2\phi_z}+1\right]^{-1/2}>0
    \label{chirality_index_eq}
\end{align}
and $0<{\cal C}_k\le 1$ shows that the transverse field in the film is ``right-handed", implying the presence of a geometric SOC. For a spin lies in the $y$-$z$ plane $(S_x=0)$,  ${\cal C}_k=1$ since the three vector $\{-\hat{\bf x},\hat{\bf S},\hat{\bf k}\}$ are orthogonal. For different physical systems, the precise value of chirality index ${\cal C}_k$ may differ, depending on the geometry under consideration.

\begin{figure}[htp!]
    \centering
    \includegraphics[width=1\linewidth]{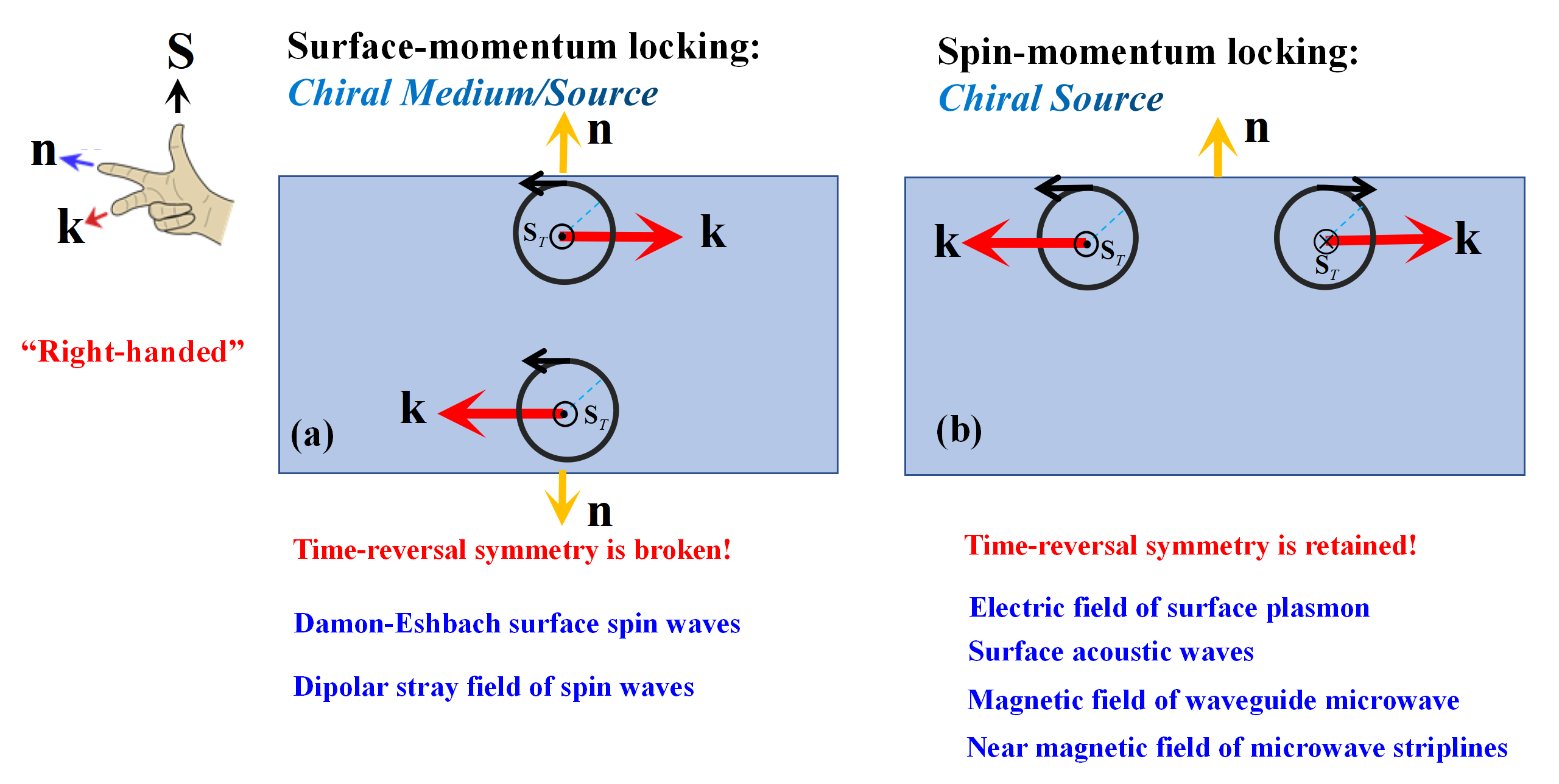}
    \caption{Two categories of wave propagation under locking relation between the wave vector {\bf k}, the spin {\bf S}, and the surface normal {\bf n} directions by the geometric SOC. (a) When the time-reversal symmetry is broken,  surface waves propagating normal to the spin are unidirectional.  (b) In a time-reversal symmetric system, the spin is locked to a momentum that can have either sign.}
    \label{generalized_SOC}
\end{figure}

Referrring to the chirality of electrons caused by the SOC of electrons in solids with broken inversion symmetry~\cite{SOC1,SOC2,SOC3,SOC4,SOC5}, the locking of the wave vector, spin, and surface normal directions described by the right-hand rule here can be interpreted as non-relativistic ``geometrical spin-orbit interaction". The associated constraints on wave propagation lead to two categories, depending on whether time-reversal symmetry is broken, as addressed in Fig.~\ref{generalized_SOC}.
When time-reversal symmetry is broken, the spin of all waves,  propagating and evanescent, is right-handed, e.g., for magnons in ferromagnets relative to the saturation magnetization. The upper surface waves in a film with wave vector normal to an in-plane magnetization can only propagate to the right direction; at the lower surface, the waves propagate to the left, as shown in Fig.~\ref{generalized_SOC}(a). This explains the chirality of the Damon-Eshbach surface spin waves~\cite{DEmode,DE2} and the dipolar stray field outside of the magnetic film emitted by the spin waves~\cite{Tao_chiral_pumping,Tao_chiral_excitation}. When the time-reversal symmetry is not broken, the spin can point in any direction, and on a given surface, all in-plane wave vectors are allowed. However, the spin is locked to the wave vector by the right-hand rule, as shown in Fig.~\ref{generalized_SOC}(b). This explains the propagation properties of several classical waves in nonmagnetic materials, including the electric field of surface plasmon~\cite{SPP1,SPP2,SPP3,SPP4,SPP5}, the surface acoustic waves~\cite{surface_acoustic_wave,surface_magnetoelastic,surface_acoustic_wave2}, the magnetic field of guided microwaves, and the near magnetic field of microwave striplines~\cite{Yu_Chiral_Coupling2021}.

\subsubsection{Chiral magnetic fields}

\label{chiral_magnetic_fields}

\begin{figure}[htp]
\includegraphics[width=0.95\linewidth]{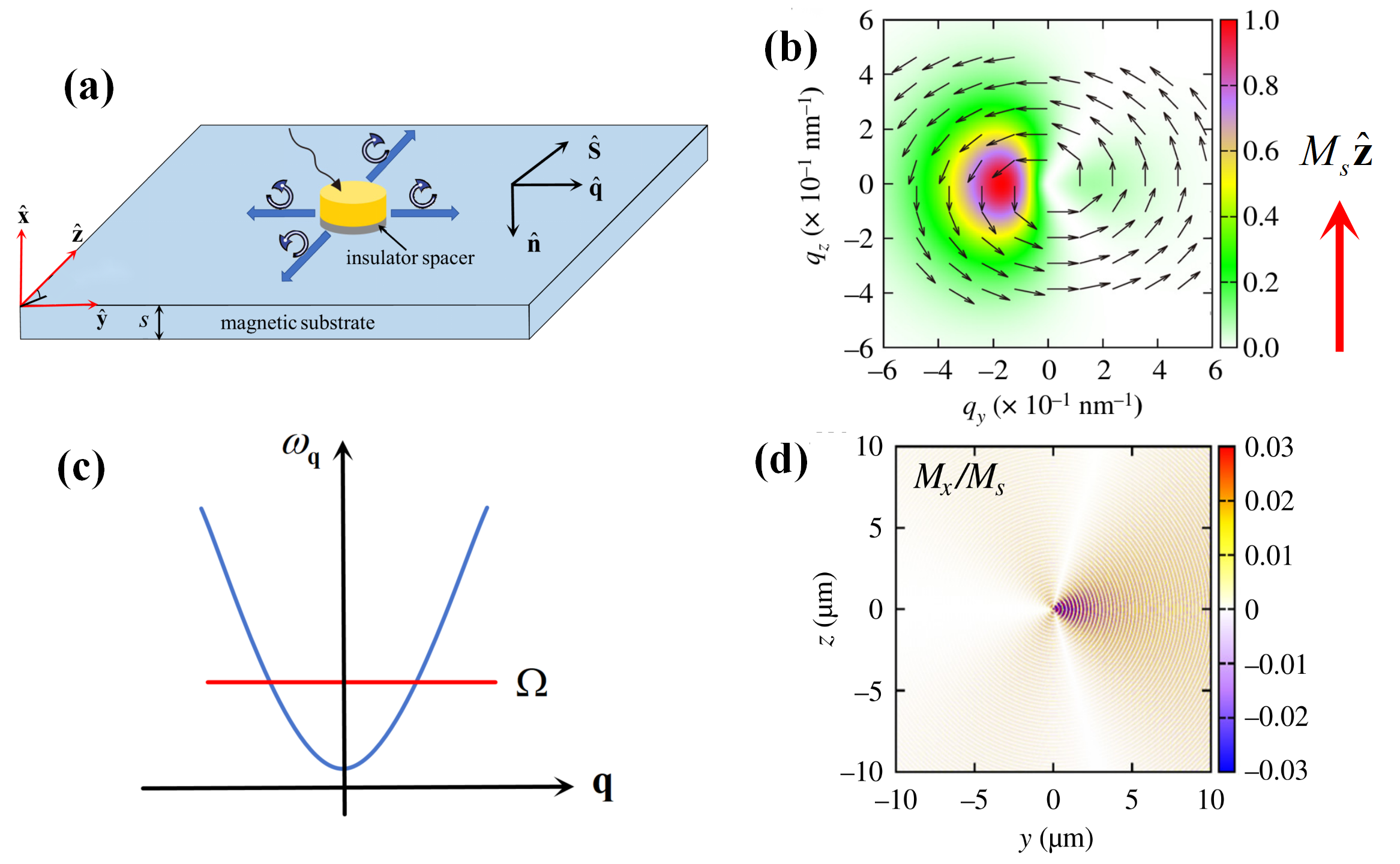}
    \caption{Geometric SOC of the stray magnetic field of an excited magnetic dipole and its detection by chiral spin pumping. (a) plots the spin-momentum locking for a field that is evanescent in the normal $-\hat{\bf x}$-direction. The photonic spin $\hat{\bf S}$ and the wave vector $\hat{\bf q}$ are locked, and constrained by the chirality index $-\hat{\bf x}\cdot(\hat{\bf S}\times\hat{\bf q})=1$. (b) In this plot of the spin density ${\bf S}({\bf q})$ of the stray field in reciprocal space arrows represent spin direction and the colors the modulus $|{\bf S}({\bf q})|$ coded according to the side bar. (c) A sketch of a resonant excitation, i.e., the resonance frequency $\Omega$ of the magnetic dipole lies above the gap $\omega({\bf q}=0)$ of the (exchange) spin waves in the magnetic substrate. (d) The anisotropy of the spin pumping into a film with equilibrium magnetization along the \(\hat{\bf z}\)-direction reflects the photonic SOC.\\
    \textit{Source:} Figures (a), (b), and (d) are taken from Ref.~\cite{PhysRevApplied.22.034042}.}
    \label{chiralexcitation}
\end{figure}

\textbf{Stray field of magnetic dipoles}.---We first illustrate the chirality of the evanescent vector field emitted by a point magnetic moment ${\bf m}(t)$, as shown in Fig.~\ref{chiralexcitation}(a). Disregarding an in-plane magnetic anisotropy, an external magnetic field $H_0\hat{\bf z}$ biases the saturation magnetization along the ${\bf z}$-direction. When weakly excited, the magnetization of the point source reads  
\begin{align}
    {\bf M}({\bf r},t)=\delta({\bf r}){\bf m}(t)=\delta({\bf r})(\delta m e^{-i\Omega t},i\xi^2\delta m e^{-i\Omega t},m_s)^T
\end{align}
where $m_s$ is the saturation magnetic moment along the $\hat{\bf z}$-direction, the fluctuation $\delta m \ll m_s$, and the ellipticity $\xi^2>0$. 
According to Coulomb's law and dropping the time variable, the magnetic stray field reads
\begin{align}
    h_{\beta}({\bf r})=\frac{1}{4\pi}\partial_{\beta}\int_{-\infty}^{\infty}d{\bf r}'\frac{\partial_{\alpha}'{\bf M}_{\alpha}({\bf r}')}{|{\bf r}-{\bf r}'|}=\sum_{q_y,q_z}e^{i(q_yy+q_zz)}h_{\beta}(x,q_y,q_z),
\end{align}
where $\{\alpha,\beta\}\in\{x,y,z\}$ are Cartesian indices. Expanding into plane-waves with wave vectors ${\bf q}=q_y\hat{\bf y}+q_z\hat{\bf z}$,  we analyze the chirality of the Fourier components $h_{\beta}(x,q_y,q_z)$, see Sec.~\ref{photonic_SOC}.

Using Weyl's identity~\cite{Yu_chirality,Weyl_identity}
 \begin{align}
     \int dq_x\frac{e^{iq_xx}}{(q_y^2+q_z^2)+q_x^2}=\frac{\pi}{\sqrt{q_y^2+q_z^2}}e^{-\sqrt{q_y^2+q_z^2}|x|},
 \end{align}
the Fourier components of the stray field in the lower half-space ($x<0$) read
 \begin{align}
&h_x(x,q_y,q_z)=F_{q_y,q_z}e^{\sqrt{q_y^2+q_z^2}x}/2,\notag\\
&h_y(x,q_y,q_z)=F_{q_y,q_z}e^{\sqrt{q_y^2+q_z^2}x}iq_y/\left(2\sqrt{q_y^2+q_z^2}\right),\notag\\&h_z(x,q_y,q_z)=F_{q_y,q_z}e^{\sqrt{q_y^2+q_z^2}x}iq_z/\left(2\sqrt{q_y^2+q_z^2}\right),
 \end{align}
 in which 
$F_{q_y,q_z}=m_x\sqrt{q_y^2+q_z^2}+m_yiq_y+m_ziq_z$. According to
Eq.~\eqref{transfield}, the momentum ${\bf q}=q_y\hat{\bf y}+q_z\hat{\bf z}$ and the field vector ${\bf h}(q_y,q_z)$ are locked by $iq_yh_y(x,q_y,q_z)+iq_zh_z(x,q_y,q_z)=-\sqrt{q_y^2+q_z^2}h_x(x,q_y,q_z)$. The  spin density ${\bf S}={\mu_0}/(4\Omega){\rm Im}[{\bf h}^*(x,q_y,q_z)\times {\bf h}(x,q_y,q_z)]$~\cite{PhysRevApplied.22.034042,spin_density,spin_density1,spin_density2,spin_density3,Yu_chirality} with
 \begin{align}
     &S_x(x,q_y,q_z)=0,\notag\\
     &S_y(x,q_y,q_z)=-\frac{\mu_0}{8\Omega}|F_{q_y,q_z}|^2q_ze^{2\sqrt{q_y^2+q_z^2}x}/\sqrt{q_y^2+q_z^2},\notag\\
     &S_z(x,q_y,q_z)=\frac{\mu_0}{8\Omega}|F_{q_y,q_z}|^2q_ye^{2\sqrt{q_y^2+q_z^2}x}/\sqrt{q_y^2+q_z^2},
 \end{align}
is evanescent for all wave vectors, transverse since ${\bf q}\cdot{\bf S}=0$, and right-handed since $\hat{\bf n}\cdot(\hat{\bf S}\times\hat{\bf q})=1$, as expected from the general rule Eq.~\eqref{chirality_index_eq}.

As illustrated by Fig.~\ref{chiralexcitation}(a), below the magnetic dipolar point source, the spin ${\bf S}({\bf q})$ of its stray magnetic field is locked to its wave vector ${\bf q}$: when the wave propagates along $\hat{\bf y}$, the photonic spin must be along $\hat{\bf z}$.
Figure~\ref{chiralexcitation}(b) shows a plot of the distribution of the microwave spin density ${\bf S}(x,{\bf q})$, in which the arrows in the figure denote the direction of the spin that is always perpendicular to the wave vector ${\bf q}$. The stray field generated by the point magnetic moment is ``right-handed" with spin-momentum perfectly locked by $C_q=1$ and obeys the right-hand rule of Eq.~(\ref{chirality_index_eq}) with fixed phase $\phi_y=\phi_z=\pi/2$. This spin-momentum locking relation is reminiscent of the Rashba SOC~\cite{Rashba_SOC,Rashba_SOC1} of electrons.   

The geometrical spin-orbit interaction affects the spin waves in thin magnetic films in proximity of a dynamic magnetic dipole formed by an excited nanomagnet~\cite{PhysRevApplied.22.034042} or nitrogen-vacancy center in diamond~\cite{xue2025}, as illustrated by Fig.~\ref{chiralexcitation}(a). When the resonance frequency of the nanomagnet $\Omega$ lies above the gap of the spin waves in the magnetic films, as illustrated for exchange magnons in Fig.~\ref{chiralexcitation}(c), its stray field ``pumps" spin waves into the film. Figure~\ref{chiralexcitation}(d) illustrates the configuration in which the saturation magnetization of the film points along the $\hat{\bf z}$-direction, as the magnon spin.  According to Fig.~\ref{chiralexcitation}(b), for $q_z=0$ only the stray field with positive $q_y$ has spin along the $\hat{\bf z}$-direction that can interact with the spin waves of the same wave vector. This implies that spin waves with positive $q_y$  [propagate to the right in Fig.~\ref{chiralexcitation}(d)] are most efficiently excited. This spin wave beam rotates with the in-plane film magnetization.

{\bf Damon-Eshbach mode}.---The Damon-Eshbach surface spin wave mode lives in thick ferromagnetic slabs when the wave vector is normal to an in-plane magnetization as in Fig.~\ref{chiral_waves}(a)~\cite{DEmode,Surface_dynamic_magnetic}. Its chirality originates from the dipolar interaction in the long-wavelength limit. 
We adopt a static applied magnetic field ${\bf H}_{\rm app}=H_{\rm app}\hat{\bf y}$, a sample thickness $d$, and consider the long-wavelength limit in which we may disregard the exchange interaction. 
For small pertubation from the equilibrium with magnetization \(M_s\), ${\bf M}=M_s\hat{\bf y}+{\bf m}$ with $M_s\gg |{\bf m}|$ and transverse fluctuations  
\begin{align}
 {\bf m}({\bf r},t)=\left({\bf a}e^{i k_z z}+{\bf b}e^{-i k_z z }\right)e^{i({\pmb \kappa }\cdot {\pmb \rho}-i\omega t)},
 \label{expand_surface_m}
\end{align}
where ${\pmb \rho}=x\hat{\bf x}+y\hat{\bf y}$  and ${\pmb \kappa }=k_x\hat{\bf x}+k_y\hat{\bf y}$ are position and wave vectors in the plane.
Substituting Eq.~\eqref{expand_surface_m} into the dipolar stray field Eq.~\eqref{dipoar_field_H} inside the magnetic slab yields
\begin{align}
 	{\bf H}^{(d)}({\bf r},t)&=e^{i({\pmb \kappa}\cdot{\pmb \rho}-\omega t)}\left\lbrace
 		\left( 	
 		\begin{array}{ccc}
 		-\dfrac{k_x^2}{k^2}	&-\dfrac{k_xk_y}{k^2}  &-\dfrac{k_xk_z}{k^2}  \\
 		-\dfrac{k_xk_y}{k^2}	& -\dfrac{k_y^2}{k^2} & -\dfrac{k_yk_z}{k^2} \\
 		-\dfrac{k_xk_z}{k^2}	& -\dfrac{k_yk_z}{k^2} & -\dfrac{k_z^2}{k^2}
 		\end{array}\right) 
 		 e^{ik_z z}\right.+
 		\left( 	\begin{array}{ccc}
 		\dfrac{k_x^2}{2\kappa }&	\dfrac{k_xk_y}{2\kappa }	& -\dfrac{ik_x}{2}	\\
 			\dfrac{k_xk_y}{2\kappa }&	\dfrac{k_y^2}{2\kappa }	& -\dfrac{ik_y}{2} \\
 			-\dfrac{ik_x}{2}&	-\dfrac{ik_y}{2}	& -\dfrac{\kappa }{2}
 		\end{array}\right) \dfrac{1}{\kappa -ik_z}e^{\kappa  z}\nonumber\\
 		&+\left. \left( \begin{array}{ccc}
 			\dfrac{k_x^2}{2\kappa }	&\dfrac{k_xk_y}{2\kappa } &\dfrac{ik_x}{2} \\
 			\dfrac{k_xk_y}{2\kappa }	&\dfrac{k_y^2}{2\kappa } &\dfrac{ik_y}{2} \\
 			\dfrac{ik_x}{2}	& \dfrac{ik_y}{2}&
 			-\dfrac{\kappa }{2}
 		\end{array}\right) \dfrac{e^{-d(\kappa +ik_z)}}{\kappa +ik_z}e^{-\kappa  z} \right\rbrace
 		\left( \begin{array}{c}
 			a_x \\
 			a_y \\
 			a_z
 		\end{array}\right) +e^{i({\pmb \kappa}\cdot{\pmb \rho}-\omega t)}\{ k_z\rightarrow -k_z \}\left( \begin{array}{c}
 			b_x \\
 			b_y \\
 			b_z
 		\end{array}\right).
 		\label{dipolar_field_DE}
 	\end{align}
The terms $\propto e^{\pm i k_z z}$ and $\propto e^{\pm \kappa z}$ represent the bulk and surface contributions, respectively. With Eq.~(\ref{dipolar_field_DE}) the (linearized) LL equation becomes 
 \begin{align}
		& -i\omega{m}_x({\bf r})=-\mu_0 \gamma M_s {H}_z^{(d)}({\bf r}) +\mu_0 \gamma H_{\rm app}  {m}_z({\bf r}), \nonumber\\
		& -i\omega {m}_z({\bf r})=\mu_0 \gamma M_s {H}_x^{(d)}({\bf r})-\mu_0 \gamma H_{\rm app}  {m}_x({\bf r}).
        \label{lin_LLEQ}
\end{align}
At resonance, the coefficients of $e^{\pm i k_z z}$ and $e^{\pm \kappa z}$ vanish,  which leads to two characteristic equations that govern the mode dispersion $\omega(k_x,k_y)$ and $k_z$.

The first characteristic equation leads to the dispersion relation
\begin{align}
    \omega^2 = \omega_H^2+\omega_H\omega_M\frac{k_z^2 + k_x^2}{k_z^2+\kappa^2},
\end{align}
where $\omega_H = \mu_0 \gamma H_{\rm app}$ and $\omega_M = \mu_0 \gamma M_s$. On the other hand,
$k_z$ solves the second characteristic equation
\begin{align}
    (\beta_1k_x)^2 + k_z^2(\beta_2 + 1)^2-\kappa^2-2k_z\kappa(\beta_2 + 1)\cot(k_z d)=0,
    \label{charasitc_surface_magnon2}
\end{align}
where $\beta_1 = \omega\omega_M/(\omega_H^2 - \omega^2)$ and $ \beta_2 = \omega_H\omega_M/(\omega_H^2 - \omega^2)$. 
When propagating normal to the saturation magnetization with $k_y=0$,  $k_z=i\eta_2$ is purely imaginary, and the mode amplitude decays exponentially as a function of distance from the surface. The in-plane wave vector \(k_x\) obeys ~\cite{DEmode,Surface_dynamic_magnetic}
    \begin{align}
    (\beta_1k_x)^2 -\eta_2^2(\beta_2 + 1)^2-\kappa^2-2\eta_2\kappa(\beta_2 + 1)\coth(\eta_2 d)=0.
    \label{characeristic_2}
\end{align}

These modes are localized at the upper or the lower surfaces of the slab, depending on the sign of the wave number \(k_x\), i.e. the propagation direction. 
 For a thick slab with $e^{\eta_2 d}\gg1$, $\coth{(\eta_2 d)}\rightarrow1$, so according to Eq.~\eqref{characeristic_2}  
 \begin{align}
     \eta_2(\beta_2+1)+\kappa\approx |\beta_1 k_x|.
         \label{charasitc_surface_magnon2_eta}
 \end{align} 
 Combining it with Eq.~\eqref{lin_LLEQ} yields 
\begin{align}
	m_z(z)=&{\cal C} [e^{-\eta_2 z} (-\beta_2 \eta_2+\beta_1 k_x) +{\cal D} e^{\eta_2(z+d)}(\beta_2 \eta_2+\beta_1 k_x)],\nonumber\\
	m_x(z)=&i{\cal C} [e^{-\eta_2 z} (-\beta_1 \eta_2+\beta_2 k_x) +{\cal D} e^{\eta_2(z+d)}(\beta_1 \eta_2+\beta_2 k_x)],
\end{align}
where ${\cal C}$ is a normalization constant and ${\cal D}=(\eta_2(\beta_2 +1)-\beta_1 k_x +\kappa)/(\eta_2(\beta_2 +1)+\beta_1 k_x-\kappa)$.
When $k_x<0$, ${\cal D}\rightarrow0$ according to Eq.~\eqref{charasitc_surface_magnon2_eta}, so $m_{x,z}\propto e^{-\eta_2 z}$ is localized at the lower surface of the slab. When $k_x>0$, the mode becomes localized at the upper surface. When $k_x<0$ and $k_y\rightarrow 0$ with ${\cal D}\rightarrow0$, $m_x+i (k_x/\kappa) m_z\approx0$, Damon-Eshbach modes are nearly right 
 circularly polarized with $m_x=im_z$, i.e., fixed by the direction of the saturation magnetization, as illustrated by Fig.~\ref{chiral_waves}(a).

Circularly polarized modes generate circularly polarized magnetic stray fields above or below the film. When $k_x<0$, according to Eq.~\eqref{dipoar_field_H}, the dipolar field below the film ($z<-d$) with $m_x=im_z$
\begin{align}
    H_z^{(d)}&=-k_xe^{|k_x|z}e^{i k_x x}\int_{-d}^0 dz' e^{-|k_x|z'}m_z(z'),\nonumber\\
    H_x^{(d)}&=ik_xe^{|k_x|z}e^{i k_x x}\int_{-d}^0 dz'e^{-|k_x|z'}m_z(z'),
\end{align}
is left circularly polarized with $H_x^{(d)}=-iH_z^{(d)}$, opposite to that of the spin wave. Figure~\ref{chiral_waves}(a) shows the special chirality expected from the generalized SOC in Fig.~\ref{generalized_SOC}.

\subsubsection{Chiral electric fields}
\label{chiral_electric_fields}

\begin{figure}[htp!]
    \centering
    \includegraphics[width=0.88\linewidth]{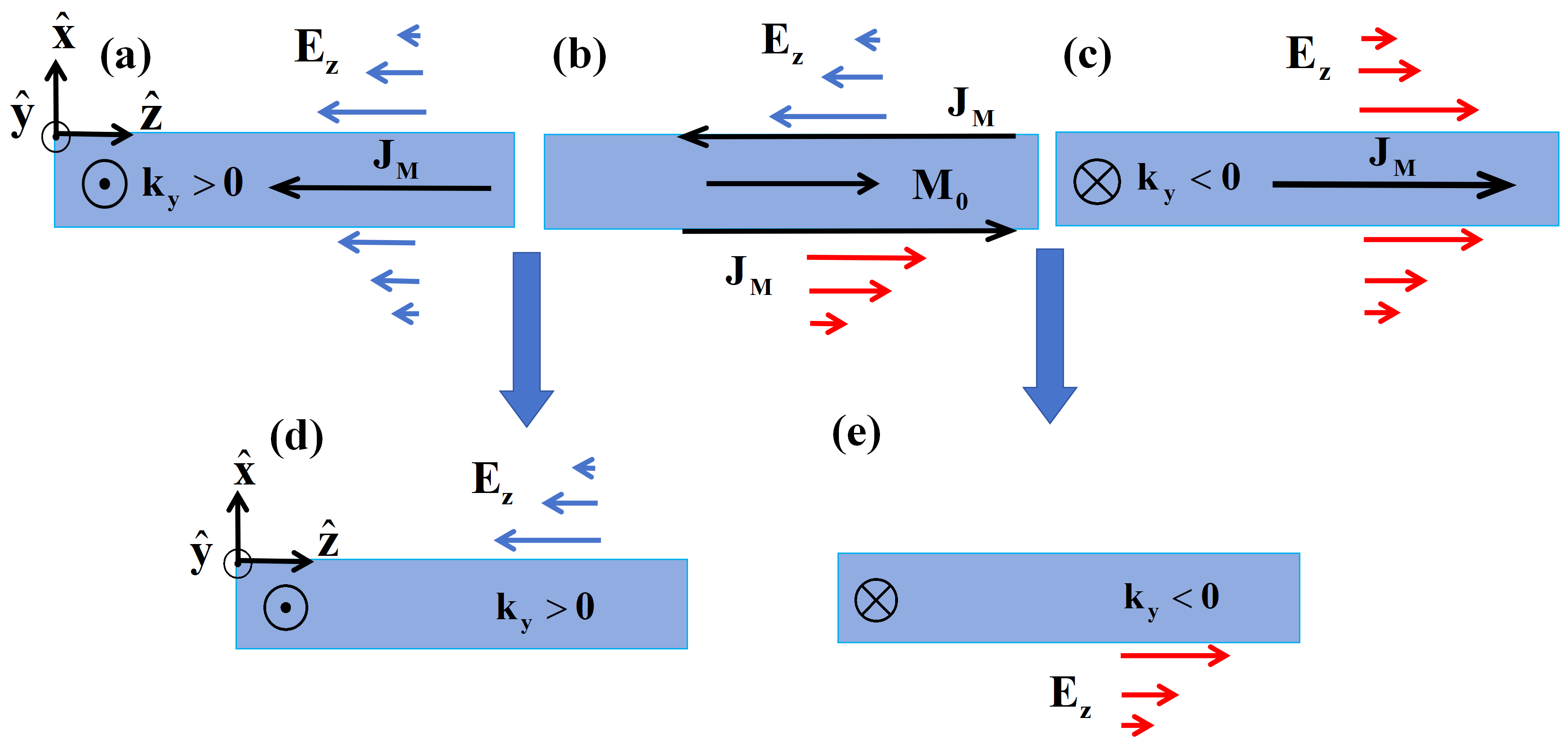}
    \caption{Snapshot of the stray electric field distribution radiated by the volume magnetization current for $k_y>0$ [(a)] and for $k_y<0$ [(c)], and the surface magnetization current [(b)], while (d) and (e) illustrate the total fields.}
    \label{chiral_electric_field}
\end{figure}

{\bf Chiral electric fields by magnetization dynamics}.---The dynamics of magnetic dipoles also generates electric fields that we here show to be chiral.
According to Maxwell's equations \eqref{maxwell_equation_2} a (bound) magnetization current \(\bf{J}_M=\nabla\times\mathbf{M}\) generates a magnetic induction with rotation
\begin{align}
\nabla\times{\bf B}=\mu_0{\bf J}_M+\mu_0\varepsilon_r\partial_t{\bf E}.
\label{magnetization_current_relation}
\end{align}
By taking the curl of the first equation in Eq.~\eqref{maxwell_equation_2} and inserting Eq.~\eqref{magnetization_current_relation}, the stray electric field obeys the wave equation
\begin{align}
    (\mu_0\varepsilon_r\partial_t^2-\nabla^2){\bf E}=-\mu_0\partial_t{\bf J}_M.
    \label{jm_radiation}
\end{align}
Figure~\ref{chiral_electric_field} illustrates this result for  circularly polarized spin waves with $M_y=iM_x$ propagating along the $\hat{\bf y}$-direction (normal to the saturation magnetization) in a thin ferromagnetic film with thickness $2d_F$ as
\begin{align}
M_x(\mathbf{r}, t) &= M_x(t)\mathrm{e}^{ik_y y}[\Theta(x + d_{{F}})-\Theta(x - d_{{F}})],\nonumber\\
 M_y(\mathbf{r}, t) &= M_y(t)\mathrm{e}^{ik_y y}[\Theta(x + d_{{F}})-\Theta(x - d_{{F}})],
\end{align} 
where the Heavyside step function $\Theta(x)=\begin{cases}
1, & x > 0 \\
0, & x < 0
\end{cases}$,
i.e., the magnetization is uniform over the film thickness.
The associated  (non-equilibrium) magnetization current 
\begin{align}
\mathbf{J}_M &= \nabla\times\mathbf{M} = (\partial_x M_y - \partial_y M_x)\hat{\bf z}\nonumber\\
&=  \begin{cases}
-k_y M_y(t)\mathrm{e}^{ik_y y}[\Theta(x + d_{\mathrm{F}})-\Theta(x - d_{\mathrm{F}})]\hat{\bf z}, & \text{volume magnetization current}, \\
M_y(t)\mathrm{e}^{ik_y y}[\delta(x + d_{\mathrm{F}})-\delta(x - d_{\mathrm{F}})]\hat{\bf z}, & \text{surface magnetization current},
\end{cases}
\label{magnetization_current}
\end{align}
contains contributions from both the volume and surface magnetization currents. The sign of the former is governed by the wave propagation direction $k_y$, as shown in Fig.~\ref{chiral_electric_field}(a) and (c). On the other hand, the directions of the surface magnetization currents 
$-M_y(t)\delta(x-d_F)$ and $M_y(t)\delta(x+d_F)$
located at the upper and lower surfaces with opposite directions do not depend on $k_y$, as shown in Fig.~\ref{chiral_electric_field}(b).

 Since the sign of the magnetization current governs that of the stray electric field, the fields generated by the volume magnetization current change sign with the propagation direction $k_y$, as shown in Fig.~\ref{chiral_electric_field}(a) and (c) at both sides of the ferromagnetic film.
The magnetization currents at the two surfaces flow in opposite directions but do not depend on the spin wave vector, so the associated electric fields are opposite above and below the film, as shown in Fig.~\ref{chiral_electric_field}(b). 
The total stray electric field exposes the same chirality as the magnetic field in this case: When $k_y>0$, it exists only above the film as shown in Fig.~\ref{chiral_electric_field}(d) [the sum of those in Fig.~\ref{chiral_electric_field}(a) and (b)]; when $k_y<0$, the field is finite only below the film as shown in Fig.~\ref{chiral_electric_field}(e).

\begin{figure}[htp!]
    \centering
    \includegraphics[width=0.9\linewidth]{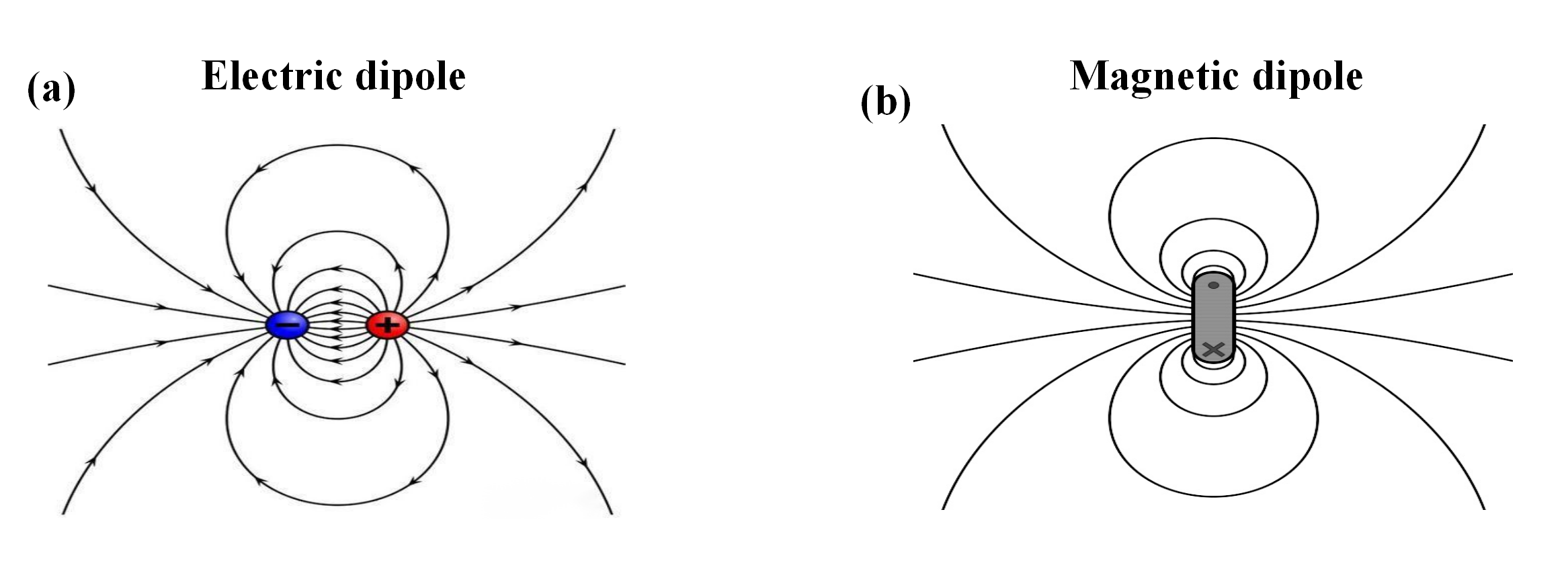}
    \caption{Duality between electric and magnetic dipoles. (a) The electric dipole and its electric stray field. (b) The magnetic dipole and its magnetic stray field.}
    \label{dual_E_H}
\end{figure}

{\bf Chiral electric field of electric dipoles}.---Electric and magnetic dipoles are dual when the associated ${\bf E}$ and ${\bf B}$ fields obey the same equation of motion. In the quasi-static limit, the magnetic and electric polarization generate electric and magnetic fields:
\begin{align}
 \mathbf{E}^{(d)}(\mathbf{r}, t) = \frac{1}{4\pi\varepsilon_0} \nabla \int \frac{\nabla' \cdot \mathbf{P}(\mathbf{r'}, t)}{|\mathbf{r} - \mathbf{r'}|} d\mathbf{r'},&&   \mathbf{H}^{(d)}(\mathbf{r}, t) = \frac{1}{4\pi} \nabla \int \frac{\nabla' \cdot \mathbf{M}(\mathbf{r'}, t)}{|\mathbf{r} - \mathbf{r'}|} \, d\mathbf{r'},
 \label{dipoleEH}
\end{align}
as illustrated in Fig.~\ref{dual_E_H}. The duality motivates the comparison of the electric and magnetic polarization waves in magnets and ferroelectrics, respectively~\cite{ferron,bauer_ferron_trans}.

The collective excitations of the magnetic order or spin waves are quantized as ``magnons", a quasiparticle that carries spin, energy, momentum, etc. 
Analogously, the excitation of the ferroelectric order carrying electric polarization is quantized as ``ferrons"~\cite{ferroelectric,ferron}, another quasiparticle that also carries energy, momentum, etc.
and are conduits for transport of the electric polarization and heat~\cite{ferroelectric,ferron,bauer_ferron_trans,Capacitors,Point_Contacts}. The observation of electric field–dependent thermal conductivity and diffusivity in a bulk lead zirconium titanate–based ferroelectrics~\cite{ferron_transport_exp}  provided the first experimental indications of electric dipole carrying excitations. Two recent experiments provide direct evidence of propagating ferrons~\cite{Coherent_exp,Diffuse_exp}.

Tang \textit{et al.} showed that bulk ferrons couple with optical photons to form ``ferron-polaritons"~\cite{fe_polarition}, but did not take the electric dipolar interaction into account. Zhuang and Hu~\cite{bulkmode_ferron2} predicted that the radiative loss due to electric stray fields reduces the resonance frequencies and significantly shortens the relaxation time of the fundamental mode in thin ferroelectric films.  Zhou \textit{et al}.~\cite
{surface_ferron} and Rodr\'iguez-Su\'arez \textit{et al}.~\cite{bulkmode_ferron} report important dipolar interaction effects on both surface and bulk polarization wave dispersions in the long-wavelength limit.

We focus on the most common displacive ferroelectrics in which the phase transition to the ordered state is triggered by a soft optical phonon. We show that the dipolar interaction importantly dominates the fluctuations of the free energy density
\begin{align}
{\cal F}=\dfrac{g}{2}(\nabla {\bf P})^2+{\cal G}({\bf P})-\dfrac{1}{2}{\bf E}^{(d)}\cdot {\bf P}-{\bf E}^{(e)}\cdot {\bf P},
\end{align}
where $g$ is the Ginzburg-type parameter, ${\cal G}({\bf P})$ is the Landau–Devonshire free energy~\cite{ferroelectric,ferron_landau,ferron_landau1}, ${\bf E}^{(d)}$ is the dipolar field  of the electric polarization $\bf P$, and ${\bf E}^{(e)}$ is an external electric field. 
Table.~\ref{Landau_energy} lists ${\cal G}({\bf P})$ expands up to the fourth order in ${\bf P}$ for typical ferroelectric materials. 
We consider a uniaxial ferroelectric crystal with easy axis along the $\hat{\bf y}$-direction with free energy density
\begin{align}
   {\cal F}= \dfrac{g}{2} (\nabla {\bf P})^2+ \frac{\alpha_1}{2}P_y^2+\frac{\alpha_2}{4}P_y^4+\frac{\alpha_3}{2}(P_x^2+P_z^2)-\frac{1}{2}{\bf E}^{(d)}\cdot {\bf P}-{\bf E}^{(e)}\cdot {\bf P},
   \label{free_energy}
\end{align}
where $\alpha_{1,2,3}$ are the Landau parameters with $\alpha_1<0$, $\alpha_2>0$, and $\alpha_3>0$. 
At equilibrium, the spontaneous polarization ${\bf P}_0$ minimizes the free energy by  $\delta {\cal F}/\delta {\bf P}|_{{\bf P}_0}=0$ by  pointing along the easy axis direction ${\bf P}_0=\{0,P_0=\sqrt{-\alpha_1/\alpha_2},0\}$ when ${\bf E}^{(e)}\rightarrow 0$.
In an excited state, the electric polarization fluctuates around ${\bf P}_0$, i.e.,  ${\bf P}=\{p_x,P_0+p_y,p_z \}$, where $P_0\gg p_{x,y,z}$ in the linear response regime. For plane waves ${\bf p}=\delta {\bf p}e^{i {\bf k}\cdot {\bf r}}$
\begin{align}
	\delta \mathcal{F} & =\mathcal{F}\left({\bf P}_0+\delta {\bf p}\right)-\mathcal{F}\left({\bf P}_0\right)\nonumber \\
	& =\left( -\dfrac{g}{2} k^2+\left(\frac{\alpha_1}{2}+\frac{3}{2} \alpha_2 P_0^2+\alpha_3\right)+\frac{1}{2 \varepsilon_0} \frac{(k_x+k_y+k_z)^2}{k^2}\right) \delta p^2,
\end{align}
where $k=(k_x+k_y+k_z)^{1/2}$. 
The third term is the dipolar interaction, which with  $\sim 5.6\times 10^{10}~{\rm Nm}^2/{\rm C}^2$  for the ferroelectric LiNbO$_3$ is larger than the second one $\sim 4.3\times 10^9~{\rm Nm}^2/{\rm C}^2$~\cite{surface_ferron,LiNbO3,LiNbO3_g}. The gradient term becomes important only for very large wave numbers $k\sim 10^{10}~{\rm m}^{-1}$. Thus, at long wavelengths, the dipolar interaction plays an important role in the excitations of the ferroelectric order.

\begin{table}[htp!]
\caption{
The Landau free energy for typical ferroelectric materials~\cite{ferroelectric}.}
	\renewcommand{\arraystretch}{1.7}
	\begin{tabular}{|c|c|}
		\hline
		Typical material                               & Landau free energy  (fourth order)\\ \hline
		\multirow{2}{*}{\begin{tabular}[c]{@{}c@{}}BaTiO$_3 $\\ PbZr$_{1-x} $Ti$ _x $O$ _3 $\end{tabular}} & \multirow{2}{*}{$ \alpha_1 (P_x^2+P_y^2+P_z^2)+\alpha_{11}(P_x^4+P_y^4+P_z^4)+\alpha_{12} (P_x^2P_y^2+P_z^2P_y^2+P_x^2P_z^2) $}\\
		&                                               \\ \hline
		\multirow{2}{*}{\begin{tabular}[c]{@{}c@{}}LiTaO$ _3 $\\ LiNbO$_3 $\end{tabular}}                  & \multirow{2}{*}{$  \frac{\alpha_1}{2} P_y^2 +\frac{\alpha_2}{4} P_y^4+\frac{\alpha_3}{2}(P_x^2+P_z^2) $}                    \\
		&                                               \\ \hline
		SrBi$ _2 $Nb$ _2 $O$ _9 $                      & $ \alpha_1(P_x^2+P_y^2)+\alpha_{11}(P_x^4+P_y^4)+\alpha_{12}P_x^2P_y^2 $               \\ \hline
	\end{tabular}
\label{Landau_energy}
\end{table}

\begin{figure}[htp!]
    \centering
    \includegraphics[width=0.9\linewidth]{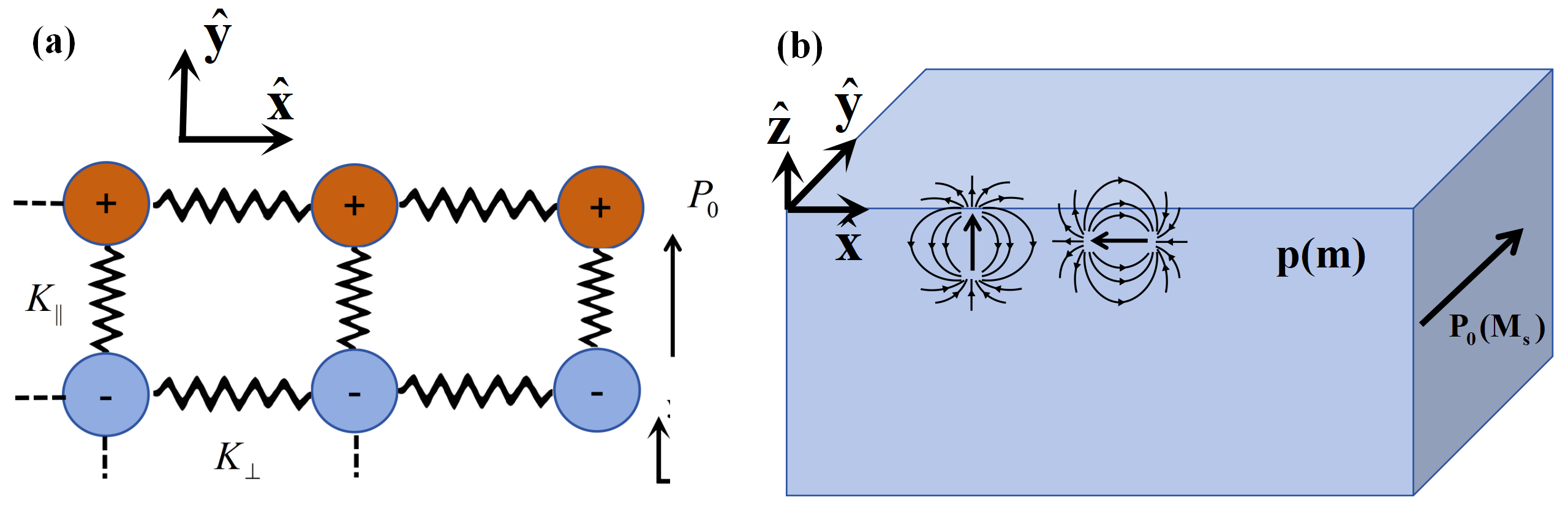}
    \caption{(a) Ball-spring harmonic oscillator model for the dynamics of electric polarization. (b) The surface ferroelectric (ferromagnetic) mode in ferroelectric (ferromagnetic) slabs with wave vectors normal to the in-plane equilibrium electric polarization $P_0$ (magnetization $M_s$). The associated stray electric and magnetic fields are similar.}
    \label{spring}
\end{figure}

In ferromagnetic crystals, the electric polarization ${\bf P}=\sum _i Q_i {\bf r}_i/V$ depends on the Bohr effective ionic charges $Q_i$ at position ${\bf r}_i$ in a microscopic volume $V$.  
An effective field $ {\bf E}_{\rm eff}$ drives the ionic motion according to Newton's Law
$\ddot{\bf r}_j=({Q_j}/{  m_j}) {\bf E}_{\rm eff}$, where $m_j$ is the mass of the $j$-th ion and
\begin{align}
    \dfrac{1}{\varepsilon _0 \Omega_p^2}\ddot{\bf P}={\bf E}_{\rm eff},
\end{align}
where $\Omega_p^2={(\varepsilon_0 V)}^{-1} \sum_j {Q_j^2}/{m_j}$ is the  ionic plasma frequency. The effective electric field 
\begin{align}
    {\bf E}_{\rm eff} =-
	\left(\dfrac{\delta { F}}{\delta {\bf P}}\right)_T-\gamma \left(\dfrac{\partial {\bf P}}{\partial t}\right)
\end{align}
is governed by the total free energy  $F=\int d{\bf r} {\cal F}({\bf r})$ and the phenomenological damping constant $\gamma $. We thus derive the Landau-Khalatnikov-Tani (LKT) equation of motion for the electric polarization~\cite{LKT1,LKT2,LKT3,LKT4}
\begin{align}
    m_p \dfrac{\partial^2{\bf P}}{\partial t^2}+{\gamma} \dfrac{\partial {\bf P}}{\partial t}=-\dfrac{\delta F}{\delta {\bf P}},
    \label{LKT}
\end{align}
where $m_p={1}/{(\varepsilon_0\Omega_p^2)}$. 
The equation of motion of the vector fluctuations follows from substituting Eq.~(\ref{free_energy}) into (\ref{LKT}). When linearized
\begin{align}
	&({1}/{\Omega_p^2})\partial_t^{2}p_{x,z}+K_{\perp} p_{x,z} =\varepsilon_0E_{x,z}^{(d)},\nonumber\\
&({1}/{\Omega_p^2}) \partial_t^{2}p_y+K_{\parallel} p_y=\varepsilon_0E_y^{(d)},
    \label{P_eom}
\end{align}
where $K_\perp=\varepsilon_0 \alpha_3>0$ and $K_\parallel=\varepsilon_0(\alpha_1+3\alpha_2P_0^2 )=-2 \varepsilon_0 \alpha_1>0$ are dimensionless force constants.
We thus arrive at a harmonic oscillator with anisotropic force constants and driving fields, as illustrated by Fig.~\ref{spring}(a). 

It is instructive to compare the electric and magnetic polarization waves. The surface spin waves in thick ferromagnetic films (Damon-Eshbach mode) distinguish themselves by the intrinsic chirality that originates in the broken time-reversal symmetry (Sec.~\ref{chiral_magnetic_fields}).
However, the electric and magnetic dipoles obey different equations of motion due to their different symmetries under time reversal and kinetic energies. 
As a consequence, the magnetization processes counter-clockwise around the effective magnetic field according to the LLG equation of motion, while lattice vibrations govern the dynamics of the electric polarization. Table~\ref{magnon_ferron_compare} compares the basis equation for the magnon and ferron modes in the bulk and at surfaces.

\begin{table}[htp!]
 \caption{Comparison of the Damon-Eshbach mode and the surface ferron mode, noting that the dynamics is governed not only by the potentials, but also the kinetic energy.}
		\renewcommand{\arraystretch}{2}
		\begin{tabular}{|>{\centering\arraybackslash}m{2.6cm}|>{\centering\arraybackslash}m{6cm}|>{\centering\arraybackslash}m{6cm}|}
		\hline
		\makecell{} & Damon-Eshbach Mode & Surface ferron Mode \\
		\hline
		\makecell{free energy\\density} & $\mathcal{F}=\frac{\mu_0\alpha_{ex}}{2}(\nabla\mathbf{M})^2-\frac{1}{2}\mathbf{H}^{(d)}\cdot\mathbf{M}-\mathbf{H}^{(e)}\cdot\mathbf{M}$ & $\mathcal{F}=\frac{g}{2}(\nabla\mathbf{P})^2+\mathcal{G}(\mathbf{p})-\frac{1}{2}\mathbf{E}^{(d)}\cdot\mathbf{P}-\mathbf{E}^{(e)}\cdot\mathbf{P}$ \\
		\hline
		\makecell{kinetic\\equation} & $\partial_t\mathbf{M}=-\mu_0\gamma\mathbf{M}\times\mathbf{H}_{\rm eff}+\frac{\alpha_G}{M_s}\mathbf{M}\times\partial_t\mathbf{M}$ & $m_p{\partial^2\mathbf{P}}/{\partial t^2}+\gamma{\partial\mathbf{P}}/{\partial t}=-{\delta F}/{\delta\mathbf{P}}$ \\
		\hline
		\makecell{characteristic\\equation (Bulk)} & $\omega^2 = (\mu_0\gamma H_{\rm app})^2+(\mu_0\gamma)^2H_{\rm app}M_s\frac{k_z^2 + k_x^2}{k_z^2+\kappa^2}$ & $c_1c_2k^2 + c_1k_y^2 + c_2(k_x^2 + k_z^2)=0$ \\
		\hline
		\makecell{characteristic\\equation (surface)} & $(\beta_1k_x)^2 + k_z^2(\beta_2 + 1)^2-\kappa^2-2k_z\kappa(\beta_2 + 1)\cot(k_z d)=0$ & $[(c_1 + 1)k_z\sin(k_z d/2)-c_1\kappa\cos(k_z d/2)][(c_1 + 1)k_z\cos(k_z d/2)+c_1\kappa\sin(k_z d/2)] = 0$ \\
		\hline
	\end{tabular}
    \label{magnon_ferron_compare}
\end{table}

We consider the surface polarization mode for a thick ferroelectric slab with easy axis along the $\hat{\bf y}$-direction, as shown in Fig.~\ref{spring}(b). Analogous to Sec.~\ref{chiral_magnetic_fields}, we expand the electric polarization into the plane-wave mode with in-plane wave vector ${\pmb \kappa }=k_x\hat{\bf x}+k_y\hat{\bf y}$ while the scattering at the surfaces generates standing waves with $\pm k_z\hat{\bf z}$ components. Hence 
\begin{align}
 {\bf p}({\bf r},t)=({\bf a}e^{i k_z z}+{\bf b}e^{-i k_z z })e^{i({\pmb \kappa }\cdot {\pmb \rho}-i\omega t)},
 \label{expand_surface_p}
\end{align}
where the in-plane coordinate ${\pmb \rho}=x\hat{\bf x}+y\hat{\bf y}$ and ${\bf a}$ and ${\bf b}$ are constant amplitudes.
According to Eq.~\eqref{dipoleEH}, the stray electric dipolar field has the same form as Eq.~\eqref{dipolar_field_DE} inside the slab, i.e.,  
	\begin{align}
 		{\bf E}^{(d)}({\bf r},t)&=\dfrac{1}{\varepsilon_0}e^{i({\pmb \kappa}\cdot{\pmb \rho}-\omega t)}\left\lbrace
 		\left( 	
 		\begin{array}{ccc}
 		-\dfrac{k_x^2}{k^2}	&-\dfrac{k_xk_y}{k^2}  &-\dfrac{k_xk_z}{k^2}  \\
 		-\dfrac{k_xk_y}{k^2}	& -\dfrac{k_y^2}{k^2} & -\dfrac{k_yk_z}{k^2} \\
 		-\dfrac{k_xk_z}{k^2}	& -\dfrac{k_yk_z}{k^2} & -\dfrac{k_z^2}{k^2}
 		\end{array}\right) 
 		 e^{ik_z z}\right.+
 		\left( 	\begin{array}{ccc}
 		\dfrac{k_x^2}{2\kappa }&	\dfrac{k_xk_y}{2\kappa }	& -\dfrac{ik_x}{2}	\\
 			\dfrac{k_xk_y}{2\kappa }&	\dfrac{k_y^2}{2\kappa }	& -\dfrac{ik_y}{2} \\
 			-\dfrac{ik_x}{2}&	-\dfrac{ik_y}{2}	& -\dfrac{\kappa }{2}
 		\end{array}\right) \dfrac{1}{\kappa -ik_z}e^{\kappa  z}\nonumber\\
 		&+\left. \left( \begin{array}{ccc}
 			\dfrac{k_x^2}{2\kappa }	&\dfrac{k_xk_y}{2\kappa } &\dfrac{ik_x}{2} \\
 			\dfrac{k_xk_y}{2\kappa }	&\dfrac{k_y^2}{2\kappa } &\dfrac{ik_y}{2} \\
 			\dfrac{ik_x}{2}	& \dfrac{ik_y}{2}&
 			-\dfrac{\kappa }{2}
 		\end{array}\right) \dfrac{e^{-d(\kappa +ik_z)}}{\kappa +ik_z}e^{-\kappa  z} \right\rbrace
 		\left( \begin{array}{c}
 			a_x \\
 			a_y \\
 			a_z
 		\end{array}\right) +\dfrac{1}{\varepsilon_0}e^{i({\pmb \kappa}\cdot{\pmb \rho}-\omega t)}\{ k_z\rightarrow -k_z \}\left( \begin{array}{c}
 			b_x \\
 			b_y \\
 			b_z
 		\end{array}\right),
 		\label{dipolar_field_2}
 	\end{align}
 which includes the bulk contribution $\propto e^{\pm i k_z z}$ and the evanescent surface contribution $\propto e^{\pm \kappa z}$. 
 Similar to the Damon-Eshbach mode, two equations govern the mode dispersion and $k_z$. By combining Eqs.~\eqref{dipolar_field_2}, \eqref{expand_surface_p}, and \eqref{P_eom}, we arrive at the first characteristic equation that governs the relation between the frequency and momentum
\begin{align}
	c_1 c_2k^2+c_1 k_y^2+c_2 (k_x^2+k_z^2)=0,
	\label{characteristic_1}
\end{align}
where
  $c_1=-(\omega/\Omega_p)^2 +K_\perp$ and $c_2=-(\omega/\Omega_p)^2 +K_\parallel$ are dimensionless. The second characteristic equation 
  \begin{align}
 		\left[
 	(c_1+1)k_z\sin({k_zd}/{2})-c_1\kappa\cos(k_zd/2)
 	 \right] \left[
 	(c_1+1)k_z\cos(k_zd/2)+ c_1\kappa\sin(k_zd/2)
 	 \right]=0
 	 \label{characteristic_2}
 	\end{align}
fixes $k_z$. 
The solution of 
Eqs.~(\ref{characteristic_1}) and (\ref{characteristic_2}) with complex  $k_z=\eta_1+i\eta_2$ are surface polarization waves.
For slabs with large thickness $d$, $\coth{(d\eta_2/2)}\rightarrow1$ and $\tanh{(d\eta_2/2)}\rightarrow 1$, we find $\eta_1=0$ and $\eta_2=-c_1\kappa/(1+c_1)$. Substituting $k_z= i\eta_2$ into the first characteristic Eq.~\eqref{characteristic_1}, we solve the dispersion of polarization waves as
\begin{align}
	 \omega_{\pmb \kappa}^+&=\Omega_p\sqrt{(1+\delta\cos^2{\theta_{\pmb \kappa}})/2+K_\perp}, \nonumber\\
	 {\omega_{\pmb \kappa}^-}&={\Omega_p}\sqrt{\delta\sin^2{\theta_{\pmb \kappa}}+K_\perp},
	 \label{frequencies}
\end{align}
to leading order in $\delta=K_\parallel-K_\perp\ll 1$, which holds for many ferroelectric materials. For clarity, we emphasize that the surface ferrons are modified optical phonon modes that are very different from the low-frequency surface acoustic waves~\cite{Oliner1978,Kino_1987,Gustafsson_2014,PhysRevX.5.031031,Satzinger2018}.

\begin{figure}[htp!]
    \centering
    \includegraphics[width=0.88\linewidth]{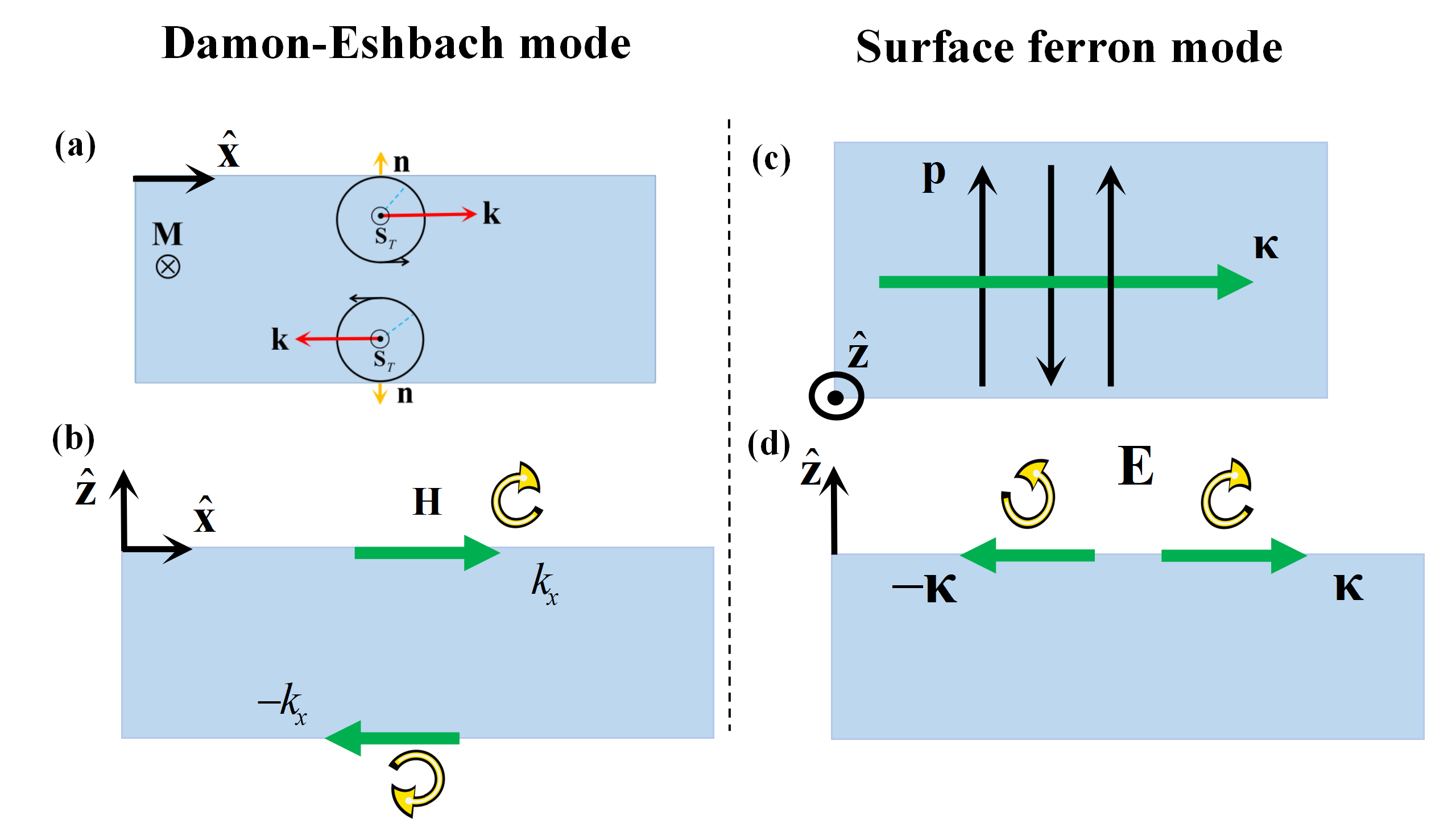}
    \caption{Comparison of the Damon-Eshbach and surface ferron modes and the associated stray fields. (a) The circularly polarized Damon-Eshbach mode at the upper (lower) surfaces is unidirectional. (b)  The circularly polarized magnetic field is generated by the Damon-Eshbach modes above (below) the film.  (c) Linearly polarized surface ferrons are not chiral. However, the associated electric fields above the film display circular polarization locked to the propagation direction [(d)].}
    \label{eigenmodes_p_model}
\end{figure}

\begin{figure}[htp!]
    \centering
\includegraphics[width=0.95\linewidth]{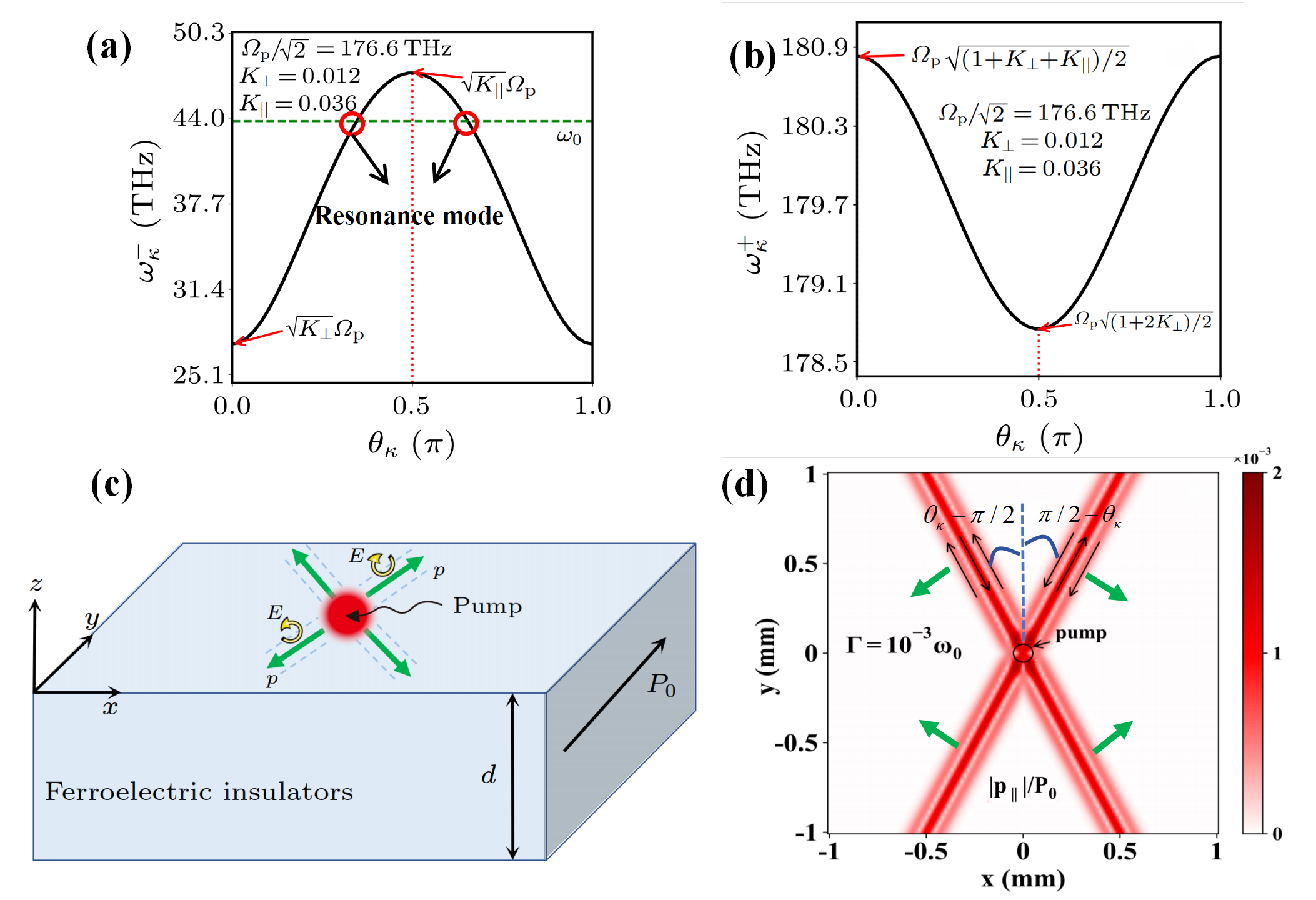}
\caption{Routing of surface ferron by focused light spot that resonantly excites the lower frequency ferron band. (a) and (b) plot the dispersion of the surface ferron branches
$\omega_{\pmb \kappa}^-$ and $\omega_{\pmb \kappa}^+$ with the room-temperature parameters of LiNbO3.
(c) Schematic of the excitation of ferrons by a focused THz laser (red dot) in a ferroelectric film with thickness $d$ and polarization poled along the $\hat{\bf y}$-direction. The green arrows illustrate the emission of directional coherent ferron beams. The yellow arrows illustrate the associated circularly polarization-locked electric stray fields. (d) The distribution of the emitted electric polarization in real space.\\
\textit{Source}: The figures are taken from Ref.~\cite{surface_ferron}.}
    \label{ferron}
\end{figure}

Magnetism and ferroelectricity break, respectively, time-reversal and inversion symmetry, and different generalized SOC (Fig.~\ref{generalized_SOC}). Figure~\ref{eigenmodes_p_model} illustrates the differences between the electric and magnetic surface modes. 
As discussed above, the Damon-Eshbach mode has only one frequency band with opposite chirality on the upper and lower surfaces, as shown in Fig.~\ref{eigenmodes_p_model}(a). 
The associated magnetic stray fields are also circularly polarized above or below the film, see Fig.~\ref{eigenmodes_p_model}(b). 
On the other hand, the surface ferrons have two branches as shown in Fig.~\ref{ferron}(a) and (b) that depend only on the angle $\theta_{\pmb \kappa}$ between the propagation direction and ${\bf P}_0$, but not the wave vector modulus [Eq.~\eqref{frequencies}]. The modes of the low- branch ${\omega_{\pmb \kappa}^-}$ are linearly polarized in the film plane since $\{|p_x({\pmb \kappa})|,|p_y({\pmb \kappa})|\}\gg |p_z({\pmb \kappa})|$~\cite{surface_ferron} and carry no angular momentum. Moreover, since $\nabla\cdot {\bf P}=0$ there are bound charges and fluctuations are purely transverse ${\bf p}\cdot {\pmb \kappa}\sim 0$.  The upper branch, however, is circularly polarized when propagating along the $\hat{\bf y}$-direction with polarization locked to its wave vector. The electric stray fields are accompanied by magnetic fields ${\bf B}=-(i/\omega)\nabla\times {\bf E}$ that are small but observable~\cite{Point_Contacts}.
Therefore, linearly polarized surface ferrons are not chiral, as shown in Fig.~\ref{eigenmodes_p_model}(c). 
However, according to Eq.~\eqref{dipoleEH},  the stray electric field of a linearly polarized surface ferron above the ferroelectric film
\begin{align}
	{\bf E}_{\rm out}^{(d)}({\pmb \kappa})\propto (\pmb{\hat{\kappa}}+i\hat {\bf z})  \kappa e^{-\kappa z}e^{i({\pmb\kappa}\cdot {\pmb \rho}-\omega_{\pmb \kappa}^- t)},
	\label{electric_stray_field}
\end{align}
is circularly polarized with $E_{\pmb \kappa}=-i E_z$.
As shown in Fig.~\ref{eigenmodes_p_model}(d), when switching the propagation direction, ${\pmb \kappa}\rightarrow -{\pmb \kappa}$, $E_{-\pmb \kappa}=-E_{\pmb \kappa}$, leading to $E_{\pmb -\pmb \kappa}=i E_z$, i.e.,  the circular polarized changes sign, the field is spin-momentum locked. We summarize the difference between the Damon-Eshbach and surface ferron modes in Fig.~\ref{eigenmodes_p_model} and Table.~\ref{magnon_ferron_compare}.

We proposed an experiment to observe surface ferrons by exciting them with a focused laser. The response should show a directional routing of the excited electric polarization in real space, as shown in Fig.~\ref{ferron}(c) that originates solely from the angle dependence of the dispersion relation [Eq.~\eqref{frequencies}]. 
In Fig.~\ref{ferron}(a) we show the ferron dispersion relation computed for the conventional ferroelectric insulator LiNbO$_3$ at room temperatures with $P_0=0.746~{\rm C/m^2}$, $ K_\perp=0.012$, and $K_\parallel=0.036$~\cite{LiNbO3,fe_polarition}.
A light frequency $\omega_0\sim 2\pi\times7$~THz, resonantly excites modes that propagate in the directions $\theta_{\pmb \kappa}$ and $\pi-\theta_{\pmb \kappa}$, as labeled by the red circles in Fig.~\ref{ferron}(a).  The plane wavefronts are then normal to the wave vectors. 
The electric polarization in real space is a superposition of the amplitudes of all excitations that interfere and form directional beams of the electric polarization along the wavefronts, see  Fig.~\ref{ferron}(d).

At long wavelengths, the dipolar interaction also affects the bulk ferron modes~\cite{surface_ferron}.  
Without it, i.e. for ${\bf E}^{(d)}=0$ in Eq.~\eqref{P_eom}, the dispersion of both bands is parabolic with $\omega_1=\sqrt{K_\perp}\Omega_p+gk^2$ and $\omega_2=\sqrt{K_\parallel}\Omega_p+gk^2$. 
Rodr\'{i}guez-Su\'{a}rez \textit{et al}.~\cite{bulkmode_ferron} calculated the effect of the dipolar interaction on both surface and volume modes for a film with thickness \textit{d}. By solving Eqs.~\eqref{characteristic_1} and~\eqref{characteristic_2}, they found two frequency bands for both the surface and the bulk modes.  For large $d$, the latter also depends only on the direction and not the modulus of the wave vector. The group velocities are strongly affected by the dipolar interaction. When $k_x=0$, the upper branch describes forward-moving waves with positive group velocity $\partial \omega(k_y)/\partial k_y>0$, while the lower branch with negative group velocity $\partial \omega(k_y)/\partial k_y<0$ corresponds to backward-moving waves.

\begin{figure}[htp!]
    \centering
    \includegraphics[width=0.95\linewidth]{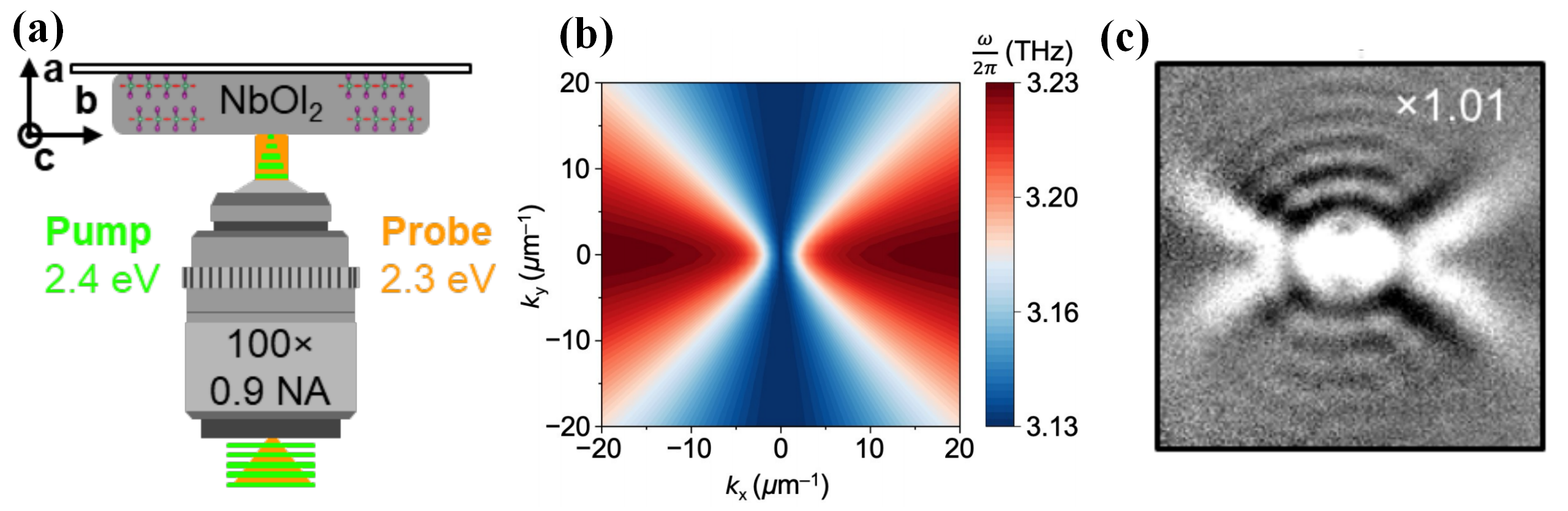}
    \caption{(a) Stroboscopic scattering microscopy for imaging ferrons. (b) Calculated ferron dispersion at $k_z=2\mu{\rm m}^{-1}$.  (c)  The images of ferron in a 240~nm thick NbOI$_2$ film at 4~K after application of a Fourier filter centered at $3.1325 \pm 0.025$~THz.\\
    \textit{Source:} These figures are taken from Ref.~\cite{Coherent_exp}.}
    \label{ferron_confirm}
\end{figure}

Recently, Choe \textit{et al}.  observed strongly anisotropic coherent ferron propagation in the van der Waals ferroelectric material  NbOI$_2$ when excited by a focused infrared laser~\cite{Coherent_exp}. Figure~\ref{ferron_confirm}(a) shows the ferron distribution excited by a femtosecond pulse polarized along the crystalline $b$-axis by a stroboscopic reflection microscopy. The authors interpreted their results by the model proposed by Zhou \textit{et al.}~\cite{surface_ferron,bulkmode_ferron} for the bulk ferron that leads to dispersion relations consistent with the predicted bulk modes in the Supplementary Material of Ref.~\cite{surface_ferron} in the limit $K_\perp=0$.  
The light frequency is much higher than that of lower-branch ferrons such that the excitation is not resonant and experiments only resolve the response of the upper branch as plotted in Figure~\ref{ferron_confirm}(b) for $k_z=2~\mu{\rm m}^{-1}$, which indeed depends more on the propagation direction than the magnitude of the wave vector. 
In spite of the differences compared to the set-up envisioned in Ref.~\cite{surface_ferron} the results plotted in Fig.~\ref{ferron_confirm}(c) show strong directionality, similar to Fig.~\ref{ferron}(d), but also additional interference patterns that may reflect the non-resonant excitation of the upper mode in the experiments.

\subsection{Chiral gating of magnons by superconductors}

\subsubsection{Theory}
\label{Chiral_gating_theory}

The gating effect is crucial in modern electronics, as it enables the creation of various electronic nanostructures for practical applications.
Magnons are promising information carriers that can avoid Joule heating in magnetic insulators~\cite{data_processing,data_processing1,data_processing2}, but they are challenging to gate since they do not directly couple to the voltage.  
The electromagnetic proximity effect was exploited to gate magnons by superconductors~\cite{46_Efficient_Gating,Golovchanskiy_gating2,Golovchanskiy_gating4,21_Fraerman,37_klos,55_Burmistrov,42_Centala,48_Giant_enhancement_of_magnon_transport,5_magnon_fluxon_Nat_Chumak,19_magnon_fluxonics_Chumak,34_Berakdar,39_Cherenkov_radiation_Buzdin,Golovchanskiy_gating1_exp,Golovchanskiy_gating3_exp,8_Pt_S_F_vortex_Blamire,13_Passamani,54_L.Tao}, metals~\cite{Seshadri_gating,Mruczkiewicz_gating1,Mruczkiewicz_gating2,metal_gate,metal_gate2_exp,45_Imaging_spin_wave_damping_underneath_metals,50_Chiral-damping-enhanced_magnon_transmission,MC_metal,Spin_Wave_Interferometer}, and other ferromagnets~\cite{magnon_trap,15_Liu,magnon_trap1,53_van_Wees,49_Chirality_enables_thermal_magnon_transistors,Fripp_gate,Liu_Magnonic_Chern_Bands,magnon_trap_Yu,magnon_trap_Yu2,Nanomagnonic_Diode,Unidirectional_spin_wave_propagation,PhysRevB.108.134436}, etc. 
In early 1970, Seshadri predicted
non-reciprocity of propagation of magnetostatic surface spin waves in a
magnetic film when covered by a perfect electric conductor with infinite
conductivity~\cite{Seshadri_gating}, such that the stray magnetic fields cannot penetrate the metal cap.  For the influence of electric conductors that cover the magnetic films with finite and infinite conductivity, Mruczkiewicz \textit{et al}. performed numerical simulations and found that for the thick metallic film, the dispersion of the spin wave is modulated~\cite{Mruczkiewicz_gating1,Mruczkiewicz_gating2}. Kostylev~\cite{metal_gate} shows that metal could enhance the radiation losses of the FMR.
Golovchanskiy \textit{et al}. simulated the effect of thick superconducting films that fully screen the stray magnetic fields~\cite{Golovchanskiy_gating2,Golovchanskiy_gating4}. For the superconductor$|$ferromagnet bilayers, the superconductor can non-reciprocally modulate the dispersion of the Damon-Eshbach mode in Sec.~\ref{chiral_magnetic_fields}, which results in a significant enhancement of the phase velocity at positive wave numbers. 
For the periodic superconductor structures on top of the ferromagnet, the superconductor induces indirect band gaps that open at positive wavenumbers, demonstrating the tunability of the spin-wave dispersion in superconductor-ferromagnetic structures.

\begin{figure}[htp!]
    \centering
    \includegraphics[width=0.95\linewidth]{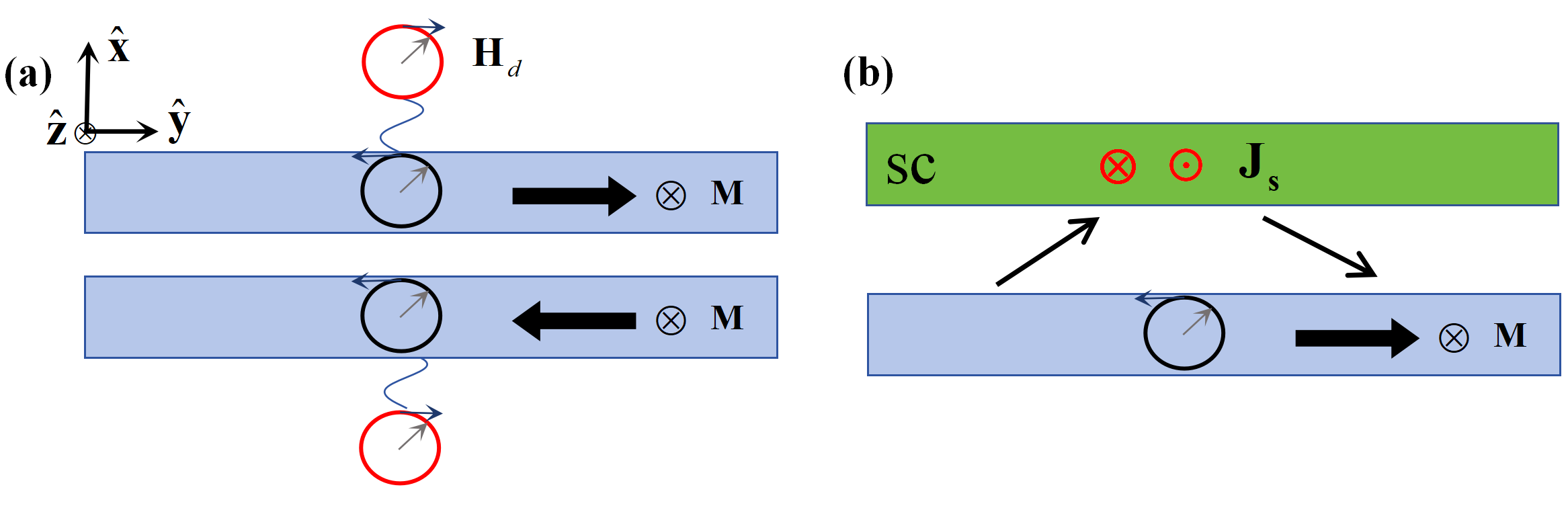}
    \caption{Gating magnons by superconductors in a chiral fashion. (a) Spin waves propagating to the right (left) emit the dipolar magnetic field ${\bf H}_d$ above (below) the film.  (b) Superconductor acts back to the magnetization via the Oersted magnetic field from the supercurrent.}
    \label{gating_01}
\end{figure}

Yu and Bauer proposed to gate magnons in high-quality ferromagnetic insulators (e.g., the YIG) by thin superconductors~\cite{46_Efficient_Gating}. 
As addressed in Fig.~\ref{chiral_waves}(a), when the spin waves propagate normal to the saturation magnetization, the stray magnetic field generated by the spin waves is chiral,  i.e., only the spin waves propagating to the right ($k_y>0$) can generate the stray field above the magnetic film. 
When a superconductor is placed on top of the magnetic film, it can only interact with the spin waves propagating to the right ($k_y>0$). 
The stray magnetic field penetrating the superconductors drives a Meissner supercurrent ${\bf J}_s$. Such supercurrent then generates an Oersted magnetic field $\tilde{\bf H}$ that acts back on the magnetization, as shown in Fig.~\ref{gating_01}(b).

\begin{figure}[htp!]
    \centering
    \includegraphics[width=\linewidth]{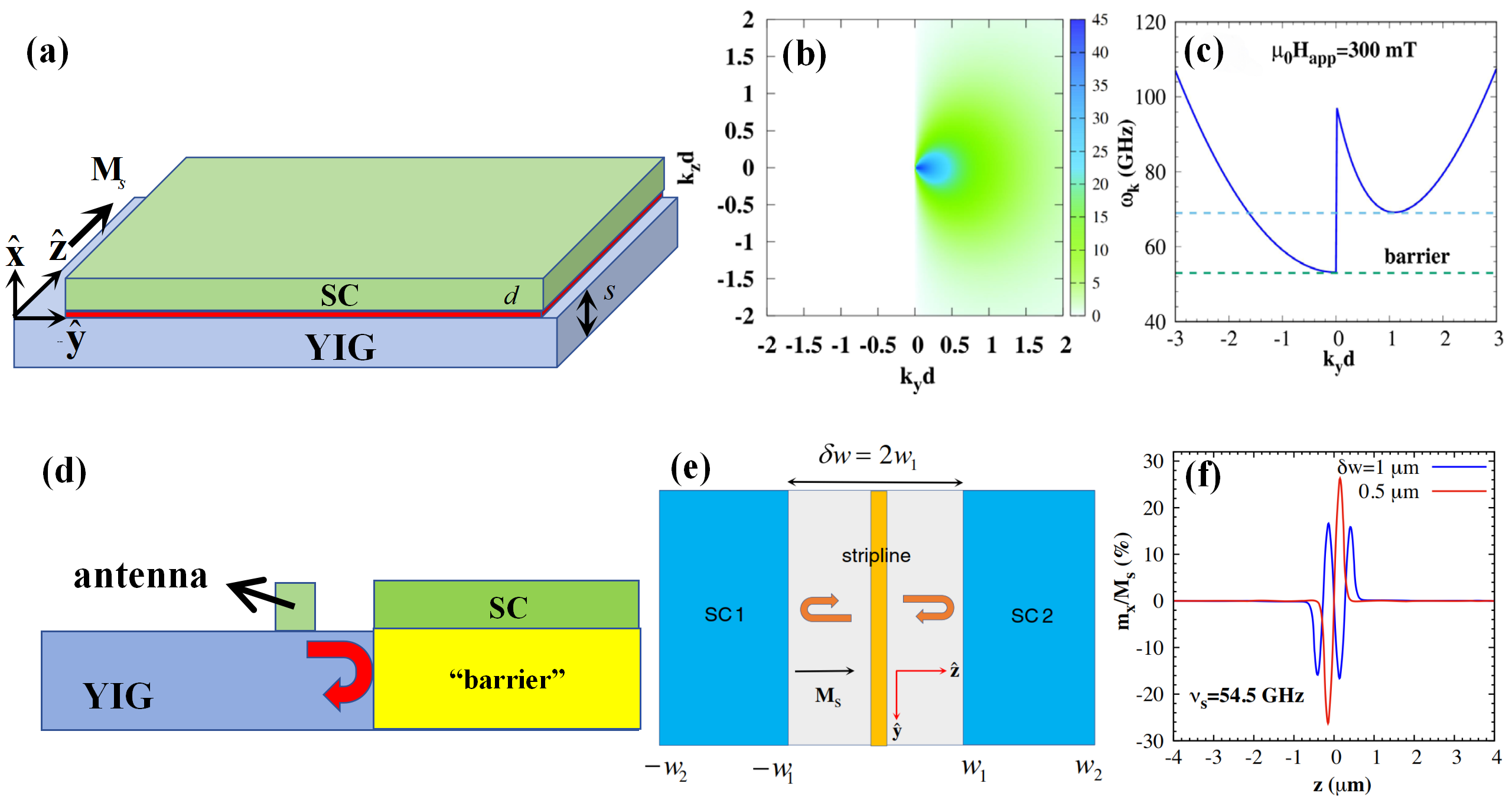}
    \caption{(a) The superconductor-magnet bilayer. (b) Chirality of the frequency shift in the reciprocal space. (c) Spin wave dispersion with momentum $k_y\hat{\bf y}$ modulated by a superconducting
gate. (d) Local gating of magnons by a superconductor. (e) Top view of the trap magnon device. (f) Snapshots of the maximum amplitudes $m_x$ for two gate distances $\delta w=\{0.5,1\}$~$\mu$m at a frequency close to the standing wave frequencies with index $n=\{3,1\}$.\\
  \textit{Source:} The figures (b), (c), (e), (f) are taken from Ref.~\cite{46_Efficient_Gating}.}
    \label{gating_02}
\end{figure}

One typical consequence of the electromagnetic proximity effect is a significant shift of the magnon frequency in a chiral fashion. 
We consider the bilayer configuration, comprising a superconductor with thickness $d$ and a magnetic film with thickness $s$, as illustrated in Fig.~\ref{gating_02}(a). 
A thin insulating spacer [the red film in Fig.~\ref{gating_02}(a)] between them suppresses the interfacial exchange interaction. 
Still, it renders the dipolar interaction unaffected.  
The static magnetic field ${\bf H}_{\rm app}=H_{\rm app}\hat{\bf z}$ biases the saturation magnetization ${\bf M}_s$ to the $\hat{\bf z}$-direction.
When the magnetic film is ultrathin, the exchange spin waves with amplitudes $m_y\approx im_x$  are circularly polarized. 
With free boundary conditions~\cite{free_boundary_conditions1,Tao_chiral_excitation}
\begin{align}
{\partial m_{x,y}}/{\partial x}|_{x = 0} = 0, \quad {\partial m_{x,y}}/{\partial x}|_{x=-s} = 0,
\end{align}
the transverse fluctuation in the wave vector space  $m_{x,y}^{(l)}(x,{\pmb{\rho};{\bf k}})\propto\cos{({l\pi x}/{s})}e^{i{\bf k}\cdot \pmb{\rho}}$, where  ${\pmb \rho}=y\hat{\bf y}+z\hat{\bf z}$ and $l$ is an integer that represents the band indices. 
The eigenmodes are normalized by Eq.~\eqref{renormalization_relation}:
\begin{align}
    m_x^{(l)}(x,{\pmb{\rho};{\bf k}})=\sqrt{\dfrac{1}{2(1+\delta_{l0})}}\dfrac{1}{\sqrt{s}}\cos{\left(\dfrac{l\pi}{s}x\right)}e^{i{\bf k}\cdot \pmb{\rho}},\quad
    m_y^{(l)}(x,{\pmb{\rho};{\bf k}})=im_x^{(l)}(x,{{\bf \rho};{\bf k}}).
    \label{eigen_modes_gating}
\end{align}

The dipolar stray fields ${\bf H}_{ d}^{(l)}({\bf r},t;{\bf k})$ of spin wave eigenmodes with momentum $\bf k$ and frequency $\omega_{\bf k}$ above the film ($x>0$) are evanescent as $\sim e^{-|{\bf k}|x}$ and nearly uniform across a superconducting gate for the wave numbers $|{\bf k}|\ll 1/d$. 
When $|{\bf k}|\ll 1/s$, only the lowest subband with $l=0$ contributes.
Thus, by substituting the eigenmodes Eq.~\eqref{eigen_modes_gating} into  Eq.~\eqref{dipoar_field_H} the dipolar field above the film
\begin{align}
    \begin{pmatrix}
H_x(\mathbf{r}, t;{\bf k}) \\
H_y(\mathbf{r}, t;{\bf k}) \\
H_z(\mathbf{r}, t;{\bf k})
\end{pmatrix}_{d}^{(0)} = \frac{1}{2}e^{-|\mathbf{k}|x}(|\mathbf{k}| + k_y)
\begin{pmatrix}
1 \\
-ik_y/|\mathbf{k}| \\
-ik_z/|\mathbf{k}|
\end{pmatrix}
e^{-i\omega_{\mathbf{k}}t} \int_{-s}^{0} m_x^{(0)}(x', \boldsymbol{\rho}; \mathbf{k})e^{|\mathbf{k}|x'} dx'.
\label{dip_fied_above_0}
\end{align}
The chirality of the dipolar field according to Eq.~\eqref{dip_fied_above_0} is most obvious for the spin wave propagating normal to the magnetization, i.e., $|{\bf k}|=|k_y|$. In this case, the term $(|{\bf k}|+k_y)/2$ in Eq.~\eqref{dip_fied_above_0} becomes $(|{k_y}|+k_y)/2$. 
The chirality is evident: for $k_y<0$, $(|{\bf k}|+k_y)/2=0$ but becomes unity when $k_y>0$, so only the spin wave propagating to the right can generate the dipolar field above the film.
According to Maxwell’s equations $\nabla\cdot {\bf E}=0$ and $\nabla\times {\bf E}=i\omega\mu_0{\bf H}$, the magnetic stray field of a wave vector $\bf k$ and frequency $\omega$ generates the electric field that has the form ${\bf E}\propto e^{-|{\bf k}|x}e^{i {\bf k}\cdot {\pmb \rho}}$.  
This electric field inside the superconductor drives the Meissner supercurrent according to the two fluid model~\cite{two_fluid_Janssonn,two_fluid_superconductor,two_fluid_Tao} with the conductivity Eq.~(\ref{conductivity}). 
Although the normal fluid is
included in the formalism, it plays a minor role when $T<0.95T_c$ since $\rho_n(T)\propto(T/T_c)^4$ rapidly decays when decreasing the temperature~\cite{46_Efficient_Gating,48_Giant_enhancement_of_magnon_transport}.
At the low temperature $T\rightarrow 0.85 T_c$, the conductivity $\sigma_c(\omega)=i\rho_ee^2/(m_e \omega)$ is purely imaginary. 
In the limit that $|{\bf k}|\ll 1/d$, the currents are uniform across the superconducting film, so we may set $x\rightarrow d/2$. 
According to the charge conservation, the divergence of the Meissner supercurrent vanishes with $\nabla\cdot {\bf J}_s=0$, implying the wave vector is normal to the supercurrent with ${\bf k}\cdot {\bf J}_s=0$. For the spin waves propagating normal to the magnetization, ${\bf J}_s$ flows in a direction parallel to the saturation magnetization, as shown by the red arrows in Fig.~\ref{gating_01}(b). 
 ${\bf J}_s$ generates the vector potential~\cite{Jackson} 
\begin{align}
    {\bf A}({\bf r},\omega)=\dfrac{\mu_0}{4\pi}\int d{\bf r}' \dfrac{{\bf J}_s(x'=d/2,{\pmb \rho}',\omega)e^{i\omega |{\bf r}-{\bf r'}|/c}}{|{\bf r}-{\bf r}'|},
\end{align}
which acts back on the spin waves.
Using Weyl identity~\cite{Yu_chirality,Weyl_identity}, we obtain for $x<0$, ${\bf A}(\mathbf{r}, \omega) = \dfrac{\mu_0 d }{2|\mathbf{k}|}{\bf J}_{s}(x = d/2, \boldsymbol{\rho}, \omega)e^{|\mathbf{k}|(x - d/2)}\propto {\bf J}_s$. $\nabla\cdot{\bf J}_s=0$ is automatically satisfied with the London gauge $\nabla\cdot {\bf A}=0$.

The transverse
($x$ and $y$) components  
\begin{align}
\tilde{H}_x(\mathbf{r}, \omega) &= i e^{|\mathbf{k}|(x - d)} \dfrac{d \omega \mu_0\sigma_c(\omega)}{4|\mathbf{k}|} (1 - e^{-|\mathbf{k}|s})
\left(m_x^{\mathrm{k}}(\boldsymbol{\rho}, \omega) - i m_y^{\mathrm{k}}(\boldsymbol{\rho}, \omega) \dfrac{k_y}{|\mathbf{k}|}\right),\nonumber\\
\tilde{H}_y(\mathbf{r}, \omega) &=\dfrac{ik_y}{|\mathbf{k}|}\tilde{H}_x(\mathbf{r}, \omega)
\end{align}
in the Oersted magnetic field $\tilde{{\bf H}}({\bf r})=(1/\mu_0)\nabla\times{\bf A}({\bf r})$
affects the magnetization dynamics. The linearized LLG equation in frequency space poses a self-consistency problem
\begin{align}
 -i\omega m_x(\mathbf{k}) &= -i\omega_{\mathbf{k}}(1 - \alpha_G)m_x(\mathbf{k}) + \mu_0 \gamma M_s \tilde{H}_y(\mathbf{k},\omega),\nonumber \\
 -i\omega m_y(\mathbf{k}) &= -i\omega_{\mathbf{k}}(1 - \alpha_G)m_y(\mathbf{k}) - \mu_0 \gamma M_s \tilde{H}_x(\mathbf{k},\omega),
\end{align}
which results in a frequency shift with 
\begin{align}
    \delta\omega_{\mathbf{k}} = \frac{d}{4}e^{-|\mathbf{k}|(\frac{s}{2}+d)}\frac{1 - e^{-|\mathbf{k}|s}}{|\mathbf{k}|}\mu_0^2\gamma M_s\frac{\rho_e e^2}{m_e}\left(\frac{k_y}{|\mathbf{k}|}+\left|\frac{k_y}{|\mathbf{k}|}\right|\right).
\end{align}
The shift is as large as ten gigahertz: for a $d=40$~nm superconducting NbN film with electron density $\rho_e=10^{29}/{\rm m}^3$~\cite{NbN_density} on top of
a $s=20$~nm YIG film, the shift $\delta \omega\approx45$~GHz when $k_y=|{\bf k}|\ll\{1/d,1/s\}$ at $\mu_0 H_{\rm app}=300$~mT. Only the magnon frequency at half of the reciprocal space is shifted, as shown in Fig.~\ref{gating_02}(b). 
For the spin waves propagating normal to the magnetization, i.e., along the $\hat{\bf y}$-direction, the eigenfrequency is shifted when $k_y>0$, but when $k_y<0$, it is not affected [Fig.~\ref{gating_02}(c)]. We can use this structure to locally create a potential barrier for magnons. As shown in Fig.~\ref{gating_02}(d), for a superconductor locally fabricated on top of the magnetic film, the frequency is shifted only in the area that is covered by the superconductor,  which acts as a potential barrier for magnons [the yellow area in Fig.~\ref{gating_02}(d)]. When the magnon is excited by an antenna, the spin waves propagating to the right feel the potential barrier and are reflected back.

The superconductor gate can trap magnons. This can be realized when the magnons propagate parallel to the saturation magnetization. 
As shown in Fig.~\ref{gating_02}(e), two superconducting
gates are located at $w_1\leq |z|\leq w_2$ on both sides of a stripline. The stripline excites spin
waves with equal amplitude to both sides that are reflected back and forth between the two superconductors. 
The magnon frequency is quantized when the spin waves form standing waves between the two superconductors. The simulation shows that they obey the quantization condition with the frequency governed by the integer number $n$, i.e., the standing wave resonance for a hard-wall potential happens when  
\begin{align}
    \nu_n=\mu_0 \gamma H_{\rm app}+\alpha_{\rm ex} \mu_0 \gamma M_s (n\pi /\delta w)^2.
    \label{magnon_cavity}
\end{align}
In Fig.~\ref{gating_02}(f), we plot snapshots of the maximum amplitudes for $m_x$ with two gate distances $\delta w=1~\mu$m (the blue curve) and $\delta w=0.5~\mu$m (the red curve) at the excitation frequency $\nu_s=54.5$~GHz close to the standing-wave frequencies with $\nu_1(\delta w=0.5\mu{\rm m})=53.6$~GHz and $\nu_3(\delta w=1\mu{\rm m})=54$~GHz. 
The steady
states in Fig.~\ref{gating_02}(f) are nearly perfectly trapped standing spin waves with odd inversion symmetry.

\begin{figure}[htp!]
    \centering
    \includegraphics[width=0.98\linewidth]{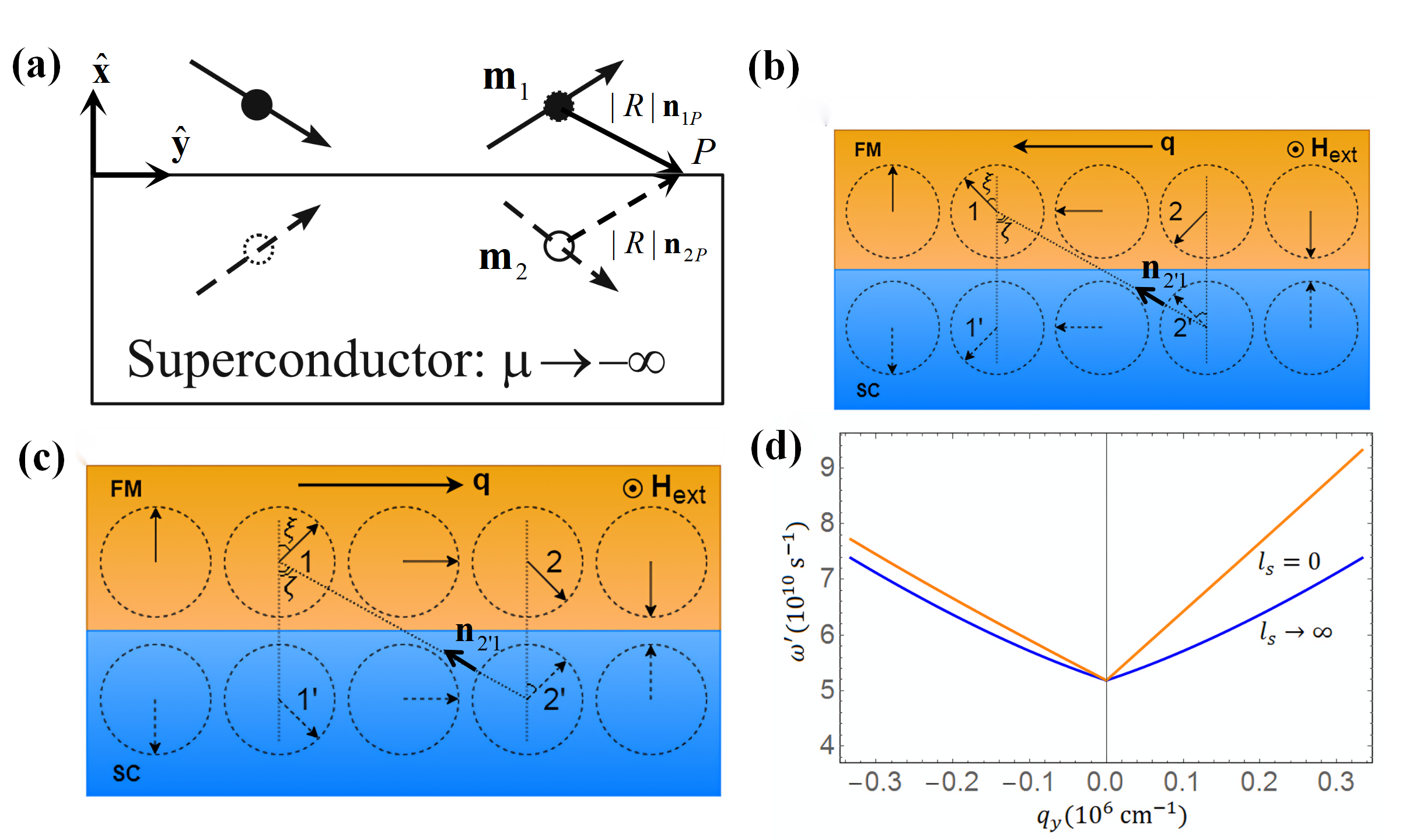}
    \caption{(a) illustrates two dipoles of different polarizations on top of a perfect conductor and their imaged dipoles that generate the magnetic induction satisfying the boundary condition.
    (b) and (c) are the snapshots of spin waves and their images propagating along opposite directions with wave vector $\pm {\bf q}\perp {\bf H}_{\rm ext}$ in the ferromagnet-superconductor bilayer biased by a magnetic field ${\bf H}_{\rm ext}$. (d) shows the spin-wave dispersion in the ferromagnet-superconductor bilayer with different London's penetration depths $l_s=\{0,\infty\}$, respectively.\\
    \textit{Source:} Figure (a) is taken from Ref.~\cite{method_of_image}; Figs.~(b), (c), (d) are taken from Ref.~\cite{21_Fraerman}. }
    \label{SF_shift_Fraerman}
\end{figure}

Kuznetsov and Fraerman also predicted the chiral gating effect of magnon by superconductors considering the dipolar interaction in a slightly different configuration Fig.~\ref{SF_shift_Fraerman}~\cite{21_Fraerman}. They considered a thin ferromagnet with thickness $h$ placed on top of the thick superconductor, with the thickness assumed to be infinite. Assuming no magnetic field penetrates the superconductor, they analyzed the electromagnetic proximity effect using the method of images.

\textbf{Method of images}---For the ferromagnet-superconductor bilayer, a convenient way to show the effect of the superconductor on the magnetization dynamics in the ferromagnet when coupled via the dipolar interaction is the method of images~\cite{method_of_image}.
The key idea of this approach is to regard the superconductor as a perfect diamagnet with the magnetic permeability $\mu\rightarrow-\infty$, as shown in Fig.~\ref{SF_shift_Fraerman}(a). 
This method disregards the finite penetration of the magnetic field due to the Meissner effect of the superconductor. Still, it qualitatively accounts for the interaction between the magnetization and the superconductor.
Under this assumption, ${\bf B}=0$ inside the superconductor. 
Combining with Maxwell's equation $\nabla\cdot {\bf B}=0$, the boundary condition that ${\bf B}_{\perp}$ is continuous requires the vanished normal component ${\bf B}_\perp$ at the interface. 
The method of images then treats the effect of the superconductor on the ferromagnet as the effective magnetization inside the superconductor, which generates the magnetic induction that obeys the boundary condition.

For a single magnetic moment ${\bf m}_{1}=\{m_x^{(1)},m_y^{(1)},m_z^{(1)}\}$ placed above the semi-infinite superconductor, the boundary condition is satisfied by introducing the image dipole ${\bf m}_{2}=\{-m_x^{(1)},m_y^{(1)},m_z^{(1)}\}$,  
as shown in Fig.~\ref{SF_shift_Fraerman}(a). 
For the arbitrary position ``$P$" at the boundary, the total magnetic induction ${\bf B}={\bf B}_{1P}+{\bf B}_{2P}$, where 
\begin{align}
   		{\bf B}_{1P}=\dfrac{\mu_0}{4\pi}\dfrac{3 {\bf n}_{1P}({\bf n}_{1P}\cdot{\bf m}_{1})-{\bf m}_{1}}{|R_{{1P}}|^3},\quad
        {\bf B}_{2P}=\dfrac{\mu_0}{4\pi}\dfrac{3 {\bf n}_{2P}({\bf n}_{2P}\cdot{\bf m}_{2})-{\bf m}_{2}}{|R_{2P}|^3},
\end{align}
are the magnetic induction generated by ${\bf m}_{1}$ and ${\bf m}_{2}$, respectively. 
From Fig.~\ref{SF_shift_Fraerman}(a), ${\bf n}_{1P}=\{n_x,n_y,n_z\}$ and ${\bf n}_{2P}=\{-n_x,n_y,n_z\}$ are the unit vector and $|R_{1P}|=|R_{2P}|=|R|$ are the distance from the magnetic moments to the position $P$ at the interface. Thus, the normal component
\begin{align}
    {\bf B}_\perp=&{\bf B}_{1P,\perp}+{\bf B}_{2P,\perp}\nonumber\\
    =&\dfrac{\mu_0}{4\pi}\left[\dfrac{3 n_x(n_xm_x^{(1)}+n_ym_y^{(1)}+n_zm_z^{(1)})-m_x^{(1)}}{|R|^3} + \dfrac{-3 n_x(n_xm_x^{(1)}+n_ym_y^{(1)}+n_zm_z^{(1)})+m_x^{(1)}}{|R|^3}\right]=0
\end{align}
obeys the boundary condition.

With the method of images to the ferromagnet-superconductor bilayer, the snapshot of the spin waves propagating in opposite directions with their imaged dipoles in the superconductors is illustrated in Fig.~\ref{SF_shift_Fraerman}(c) and (d).
For the positive and negative wave vectors $\bf q$ normal to the applied magnetic field ${\bf H}_{\rm ext}$, the distributions of the magnetic dipoles and their images differ, which induces different interactions between the dipoles and their images via the dipolar interaction. The pairwise interaction energies $E_{ij}$ of one dipole $i$ coupling with the other dipole $j$ read
\begin{align}
   E_{ij}= -{\bf m}_i\cdot {\bf B}_{j\rightarrow i}=\dfrac{\mu_0}{4\pi}\dfrac{-3 ({\bf n}_{ji}\cdot {\bf m}_i)({\bf n}_{ji}\cdot{\bf m}_j)+{\bf m}_i\cdot {\bf m}_j}{|R_{ji}|^3},
\end{align}
where ${\bf B}_{j\rightarrow i}$ is the magnetic induction at the dipole $i$ generated by the dipole $j$, $|R_{ji}|$ is the distance between the dipoles $i$ and $j$, and ${\bf n}_{ji}$ is the unit vector from  the dipole $j$ to $i$.
As seen from the system symmetry in Fig.~\ref{SF_shift_Fraerman}(b) and (c), the interaction energies between the dipoles in the ferromagnet and those in the superconductor depend on the wave propagation direction. 
The interaction energies between the dipole $1$ and dipole $2'$ with $+{\bf q}$ in Fig.~\ref{SF_shift_Fraerman}(b) are different from those between the dipole  $2$ and dipole $1'$ with $-{\bf q}$. Indeed, for the spin wave propagating to the left, 
${\bf n}_{2'1}$ is almost parallel to the magnetic dipoles ${\bf m}_{1}$ and ${\bf m}_{2'}$; for the spin wave propagating to the right, ${\bf n}_{2'1}$ is almost perpendicular to the magnetic dipoles. 
Thus, since the vector $({\bf n}_{12'}\cdot {\bf m}_1)({\bf n}_{12'}\cdot{\bf m}_2)$ is dependent on the spin wave propagation direction, generally $E_{12'}^+> E_{12'}^-$, where ``$+$" and ``$-$" correspond to $+{\bf q}$ and $-{\bf q}$ directions. 
The coupling energy for the spin waves propagating to the right is larger than that propagating to the left, implying the non-reciprocity of the spin-wave dispersion, i.e., the $+{\bf q}$ branch has an upper-frequency shift compared to the $-{\bf q}$ branch. 
Kuznetsov and Fraerman calculated the spin-wave dispersion in such a configuration using Maxwell's and London's equations with different London's penetration depths $l_s$. As shown in Fig.~\ref{SF_shift_Fraerman}(d),   clear non-reciprocity in the dispersion is revealed when $l_s=0$, while when $l_s\rightarrow\infty$, the non-reciprocity disappears due to the vanished superconductivity.

\subsubsection{Experiments}

\begin{figure}[htp!]
    \centering
    \includegraphics[width=0.98\linewidth]{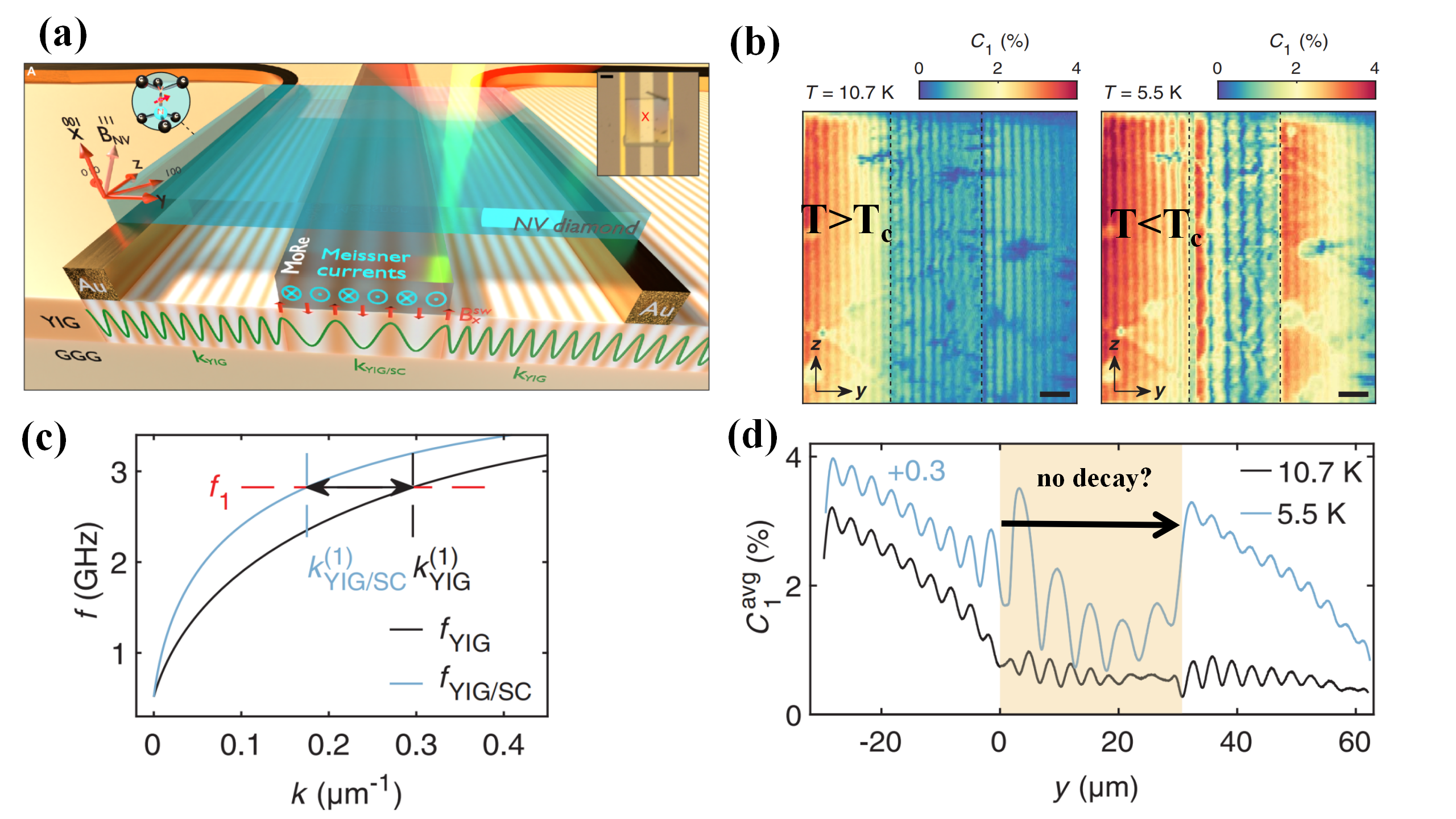}
    \caption{(a) Experimental configuration of gating spin waves by superconductors. (b) shows the spatial maps of the NV electron spin resonance contrast ``$C_1$" above and below the superconducting transition temperature $T_c$. (c) compares the spin-wave dispersion $f_{\rm YIG}$ for the bare YIG film and $f_{\rm YIG/SC}$ for the YIG film gated by a superconductor. The superconductor shifts the spin-wave dispersion upward, manifesting as an increase in wavelength at the excitation frequency $f_1$. (d) the data from (b) averaged over the $\hat{\bf z}$-direction, in which the yellow shading indicates the superconductor strip, and the amplitudes of spin wave between the superconductor seem not to decay.\\
    \textit{Source:} These figures are taken from Ref.~\cite{24_SF_shift_van_der_sar}.}
    \label{van_der_Sr_01}
\end{figure}

Recently, Borst \textit{et al.} observed the magnon frequency shift when gated by superconductors by using the NV center in diamonds that can locally measure the stray field emitted by the spin waves~\cite{24_SF_shift_van_der_sar}.
The experimental configuration 
is illustrated in Fig.~\ref{van_der_Sr_01}(a), in which two Au strips are fabricated on the YIG film as the spin wave generator.
A molybdenum-rhenium (MoRe) superconducting strip is placed in the middle of the sample as the spin wave modulator. The authors used a diamond membrane on top of the sample containing a thin layer of NV-center spins to image the spin waves. 
These NV-center spins are sensitive to magnetic fields, which can be used to detect the microwave magnetic stray fields emitted by spin waves. This enables imaging through optically opaque materials. The applied magnetic field is along one of the four possible crystallographic NV orientations to image the spin waves. 
When the applied magnetic field is much smaller
than the YIG saturation magnetization, the magnetization direction is fixed in the film plane due to the shape anisotropy, which is almost along the in-plane $\hat{\bf z}$-direction; therefore, the experimental configuration is similar to the Damon-Eshbach geometry as we discussed in Sec.~\ref{Chiral_gating_theory}.

By applying an AC electric current of microwave frequency to the gold
microstrip located at the left-hand side of the sample, the spin wave is excited with the wavevector ${\pmb\kappa}=k_y\hat{\bf y}$ that propagates to the right, as shown in Fig.~\ref{van_der_Sr_01}(b). 
From Fig.~\ref{van_der_Sr_01}(b), the spin waves propagating underneath the MoRe strip experience no change in wavelength when $T>T_c$.
However, when $T<T_c$, the spin waves underneath the MoRe strip are strongly modulated, i.e., their wavelength becomes much longer. 
The change of the wavelength may be understood by the frequency shift due to the back action of the superconducting Meissner current that is driven by the stray field of the spin waves~\cite{47_Gating_ferromagnetic_resonance}.
Indeed, when the spin waves propagate underneath the superconductor, their dispersion changes, resulting in the frequency shift as shown by the blue curve in Fig.~\ref{van_der_Sr_01}(c).  
The authors highlighted the spatial distribution of the spin waves in Fig.~\ref{van_der_Sr_01}(d). 
According to Fig.~\ref{van_der_Sr_01}(d), it is strange to observe that the amplitudes of the spin waves at the left and right-hand sides of the superconductor strip appear not to decay. This may imply that the spin waves can propagate longer under the superconductors.

\begin{figure}[htp!]
    \centering
    \includegraphics[width=\linewidth]{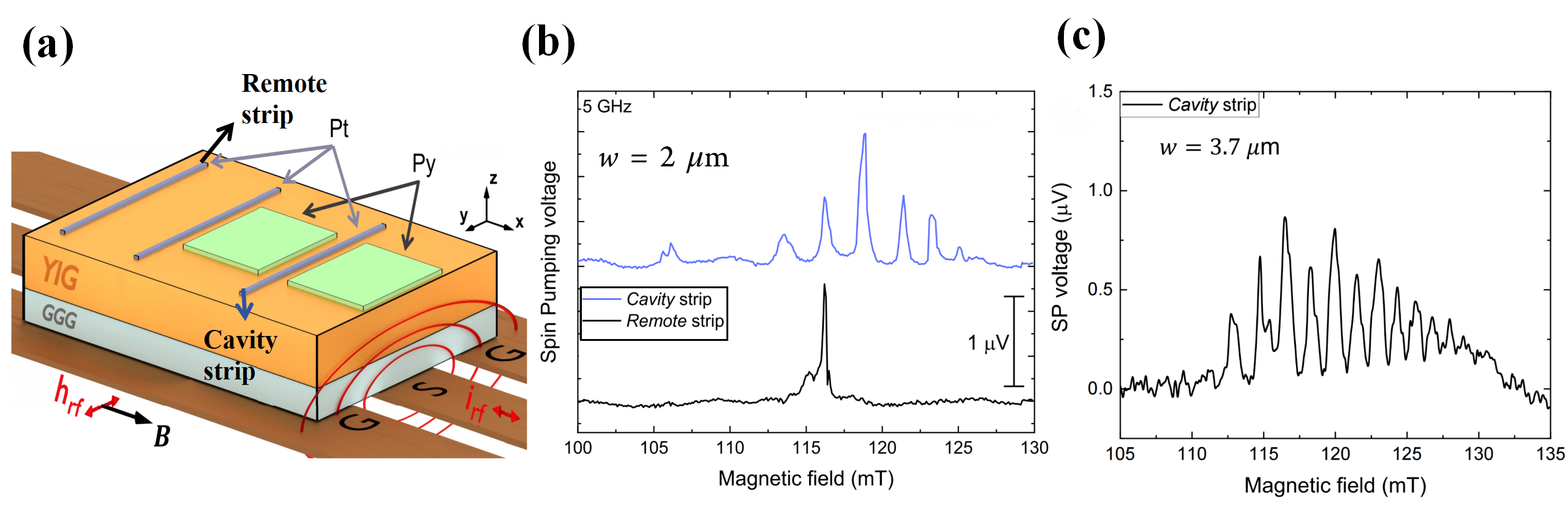}
    \caption{(a) Experimental configuration. (b) Spin pumping voltage $V^{sp}$ measured by the ``cavity strip" and ``remote strip" at the microwave frequency $f=5$~GHz by changing the magnitude of the magnetic field. (c) Spin pumping voltage $V^{sp}$ measured inside the cavity with the cavity width $w=3.7~\mu$m.\\
 \textit{Source:} These figures are taken from~\cite{magnon_trap}.
}
 \label{magnon_trap}
\end{figure}

The magnon trapping phenomenon was observed by Santos \textit{et al.}~\cite{magnon_trap} in a setup illustrated in Fig.~\ref{magnon_trap}(a). 
In this experiment, two Permalloys (Py) with the edge-to-edge distance $w$ along the $\hat{\bf x}$-direction are placed on top of the YIG film, which locally forms the YIG$|$Py bilayer regions. 
Due to exchange and dipolar interactions, these regions exhibit different magnetic properties. They act as reflecting boundaries for magnons, thus forming a magnonic cavity between the two Py layers. 
On the other hand, one Pt strip is placed in the middle of the cavity as the ``cavity strip", and two Pt strips placed outside the cavity are labeled as the ``remote strip."
The YIG film is grown on a GGG substrate, and the entire sample is placed on top of a stripe line waveguide, which is then connected to a vector network analyzer (VNA).
A static magnetic field ${\bf B}=\mu_0 H\hat{\bf x}$ is applied along the $\hat{\bf x}$-direction to bias the magnetization, so the microwave field ${\bf h}_{\rm rf}$ generated by the strip line waveguide is perpendicular to the applied magnetic field to excite the spin waves. 
Since the magnetic field ${\bf h}_{\rm rf}$ is nearly uniform over a relatively large scale when out of the cavity, only the ferromagnetic resonance (FMR) is excited; however, when inside the cavity, the frequency is expected to be quantized.
The spin-pumping voltage $V^{sp}$ detected by the Pt strip is used to identify the resonance modes, as spin-wave resonances in the YIG film can inject a spin current into the Pt strip via the spin-pumping effect. This spin current is then converted into the charge current by the inverse spin Hall effect, resulting in the spin pumping voltage $V^{sp}$.

Figure~\ref{magnon_trap}(b) shows the spin pumping voltage $V^{sp}$ measured by the cavity strip (the blue curve) and the remote strip (the black curve) at a constant microwave frequency $\sim 5$~GHz when changing the magnitude of the magnetic field.
For the $V^{sp}$ measured by the remote strip, there is only one peak, which represents the FMR since it obeys the Kittel formula~\cite{kittel_book}.
Several peaks with almost equal peak spacing were observed by the Pt strip inside the cavity, as shown by the blue curve, which implies the existence of the magnonic cavity. To identify the cavity mode,  the $V^{sp}$ by the cavity strip in the cavity is measured with different widths $w=\{1.6,2.0,3.7\}~\mu$m. 
When $w=3.7~\mu$m as in Fig.~\ref{magnon_trap}(c), the discrete peaks with smaller spacing and larger line width were observed. 
The measured cavity resonances follow the theoretical expectation \eqref{magnon_cavity} for a magnonic cavity from Refs.~\cite{46_Efficient_Gating,magnon_trap1}, leading to the spacing $\Delta H\propto1/w^2$ under the constant frequency $f$.

\begin{figure}[htp!]
    \centering
    \includegraphics[width=\linewidth]{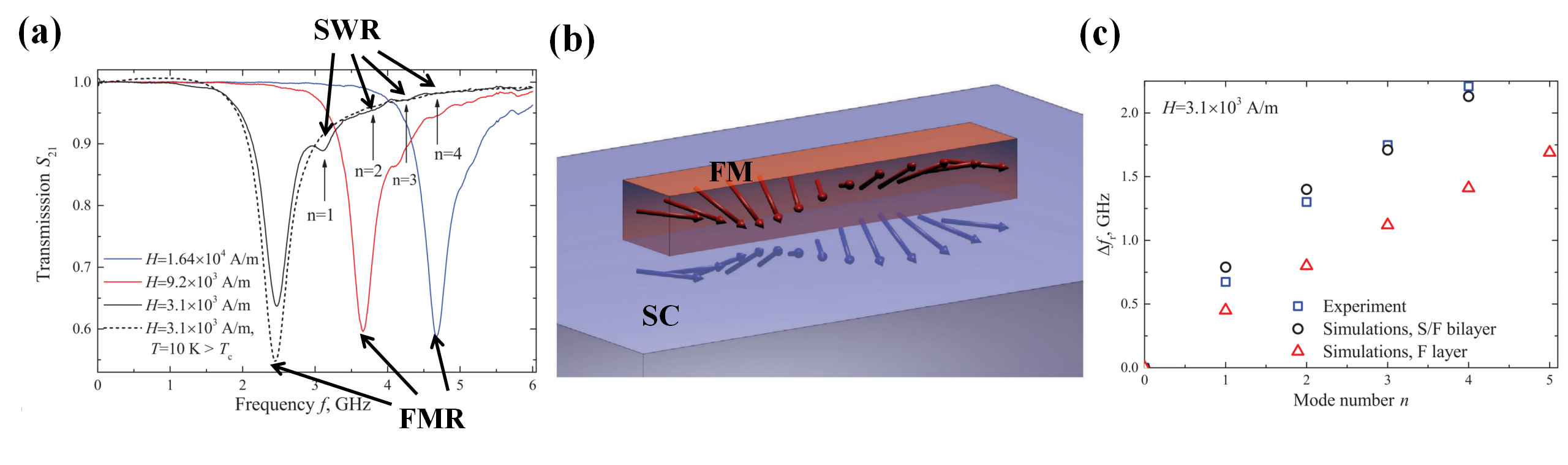}
    \caption{(a) Microwave transmission $S_{21}$ of different microwave frequencies $f$ under several fixed magnetic fields $H$. 
    (b) illustrates the method of images for a ferromagnetic film placed on an infinite superconductor, where the red arrow represents the magnetic moments within the ferromagnet, and the blue arrows indicate the imaged magnetic moments. 
    (c) Comparison of the simulated frequency difference $\Delta f$ with the experimental data.  \\
    \textit{Source:} These figures are taken from Ref.~\cite{2_FS_for_Magnonic_application_Ustinov}. } 
    \label{SF_shift}
\end{figure}

Golovchanskiy \textit{et al.} reported unconventional microwave absorptions in ferromagnet-superconductor bilayers~\cite{2_FS_for_Magnonic_application_Ustinov}. 
A permalloy film (Py) with width $W$ and thickness $d$ was placed on top of the coplanar waveguide made of Niobium (Nb). Microwave absorption spectroscopy was performed using a VNA.
As depicted by the black curve in Fig.~\ref{SF_shift}(a),  several additional weaker absorptions are observed except for the primary absorption below the superconducting transition temperature $T_c$, which are labeled by the integer $n=\{1,2,3,4\}$.
When increasing the magnetic field, the additional absorptions remain, as shown by the red and blue curves in Fig.~\ref{SF_shift}(a).
However, when $T>T_c$, the additional absorptions disappear, and the primary absorption has a narrow frequency shift compared to that when $T<T_c$, as shown by the dashed curve.
These distinct absorption features above and below $T_c$ imply the effect of superconductivity on the magnetization dynamics.
The strong absorption is attributed to the FMR mode in which its frequency $f_r$ depends on the applied field $H$, effective anisotropy field $H_{{\rm a}}$, and the magnetization saturation $M_s$ via the Kittel formula~\cite{Chen_FMR,kittel_book,Kittel_mode,Alvarez_2013}
\begin{align}
    2\pi f =\mu_0\gamma\sqrt{(H+H_a)(H+H_a+M_s)},
    \label{FMR_fit}
\end{align}
where $\gamma$ is the gyromagnetic ratio. 
By comparing the fitting parameters in Eq.~\eqref{FMR_fit} with the data when $T<T_c$ and $T>T_c$, it is found that the superconductivity increases the FMR frequency by enhancing the effective anisotropy field $H_a$ by about $8\%$: when $H=3.1\times 10^4$~A/m, the frequency shift is about $0.02$~GHz, i.e., $f_r=2.47$~GHz at $T<T_c$ and $f_r=2.45$~GHz at $T>T_c$, as shown in Fig.~\ref{SF_shift}(a).
The smaller absorptions represent the spin-wave resonance  (SWR). 
As labeled in Fig.~\ref{SF_shift}(a), the observed SWR mode is identified as the standing magnetostatic surface wave (MSSW) after excluding the possibilities of the forward (backward) volume modes and the perpendicular exchange standing spin waves~\cite{2_FS_for_Magnonic_application_Ustinov}.  Typically, the MSSW exists with an in-plane wave vector perpendicular to the direction of the in-plane magnetic field and follows the dispersion relation 
\begin{align}
    2\pi f =\mu_0\gamma\sqrt{(H+H_{\rm a})(H+H_{\rm a}+M_s)+M_s^2\left(1-e^{-2kd }\right)/4},
    \label{FMR_fitMSSW}
\end{align}
where $k=2\pi/\lambda$ with $\lambda$ quantized by the width of the sample, i.e., $(2n-1)\lambda_n/2=W=130~\mu$m with $n=\{1,2,\cdots,5\}$. 
However, Eq.~\eqref{FMR_fitMSSW} does not match the experimentally observed data, implying the impact of superconductivity on a spin-wave dispersion relation.

Golovchanskiy \textit{et al.} used the method of images~\cite{method_of_image,21_Fraerman}, as addressed in Fig.~\ref{SF_shift_Fraerman}(a), to analyze the effect of the superconductor on the spin-wave modes.
The interaction between the superconductor and the ferromagnet is treated as the magnetostatic interaction between two ferromagnets, i.e., the coupling of macro-spins in the ferromagnetic layer ${\bf M}_{\rm}=\{M_x,M_y,M_z\}$ with their mirror
images ${\bf M}_{\rm mirror}=\{M_x,M_y,-M_z\}$ in the superconductor, as shown in Fig.~\ref{SF_shift}(b), which allows to use the micromagnetic simulations to analyze the electromagnetic proximity effect.
The simulated results for the MSSWs with a single ferromagnetic layer and the superconductor$|$ferromagnet bilayer are compared with the experimental data in Fig.~\ref{SF_shift}(c), where $\Delta f_r=f_{\rm SWR}-f_{\rm FMR}$ is the frequency difference between the FMR and SWR mode. The simulation result agrees well with the experimental data. By comparing with the simulation result of the single ferromagnetic layer [the red triangles in Fig.~\ref{SF_shift}(c)], it is concluded that the superconductor affects the MSSW mode by shifting up its eigenfrequency.

\subsection{Chiral-gating-enhanced magnon transport by superconductors}
\label{Chiral_gating_enhanced_magnon_transport}

The experiments by Borst \textit {et al.} imply the enhanced transport of magnons beneath the superconductors~\cite{24_SF_shift_van_der_sar}.
 On the other hand, approximations for the boundary conditions applied in the  calculations~\cite{46_Efficient_Gating,21_Fraerman} of the modulation of spin waves by superconductors (Sec.~\ref{Chiral_gating_theory}) call for a rigorous treatment.

\begin{figure}[htp!]
    \centering
    \includegraphics[width=0.98\linewidth]{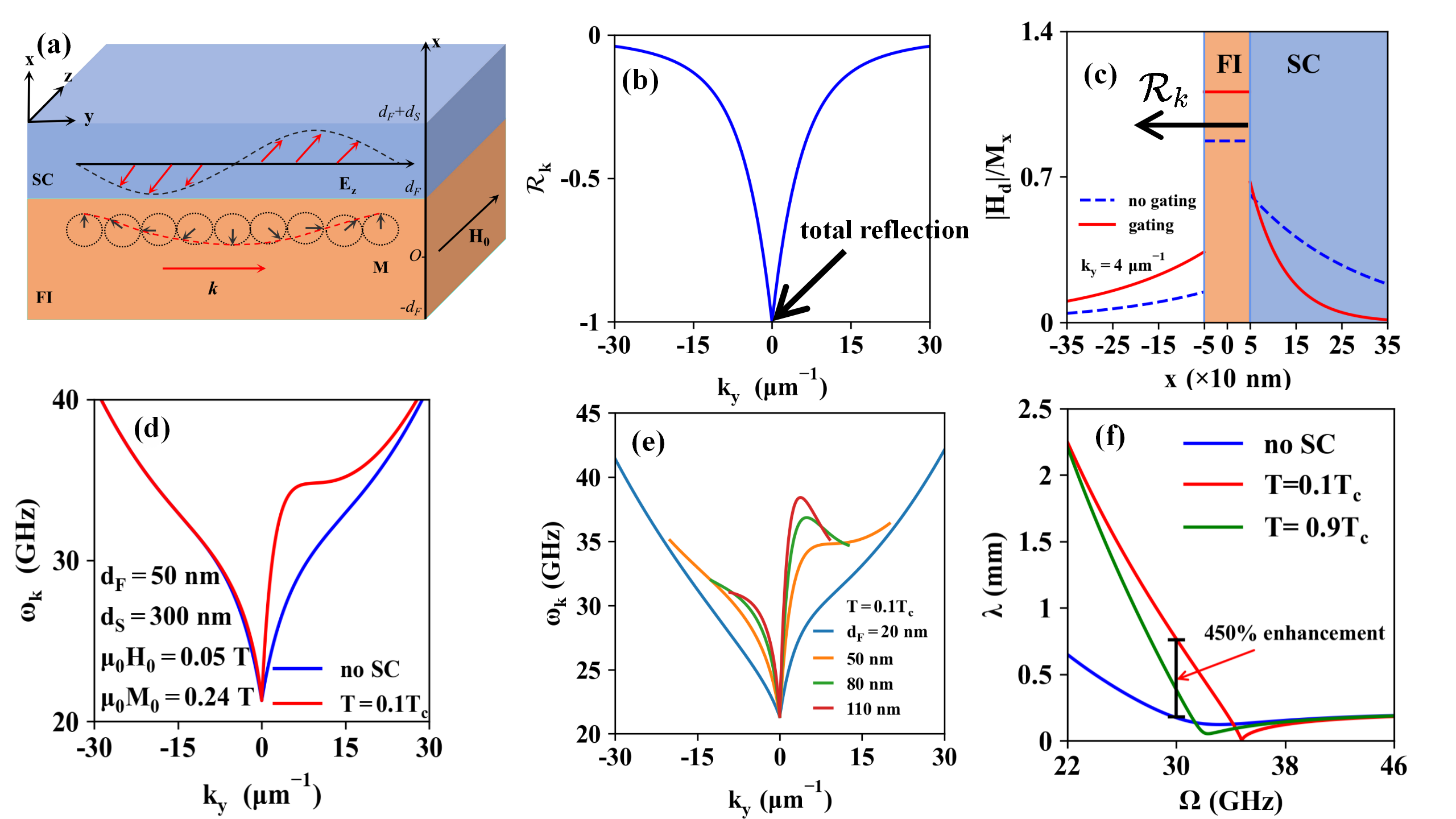}
    \caption{Enhanced magnon transport by superconductors. (a) In the superconductor$|$ferromagnetic insulator heterostructure, an in-plane external magnetic field ${\bf H}_0$ biases the saturation magnetization ${\bf M}_0$ to the $\hat{\bf z}$-direction. The spin waves propagating normally to ${\bf M}_0$ radiate the electric fields along ${\bf M}_0$. (b) Reflection coefficient ${\cal R}_k$ versus  $k_y$. For $k_y=0$, ${\cal R}_k=-1$ implies the total reflection of FMR. 
    (c) Enhancement of radiated magnetic field by the proximity superconductor with $k_y=4$~$\mu$m$^{-1}$. (d) Frequency shift by the superconductor gate when the temperature $T=0.1T_c$. (e) Dispersion relation at $T=0.1T_c$ with different ferromagnet thicknesses $d_F=\{20,50,80,110\}$~nm. (f)  Significant enhancement of magnon transport by superconductors as a function of excited frequency $\Omega$.\\
    \textit{Source:} Figures (a), (c), (d), and (f) are adopted from Ref.~\cite{48_Giant_enhancement_of_magnon_transport}. }
    \label{enhancement}
\end{figure}

Zhou \textit{et al.} rigorously consider the boundary conditions and solve consistently Maxwell's and London's equations in the superconductor$|$ferromagnet configuration~\cite{48_Giant_enhancement_of_magnon_transport}. They find that the superconductor chirally enhances the group velocity of magnons, enabling the unidirectional enhancement up to 450$\%$ to the magnon transport in a YIG film~\cite{48_Giant_enhancement_of_magnon_transport}. For the bilayer configuration~\cite{46_Efficient_Gating}, a superconductor thin film with thickness $d_S$ is fabricated above the ferromagnetic film with thickness $2d_F$. 
A static in-plane magnetic field ${\bf H}_0$ biases the saturation magnetization ${\bf M}_0$ to the $\bf \hat{z}$-direction, as shown in Fig.~\ref{enhancement}(a). 
In the bilayer, the electric field distribution is governed by Maxwell's and London's equations with their boundary conditions summarized in Table~\ref{Equations_of_E_and_H} and \ref{boundary}.
For the plane waves $M_{x,y}= {\cal M}_{x,y}e^{i ({\bf k}\cdot {\pmb \rho}-\omega t)}$, the electric field ${\bf E}({\bf r},t)={\pmb {\cal E}}(x)e^{i(\mathbf{k} \cdot {\pmb \rho}-\omega t)}$. According to Table.~\ref{Equations_of_E_and_H}, the amplitudes ${\pmb {\cal E}}(x)$ in different regions reads
 \begin{subequations}
\begin{align}
 &\text{in SC}:~~~ {\pmb{\cal E}}(x)={\pmb{\cal E}}_{1}e^{i {\cal B}_{k} x},\\
&\text{in FI}:~~~ {\pmb{\cal E}}(x)={\bf U} +{\pmb{\cal E}}_{0}e^{i {\cal A}_{ k} x}+{\pmb{\cal E}}_{0}'e^{-i {\cal A}_{ k} x}\label{sol_F},\\
&\text{in vacuum}~(x<-d_F):~~~ {\pmb{\cal E}}(x)={\pmb{\cal E}}_{2}e^{-i {\cal C}_k x}.
 \end{align}
\end{subequations}
In the ferromagnetic insulator (FI), Eq.~\eqref{sol_F} includes the general solution $\propto e^{\pm i{\cal A}_k x}$ due to the reflection by the interfaces and a special solution ${\bf U}=\omega\mu_0(-k_z {\cal M}_y/{\cal A}_{ k}^2, k_z {\cal M}_x/{\cal A}_{ k}^2, -k_y {\cal M}_x/{\cal A}_{ k}^2)$.  
The reflection by the other surface of the superconductor is disregarded when it is sufficiently thick with $d_S\gg \lambda_L$. 
For the microwave frequency $\omega\sim100$~GHz, $k_F\sim 0.01~\mu{\rm m}^{-1}$ and $k_0\sim 0.003~\mu {\rm m}^{-1}$ are tiny, such that ${\cal A}_{k}=\sqrt{k_F^2-k^2}\rightarrow ik$, ${\cal C}_k=\sqrt{k_0^2-k^2}\rightarrow ik$, and ${\cal B}_k=\sqrt{k_S^2-k^2}\rightarrow i\sqrt{k^2+1/\lambda_L^2}$. The coefficients $\{\pmb{\cal E}_0,\pmb{\cal E}_0',\pmb{\cal E}_1,\pmb{\cal E}_2\}$ are solved by the boundary conditions of the electric and magnetic fields at different interfaces (Table.~\ref{boundary}).

For the experimental Damon-Eshbach configuration with $k_z=0$, the spin waves are modulated most strongly by the superconductors~\cite{24_SF_shift_van_der_sar}. 
By solving the coefficients $\{\pmb{\cal E}_0,\pmb{\cal E}_0',\pmb{\cal E}_1,\pmb{\cal E}_2\}$ according to the boundary conditions, the electric field inside the FI reads  
\begin{align}
    E_z({\bf r},t)&=E_z^{(0)}({\bf r},t)+{\cal R}_kE_z^{(0)}({\bf r},t)|_{x=d_F}e^{-i{\cal A}_k(x-d_F)},
    \label{electric_field_h}
\end{align}
expressed as a superposition of the electric field emitted by a single FI and its reflection by the superconductor via a wave-number dependent reflection coefficient ${\cal R}_{k}=({\cal A}_k-{\cal B}_k)/({\cal A}_k+{\cal B}_k)$, where  $k=|k_y|$ in the Damon-Eshbach configuration, and $E_z^{(0)}$ is the electric field emitted by a single FI layer.

Equation~\eqref{electric_field_h} shows that the influence of the superconductor on the ferromagnet is governed by the reflection coefficient ${\cal R}_{k}$. Figure~\ref{enhancement}(b) shows how the superconductor affects the electric field by plotting the wave vector dependence of the reflection coefficient ${\cal R}_k$. 
For a large wave vector $k\gg \lambda_L^{-1}$, the emitted electric field rapidly decays within the decay length $k^{-1}\ll \lambda_L$, such that $|{\cal A}_k|\sim |{\cal B}_k|$, implying a vanished reflection ${\cal R}_k\sim ({\cal A}_k-{\cal A}_k)/(2{\cal A}_k) \rightarrow0$, i.e., the superconductor hardly affects the spin waves. For the FMR with $k=0$, $|{\cal A}_k|\ll |{\cal B}_k|$ and ${\cal R}_k\sim -{\cal B}_k/{\cal B}_k=-1$.  In this case, the radiative electric field with the wavevector ${\bf k}=\sqrt{\omega^2\mu_0\varepsilon_0}\hat{ \bf x}$ is totally reflected by the superconductor with a $\pi$ phase shift, resulting in no net supercurrent driven inside the superconductor, i.e., the Kittel magnon and Cooper-pair supercurrent decouples~\cite{47_Gating_ferromagnetic_resonance}. The superconductor modulates the spin waves for small $k$ since ${\cal R}_k$ rapidly departs from $-1$. The spin waves with $k\sim \lambda_L^{-1}$ are most strongly modulated by the superconductors.

To address the influence of the superconductor on the spin waves, it is necessary to determine the magnetic stray field inside the ferromagnet, as it directly couples with the magnetization.
According to Eq.~\eqref{electric_field_h} and Faraday's Law, the magnetic field decays within the decay length $\sim 1/k$ inside the FI. 
For a thin ferromagnetic insulator with thickness $d_F<1/k$, we are allowed to take the spatial average for the stray field along the surface normal $\hat{\bf x}$-direction, such that 
\begin{align}
H_{d,y}&=M_y(
P_{k}e^{-kd_F}-1)+{\cal R}_{k}M_y{P_{k}e^{-kd_F}}\left(1-e^{-2kd_F}\right)\frac{1+a_{k_y}\text{sgn}{(k_y)}}{2} 
\equiv \kappa_y(k_y)M_y,\nonumber\\
 H_{d,x}&=-M_x{P_ke^{-kd_F}}
  +{\cal R}_kM_x{P_ke^{-kd_F}}\left(1-e^{-2kd_F}\right)\frac{1+a_{k_y}^{-1}\text{sgn}{(k_y)}}{2}\equiv\kappa_x(k_y)M_x,
  \label{dipolar_field}
\end{align}
where the form factor $P_k=\sinh{(kd_F)}/(kd_F)$ and the amplitudes ${\cal M}_x=-ia_{k_y} {\cal M}_y$. 
The ellipticity $a_{k_y}=1$ indicates the circular polarization. Accordingly, the magnetic field is also expressed as the superposition of emission by an isolated ferromagnet and its reflection by the superconductors. When $|k_y|\gg \lambda_L^{-1}$,   ${\cal R}_k\rightarrow0$ as in Fig.~\ref{enhancement}(b), the influence of the superconductor vanishes, such that the dipolar field recovers to that of a single FI.
On the other hand, when $k_y=0$, ${\cal R}_k=-1$; however, different from the physics in Eq.~\eqref{electric_field_h},  the term $\propto {\cal R}_k(1-e^{-2k d_F})=0$ according to Eq.~\eqref{dipolar_field}, such that the superconductor does not affect the magnetic field distribution inside the FI and its FMR. 
When $k_y\in[0,\lambda_L^{-1}]$, the magnetic field is enhanced inside the FI assuming $a_{k_y}\sim 1$.  The chirality of the dipolar magnetic field is now clear: for circularly polarized spin waves with $a_{k_y}\rightarrow1$,  $(1+a_{k_y}\text{sgn}(k_y))/2\rightarrow 0$ and $(1+a^{-1}_{k_y}\text{sgn}(k_y))/2\rightarrow 0$ when $k_y<0$, but they become unity when $k_y>0$. Figure~\ref{enhancement}(c) shows the effect of the reflection by plotting the magnetic field distribution with and without the superconductor gate. When $k_y=4$~$\mu$m$^{-1}$,  the stray magnetic field mainly exists above the ferromagnetic film $x>0$ without gating; when the ferromagnetic film is in proximity to the superconductor, the magnetic field is reflected and decays faster inside the superconductor, which is then enhanced inside the ferromagnetic film.

The enhanced magnetic field inside the FI modulates the properties of spin waves. Substituting Eq.~\eqref{dipolar_field} into the LLG equation, the eigenfrequency, inverse lifetime, and ellipticity of the collective modes read 
\begin{subequations}
\begin{align}
    \omega({k_y})&=\mu_0 \gamma \sqrt{(H_{k_y}-\kappa_x({k_y})M_0)(H_{k_y}-\kappa_y({ k_y})M_0)},\label{S_F_dispersion}\\
    \Gamma({ k_y})&=\dfrac{\mu_0\gamma\alpha_G}{2}\left[2H_{k_y}-M_0\left(\kappa_x(k_y)+\kappa_y(k_y)\right)\right],\label{inverse_lifetime}\\
    a({k}_y)&=\sqrt{(H_{k_y}-\kappa_x(k_y)M_0)/(H_{k_y}-\kappa_y(k_y)M_0)},
   \label{ellipticity}
\end{align}
\label{S_F_solution}
\end{subequations}
where $H_{k_y}=H_0+\alpha_{\rm ex}k_y^2M_0$, noting in Eq.~\eqref{dipolar_field}
\begin{align}
  \kappa_y(k_y)=&  
P_{k}\left(e^{-kd_F}+{\cal R}_{k}{e^{-kd_F}}\left(1-e^{-2kd_F}\right)\frac{1+a_{k_y}\text{sgn}{(k_y)}}{2}\right)-1, \nonumber\\
\kappa_x(k_y)=&P_k\left(
  {\cal R}_k{e^{-kd_F}}\left(1-e^{-2kd_F}\right)\frac{1+a_{k_y}^{-1}\text{sgn}{(k_y)}}{2}-{e^{-kd_F}}\right).\nonumber
\end{align}
Equation~\eqref{S_F_solution} provides self-consistent equations for determining the dispersion, inverse lifetime, and ellipticity, which call for numerical calculation.
The superconductor induces an upward chiral shift in the spin wave dispersion: as in Fig.~\ref{enhancement}(d),  the dispersion shifts upward when $k_y\in[0,\lambda_L^{-1}]$ because the superconductor enhances the magnetic field inside the FI; when $k_y\gg \lambda_L^{-1}$ or $k_y \leq 0$,  the superconductor's influence on the magnetic field vanishes, and the dispersion of spin waves recovers to the case with no superconductors. Figure~\ref{enhancement}(e) displays the dispersion relations for different ferromagnet thicknesses $d_F$. The frequency shift increases with $d_F$. Such a frequency shift is theoretically predicted~\cite{46_Efficient_Gating,41_Eremin} and experimentally verified~\cite{24_SF_shift_van_der_sar}.  The magnon transport properties in this configuration have received less attention.

The superconductor enhances the magnon transport by significantly increasing the spin-wave group velocity and slightly reducing its lifetime. Using the experimental configuration in Ref.~\cite{24_SF_shift_van_der_sar}, we consider the gated spin waves excited by an external Oersted magnetic field ${\bf h}({\bf r},t)$ of frequency $\Omega$, emitted from a long stripline along the saturation magnetization $\hat{\bf z}$-direction. 
The excited magnetic field couples with the transverse magnetization via the Zeeman interaction
$\hat{H}_{\text{int}}=-\mu_0\int {\bf M(r)}\cdot{\bf h}({\bf r},t) d{\bf r}$. According to the linear response theory~\cite{spin_susceptibility}, the excited 
Fourier components of the magnetization
$M_{\alpha}(x,k_{y},\Omega)=\mu_{0}(\gamma\hbar)^{2}\int_{-d_F}^{d_F}dx^{\prime}\chi_{\alpha\beta}(x,x^{\prime},k_{y},\Omega) h_{\beta
}(x^{\prime},k_{y},\Omega)$,
where $h_\beta(x,k_y,\Omega)=(1/L)\int dy e^{-i k_y y} h_\beta(y,t)$ is the Fourier components of the Oersted magnetic field with $L$ being the sample length, and the spin susceptibility
\begin{equation}
\chi_{\alpha\beta}(x,x^{\prime},\mathbf{k},\Omega)=-\frac{2M_0}{\gamma\hbar^2
}{\cal M}_{\alpha}^{\mathbf{k}}(x){\cal M}_{\beta}^{\mathbf{k}\ast}(x^{\prime})\frac{1}{\Omega-\tilde{\omega}_{\mathbf{k}}}.
\end{equation}
The amplitudes ${\cal M}_y^{\bf k}=\dfrac{1}{2\sqrt{2d_F}}\dfrac{1}{\sqrt{a_{\bf k}}}$ and ${\cal M}_x^{\bf k}=-\dfrac{i}{2\sqrt{2d_F}}{\sqrt{a_{\bf k}}}$ are normalized according to the normalization condition Eq.~\eqref{renormalization_relation}.
Summing over the Fourier components leads to the excited magnetization in real space
\begin{align}
  M_{\alpha}(y,t)&=\sum_{k_y}e^{ik_yy}M_{\alpha}(k_y,t)
  =4iL\mu_0\gamma d_F M_0{\cal M}_{\alpha}^{k_\Omega}{\cal M}_{\beta}^{k_\Omega*}\dfrac{1}{v_{k_\Omega}}e^{i (k_\Omega y-\Omega t)}e^{-\tilde{k}_{_\Omega}y}h_\beta(k_\Omega),
\end{align}
where $\tilde{k}_y=k_{\Omega}+i\tilde{k}_{\Omega}$ is the positive root of $\tilde{\omega}(\tilde{ k}_y)=\Omega$ and $v_{\bf k}=\partial\omega({\bf k})/\partial {\bf k}$ is the group velocity of spin waves. The decay length $\lambda=1/\tilde{k}_{\Omega}$ governs the magnon transport property. The solution of equation  $\tilde{\omega}(\tilde{ k}_y)=\Omega$ shows that the characteristic decay length obeys a simple relation with
\begin{align}
\lambda=1/\tilde{k}_{\Omega}={v_{k_\Omega}}/\Gamma(k_\Omega)=v_{k_\Omega}\tau(k_\Omega),
\label{decay_length}
\end{align} 
governed by the group velocity and the magnon lifetime $\tau(k)=1/\Gamma(k)$. According to Fig.~\ref{enhancement}(d) and (e), the magnon group velocity is strongly enhanced when $k_y\in[0,\lambda_L^{-1}]$.
The magnon lifetime is only slightly affected~\cite{46_Efficient_Gating}.
As plotted in Fig.~\ref{enhancement}(f), the decay length of the excited magnetization becomes much larger due to the screening of the stray field by superconductors. The enhancement of magnon transport for YIG can be as large as $450\%$, resulting in the decay length exceeding millimeters.

\textbf{Absence of equilibrium current}.---The asymmetric dispersion $\omega({\bf k})\neq \omega({-\bf k})$ could raise the issue whether an equilibrium current ${\bf J}^{(0)}=\sum_{\bf k}{ v}_{\bf k} n(\omega({\bf k}))$ exists in such systems, in which $n(\omega({\bf k}))=({e^{\beta \hbar\omega({\bf k})}-1})^{-1}$ is the equilibrium magnon  Bose-Einstein distribution. Here, we demonstrate that the equilibrium magnon current vanishes even with the asymmetric dispersion modulated by the superconductors.

We first consider the one-dimensional case. The group velocity $v_k=d\omega(k)/dk$; the magnon current
\begin{align}
	J^{(0)}=\sum_k v_k n(\omega({ k}))\rightarrow\dfrac{1}{2\pi} \lim_{k_1,k_2\rightarrow\infty}
	\int_{-k_2}^{k_1} \dfrac{d\omega(k)}{dk}n(\omega({ k}))dk.
\end{align}
We calculate such an integral analytically by dividing it into two parts
\begin{align}
\lim_{k_1,k_2\rightarrow\infty}\int_{-k_2}^{k_1} \dfrac{d\omega(k)}{dk}n(\omega({ k}))dk&=\lim_{k_1,k_2\rightarrow\infty}\left(	\int_{-k_2}^{0} \dfrac{d\omega(k)}{dk}n(\omega({ k}))dk+	\int_{0}^{k_1} \dfrac{d\omega(k)}{dk}n(\omega({ k}))dk\right)\nonumber\\
&=\lim_{k_1,k_2\rightarrow\infty}\left(\int_{\omega(-k_2)}^{\omega(0)} n(\omega({ k}))d\omega(k)+\int_{\omega{(0)}}^{\omega{(k_1)}} n(\omega({ k}))d\omega(k)\right)\nonumber\\
 &=\lim_{k_1,k_2\rightarrow\infty}\left(\dfrac{1}{\beta\hbar} \left(\ln{\frac{|1-e^{\beta\hbar\omega{(k_1)}}|}{|1-e^{\beta\hbar\omega{(-k_2)}}|}}\right)+(\omega{(-k_2)}-\omega{(k_1)})\right).
	\label{tak}
\end{align}
In the continuum model, when $\{k_1,k_2\}\rightarrow\infty$, $\{\omega(k_1),\omega(-k_2)\}\rightarrow\infty$, such that $e^{\beta\hbar\omega(k_1)}\gg 1$ and $ e^{\beta\hbar\omega(-k_2)}\gg 1$. The limit \eqref{tak} implies the absence of an equilibrium current:
\begin{align}
&\lim_{k_{1,2}\rightarrow\infty} \left(\dfrac{1}{\beta\hbar} \left(\ln{\frac{|1-e^{\beta\hbar\omega(k_1)}|}{|1-e^{\beta\hbar\omega(-k_2)}|}}\right)+(\omega(-k_2)-\omega(k_1))\right)=\lim_{k_{1,2}\rightarrow\infty}\left(\dfrac{1}{\beta\hbar} \left(\ln{\frac{e^{\beta\hbar\omega(k_1)}}{e^{\beta\hbar\omega(-k_2)}}}\right)+(\omega(-k_2)-\omega(k_1))\right)\nonumber\\
&=\lim_{k_{1,2}\rightarrow\infty}\left(\omega(k_1)-\omega(-k_2)+(\omega(-k_2)-\omega(k_1))\right)=0.
    \label{limit}
\end{align}

Equilibrium current is also absent for the two-dimensional case, \textit{i.e.,} $ {\bf k}=k_y \hat{\bf y}+k_z \hat{\bf z}$, with
\begin{align}
	{\bf J}^{(0)}=\sum_{\bf k} v_{\bf k} n(\omega({\bf k}))\rightarrow\dfrac{1}{(2\pi)^2} 
	\iint_{-\infty}^{\infty}\dfrac{\partial\omega({\bf k})}{\partial {\bf k}}n(\omega({\bf k}))dk_ydk_z.
\end{align}
To prove that, we use Green's formula
\begin{align}
	\oint_L (Pdk_y+Qdk_z)=\iint_\infty\left(\dfrac{\partial Q}{\partial k_y}-\dfrac{\partial P}{\partial k_z}\right)dk_ydk_z.
\end{align}
By taking $Q=0$ and $P=-\left(\frac{\ln{|1-e^{\beta\hbar\omega({\bf k})}|}}{\beta\hbar}-\omega({\bf k})\right) $ such that $-\partial P/\partial \omega({\bf k})=n(\omega({\bf k}))$, we obtain
\begin{align}
J^{(0)}_z&\propto
\iint_{-\infty}^\infty \dfrac{\partial\omega({\bf k})}{\partial k_z}n(\omega({\bf k}))dk_ydk_z= 
-\iint_{-\infty}^\infty \dfrac{\partial\omega({\bf k})}{\partial k_z}\dfrac{\partial P}{\partial \omega({\bf k})}dk_ydk_z	\nonumber\\
&=-\iint_{-\infty}^\infty \dfrac{\partial P}{\partial k_z}dk_ydk_z=\oint_L Pdk_y,
\end{align}
where the integral path of $\oint_L Pdk_y$ is along the edge of the integration region. Again $\omega({\bf k})\rightarrow\infty$ when $|{\bf k}|\rightarrow\infty$, so
\begin{align}
	\lim_{|{\bf k}|\rightarrow\infty}P=&-\lim_{|{\bf k}|\rightarrow\infty}\left(\frac{\ln{|1-e^{\beta\hbar\omega({\bf k})}|}}{\beta\hbar}-\omega({\bf k})\right)
	=0;
\end{align}
thus $J^{(0)}_z\propto\oint_L P dk_y=0$. 
Similarly, for ${ J}^{(0)}_y$, 
we take  $P=0$ and $Q=\left(\frac{\ln{|1-e^{\beta\hbar\omega({\bf k})}|}}{\beta\hbar}-\omega({\bf k})\right)$ such that $\partial Q/\partial \omega({\bf k})=n(\omega({\bf k}))$. 
Accordingly, 
\begin{align}
J^{(0)}_y=\dfrac{1}{(2\pi)^2} 
\iint_{-\infty}^\infty \dfrac{\partial\omega({\bf k})}{\partial k_y}n(\omega({\bf k}))dk_ydk_z = \dfrac{1}{(2\pi)^2}	\oint_L Q dk_z=0.
\end{align}
Thus, even with the asymmetric dispersion $\omega({\bf k})\neq \omega({-\bf k})$, the equilibrium magnon current ${\bf J}=\sum_{\bf k}{ v}_{\bf k} n_{\bf k}(\omega({\bf k}))$ vanishes.

\subsection{Magnonic crystals created by microstructured and type II superconductors}

\subsubsection{Theory}

Magnonic crystals are artificial magnetic materials with a periodic modulation of magnetic parameters that can alter the dispersion and properties of spin waves~\cite{MC1,MC2,MC3,MC4,MC5,MC6,MC7}.

\begin{figure}[htp!]
    \centering
    \includegraphics[width=0.95\linewidth]{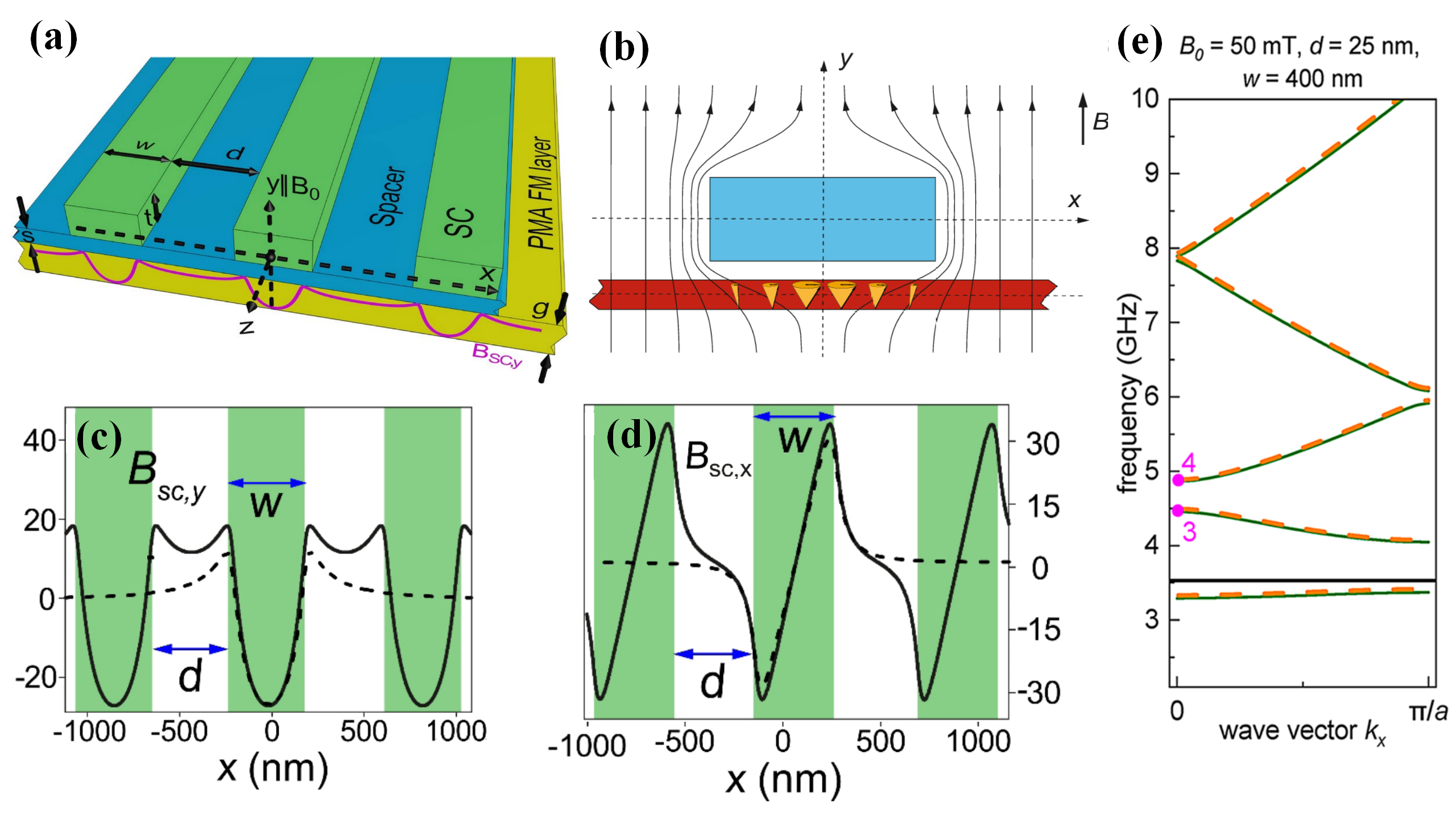}
    \caption{(a) shows the configuration of the magnonic crystal, in which the periodic magnetic field $B_{{\rm SC},y}$ (the magenta curve) inside the ferromagnetic layer strongly modulates the dispersion of the spin wave. (b) shows the magnetic field distribution of a single superconductor strip. (c) and (d) display the induced magnetic field $B_{sc,y}$ and $B_{sc,x}$ in the ferromagnetic film by a single superconductor strip (the dashed curves)  and the superconductor strip array (the solid curves), respectively. (e) Spin wave dispersion of the magnonic crystal.\\
   \textit{ Source:} Figures (a), (c), (d), (e) are taken from Ref.~\cite{42_Centala}; Fig.~(b) is taken from Ref.~\cite{37_klos}.
    }
    \label{MC_Centala}
\end{figure}

Recently, Kharlan \textit{et al.} proposed using the diamagnetism of superconductors to realize a periodic magnetic field inside the ferromagnetic film generated by an array of superconducting strips~\cite{42_Centala}. 
Figure~\ref{MC_Centala}(a) illustrates a magnonic crystal configuration that includes the gallium-doped YIG (Ga:YIG) ferromagnetic film and an array of Nb superconducting strips.
A nonmagnetic spacer, with a thickness of $s$, is inserted between the superconductor strips and the ferromagnet to avoid the exchange proximity effect at the superconductor$|$ferromagnet interface, such that only the electromagnetic coupling is allowed in this configuration. 
Different from the experimental configuration in Ref.~\cite{Golovchanskiy_gating3_exp}, the external static magnetic field ${\bf B}_0$ is applied \textit{perpendicular} to the superconducting strips along the normal $\hat{\bf y}$-direction, with the spatial distribution of the static magnetic field is already modulated by the superconductors. The dynamical electromagnetic proximity effect is disregarded in interpreting the experimental data.
The Ga:YIG ferromagnetic film exhibits strong perpendicular magnetic anisotropy, resulting in an out-of-plane magnetization ground state.

The static magnetic field inside the superconductors generates the persistent Meissner supercurrent, which induces the Oersted magnetic field that affects the distribution of the magnetic field near the superconductor strips. 
The magnetic field near the superconductor is strongly modulated by the superconductor, as described in Fig.~\ref{MC_Centala}(b): at $x=0$ below the center of the superconductor, the magnetic field is strongly suppressed and becomes along the $\pm \hat{\bf x}$-direction. 
When close to the left and right edges of the superconductor, the magnetic field along the $\hat{\bf y}$-direction is enhanced.
The distribution of $B_{{\rm SC},y}$ generated by a single superconductor stripe is shown by the dashed curve in Fig.~\ref{MC_Centala}(c).
 On the other hand, the component $B_{{\rm SC},x}$  has the opposite sign to $x=0$ at the left and right edges of the superconductor, as shown by the dashed curve in Fig.~\ref{MC_Centala}(d). 
 For an array of superconducting strips spaced at an interval $d$, both $B_{{\rm SC},x}$ and $B_{{\rm SC},y}$ exhibit periodic distributions, as illustrated by the solid curves in Figs.~\ref{MC_Centala}(c) and (d). 

The periodic magnetic field has a strong effect on the characteristics of spin waves. The dispersion of the spin wave along the $\hat{\bf x}$-direction is shown in Fig.~\ref{MC_Centala}(e), where the black curve represents the FMR frequency and the dashed orange and solid green curves represent the analytical and simulation results, respectively. The frequency gap is observed. The analytical result agrees well with the simulation result.

\subsubsection{Experiments}

\begin{figure}[htp!]
    \centering
    \includegraphics[width=0.95\linewidth]{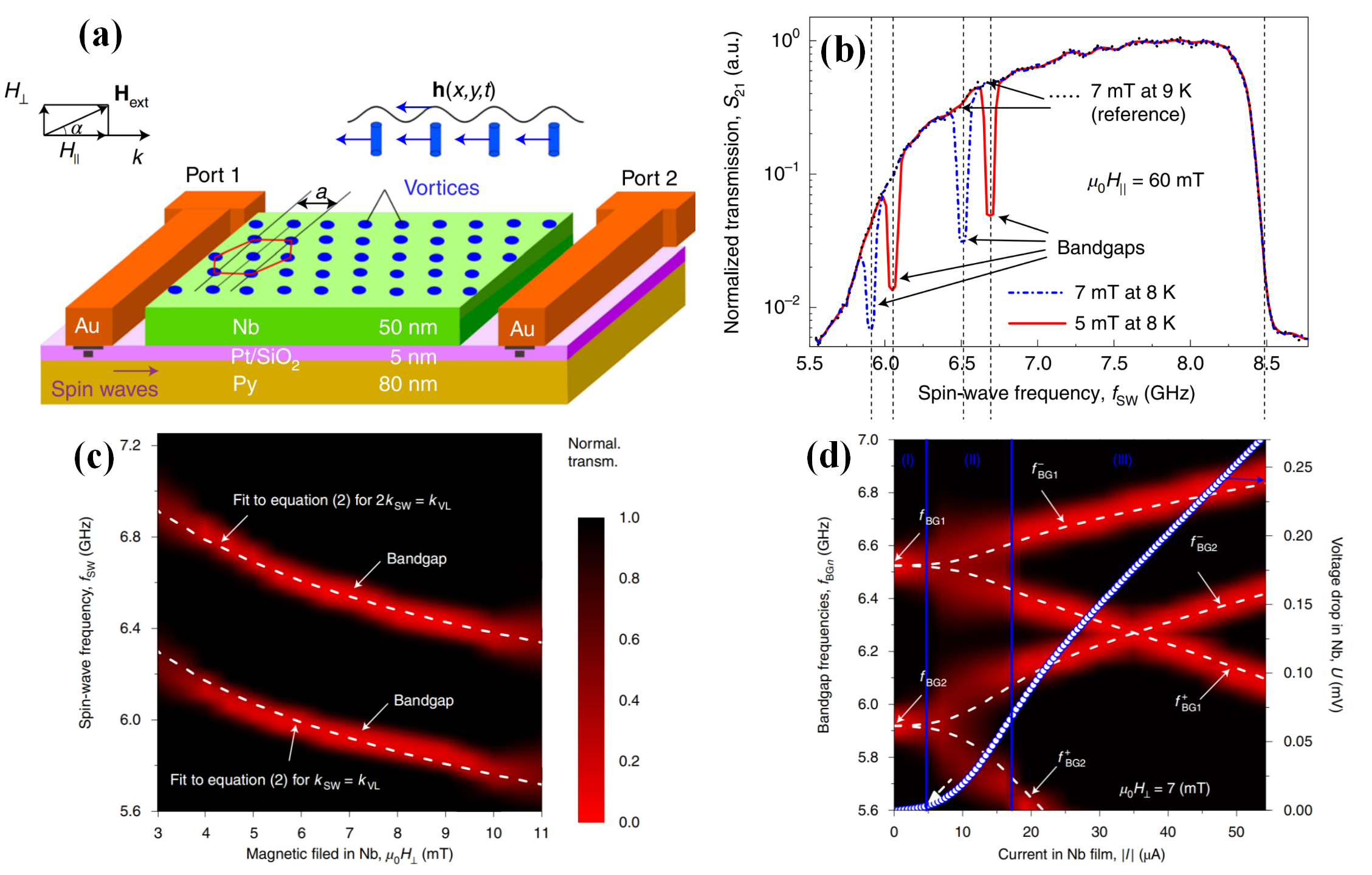}
    \caption{(a) Experimental configuration. 
    (b) Measured band gaps in the microwave transmission with a series of out-of-plane magnetic field $\mu_0 H_{\perp}$.  (c) Normalized microwave transmission through the Py$|$Nb bilayer as a function of the magnetic field $\mu_0 H_{\perp}$. The dashed curves are calculated according to Eq.~\eqref{chumark_Bragg} for the central band gap frequencies. (d) Normalized microwave transmission as a function of 
the current in the Nb layer. The dashed curves are plotted according to Eq.~\eqref{chumark_doppler}.\\
\textit{Source:} These figures are taken from Ref.~\cite{5_magnon_fluxon_Nat_Chumak}.
}
\label{chumark}
\end{figure}

In 2019, the Chumak group demonstrated that the type II superconductor can gate the spin-wave dispersions~\cite{5_magnon_fluxon_Nat_Chumak,19_magnon_fluxonics_Chumak,34_Berakdar,39_Cherenkov_radiation_Buzdin}. In this device, an out-of-plane magnetic field creates an array of vortices that provides a periodic static magnetic field, which tunes the band structure of magnons. As in Fig.~\ref{chumark}(a), a superconducting Nb film is placed above the ferromagnetic Py waveguide. The Py waveguide is covered with a 5-nm-thick Pt or SiO$_2$ cap layer to prevent the degradation of the magnetic properties on the surface. 
Two Au microwave antennas were produced on top of the waveguide, separated by 5.5~$\mu$m and positioned perpendicular to the Py spin-wave waveguide. The external magnetic field includes two components that serve different purposes. The in-plane component of the field is set as $\mu_0H_\parallel=59.5$~mT in all measurements, which is aligned along the Py waveguide and magnetizes it along its long axis. 
This field component sets the spin wave propagation to the backward volume magnetostatic spin wave configuration. The vertical component $\mu_0H_\perp$ creates a vortex lattice in the Nb superconducting film. 
The vortex lattice can be viewed as small cylinders whose cores are in the normal state, arranged in a hexagonal lattice. Each vortex can be regarded as a tiny whirl of supercurrent, producing a local magnetic field. Such a magnetic field enters the Py film and couples with the spin waves. 
Since the vortices are arranged in a hexagonal lattice, the spin wave can experience a similarly periodic potential. 
Such a bilayer forms a magnonic crystal with a specific band gap in the spin wave transport spectrum.

To detect the band gap, the microwave is applied to the excitation antenna (port 1) and detected at the detection antenna (port 2). Figure~\ref{chumark}(b) shows the transmission spectrum $S_{21}$ of microwaves from port 1 to port 2. 
The black curve is measured above $T_c$, at which there is no vortex lattice, and the red and blue curves are measured in the presence of the vortex lattice ($T<T_c$) with $\mu_0 H_\perp=5$~mT and $7$~mT, respectively. According to the blue curves, several characteristic dips are visible in the spin wave transmission, which implies the frequency band gaps.
When changing the strength of $\mu_0 H_\perp$, the characteristic dips shift, as shown by the red curve. The authors interpreted such dips as the Bragg scattering of spin waves at the vortex lattice. That is, the frequency band gap is opened at the Bragg condition $2a_{\rm VL}=n\lambda_{\rm SW}$, where $a_{\rm VL}=(2 \Phi_0/\sqrt{3}H_\perp)^{1/2}$ is lattice constant of the vortices and 
$\Phi_0=h/(2e)$ is the quantum of magnetic flux. 
Thus, the matching condition for the wave vectors of spin waves reads
 \begin{align}
     k_{\rm SW}(f_{{\rm BG},n})={n \pi (2\sqrt{3} H_{\perp}\Phi_0)^{1/2}}/{2},
     \label{chumark_Bragg}
 \end{align}
 where $n=\{1,2,\cdots\}$ is the order of the Bragg resonance and $f_{{\rm BG},n}$ is the bandgap frequency. Figure~\ref{chumark}(c) shows a series of normalized transmission spectra through the Py waveguide as a function of $\mu_0 H_{\perp}$. The dashed curves follow Eq.~\eqref{chumark_Bragg}, showing excellent agreement with the experimental data for the first two resonance conditions ($n=1,2$).

 When a current ${\bf I}$ is applied perpendicular to the Py spin-wave waveguide in the Nb layer, the vortex lattice experiences a Lorentz force ${\bf F}_{\rm L}={\bf I}\times {\bf B}_{\rm VL}$ and is driven into motion, where ${\bf B}_{\rm VL}$ represents the magnetic field of the vortex lattice. The coupling of magnons with this moving lattice allows for the observation of the spin-wave Doppler effect~\cite{Spin_Wave_Doppler_Yu,Spin_Wave_Doppler1,Spin_Wave_Doppler2,Spin_Wave_Doppler3,Spin_Wave_Doppler4,Spin_Wave_Doppler5,Spin_Wave_Doppler6}. As shown in Fig.~\ref{chumark}(d), when the vortices move in the same direction as the spin 
waves propagate, $f_{\rm BG1}$ and $f_{\rm BG2}$ shift towards higher frequencies, which are labeled as $f_{\rm BG1}^-$ and $f_{\rm BG2}^-$. In contrast, by reversing the direction of the current, the vortices and the spin waves travel in 
opposite directions and $f_{\rm BG1}$ and $f_{\rm BG2}$ shift towards lower frequencies, which are labeled as $f_{\rm BG1}^+$ and $f_{\rm BG2}^+$. 
As shown by the dashed curves in Fig.~\ref{chumark}(d), the observed band gap agrees well with the expected position of the Doppler-shifted band gap
\begin{align}
    f_{{\rm BG},n}^{\pm}(k)=f(k_{{\rm BG},n}\pm \Delta k_n^\pm),&&\Delta k_n^\pm=k(v_{\rm SW})-k(v_{\rm SW}\pm v_{\rm VL}),
    \label{chumark_doppler}
\end{align}
where 
$v_{\rm SW}$ is the corresponding spin-wave velocity at frequency $f_{{\rm BG},n}$ and $v_{\rm VL}\propto |I|$ is the vortex lattice velocity.
Note that for lower currents [see region (I) in Fig.~\ref{chumark}(d)], the spin wave spectrum is unaffected. This is because the vortices are pinned, and only when the Lorentz force ${\bf F}_{\rm L}$ is larger than the pinning force ${\bf F}_{\rm p}$, i.e., $|I|>I_d$ with $I_d$ the critical current (depinning current), the vortex lattice starts to move. 
Using this property, in 2022, Dobrovolskiy and Chumak~\cite{19_magnon_fluxonics_Chumak} further considered the Nb microstrips with asymmetric pinning, i.e., the positive critical current $I_d^+$ is not equal to the negative critical current $I_d^-$. They apply an AC current with a maximum value satisfying $|I_d^+|<|I_{\rm max}|<|I_d^-|$, then the band gap shifts upward for $I>I_d^+$ but is not affected when $I<0$. Thereby, they realized the rectification (vortex diode or ratchet) effect, which can be used to create highly sensitive, dynamically tunable radio-frequency filters with ultralow energy consumption.

\begin{figure}[htp!]
      \centering
      \includegraphics[width=\linewidth]{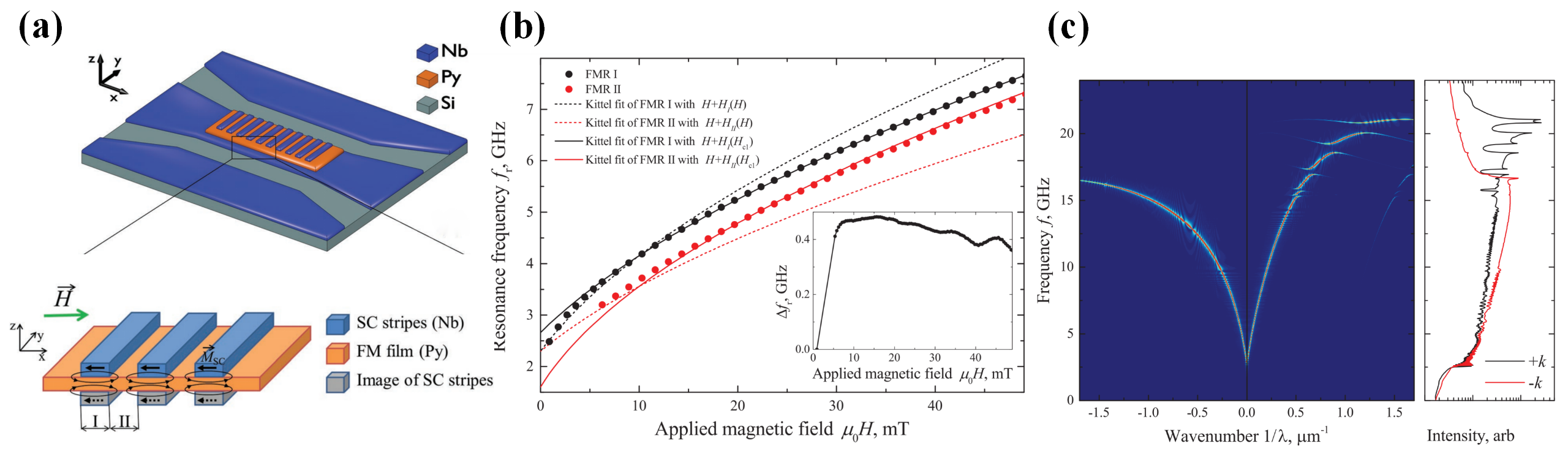}
      \caption{(a) Schematic illustration of the magnonic crystal and its phenomenological model. (b) Dependencies of the FMR frequency on the external magnetic field $H$. (c) Band structure of the hybrid magnonic crystal. Here, the right panel shows the intensity of the MSSW for $+k$ and $-k$ versus frequency.\\
      \textit{Source:} These figures are taken from Ref.~\cite{Golovchanskiy_gating3_exp}.}
      \label{MC_Golovchanskiy}
  \end{figure}

 Golovchanskiy \textit{et al.} fabricated the magnonic crystal using the superconductor strips on top of a ferromagnetic film~\cite{Golovchanskiy_gating3_exp}, as shown in the top panel of Fig.~\ref{MC_Golovchanskiy}(a). 
The Nb strips of the triangular cross-section are uniformly spaced on top of the Py film, which is fabricated on top of the superconducting coplanar waveguide made of Nb. 
The FMR absorption spectroscopy is measured using the VNA. 
The measured FMR spectrum is shown in Fig.~\ref{SF_shift}(e) when $T<T_c$.  
Two FMR signals are unexpectedly observed, which are labeled as ``FMR I" by the black dots and ``FMR II" by the red dots.
It is found that such two modes are merged into one signal when $T>T_c$, implying the unique effect of the superconductors. 

Golovchanskiy \textit{et al.} explained such an effect by the method of images, i.e., the effect of the superconductor on the FMR of Py is equivalent to the impact of the imaged magnetization inside the superconductors.  
Here, the superconductors are treated as ideal diamagnets. Thus, inside the superconductors, ${\bf B}=0$, such that the applied external magnetic field ${\bf H}$ induces the effective magnetization ${\bf M}_{\rm sc}(H)=-{\bf H}$ everywhere inside the superconductor strips.
Since the sample is placed on top of the superconducting Nb waveguide, each superconducting stripe with induced magnetization ${\bf M}$ is accompanied by its mirrored image with respect to the surface of the superconducting waveguide, as shown in the bottom panel of Fig.~\ref{MC_Golovchanskiy}(a). 
These induced and imaged magnetizations ${\bf M}$ emit the 
\textit{static} stray magnetic field in the Py film along the $\hat{\bf x}$-direction as shown by the black arrows in the bottom panel of Fig.~\ref{MC_Golovchanskiy}(a). In this analysis, the effect of the superconductors on the dynamical magnetization in the Py film is disregarded. 
The periodic superconductor strips divide the ferromagnetic film into two regions: Region I, where the Py is covered by the superconductor strip with its image, and Region II, where no superconductor is covered. 
The different environments of Region I and Region II imply different magnetization dynamics.
In particular, the average magnetic field in the Py film along the $\hat{\bf x}$-direction that is induced by the superconducting stripes in Region I is $H_{\rm I}(H)\approx0.18 H$ and in Region II is $H_{\rm II}(H)\approx-0.27H$. 
Thereby, the experimental data may be fit using Eq.~\eqref{FMR_fit} by replacing $H$ to $H+H_{\rm I}(H)$ and $H+H_{\rm II}(H)$, corresponding to two FMR.
As shown by the dashed curves in Fig.~\ref{MC_Golovchanskiy}(b), the Kittel formula \eqref{FMR_fit} with renormalized magnetic field fits the measurement well at low magnetic fields $\mu_0H<10$~mT. 
However, the fitting deviates from the experiment when $\mu_0 H>10$~mT. 
This could be due to the destruction of complete diamagnetism of the superconductor above the first superconducting critical field $H_{\rm c1}$. 
In this case, the magnetic 
flux starts to penetrate the Nb strip in the form of Abrikosov vortices, and the ideal diamagnetic response of Nb stripes to the external magnetic field ceases. 
Thus, when $\mu_0 H>\mu_0 H_{\rm c1}\approx10$~mT, the authors adjust the fitting by fixing the diamagnetic response of superconductors at $H_{\rm c1}$, i.e., using Eq.~\eqref{FMR_fit} by replacing $H$ with $H+H_{\rm I}(H_{\rm c1})$ and $H+H_{\rm II}(H_{\rm c1})$. 
Both fit well at a high magnetic field, which is shown by the solid curves in Fig.~\ref{MC_Golovchanskiy}(b). 
The authors emphasized that such FMR spectrum is nonlinear since it has different features when $H<H_{\rm c1}$ and $H>H_{\rm c1}$, which is also shown in the inset of Fig.~\ref{MC_Golovchanskiy}(b), where the FMR difference of the two regions $\Delta f_r=f_{\rm I}-f_{\rm II}$ increases rapidly when magnetic field is changed from $0$ to $10$~mT, and then decreases slowly at higher magnetic field. Golovchanskiy \textit{et al.}~\cite{Golovchanskiy_gating3_exp} performed numerical simulations for the MSSW in the magnonic crystal geometry where the superconducting strips are aligned along the $\hat{\bf x}$-direction in Fig.~\ref{MC_Golovchanskiy}(a) but without the superconductor film below the Py film. 
The simulation result shows a clear non-reciprocity of the dispersions as shown in Fig.~\ref{MC_Golovchanskiy}(c). For the positive wave numbers $1/\lambda>0$, the dispersions reveal forbidden bands at a high frequency $f>15$~GHz. However, the dispersion is continuous for the negative wave numbers $1/\lambda<0$.

\section{Ultrastrong coupling between magnons and photons in SC$|$FM$|$SC Josephson junctions}
\label{Ultrastrong_coupling_in_SC_FM_SC_Josephson_junctions}

\subsection{Giant shift of ferromagnetic resonance by Meissner effect}

\subsubsection{Experiments}

\begin{figure}[htp!]
    \centering
    \includegraphics[width=\linewidth]{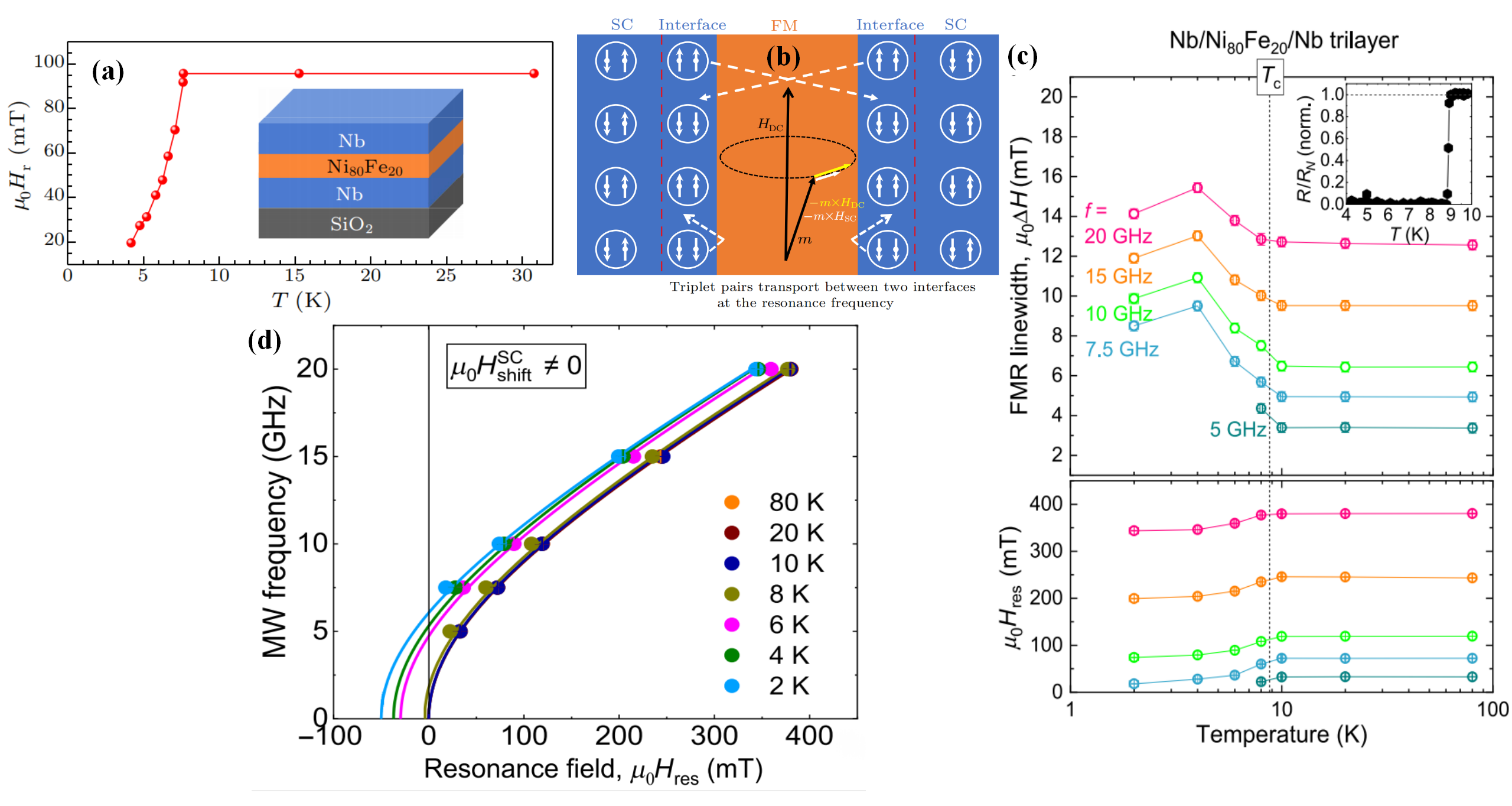}
    \caption{FMR shifts in the superconductor$|$ferromagnet$|$superconductor Josephson junctions. (a) Resonance field shift in Nb(100~nm)$|$Ni$_{80}$Fe$_{20}$(20~nm)$|$Nb(100~nm) trilayer. (b) Possible mechanism that the spin-triplet Cooper pairing is induced at the superconductor$|$ferromagnet interfaces. Such Cooper pairs with the same spin to the magnetization can pass through the ferromagnet and provide the extra torque. (c) FMR linewidth and resonance field $\mu_0 H_{\rm res}$ above and below $T_c$ in the  Nb(100~nm)$|$Ni$_{80}$Fe$_{20}$(15~nm)$|$Nb(100~nm) trilayer. (d) Numerical fitting with $\mu_0 H_{\rm shift}^{\rm sc}\neq 0$ for the experimental data.\\   
\textit{Source:} The figures are taken from Refs.~\cite{4_sun} [(a) and (b)], and~\cite{7_SFS_shift_Blamire} [(c) and (d)].  }
    \label{shift_exp}
\end{figure}

In 2018, Li \textit{et. al.} measured the FMR in a series of superconductor$|$ferromagnet heterostructures~\cite{4_sun}. 
As shown in Fig.~\ref{shift_exp}(a), they fabricated the ferromagnetic Josephson junction with a ferromagnet layer of Ni$_{80}$Fe$_{20}$ (20~nm in thickness) sandwiched by two superconducting layers of Nb (100~nm in thickness); the heterostructure is fabricated on a SiO$_2$ substrate. The FMR is measured at X-band ($\sim$ 9~GHz) in a cavity with the in-plane resonance magnetic field $H_r$.
In such a heterostructure, the superconducting transition temperature $T_c\sim 8.7$~K. According to Fig.~\ref{shift_exp}(a), the resonance field changes slightly when the temperature is above $T_c$; however, when the temperature is decreased below $T_c$, the resonance field $H_r$ rapidly shifts downward. 
For example, when the temperature is $\sim 4.2~$K, the resonance field has a giant shift $\sim 70$~mT compared to the $H_r$ above $T_c$, comparable to half of the bare FMR resonance field. The giant shift indicates a strong modulation of the magnetization dynamics by the superconductor. Li \textit{et. al.} explained the effect via the spin-transfer torque induced by the spin-triplet supercurrents, as shown in Fig.~\ref{shift_exp}(b). When the temperature $T<T_c$, the spin-triplet Cooper pairs are induced by the spin-mixing and spin-flip scattering at the interfaces between the ferromagnet and superconductor; the spin-singlet Cooper pairs only exist away from the SC$|$FM interfaces. The spin-triplet Cooper pairs are spin-polarized, so the coherent charge and spin transport occur due to the total angular momentum conservation between magnons and their induced spin-triplet Cooper pairs. 
The spin-triplet Cooper pairs with up spins can spread to the ferromagnet layer; however, those with down spins are reflected by the ferromagnet. The triplet Cooper pairs passing through the ferromagnet with the same spins exert a strong torque on the precessing magnetization, which has the same direction as the torque generated by the applied magnetic field, resulting in the resonance field $H_r$ shifting to a lower field.

In 2019, a similar experiment was performed by Jeon \textit{et al.}~\cite{7_SFS_shift_Blamire}. However, they came up with a different explanation: the strong Meissner effect and the magnetic flux pinning of the thick adjacent superconductor layers cause the frequency shift. The experiment is organized as follows: the Nb$|$Ni$_{80}$Fe$_{20}|$Nb trilayers are deposited on thermally oxidized Si substrates. The thickness of the Nb film, $100$~nm, is comparable to its London's penetration depth, and the thickness of Ni$_{80}$Fe$_{20}$ is $15$~nm.
The sample is attached to the broadband coplanar waveguide with a static magnetic field $\mu_0H$ parallel to the film plane. The FMR spectrum is measured by sweeping $\mu_0 H$ at the fixed microwave frequencies $f\sim 5-20$~GHz, around which the microwaves are absorbed by the sample at the resonance frequencies.  
The FMR data are well fitted by the Lorentzian functions, with the FMR linewidth $\mu_0\Delta H$ and the resonance field $\mu_0H_{\rm res}$ accurately determined.
The temperature dependence of $\mu_0 \Delta H$ and $\mu_0 H_{\rm res}$ are summarized in Fig.~\ref{shift_exp}(c).
When $T>T_c$, both the linewidth and resonance field are almost independent of temperature; however, when $T<T_c$, $\mu_0 \Delta H$ broadens largely and the resonance field $\mu_0 H_{res}$ shifts significantly to a lower field. The resonance field shifts about $30~$mT when $T$ decreases to 2~K at $f=20$~GHz and shifts 45~mT at $f=10$~GHz; the effect is more pronounced for a lower $f$. 
For a quantitative analysis, the linewidth $\mu_0 \Delta H$ at different frequencies $f$ relates to the Gilbert damping via $\mu_0 \Delta H(f)=\mu_0 \Delta H_0 +4\pi \alpha f /(\sqrt{3} \gamma) $, where $\mu_0 \Delta H_0$ is the zero frequency line broadening and $\gamma$ is the gyromagnetic ratio. 
The extracted $\alpha$ decreases when the temperature $T<T_c$ decreases. The authors explained it by the suppressed outflow of spin currents from the precessing magnetization in Ni$_{80}$Fe$_{20}$ due to the development of singlet Cooper pairing in the adjacent Nb layers. 
The resonance-field shift could be explained by the trapped magnetic flux inside the ferromagnet due to the screening of the neighboring Nb layers, which results in the modified Kittel formula 
\begin{align}
  2\pi  f=\mu_0\gamma\sqrt{ \left(H_{\rm res}+H_{\rm shift}^{\rm SC} +M_{\rm eff}\right) \left(H_{\rm res}+H_{\rm shift}^{\rm SC}\right)}.
  \label{FMR_jeon}
\end{align}
Here, $ M_{\rm eff}$ is the effective saturation magnetization and $\ H_{\rm shift}^{\rm SC}$ is the correction due to the superconductors. As shown in Fig.~\ref{shift_exp}(d), all the experiment data fits well by a suitable $H_{\rm shift}^{\rm SC}\neq 0$.

\begin{figure}
    \centering
    \includegraphics[width=0.75\linewidth]{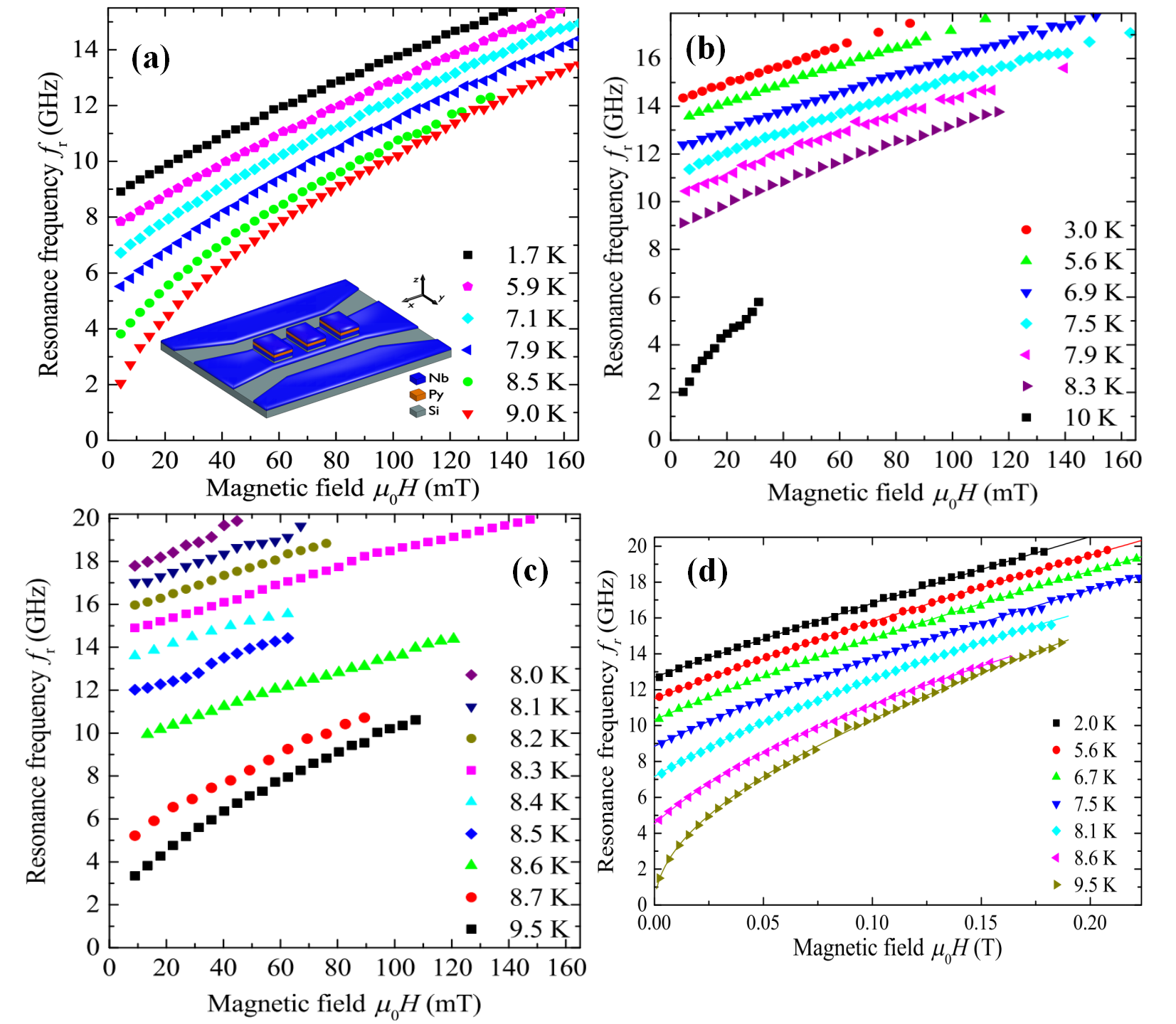}
    \caption{FMR frequencies as a function of external magnetic fields at different temperatures and in different samples Nb(110~nm)$|$Py(19~nm)$|$Nb(110~nm) [(a)],  Nb(140~nm)$|$Py(45~nm)$|$Nb(140~nm) [(b)],  Nb(110~nm)$|$Py(350~nm)$|$Nb(110~nm) [(c)], and Nb(100~nm)$|$Py(35~nm)$|$Nb(100~nm) [(d)]. 
    The solid curves at $T<T_c$ in (d) show the corresponding fitting with Eq.~\eqref{FMR_shift_fit}.  \\
    \textit{Source:} Figures (a), (b), and (c) are taken from Ref.~\cite{10_SFS_shift_Ustinov}; Fig.~(d) is taken from Ref.~\cite{25_SFS_shift_Stolyarov}.  }
    \label{shift_exp1}
\end{figure}

Golovchanskiy \textit{et. al.} performed similar measurements~\cite{10_SFS_shift_Ustinov,25_SFS_shift_Stolyarov}. As shown in Fig.~\ref{shift_exp1}(a), the coplanar waveguide is made by the superconductor Nb of thickness 150~nm. A series of Nb$|$Py$|$Nb films with a spacing of 25~$\mu$m along the ${\bf \hat{x}}$-direction are placed directly on the waveguide. A Si spacer of 20-nm thickness is deposited between the waveguide and the three Nb$|$Py$|$Nb heterostructures to suppress their interlayer exchange interaction and ensure their electrical insulation.
The measured resonance frequency as a function of magnetic fields \(H\) is shown in Fig.~\ref{shift_exp1} for the three samples Nb(110~nm)$|$Py(19~nm)$|$Nb(110~nm) [Fig.~\ref{shift_exp1}(a)], Nb(140~nm)$|$Py(45~nm)$|$Nb(140~nm)  [Fig.~\ref{shift_exp1}(b)], and Nb(110~nm)$|$Py(350~nm)$|$Nb(110~nm) [Fig.~\ref{shift_exp1}(c)].
When the temperature decreases below $T_c$, the FMR shifts to a higher frequency for all three samples; the FMR shift is enhanced when increasing the thickness of Nb and Py~\cite{10_SFS_shift_Ustinov}. As shown in Fig.~\ref{shift_exp1}(a), the frequency shifts about 6~GHz when $\mu_0 H=20~$mT at $T\sim2$~K compared to the frequency when $T>T_c$. While in Fig.~\ref{shift_exp1}(b), the frequency shift becomes $\sim 10$~GHz with the thicker Nb and Py films. In Fig.~\ref{shift_exp1}(c), as the thickness of Py is further increased, the frequency shift reaches about 14~GHz at $T\sim 8$~K.
 In Ref.~\cite{25_SFS_shift_Stolyarov}, Golovchanskiy \textit{et al}. further measured a series of Nb$|$Py$|$Nb samples with different thicknesses. 
 As shown by the solid curves in Fig.~\ref{shift_exp1}(d) for the Nb(100~nm)$|$ Py(35~nm)$|$Nb(100~nm) heterostructure, the experimental data fits well with the modified Kittel formula 
 \begin{equation}
    2\pi f_r=\mu_0 \gamma\sqrt{\left[H+H_a+H_s(1-\alpha_s H^2)\right]\left(H+H_a+M_{\rm eff}\right)},
     \label{FMR_shift_fit}
 \end{equation}
 where $H_a$ and $M_{\rm eff}$ are the uniaxial anisotropy field and the effective saturation magnetization, respectively,  $H_s$ is the effective torque by the superconductors that acts as the fitting parameter, and  $\alpha_s$ considers the dependence of the torque on the applied field. From Eq.~\eqref{FMR_shift_fit}, the superconductor provides a field-like torque $H_s(1-\alpha_sH^2)$ that modifies the resonance frequency. The authors concluded that the mechanism behind the superconducting torque is
 not the spin transfer torque induced by the spin-triplet Cooper pairs~\cite{4_sun} or the giant demagnetization effect~\cite{16_Giant_demagnetization_effects_Buzdin}, but is purely electromagnetic~\cite{20_shift_Silaev}, i.e., the magnetization precession at the SC$|$F interface induces macroscopic superconducting currents in superconductor layers. These currents induce a magnetic field in the opposite phase to the precession of $M$ that shifts the frequency.

\begin{figure}[htp!]
    \centering
    \includegraphics[width=1.05\linewidth]{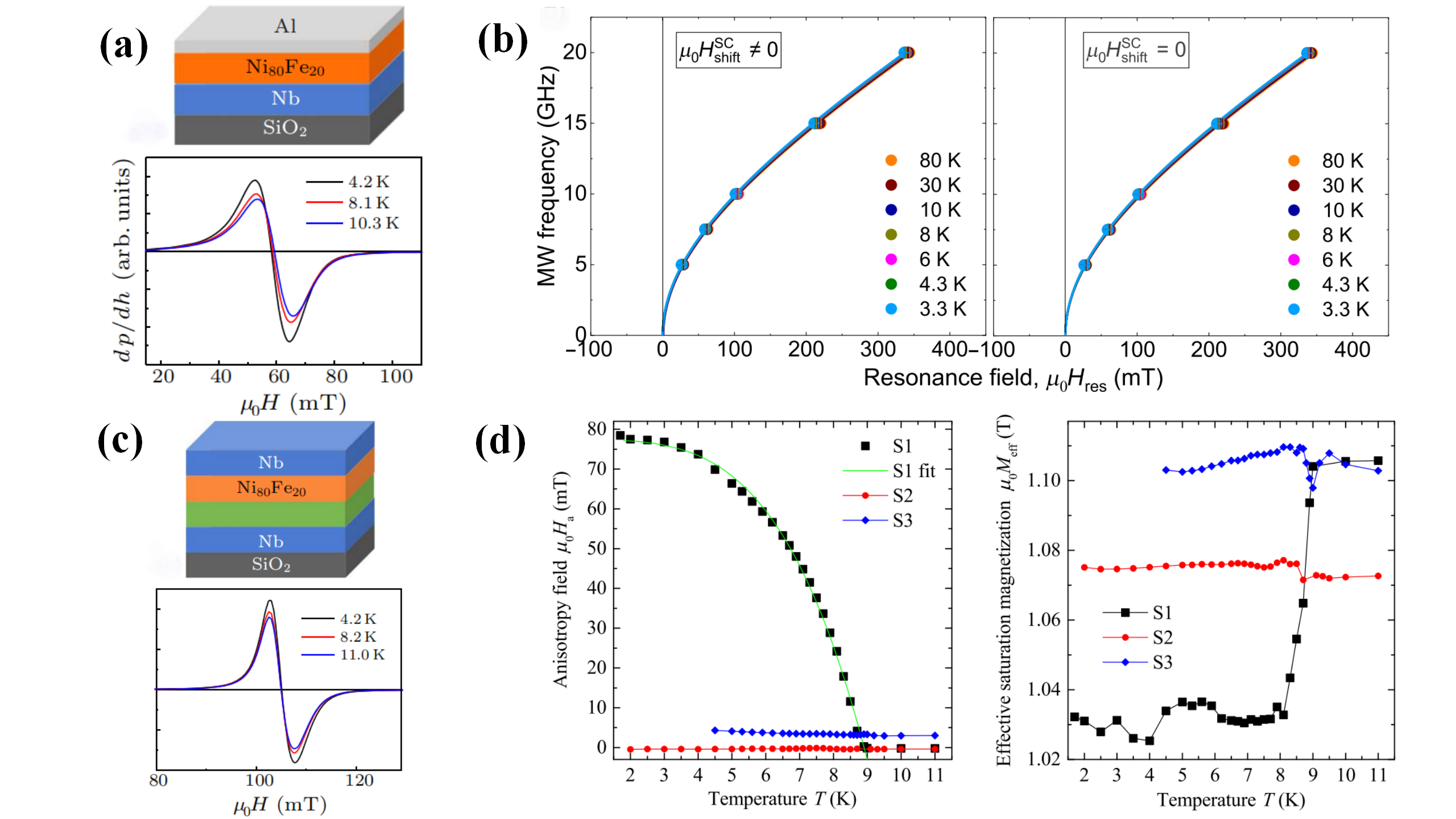}
    \caption{Absence of frequency shift in different configurations. (a) FMR spectra of a Nb(100~nm)$|$Ni$_{80}$Fe$_{20}$(20~nm) bilayer. The resonance field $H_r$ shifts slightly below $T_c$. 
    (b) Data fitting with and without the correction term $\mu_0 H_{\rm shift}^{SC}$ for the measurements of the Nb(100~nm)$|$Ni$_{80}$Fe$_{20}$(15~nm) bilayer. (c) FMR spectra of a Nb(100~nm)$|$MgO(10~nm)$|$Ni$_{80}$Fe$_{20}$(20~nm) multilayer. (d) Temperature dependence of the anisotropy field $\mu_0 H_a$ and effective saturation magnetization $\mu_0 M_{\rm eff}$. \\
    \textit{Source:} The figures are taken from Refs.~\cite{4_sun} [(a) and (c)], ~\cite{7_SFS_shift_Blamire} [(b)], and~\cite{10_SFS_shift_Ustinov} [(d)].}
    \label{shift_exp2}
\end{figure}

We reviewed several experiments~\cite{10_SFS_shift_Ustinov,7_SFS_shift_Blamire,25_SFS_shift_Stolyarov,4_sun} above that observed the giant shift of FMR in SC$|$FM$|$SC sandwiched structures. Such an effect turns out to be sensitive to the configuration. The most obvious phenomenon is that the FMR shift almost vanishes in the superconductor-ferromagnet bilayers, which is reported by Li \textit{et al.}~\cite{4_sun} and Jeon \textit{et al.}~\cite{7_SFS_shift_Blamire}. 
As shown in Fig.~\ref{shift_exp2}(a), the resonance field only has a slight shift below $T_c$ from $T=10.3$~K to $T=4.2$~K in the Nb(100~nm)$|$Ni$_{80}$Fe$_{20}$(20~nm) bilayer; the resonance field only shifts about 1~mT, far less than that of the Nb(100~nm)$|$Ni$_{80}$Fe$_{20}$(20~nm)$|$Nb(100~nm) trilayer~\cite{4_sun}. 
In Fig.~\ref{shift_exp2}(b), Jeon \textit{et al.}~\cite{7_SFS_shift_Blamire} investigated the FMR frequency in the Nb(100~nm)$|$Ni$_{80}$Fe$_{20}$(15~nm) bilayer and fit their measured data with the renormalized FMR~Eq.~\eqref{FMR_jeon}.  
The resonance frequency is almost independent of temperature, and the fitted data show agreement with $H_{\rm shift}^{\rm SC}=0$, i.e., the superconductor has a negligible effect on the resonance frequency in the SC$|$FM bilayer.
On the other hand, the frequency shift appears to vanish as well when an insulating layer is inserted. 
As in Fig.~\ref{shift_exp2}(c), the Nb(100~nm)$|$MgO(10~nm)$|$Ni$_{80}$Fe$_{20}$(20~nm)$|$Nb(100~nm) multilayer also shows the FMR frequency independent of temperature~\cite{4_sun}. A similar phenomenon is obtained by Golovchanskiy \textit{et al.}~\cite{10_SFS_shift_Ustinov}: Fig.~\ref{shift_exp2}(d) shows the fitting parameters $\mu_0 H_a$ from Eq.~\eqref{FMR_fit} 
with different samples, i.e.,  sample S1: Nb(110~nm)$|$Py(19~nm)$|$Nb(110~nm), sample S2: Nb(110~nm)$|$Py(19~nm)$|$Nb(7~nm), and sample S3: Nb(110~nm)$|$Py(19~nm)$|$AlO$_x$(10~nm)$|$Nb(110~nm). From Fig.~\ref{shift_exp2}(d), the parameters with sample S3 slightly change when decreasing the temperature, indicating the independence of the FMR frequency with the temperature. 
Sample S2 shares the same phenomenon. Note that the superconductor film with a thickness of 7~nm is ultrathin in sample S2. 
In this case, the authors explained the vanished frequency shift by the vanished superconductivity of the ultrathin film when its thickness is far below the superconducting coherence length and London's penetration depth~\cite{10_SFS_shift_Ustinov}.

\begin{figure}[htp!]
    \centering
    \includegraphics[width=\linewidth]{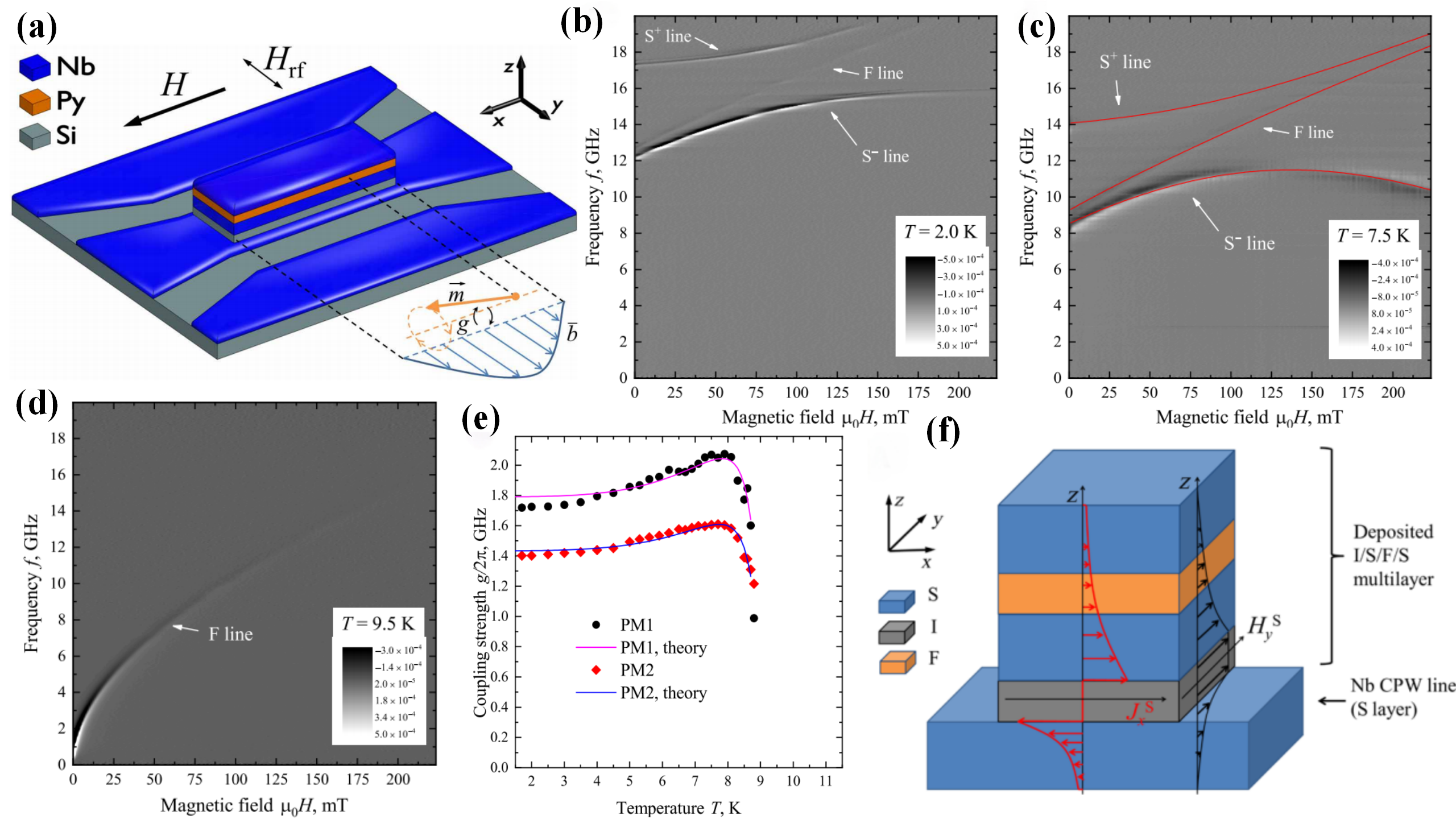}
    \caption{(a) Experimental configuration. The I$|$SC$|$FM$|$SC multilayers of length \textit{L} = 1.1~mm along the $\bf \hat{x}$-axis and width \textit{W} = 130 $\mu$m along the $\bf \hat{y}$-axis is placed directly on top of the central line of the superconducting coplanar waveguide. The magnetic field $H$ is applied along the $\bf \hat{x}$-direction. The radio-frequency magnetic field $H_{\rm rf}$ oscillates along the $\bf \hat{y}$-axis. The orange arrow shows the magnetization $\bf M$, and the blue arrow indicates the magnetic field component of the Swihart electromagnetic standing waves in the SC$|$I$|$SC subsystem. (b)-(d) Microwave transmission spectra $S_{21}$ of the PM1 sample measured at different temperatures below [(b) and (c)] and above (d) the superconducting critical temperature. (e) Temperature dependence of the coupling strength $g/(2\pi)$ of samples PM1 and PM2. (f) Microscopic mechanism of the magnon-photon coupling. The black and red arrows indicate the field and current distributions that result in the magnon-photon coupling. \\
    \textit{Source: }The figures are taken from Ref.~\cite{18_photon-to-magnon_coupling_Ustinov}.}
    \label{shift_exp3}
\end{figure}

The giant shift of FMR in the SC$|$FM$|$SC heterostructure indicates an ultrastrong coupling between the magnon and superconductor. 
Besides that, it is reported that ultrastrong magnon-photon coupling can also be achieved in the multilayered SC$|$FM heterostructures~\cite{18_photon-to-magnon_coupling_Ustinov,17_On-Chip_Photon-To-Magnon_Coupling_Ustinov,35_magnon_photon_with_SC_resonator_Affronte,40_magnon-photon_bilayer_Affronte}. 
In 2021, Golovchanskiy \textit{et al}. designed a series of samples that consists of the I$|$SC$|$FM$|$SC multilayer on top of the Nb waveguide of thickness 140~nm and measured the transmission of microwaves $|S_{21}|(f,H)$ by the VNA~\cite{18_photon-to-magnon_coupling_Ustinov}.
An external bias magnetic field $H\hat{\bf x}$ is applied, as shown in Fig.~\ref{shift_exp3}(a). 
Figures~\ref{shift_exp3}(b)-(d) addresses the microwave transmission $S_{21}$ across the sample Si(30~nm)$|$Nb(102~nm)$|$Py(35~nm)$|$Nb(103~nm)  (which is named ``PM1") below $T_c$ [(b,c)] and above $T_c$ [(d)] at different $H$.
When $T>T_c$, they find a single absorption line, which is indicated as the ``F" line in Fig.~\ref{shift_exp3}(d), which represents the conventional FMR absorption by the ferromagnetic layer since it obeys the Kittel formula Eq.~\eqref{FMR_fit}. 
The fitting data yields negligible $\mu_0H_a\sim 10^{-4}$~T, and $\mu_0 M_{\rm s}=\mu_0M_{\rm eff}=1.13$~T, which are typical for Py thin films.

The transmission spectra change dramatically when $T<T_c$. As shown in Fig.~\ref{shift_exp3}(b), there exist three absorption lines, in which the straight line (F line) corresponds to the FMR absorption of the hybrid SC$|$FM$|$SC subsystem, consistent with the other experiments~\cite{10_SFS_shift_Ustinov,4_sun} that the FMR frequency has an upward shift in the SC$|$FM$|$SC heterostructure when $T<T_c$. On the other hand, the $S^+$ and $S^-$ lines exhibit an anticrossing phenomenon, indicating a strong interaction between two modes. The whole SC$|$I$|$SC$|$FM$|$SC system in Fig.~\ref{shift_exp3}(a) may be divided into two subsystems, i.e., the SC$|$I$|$SC subsystem and the  SC$|$FM$|$SC subsystem.
In the SC$|$I$|$SC subsystem, the photon velocity is reduced according to Swihart~\cite{Swihart_mode}  $\tilde{c}=c\sqrt{d_I/[\varepsilon_I (2\lambda_L+d_I)]}$ (as reviewed in Sec.~\ref{magnon_polariton_theory}), where $c$ is the velocity of light in the vacuum, $d_I$ and $\varepsilon_I$ are the thickness and the dielectric constant of the insulator layer, and $\lambda_L$ is London's penetration depth of the SC layer. 
For the sample with $d_I=30$~nm, $\varepsilon_I\sim 10$, and $\lambda_L=90$~nm, $\tilde{c}=0.12c$, corresponding to the resonance frequency $f^S_r=\tilde{c}/(2L)\approx17$~GHz for a $\lambda/2$ resonator with the length $L$ = 1.1~mm. 
This frequency matches the microwave resonator's resonance frequency at zero field as in Fig.~\ref{shift_exp3}(b). 
Accordingly, the anticrossing of $S^+$ and $S^-$ curves originate from the coupling of Swihart mode in the S$|$I$|$S  and FMR in the ferromagnetic S$|$F$|$S resonators [Fig.~\ref{shift_exp3}(b) and (c)]~\cite{Hopfield_model,Hopfield_model2,Hopfield_model3}. 
Compared with the absorption spectrum in Fig.~\ref{shift_exp3}(b) at $T=2.0$~K, Fig.~\ref{shift_exp3}(c) show the shift of the $S^+$, $S^-$, and F curves when increasing the temperature to $T=7.5$~K.  This shift may be explained by the temperature dependence of London's penetration depth, which affects the resonance frequency of the Swihart mode. The coupling strength $g$ is half the minimum split between the spectral lines. 
As shown in Fig.~\ref{shift_exp3}(e), the coupling strength $g$ for both samples ``PM1'' and ``PM2'' [Si(15~nm)$|$Nb(110~nm)$|$Py(20~nm)$|$Nb(110~nm)] increases substantially at higher temperatures; however, it decreases rapidly when approaching the superconducting critical temperature $T_c$.
The large coupling strength below $T_c$ indicates the hybrid system operates in the ultrastrong coupling regime. 
In Fig.~\ref{shift_exp3}(f), Golovchanskiy \textit{et al.}~\cite{18_photon-to-magnon_coupling_Ustinov} explained the coupling in terms of the FMR mode in the upper SC$|$FM$|$SC trilayer that generates the magnetic field $H_y^S$ and supercurrent $J_x^S$ in the lower SC$|$I$|$SC trilayer, thereby mediating a coupling with the Swihart mode with the  coupling strength
\begin{align}
    g=2\pi\alpha \lambda^{-3/4}\sinh{\left( \dfrac{d_S+d_F/2}{\lambda}\right)}/\sinh{\left( \dfrac{2d_S+d_F}{\lambda}\right)},
    \label{coupling_strength_18}
\end{align}
where $\alpha$ is a fitting parameter and $\lambda$ is London's penetration depth. Equation~\eqref{coupling_strength_18} agrees well with the experimental data of samples PM1 and PM2, as shown by the red and blue curves in Fig.~\ref{shift_exp3}(e).

\subsubsection{Theory}

As addressed above, the experiments observed a giant frequency shift comparable to the bare Kittel frequency when $T<T_c$ in the SC$|$FM$|$SC heterostructures, which, however, vanishes in the SC$|$FM bilayers~\cite{4_sun}. Later, several independent experiments observed a similar phenomenon~\cite{7_SFS_shift_Blamire,10_SFS_shift_Ustinov,25_SFS_shift_Stolyarov,bai_shift}. Theorists explained these experiments in terms of the dipolar interaction between magnons and Cooper pairs~\cite{16_Giant_demagnetization_effects_Buzdin,20_shift_Silaev,47_Gating_ferromagnetic_resonance}.

\begin{figure}[htp!]
    \centering
    \includegraphics[width=\linewidth]{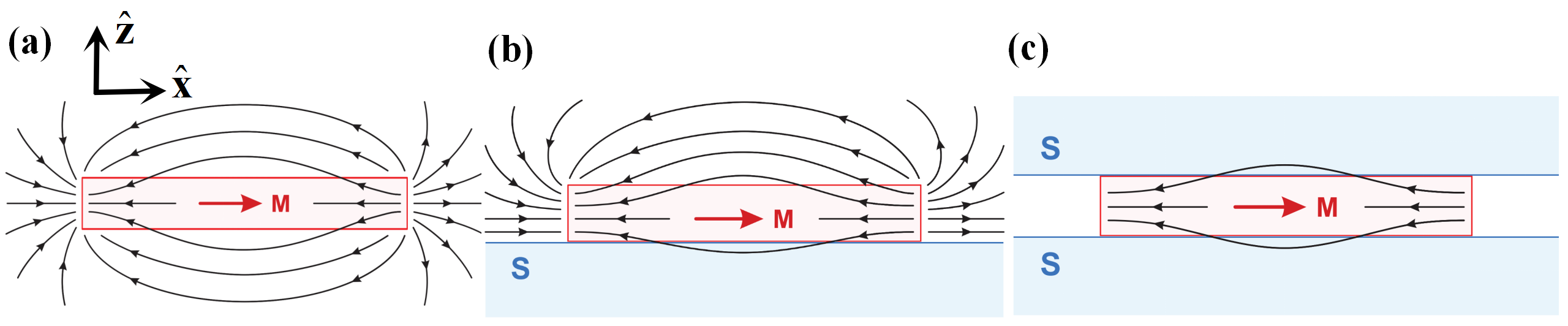}
    \caption{Magnetic field profiles for an isolated ferromagnetic slab [(a)], ferromagnet-superconductor bilayer [(b)], and a ferromagnetic slab sandwiched between two superconductors [(c)].\\
    \textit{Source:} These figures are taken from Ref.~\cite{16_Giant_demagnetization_effects_Buzdin}.}
    \label{Buzdin}
\end{figure}

Mironov and Buzdin proposed that in the SC$|$FM$|$SC heterostructures, the superconductor could screen the \textit{static} stray field of the ferromagnet slab, resulting in the giant demagnetization effect and providing a hint of the experimentally observed frequency shift~\cite{16_Giant_demagnetization_effects_Buzdin}.
To demonstrate the demagnetization effect, the authors analyzed a single ferromagnetic insulator with a length of $2L$ and a thickness of $2h$, using the coordinate system shown in Fig.~\ref{Buzdin}(a). 
The magnetization ${\bf M}$ directed to the $\hat{\bf x}$-direction inside the slab is assumed to be uniform. The demagnetization effect only arises for a finite $L$. 
As shown in Fig.~\ref{Buzdin}(a), the demagnetization field could be regarded as caused by the surface magnetic charge $\rho_M=-\nabla\cdot {\bf M}$. The magnetic charge density $\sigma=\pm M$ is distributed along the boundaries $x=\pm L$, and the induced magnetic stray field $H$ lies in the $x$-$z$ plane. 
The demagnetization field $H\propto h/r\rightarrow0$ at distances $r\gg h$; the demagnetization field is weak at the central part of the slab.
When the ferromagnetic film is placed on the superconductor with a thickness well exceeding London's penetration depth $\lambda$, as shown in Fig.~\ref{Buzdin}(b), the authors used the method of image to estimate the demagnetization field, i.e., each real magnetic charge $\rho_m$ located at the point $(x, y, z)$ above the edge of the superconductor (i.e., for $z> -h$) is accompanied by the image charge of the same sign located at the point \(( x, y,  -2h-z)\). 
The field \(H\) in the region \(z > -h\) then coincides with the field produced by both real and image charges; as a result, the field $H$ is doubled in the central part of the ferromagnetic film compared to the case of the isolated ferromagnet. 
However, the demagnetization is still negligibly small and could not shift the FMR frequency. 
Remarkably, the demagnetization effect becomes dramatically enhanced when the ferromagnet slab is sandwiched between two superconductors, as shown in Fig.~\ref{Buzdin}(c). In this case, the magnetic charges at the ferromagnet edges $x =\pm L$ induce an infinite series of image charges in the superconductors, and the field $H$ induced inside the ferromagnet coincides with the field from two parallel magnetic planes
(``magnetic capacitor'') perpendicular to the $\hat{\bf x}$ axis and located at $x=\pm L$, resulting in the remarkable demagnetization field ${\bf H}=-{\bf M} $ that causes the giant FMR shift. It is noted that this effect vanishes for thin magnetic films, as when the sample length $L\rightarrow\infty$, the magnetic stray field $H=0$ outside the slab and the demagnetization effect vanishes.

Silaev~\cite{20_shift_Silaev}, Zhou and Yu~\cite{47_Gating_ferromagnetic_resonance} interpreted such a phenomenon from a different point of view. Silaev explained it by the Anderson-Higgs mechanism of mass generation~\cite{20_shift_Silaev}. Zhou and Yu go beyond the quasi-static approximation and demonstrate that the superconductor can reflect the electromagnetic waves radiated from the ferromagnetic insulator in the magnetization dynamics~\cite{47_Gating_ferromagnetic_resonance}.
The electromagnetic fields are reflected back and forth in the SC$|$FM$|$SC heterostructure, so the electric field penetrates both superconductors. This electric field inside the superconductors drives the oscillating Meissner supercurrent, which, in turn, generates the Oersted magnetic field that shifts the FMR frequency. 
When the temperature $T>T_c$, the superconductor becomes a normal metal, and the back-and-forth reflection results in the normal current inside the metal, which provides considerable extra damping but does not shift the FMR frequency.

\textbf{Electromagnetic radiation of a single ferromagnet}.---We first address the electromagnetic radiation of a single ferromagnetic insulator (FI) by calculating the electromagnetic field emitted by the FMR of the FI film, which goes beyond the quasistatic approximation.
An in-plane external magnetic field biases the static magnetization of a thin FI film of thickness $2d_F$, such that the in-plane static magnetization brings zero static demagnetization field.

\begin{figure}[htp!] 
  \centering 
\includegraphics[width=1\linewidth]{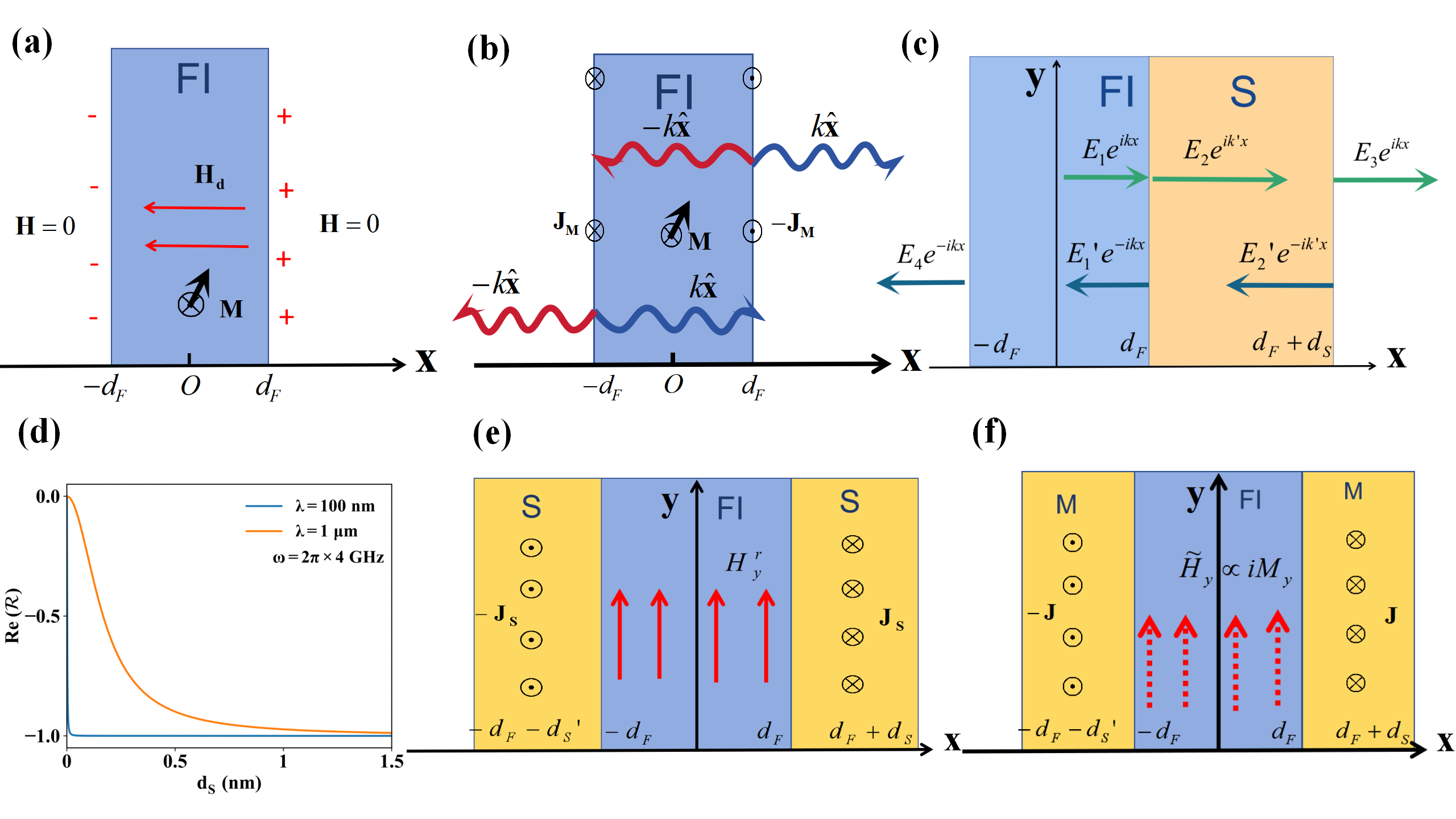}
 \caption{Electromagnetic radiation and its frequency shift of FMR in different configurations.
 (a) Quasi-static description of the radiated magnetic field by FMR of a single ferromagnetic layer. 
 (b) Electric fields radiated by the surface magnetization current at two surfaces of the magnetic insulator. 
 (c) Radiated electric field of the FI$|$SC  heterostructure. (d) Reflection coefficient Re($\cal R$) as a function of the superconductor thickness $d_S$ with different London's penetration depth $\lambda_L=100$~nm and $1$~$\mu$m. (e) Supercurrent distribution in superconductors of the SC$|$FM$|$SC trilayer and its generated magnetic field $H_y^r$ that affects the magnetization dynamics. (h) shows the situation when the superconductor becomes a normal metal. \\
 \textit{Source:} Figures (b), (c), (d) are taken from Ref.~\cite{47_Gating_ferromagnetic_resonance}.}
 \label{shift} 
\end{figure}

The quasistatic approximation is often exploited when calculating the electromagnetic field distribution of magnetization for low frequencies and near fields.
\textit{In the quasistatic approximation}, the radiation term of the magnetic field is disregarded, i.e., $\nabla\times {\bf H}=0$, and the magnetic field generated by magnetization satisfies the Coulomb's Law $\nabla^2{\bf H}=-\nabla(\nabla\cdot {\bf M}) $ \eqref{quansi_M_H}. In this viewpoint, even for the dynamical magnetization, the stray field can be regarded as the field generated by the instantaneous magnetization, which allows for analysis using the magnetic charge $\rho_M=-\nabla\cdot {\bf M}$ in the calculation.
As shown in Fig.~\ref{shift}(a), the $x$-component of the magnetization leads to the surface magnetic charge $\sigma_m=\pm M_x$ at two surfaces $x=\pm d_F$ of the film, which behaves like a ``magnetic capacitor''.  These magnetic charges generate the demagnetization field $H_d \hat{\bf x}=-M_x\hat{\bf x}$ inside the ferromagnetic film. However, the magnetic field vanishes outside the film since the fields from the two surfaces with opposite magnetic charges cancel. The vanished magnetic field outside the slab implies the superconductor could not affect the magnetization, whether for the FM$|$SC bialyer or the S$|$F$|$S trilayer, which is inconsistent with the experiments~\cite{4_sun,7_SFS_shift_Blamire,10_SFS_shift_Ustinov,25_SFS_shift_Stolyarov,bai_shift,shift_yangguang}.

From the radiative point of view, the situation becomes different. Indeed, the magnetization leads to an oscillating ``magnetization current'', e.g., for the FMR, 
 \begin{align}
{\bf J}_M=\nabla\times {\bf M}=[\delta(x+d_F)-\delta(x-d_F)]M_y\hat{\bf z}\propto M_y,
\end{align}
which only has the $z$-component and is located at two film surfaces with opposite flow directions, as shown in Fig.~\ref{shift}(b). 
This oscillating current radiates the electromagnetic waves outside the slab according to Maxwell's equation $ \nabla^2 {\bf E}+k^2{\bf E}=-i\omega\mu_0 \nabla\times {\bf M}$ with the solution 
\begin{align}
    {\bf E}({\bf r})=\dfrac{i\omega\mu_0 }{4\pi}\int \dfrac{[\nabla'\times {\bf M}({\bf r'})]e^{i k|{\bf r-r'}|}}{|{\bf r-r'}|}d{\bf r'}=\dfrac{i\omega\mu_0 }{4\pi}\int \dfrac{{\bf J}_M({\bf r'})e^{i k|{\bf r-r'}|}}{|{\bf r-r'}|}d{\bf r'},
\end{align}
where $k=\sqrt{\omega^2\mu_0\varepsilon_0}$. Since the magnetization current only has the $z$-component, only $E_z$ exists, which reads
\begin{align}
      E_z=\dfrac{\mu_0 \omega M_y}{2 k}\begin{cases}
       e^{-ik(x-d_F)}-e^{ik(x+d_F)},  & -d_F<x<d_F \\
       e^{ik(x-d_F)}-e^{ik(x+d_F)},   &     x>d_F\\
       e^{-ik(x-d_F)}-e^{-ik(x+d_F)}, &     x<-d_F
    \end{cases}.
\label{full_solution_single_layer}
\end{align}
As shown in Fig.~\ref{shift}(b), the oscillating magnetization currents radiate four plane waves, i.e., two waves with positive wavevector $k\hat{\bf x}$  and two waves with negative wavevector $-k\hat{\bf x}$ from the surfaces $x=\pm  d_F$ with opposite amplitudes. According to Maxwell's equation $\nabla\times{\bf E}=-\partial_t{\bf B}$, the associated magnetic induction $B_x=0$, $B_z=\mu_0(H_0+M_0)$ is static, and 
\begin{align}
      B_y=\dfrac{\mu_0  M_y}{2}\begin{cases}
    e^{ik(x+d_F)}+e^{-ik(x-d_F)},  & -d_F<x<d_F \\
      e^{ik(x+d_F)}-e^{ik(x-d_F)},  &     x>d_F\\
       -e^{-ik(x+d_F)}+e^{-ik(x-d_F)},  &     x<-d_F
    \end{cases}
    \label{magnetic_field}
\end{align}
does not vanish outside the slab. 
Only for the near field distribution and low frequency that satisfies $kx\rightarrow0$, $B_y=0$ outside the slab, which recovers the results by the quasistatic approximation.

\textbf{Electromagnetic radiation of ferromagnet$|$superconductor bilayer}.---When a superconductor film covers the ferromagnetic insulator film, as shown in Fig.~\ref{shift}(b),  the radiated electric field is strongly reflected by the adjacent superconductors. The reflection process of the ferromagnet$|$superconductor bilayer is shown in Fig.~\ref{shift}(c), where the electromagnetic wave is reflected by the vacuum$|$FI, FI$|$SC, and SC$|$vacuum interfaces. In the FI or vacuum, the electric field satisfies
$\nabla^2 {\bf E}+k^2{\bf E}=0$, which has the radiative solutions
\begin{align}
      E_z(x)=\begin{cases}
    E_1e^{ikx}+E_1'e^{-ikx},  & -d_F<x<d_F \\
      E_3e^{ikx},  &     x>d_F+d_S\\
       E_4e^{-ikx},  &     x<-d_F
           \end{cases}.
\end{align}
Inside the superconductor, the electric field obeys $\nabla^2 {\bf E}+k'^2{\bf E}=0$, so the solution $E_z(x)=E_2 e^{ik' x}+E_2'e^{-i k' x}$, where 
$k'=\sqrt{\omega^2 \mu_0 \varepsilon_0-1/\lambda_L^2}\sim i/\lambda_L$ when $\omega^2 \mu_0 \varepsilon_0\ll1/\lambda_L^2$ (e.g., $k\sim 83.8$~m$^{-1}$ is much smaller than $1/\lambda_L\sim 10^7$~m$^{-1}$ with  frequency $\omega\sim 2\pi\times 4$~GHz and $\lambda_L\sim100$~nm). Accordingly, electromagnetic waves no longer propagate inside the superconductor but decay with the penetration length $\lambda_L$ due to the Meissner effect.

$E_z$ and $H_y$ are continuous at the interfaces, matching which we can solve all the amplitudes $\{E_1,E_1',E_2,E_2',E_3,E_4 \}$. 
The electric field inside the FI reads 
\begin{align}
    E_z(-d_F<x<d_F)={\cal R}E_{\text{single}}|_{x=d_F}e^{-ik_0(x-d_F)}
    +E_{\text{single}}(x),
    \label{E_bilayer_in_f}
\end{align}
which is interpreted as the radiated electric field from a single magnetic insulator  $E_{\text{single}}(x)$  [Eq.~(\ref{full_solution_single_layer})] plus the field reflected by the superconductor at the interface with the reflection coefficient
\begin{align}
    {\cal R}=\frac{e^{ik' d_S}(k^2-k'^{2})+e^{-ik' d_S}(k'^{2}-k^2)}{e^{ik' d_S}(k-k')^2-e^{-ik' d_S}(k+k')^2}.
\end{align}
When $d_S=0$, $e^{\pm i k' d_S}=1$, such that ${\cal R}=0$ and Eq,~(\ref{E_bilayer_in_f}) recovers to the solution (\ref{full_solution_single_layer}) of the single-layer case. 
However, as shown in Fig.~\ref{shift}(d), even with a small $d_S\ll \lambda$, the reflection coefficient  ${\cal R}\rightarrow -1$ since $k=\omega\sqrt{\mu_0\varepsilon_0}$ is much smaller than $|k'|\approx 1/\lambda$ at microwave frequencies. This implies total reflection with a $\pi$-phase shift of the electric fields at the FI$|$SC interface, even with an ultrathin conventional superconductor layer.  Therefore, at the interface between the ferromagnet and superconductor, the electric field $E_z=(-1) E_{\text{single}}|_{x=d_F}+E_{\text{single}}|_{x=d_F}\sim 0$ due to the  $\pi$-phase shift  ${\cal R}=-1$. Since $E_z$ is continuous at interfaces, the electric field vanishes in the superconductor that generates no supercurrent and thereby leads to no modulation on the FMR, explaining the absence of FMR shift in all the available experiments with thick superconductors~\cite{4_sun,7_SFS_shift_Blamire,10_SFS_shift_Ustinov,25_SFS_shift_Stolyarov}.

\textbf{Electromagnetic radiation of superconductor$|$ferromagnet$|$superconductor trilayer}.---The electromagnetic field distribution becomes very different for the  SC$|$FI$|$SC heterostructure. In this case, the electric field reflected by one superconductor is reflected by the other, such that the reflection becomes back and forth, resulting in the electric field penetrating both superconductors and driving the supercurrent as illustrated in Fig.~\ref{shift}(e). 
The supercurrent, in turn, generates the Oersted magnetic field $H_y^r\hat{\bf y}$ normal to the saturation magnetization $M_s\hat{\bf z}$ inside the ferromagnetic insulator, which causes the FMR frequency shift.

Similar to the SC$|$FI heterostructure, the electric field distribution is calculated by matching the boundary conditions. For the ferromagnetic film sandwiched by two superconductors with thickness $d_S$ and $d_S'$, the magnetic field inside the ferromagnet reads 
\begin{align}
   H_y^r=\frac{kM_y(Ge^{ikx}-e^{-ikx})}{k(Ge^{ikd_F}-e^{-ikd_F})-k'f(u)(Ge^{ikd_F}+e^{-ikd_F}) }-M_y,\quad H^r_x=-M_x,
   \label{general_magnetic_field}
\end{align}
where $G=-\frac{-2k\sinh(ikd_F)+k'(f(u)e^{-ikd_F}+f(u')e^{ikd_F})}{-2k\sinh(ikd_F)+k'(f(u)e^{ikd_F}+f(u')e^{-ikd_F})}$ with $u=-[(k+k')/(k-k')]\exp(-2ik'd_S)$ and $ u'=-[(k+k')/(k-k')]\exp(-2ik'd_S')$. It affects the precession of the magnetization. 
In Eq.~\eqref{general_magnetic_field}, the real part of the radiated magnetic field $H_y^r$ is in the same phase as $M_y$, i.e., $\Re(H_y^r)=\kappa_y M_y$, which acts as the effective ``anisotropy field'' that provides a field-like torque for the magnetization and shift the frequency. This explains the changes of the ``anisotropy field'' observed in the experiment~\cite{10_SFS_shift_Ustinov} [see the ``S1" data from Fig.~\ref{shift_exp2}(d)]. 
On the other hand, the imaginary part of $H_y$ has a $\pi/2$ phase difference with $M_y$, which provides a damping-like torque that results in the extra damping coefficient $\tilde{\alpha}_G=\mu_0\gamma M_0 \Im(H_y^r)/(\omega M_y)$. 
In terms of the (linearized) LLG equation,  
\begin{align}
     -i\omega M_x+\mu_0 \gamma M_y H_0&=\mu_0 \gamma M_0\Re({H_{y}^r}/M_y)M_y+i (\alpha_G +\tilde{\alpha}_G)\omega M_y,\nonumber\\
\mu_0 \gamma H_0 M_x+i\omega M_y &=-\mu_0 \gamma M_0 M_{x}+i\alpha_G \omega M_x.
\label{linearized_LLG_Equation}
 \end{align}
Accordingly, the real part of $H_y^r$ 
\[
     \Re(H_y^r)\approx- {2 d_FM_y \tanh(d_S/\lambda)\tanh(d_S'/\lambda)}\left[\lambda(\tanh(d_S/\lambda)+\tanh(d_S'/\lambda))\right.
\left.+2d_F\tanh(d_S/\lambda)\tanh(d_S'/\lambda)\right]^{-1}
\]
renormalizes the FMR frequency to be
 \begin{align}
    \omega_{\rm K}&=\mu_0\gamma\sqrt{H_0+M_0}
     \sqrt{H_0-M_0\Re(H_y^r)/M_y},
     \label{general_omega_K}
\end{align}
while the imaginary part  
\begin{align}
      \Im({ H}_y^r)\approx&2kd_FM_y\left(
    \frac{\tanh^2(d_S'/\lambda)}{\cosh^2(d_S/\lambda)}+\frac{\tanh^2(d_S/\lambda)}{\cosh^2(d_S'/\lambda)} \right)\nonumber\\
\times&\left[\tanh(d_S/\lambda)+\tanh(d_S'/\lambda)\right.\left.+2d_F/\lambda\tanh(d_S/\lambda)\tanh(d_S'/\lambda)\right]^{-2},\nonumber
\end{align}
contributes to the additional damping coefficient $\tilde{\alpha}_G$. Silaev solved a similar issue in the situation with $d_S\gg \lambda_L$~\cite{23_magnon_photon_coupling_Silaev}. When $d_S\gg \lambda_L$, the solution Eq.~\eqref{general_omega_K} recovers to that in Ref.~\cite{23_magnon_photon_coupling_Silaev}.

In particular, for the symmetric configuration with $d_S=d_S'$, the FMR frequency
 \begin{align}
     \omega_{\rm K}=\mu_0\gamma\sqrt{(H_0+M_0)\left(H_0+\dfrac{d_F\tanh(d_S/\lambda)}{\lambda+d_F\tanh(d_S/\lambda)}M_0\right)}
     \label{FMR_frequency}
 \end{align}
shifts the bare Kittel frequency $\tilde{\omega}_K=\mu_0\gamma\sqrt{H_0(H_0+M_0)}$.
For example, for the Nb(60~nm)$|$YIG(120~nm)$|$Nb(60~nm) heterostructure,  the FMR frequency has a giant shift $\delta \omega=2\pi\times 1.6~{\rm GHz}\sim\tilde{\omega}_K/2$ with $\mu_0M_0=0.2$~T, $\lambda(T=0.5T_c)=87.8$~nm, and the bias field $\mu_0H_0=0.05$~T. 
This demonstrates the potential for an ultrastrong interaction between magnons and a Cooper-pair supercurrent, even in the presence of magnetic insulators.

The temperature efficiently modulates the FMR frequency. In the experiments~\cite{4_sun,7_SFS_shift_Blamire,10_SFS_shift_Ustinov,25_SFS_shift_Stolyarov}, the FMR shift is enhanced when decreasing the temperature when $T<T_c$, but becomes the bare Kittel mode when $T>T_c$. By the two-fluid model, i.e., the temperature affects both normal current and supercurrent via the conductivity $\sigma_n(T)=\sigma_n(T/T_c)^4$ and $\lambda_L(T)=\lambda_L(T=0{\rm K})\sqrt{1-(T/T_c)^4}^{-1}$, modulating $k'=\sqrt{\omega^2 \mu_0\varepsilon_0+i \omega \mu_0 \sigma_n(T)-1/\lambda_L^2(T)}$. When decreasing the temperature below $T_c$, a decrease in $\lambda_L(T)$ enhances the shift of the FMR frequency. When $T>T_c$, the superconductor becomes a normal metal, such that $k'=\sqrt{\omega^2 \mu_0\varepsilon_0+i \omega \mu_0 \sigma_n}\approx i/\delta$, where $\delta=(1+i)/\sqrt{2 \omega \mu_0 \sigma_n}$ is the penetration depth of microwaves in the normal metal. As in  Fig.~\ref{shift}(f), the electric field drives the normal current, generating the Oersted magnetic field inside the ferromagnetic insulator. This magnetic field $\tilde{H}_y\propto iM_y$ has a $\pi/2$ phase difference with $M_y$, thus providing a damping-like torque to the magnetization, resulting in the additional damping rather than the frequency shift.

\begin{figure}[htp!]
    \centering
    \includegraphics[width=0.86\linewidth]{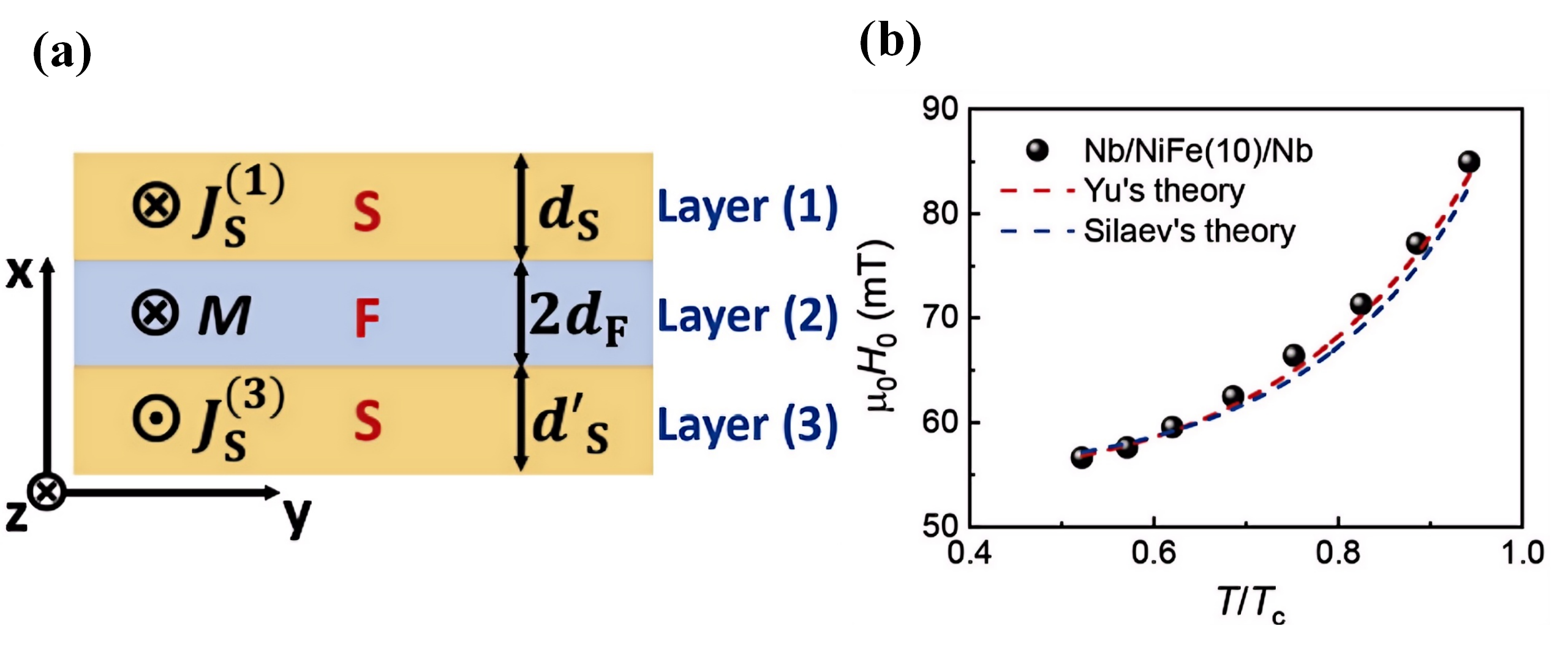}
    \caption{(a) Experiemntal configuration of the SC$|$FM$|$SC heterostructure. (b) Comparison of the measured resonance-field shift with Yu's theory~\cite{47_Gating_ferromagnetic_resonance} and Silaev's theory~\cite{20_shift_Silaev}. \\
    \textit{Source:} These figures are taken from Ref.~\cite{shift_yangguang}.}
    \label{yangguang_exp}
\end{figure}

Recently, Li \textit{et al.}~\cite{shift_yangguang} measured the resonance field shift (corresponding to the frequency shift) in the Nb$|$NiFe$|$Nb device as shown in Fig.~\ref{yangguang_exp}(a) with different thicknesses of NiFe and temperatures. The experiment results for the Nb(100~nm)$|$NiFe(10~nm)$|$Nb(100~nm) heterostructure agree excellently with the theory by Zhou and  Yu~\cite{47_Gating_ferromagnetic_resonance}, and Silaev~\cite{20_shift_Silaev}, as shown in Fig.~\ref{yangguang_exp}(b).

\subsection{Magnon-photon polariton under ultrastrong coupling}

\subsubsection{Magnon-photon ultrastrong coupling}

\label{magnon_polariton_theory}

\textbf{Swihart mode}.---We first review the electromagnetic modes in the SC$|$insulator(I)$|$SC heterostructures with surface normal along the $\hat{\bf x}$-direction, in which the thickness of the middle non-magnetic insulator is $2d_I$.
Swihart found that the superconductor strongly modulates the distribution of the electromagnetic field, significantly slowing down the speed of electromagnetic waves~\cite{Swihart_mode}. This mode is now known as the Swihart mode.

According to Maxwell's equations~\cite{Jackson},  the electric field in different regions obeys 
\begin{align}
\text{in I}:& ~~~\nabla^2 \mathbf{E}({\bf r}, t)+k^2_0 \mathbf{E}({\bf r}, t)
=0,\nonumber\\
\text{in SC}:& ~~~\nabla^2 \mathbf{E}({\bf r}, t)+k^2_s \mathbf{E}({\bf r}, t)
=0,
\label{waveequ}
\end{align}
where $k_0=\omega\sqrt{\mu_0\varepsilon_{I}}$ with $\varepsilon_{I}$ being the dielectric constant in the insulators. Since this system is isotropic in the $y$-$z$ plane, we consider the electromagnetic waves propagating along the $\bf \hat{z}$-direction with wavevector $k$ without losing generality. We assume the formal solution ${\bf E}(x,z,t)=\tilde{\bf E}(x)e^{i(kz-\omega t)}+{\rm H.c.}$, where the amplitudes
\begin{align}
    &\text{in I}:~~{\tilde {\bf E}}(x)={\pmb{\cal E}}_{0}e^{i {\cal A}_k x}+{\pmb{\cal E}}_{0}'e^{-i {\cal A}_k x},\nonumber\\
    &\text{in SC~``1"}~(x>d_I):~~{\tilde {\bf E}}(x)={\pmb{\cal E}}_{1}e^{i {\cal B}_k x},\nonumber\\
  &\text{in SC~``2"}~(x<-d_I):~~{\tilde {\bf E}}(x)={\pmb{\cal E}}_{2}e^{-i {\cal B}_k x},\nonumber
 \end{align}
in which ${\cal A}_k=\sqrt{k_0^2-k^2}$ and ${\cal B}_k=\sqrt{\omega^2\mu_0\varepsilon_0-1/\lambda^2-k^2}$ with $\lambda$ being London's penetration depth. 
In the non-magnetic insulator, the magnetic field is governed by Faraday's Law, i.e., $i\mu_0\omega {\bf H}({\bf r},t)=\nabla\times {\bf E}({\bf r},t)$, leading to
     \begin{align}
     &H_x=1/(i \omega\mu_0)(\partial_y E_z-\partial_z E_y),\nonumber\\
     &H_y=1/(i \omega \mu_0)(\partial_z E_x-\partial_x E_z),\nonumber\\
     &H_z=1/(i \omega\mu_0)(\partial_x E_y-\partial_y E_x).
     \end{align}

The boundary conditions described in Table.~\ref{boundary} couple the amplitudes $\{{\cal E}_{0x},  {\cal E}_{0x}',    {\cal E}_{0y},  {\cal E}_{0y}'\}$ by the matrix equation
\begin{equation}
    \left(\begin{matrix}
         (R_k-1)e^{i{\cal A}_k d_I} & (R_k+1)e^{-i{\cal A}_k d_I} & 0 & 0 \\
         (R_k+1)e^{-i{\cal A}_k d_I} & (R_k-1)e^{i{\cal A}_k d_I} & 0 & 0 \\
         0 & 0 & ({\cal A}_k-{\cal B}_k)e^{i{\cal A}_k d_I} & -({\cal A}_k+{\cal B}_k)e^{-i{\cal A}_k d_I} \\
         0 & 0 & ({\cal A}_k+{\cal B}_k)e^{-i{\cal A}_kd_I} & -({\cal A}_k-{\cal B}_k)e^{i{\cal A}_kd_I} \\
    \end{matrix}\right)\left(
    \begin{matrix}
        {\cal E}_{0x}\\
        {\cal E}_{0x}'\\
        {\cal E}_{0y}\\
        {\cal E}_{0y}'
    \end{matrix}
    \right)=0.
\end{equation}
This implies that the $x$- and $y$-components of the electric fields decouple, forming two distinct modes. The non-zero solutions require the determinant of the coefficient matrix to vanish.

The first mode is the TE mode with $E_x=E_z=H_y=0$. It has three components $\{H_z,H_x,E_y\}$. From the secular equation 
\begin{align}
\left| \begin{matrix}
({\cal A}_k-{\cal B}_k)e^{i{\cal A}_k d_I} & -({\cal A}_k+{\cal B}_k)e^{-i{\cal A}_k d_I} \\
({\cal A}_k+{\cal B}_k)e^{-i{\cal A}_kd_I} & -({\cal A}_k-{\cal B}_k)e^{i{\cal A}_kd_I} \\
\end{matrix} \right|=0,
\nonumber
\end{align}
we find its dispersion relation 
\begin{equation}
    \omega=\frac{1}{\sqrt{\mu_0\varepsilon_I}}\sqrt{k^2+\frac{1}{2}\left(\frac{1}{\lambda^2}+\frac{1}{d_I\lambda}\right)},
\end{equation}
which turns out to be a high frequency and hence mismatches with the FMR in conventional magnetic films.

The second mode is the TM mode with $E_y=H_x=H_z=0$. This mode has three components $\{E_z,E_x,H_y\}$. From the secular equation
\begin{align}
    \left|
    \begin{matrix}
    (R_k-1)e^{i{\cal A}_k d_I} & (R_k+1)e^{-i{\cal A}_k d_I}  \\
    (R_k+1)e^{-i{\cal A}_k d_I} & (R_k-1)e^{i{\cal A}_k d_I} 
    \end{matrix}\right|=0, 
    \nonumber
\end{align}
we find the Swihart-mode frequency~\cite{Swihart_mode}  
\begin{equation}
\Omega_s=\sqrt{\frac{d_I}{d_I+\lambda}}\frac{1}{\sqrt{\mu_0\varepsilon_I}}|{\bf k}|.
\label{Swihart_mode}
\end{equation}

\begin{figure}[htp]
    \centering
    \includegraphics[width=0.95\linewidth]{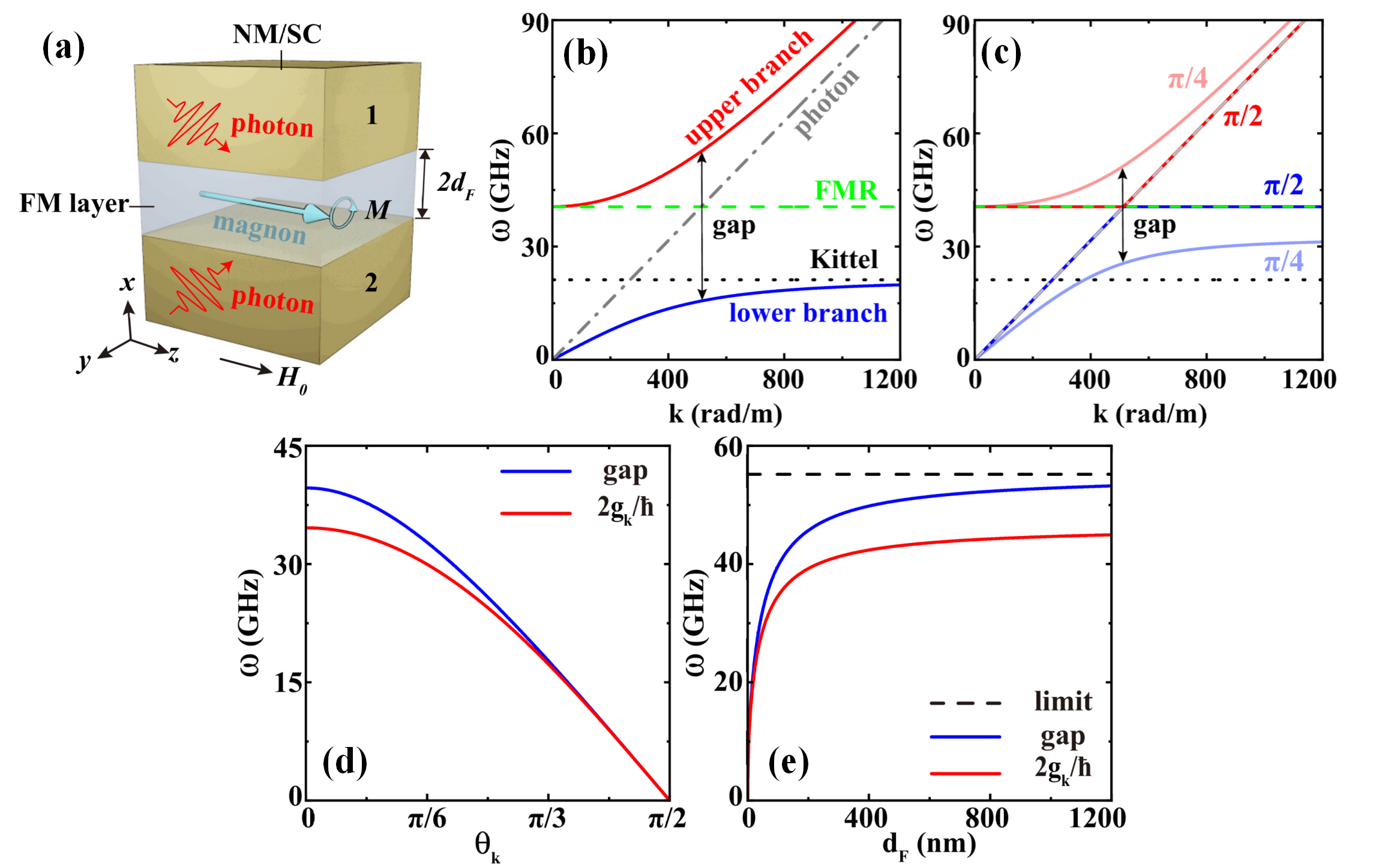}
    \caption{(a) Schematic illustration of the SC$|$FI$|$SC heterostructure. The system is subjected to an in-plane external magnetic field $H_0\hat{\bf z}$. The thickness of the FI layer is $2d_F$. (b) Dispersion relation of the magnon-photon polariton as a function of wavevector ${\bf k}\parallel H_0\hat{\bf z}$. (c) Dispersion of collective modes with two propagation directions $\theta_{\bf k}=\{\pi/4,\pi/2\}$. (d) Frequency gap at the intersection of the FMR and photon modes and the coupling strength $2g_{\bf k}/\hbar$ as a function of $\theta_{\bf k}$. (e) Dependence of the frequency gap and coupling strength on the FI thickness $d_F$, with the black dashed line indicating the upper limit of the frequency gap.\\
    \textit{Source:} The figures are taken from Ref.~\cite{52_Persistent_nodal_magnon-photon_polariton_in_ferromagnetic_heterostructures}.}
    \label{MP polariton_SC}
\end{figure}

\textbf{Magnon-photon polariton}.---In the SC$|$ferromagnetic insulator(FI)$|$SC heterostructure illustrated in Fig.~\ref{MP polariton_SC}(a), the magnons and photons interact strongly through the Zeeman interaction $-\mu_0{\bf H}\cdot{\bf M}$, with the coupling strength becoming comparable to the magnon bare frequency~\cite{52_Persistent_nodal_magnon-photon_polariton_in_ferromagnetic_heterostructures,20_shift_Silaev,23_magnon_photon_coupling_Silaev}. 

According to the normalization condition Eq.~\eqref{normalization_22}~\cite{Walkermode,spinwaveExcitation}, the  magnetization 
\begin{align}
    \hat{M}_x&=\int\frac{d{\bf k}}{2\pi} \left(-i\sqrt{\frac{\hbar M_s\omega_F}{4d_F\mu_0(H_0+M_s)}}e^{i{\bf k}\cdot{\pmb \rho}}\hat{m}_{\bf k}+{\rm H.c.}\right),\nonumber\\
    \hat{M}_y&=\int\frac{d{\bf k}}{2\pi} \left(\sqrt{\frac{\hbar\mu_0\gamma^2(H_0+M_s)M_s}{4d_F\omega_F}}e^{i{\bf k}\cdot{\pmb \rho}}\hat{m}_{\bf k}+{\rm H.c.}\right),
    \label{magnetization_operator_SC}
\end{align}
where $M_s$ is the saturation magnetization and $H_0$ is the bias magnetic field along the $\hat{\bf z}$-direction. The FMR frequency 
\begin{align}
\omega_{F}=\mu_0\gamma\sqrt{(H_0+M_s)(H_0+M_sd_F/(d_F+\lambda))}
\label{frequency_SC}
\end{align}
is renormalized by the superconductor layers taking $d_S\rightarrow+\infty$ from Eq.~\eqref{FMR_frequency}.

The ferromagnetic insulator layer sandwiched by two adjacent superconducting layers behaves like a microwave waveguide. The waveguide photon mode is the Swihart mode~\cite{Swihart_mode} as addressed by Eq.~\eqref{Swihart_mode}. 
For the ferromagnetic insulator of thickness $2d_F$ with dielectric constant $\varepsilon_{\rm FI}$, the photon mode frequency
\[
\Omega_s=\sqrt{\frac{d_F}{d_F+\lambda}}\frac{1}{\sqrt{\mu_0\varepsilon_{\rm FI}}}|{\bf k}|. 
\]
The Hamiltonian of the Swihart photons includes two contributions including the energy of the electromagnetic field and the supercurrents induced in the superconductors,  given by $ \hat{H}_{\rm Sw}=\int{\rm d}{\bf r}\left(\varepsilon_r\hat{\bf E}^2+\mu_0\hat{\bf H}^2+\mu_0\lambda^2\hat{\bf J}_{\rm s}^2\right)/2$. In terms of the photon creation and annihilation operators $\{\hat{p}^{\dagger}_k,\hat{p}_k\}$ that obey the commutation relation $[\hat{p}_k, \hat{p}^{\dagger}_{k'} ]=i\hbar\delta(k-k')$, the quantized fields
\begin{align}
    \hat{\bf E}&=\int^{+\infty}_{-\infty}\frac{{\rm d}k}{\sqrt{2\pi}}\left(\tilde{\bf E}_{k}(x)e^{ikz}\hat{p}_k+{\rm H.c.}\right),\nonumber\\
    \hat{\bf H}&=\int^{+\infty}_{-\infty}\frac{{\rm d}k}{\sqrt{2\pi}}\left(\tilde{\bf H}_{k}(x)e^{ikz}\hat{p}_k+{\rm H.c.}\right),\nonumber\\
    \hat{\bf J}_s&=\int^{+\infty}_{-\infty}\frac{{\rm d}k}{\sqrt{2\pi}}\left(\tilde{\bf J}_{s,k}(x)e^{ikz}\hat{p}_k+{\rm H.c.}\right),
\end{align}
where for $-d_F<x<d_F$, the amplitudes 
\begin{align}
    \tilde{E}_{x,k}(x)&={\cal E}_{0x}e^{i{\cal A}_k x}+{\cal E}_{0x}'e^{-i{\cal A}_k x},\nonumber\\
    \tilde{E}_{z,k}(x)&=\frac{{\cal A}_k}{k}({\cal E}_{0x}e^{i{\cal A}_k x}-{\cal E}_{0x}'e^{-i{\cal A}_k x}),\nonumber\\
    \tilde{H}_{y,k}(x)&=\frac{{\omega \varepsilon_{\rm FI} }}{k}({\cal E}_{0x}e^{i{\cal A}_k x}+{\cal E}_{0x}'e^{-i{\cal A}_k x}),\nonumber\\
    \tilde{J}_{s,k}(x)&=0,
    \label{distribution1}
\end{align}
while for $x>d_F$ and $x<-d_F$, the amplitudes  
\begin{align}
    \tilde{E}_{x,k}(x)&=\frac{\varepsilon_{\rm FI}}{\varepsilon_0+i\sigma_s/\omega}\tilde{E}_{x,k}(\pm d_F^{\mp})e^{\pm i{\cal B}_k (x\mp d_F)},\nonumber\\     \tilde{E}_{z,k}(x)&=\tilde{E}_{z,k}(\pm d_F)e^{\pm i{\cal B}_k (x\mp d_F)},\nonumber\\
    \tilde{H}_{y,k}(x)&=\tilde{H}_{y,k}(\pm d_F)e^{\pm i{\cal B}_k (x\mp d_F)},\nonumber\\
    \tilde{J}_{s,k}(x)&=\sigma_s\tilde{E}_{z,k}(x)=\frac{i}{\omega\mu_0\lambda^2}\tilde{E}_{z,k}(\pm d_F)e^{\pm i{\cal B}_k (x\mp d_F)}.
    \label{distribution2}
\end{align}
The amplitudes are solved by the boundary conditions Table~\ref{boundary}~\cite{52_Persistent_nodal_magnon-photon_polariton_in_ferromagnetic_heterostructures}:
\begin{align}
    {\cal E}_{0x}'&=\frac{({\cal B}_k+k^2){\cal A}_k-({\cal A}_k+k^2){\cal B}_k}{({\cal B}_k^2+k^2){\cal A}_k+({\cal A}_k^2+k^2){\cal B}_k}e^{2i{\cal A}_kd_F}{\cal E}_{0x}
    \approx \frac{1+\frac{d_F}{d_F+\lambda}k\lambda}{1-\frac{d_F}{d_F+\lambda}k\lambda}{\cal E}_{0x}\approx{\cal E}_{0x},
\end{align}
in which with the dispersion of the Swihart mode, ${\cal A}_k=ik\sqrt{\lambda/(d_F+\lambda)}$ and ${\cal B}_k=i\sqrt{1/\lambda^2+k^2\lambda/(d_F+\lambda)}\approx i/\lambda$.

Using the normalization relation
\begin{align}
    &\frac{1}{2}\int {\rm d}x [{\varepsilon_r}\left(\tilde{E}_{x,k}\tilde{E}^*_{x,k}+\tilde{E}_{z,k}\tilde{E}^*_{z,k}\right)+{\mu_0}\tilde{H}_{y,k}\tilde{H}^*_{y,k}+{\mu_0}\lambda^2\tilde{J}_{s,k}\tilde{J}_{s,k}^*]\approx8\varepsilon_{\rm FI}{\cal E}_{0x}{\cal E}_{0x}^{*}d_F={\hbar\Omega_s}/{2},
    \label{Hkk}
\end{align}
the magnetic field is quantized as 
\begin{align}
    \hat{\bf H}_{Sw}&=\hat{H}_{Sw,y}\hat{\bf y} =\int \frac{{\rm d}k}{\sqrt{2\pi}}\left(\frac{\Omega_s}{2k}\sqrt{\frac{\varepsilon_{\rm FI}\hbar\Omega_s}{d_F}}\hat{p}_ke^{ikz}+{\rm H.c.}\right)\hat{\bf y}.
\end{align}
For the wave propagating with the general wave vector ${\bf k}=k_y\hat{\bf y}+k_z\hat{\bf z}$, the magnetic field $\hat{\bf H}_{Sw}=\hat{H}_{Sw,y}\hat{\bf y}+\hat{H}_{Sw,z}\hat{\bf z}$ for the Swihart photons has components
\begin{align}
     \hat{H}_{{\rm Sw},y}&=\int \frac{d{\bf k}}{2\pi}~ \left(\frac{\Omega_s}{2k}\sqrt{\frac{\varepsilon_{\rm FI}\hbar\Omega_s}{d_F}}\cos\theta_{\bf k} e^{i{\bf k}\cdot{\pmb \rho}}\hat{p}_{\bf k}+{\rm H.c.}\right),\nonumber\\
     \hat{H}_{{\rm Sw},z}&=\int \frac{d{\bf k}}{2\pi}~ \left(-\frac{\Omega_s}{2k}\sqrt{\frac{\varepsilon_{\rm FI}\hbar\Omega_s}{d_F}}\sin\theta_{\bf k} e^{i{\bf k}\cdot{\pmb \rho}}\hat{p}_{\bf k}+{\rm H.c.}\right),
\end{align}
where $\theta_{\bf k}$ is the angle between the wave vector ${\bf k}$ and the bias magnetic field $H_0\hat{\bf z}$.

The coupling between magnons and photons occurs through the Zeeman interaction $\hat{H}_{\rm int}=-\mu_0\int d{\bf r}\hat{\bf M}\cdot\hat{\bf H}=\int d{\bf k}~g_{\bf k}\left(\hat{m}_{\bf k}\hat{p}_{\bf k}^{\dagger}+\hat{m}_{\bf k}^{\dagger}\hat{p}_{\bf k}-\hat{m}_{\bf k}^{\dagger}\hat{p}_{\bf -k}^{\dagger}-\hat{m}_{\bf k}\hat{p}_{\bf -k}\right)$, where the coupling constant 
\[
g_{\bf k}/\hbar=\cos\theta_{\bf k}\mu_0\gamma\Omega_s/(2|{\bf k}|)\sqrt{(H_0+M_s)M_s\Omega_s\mu_0\varepsilon_{\rm FI}/\omega_S}
\]
depends on the propagation direction. The total Hamiltonian of the system 
\begin{align}
    \hat{H}_{\rm tot}=\int d{\bf k} \left(\hbar\omega_F\left(\hat{m}_{\bf k}^{\dagger}\hat{m}_{\bf k}+\frac{1}{2}\right)+\hbar\Omega_s\left(\hat{p}_{\bf k}^{\dagger}\hat{p}_{\bf k}+\frac{1}{2}\right)+g_{\bf k}\left(\hat{m}_{\bf k}\hat{p}_{\bf k}^{\dagger}+\hat{m}_{\bf k}^{\dagger}\hat{p}_{\bf k}-\hat{m}_{\bf k}^{\dagger}\hat{p}_{\bf -k}^{\dagger}-\hat{m}_{\bf k}\hat{p}_{\bf -k}\right)\right)
\label{SC_hamiltonian}
\end{align}
goes beyond the perturbation theory. By using the Bogoliubov transformation, the dispersion relation of the collective modes for the SC$|$FI$|$SC heterostructure is governed by
\begin{equation}
    \omega^4-\omega^2\left(\omega_F^2+\Omega_s^2\right)+\omega_F^2\Omega_s^2-4\omega_F\Omega_s{g_{\bf k}^2}/{\hbar^2}=0.
    \label{Qum_dispersion_SC}
\end{equation}
Solving this equation yields the dispersions of two collective modes
\begin{equation}
    \omega_{u(l)}^2=\frac{1}{2}\left(\Omega_s^2+\omega_F^2\pm\sqrt{\left(\Omega_s^2-\omega_F^2\right)^2+16\Omega_s\omega_F\frac{g_{\bf k}^2}{\hbar^2}}\right).
\end{equation}

Figures~\ref{MP polariton_SC}(b)-(e) address the ultrastrong coupling between magnon and Swihart photon modes in various propagation directions. The calculation assumes a YIG film with thickness $2d_F=200$~nm, saturation magnetization $\mu_0M_s=0.24$~T~\cite{24_SF_shift_van_der_sar,saturation_magnetization2} at low temperatures, external magnetic field $\mu_0H_0=50$~mT, and dielectric constant $\varepsilon_{\rm FI}=8\varepsilon_0$~\cite{YIG_parameters2}. The superconductor film is NbN, holding London’s penetration depth $\lambda=80$ nm at $T=0.1 T_c\sim 1$ K~\cite{Temperature_dependence1,Temperature_dependence2,Temperature_dependence3}. When the mode propagates parallel to the magnetization, as in Fig.~\ref{MP polariton_SC}(b), the upper branch converges to the high-frequency Swihart photon mode for large $k$, while the lower branch approaches the Kittel frequency. The frequency gap $\Delta\omega=\omega_u-\omega_l$ at the crossing point of the renormalized FMR and Swihart modes is approximately 39.7~GHz, comparable to the renormalized FMR frequency, indicating ultrastrong coupling. For the arbitrary propagation directions, the anticrossings are illustrated in Fig.~\ref{MP polariton_SC}(c); the gap $\Delta\omega$ decreases with $\theta_{\bf k}=\pi/4$.
When $\theta_{\mathbf{k}}=\pi/2$, the modes decouple since $g_{\bf k}=0$ as shown in Fig.~\ref{MP polariton_SC}(d). Therefore, the anti-crossing gap depends on the propagation directions $\theta_{\bf k}$, reaching a maximum when $\theta_{\bf k}=0$ and reducing to zero at $\theta_{\bf k}=\pi/2$, illustrating anisotropic coupling. The ferromagnetic film thickness $d_F$ also influences the coupling. When $d_{F}<\lambda$, the coupling strength varies, but it saturates when $d_F\gg\lambda$ [Fig.~\ref{MP polariton_SC}(e)]. A thick YIG film $d_F\gg\lambda$ yields a maximum anti-crossing gap of around 55~GHz.

\subsubsection{Magnon-magnon ultrastrong coupling}

Ultrastrong magnon-magnon coupling can be realized when a multilayer ferromagnetic structure is sandwiched between two superconductor layers~\cite{Gordeeva2025,26_AFM_Resonances_in_SFS_Stolyarov}.  
In the range of magnon wave numbers of the order of microwave photons $k \lesssim 1000~\rm{m^{-1}}$, the strength of this interaction is many times larger than the strength of the usual dipolar interaction between ferromagnetic layers having thicknesses of tens of nanometers, separated by an insulator layer. 
Similar to the magnon-photon coupling~\cite{52_Persistent_nodal_magnon-photon_polariton_in_ferromagnetic_heterostructures,20_shift_Silaev,23_magnon_photon_coupling_Silaev}, the strength of the magnon-magnon coupling is also anisotropic. Still, its anisotropy is opposite to that of the magnon-photon coupling: while the strength of the magnon-photon interaction is maximum when the magnon propagates along the equilibrium magnetization direction $\theta_{\mathbf k} = 0$ and vanishes when $\theta_{\mathbf k} = \pi/2$ (Sec.~\ref{magnon_polariton_theory}), the strength of the magnon-magnon coupling is maximal at $\theta_{\mathbf k} = \pi/2$ and tends to zero at $\theta_{\mathbf k} = 0$. 
This feature enables the experimental separation of magnon-magnon and magnon-photon couplings. Additionally, the superconductor layers introduce extra functionality to the system, enabling the adjustment of the dispersions of the coupled magnons and photons by varying the temperature, e.g., an effective temperature-controlled tuning of the wave number over a wide range of frequencies, tens of GHz~\cite{Gordeeva2025}.

\begin{figure}[htp]
    \centering
    \includegraphics[width=0.6\linewidth]{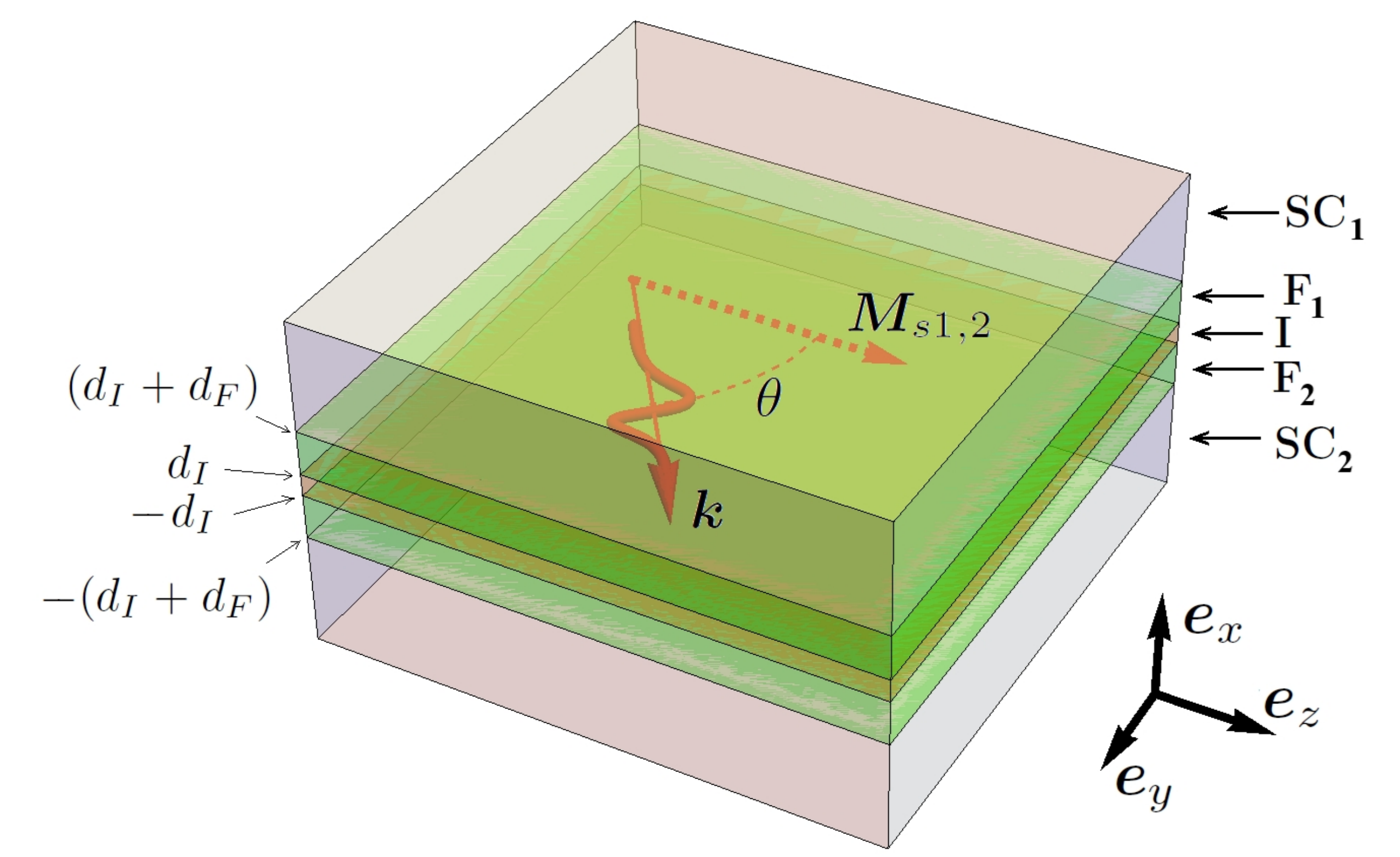}
    \caption{SC/F/I/F/SC heterostructure for the realization of ultra-strong magnon-magnon coupling. The equilibrium magnetizations $\mathbf M_{s1,2}$ of the $\rm F_{1,2}$ layers are aligned with the applied magnetic field $H_0 \mathbf e_z$.\\     
    \textit{Source:} The figure is taken from Ref.~\cite{Gordeeva2025}.}
    \label{magnon_magnon_sketch}
\end{figure}

In Ref.~\cite{Gordeeva2025}, the ultrastrong magnon-magnon coupling was proposed in a SC$|$F$|$I$|$F$|$SC heterostructure, containing two ferromagnetic insulators $F_{1,2}$ separated by a non-magnetic insulating layer $I$, as depicted in Fig.~\ref{magnon_magnon_sketch}. The material parameters, assumptions, and range of magnon wave numbers of interest are almost identical to those in Sec.~\ref{magnon_polariton_theory}. 
The in-plane static external magnetic field $\mathbf{H}_0=H_0 \mathbf e_z$ biases the equilibrium magnetizations $\mathbf M_{s1}=M_{1}\mathbf e_z $ and $\mathbf M_{s2}=M_{2}\mathbf e_z $ in the $F_1$ and $F_2$ layers, respectively. In the linear response regime magnons are described by the transverse fluctuations of the magnetizations in F$_1$ and F$_2$  as $\mathbf M_{\omega 1(2)}(\pmb \rho, t)=\tilde {\mathbf M}_{1(2)} e^{i{\mathbf k}\cdot{\pmb \rho}-i\omega t}$ with amplitudes $\tilde {\mathbf M}_{1(2)}=\tilde M_{1(2)x}\mathbf e_x+\tilde M_{1(2)y}\mathbf e_y$, wavevector $\mathbf k=k_y \mathbf e_y+k_z \mathbf e_z$, in-plane radius vector $\pmb \rho=y \mathbf e_y+z \mathbf e_z$, and frequency $\omega$. The full magnetizations of the $F_{1,2}$ layers are $\mathbf M_{1(2)}(\pmb \rho, t) = \mathbf M_{s1(2)} + \mathbf M_{\omega 1(2)}(\pmb \rho, t)$. $\theta_{\mathbf k}$ is the angle between the wave vector $\mathbf k$ and both equilibrium magnetizations $\mathbf M_{s1}$ and $\mathbf M_{s2}$.

The dispersion relations of the magnonic and photonic excitations are derived by solving the coupled Maxwell's equations for electric and magnetic fields and the Landau-Lifshitz-Gilbert (LLG) equation for the magnetizations. The Meissner currents, induced in the superconductors by the stray fields of magnons, on the magnetization dynamics, are taken into account via the renormalized demagnetization field produced by the magnetic excitations in ferromagnets:
 \begin{align}
     \hat H=-\hat N \hat {\tilde M}, ~~\hat H=\left( \begin{array}{c}
          \tilde H_{x1}  \\
          \tilde H_{x2}\\
          \tilde H_{y1}\\
          \tilde H_{y2}
     \end{array}\right),~~\hat {\tilde M}=\left( \begin{array}{c}
          \tilde M_{1x}  \\
          \tilde M_{2x}\\
          \tilde M_{1y}\\
          \tilde M_{2y}
     \end{array}\right),
     \label{H_M}
 \end{align}
where $\tilde H_{x,y1(2)}$ are the amplitudes in ${\mathbf H}_{1(2)}({\mathbf r},t)={\mathbf H}_0 + \tilde {\mathbf H}_{1(2)}e^{i{\mathbf k} \cdot \pmb{\rho}-i\omega t}$, in which the indices ``1" (``2") denote the magnetic field in $F_1$ and $F_2$, respectively;
  \begin{align}
     &\hat N=\left(\begin{array}{cccc}
        1  & 0&0&0 \\
         0 &1&0&0\\
         0&0&N&N\\
         0&0&N&N
     \end{array}\right),~~~~N=\dfrac{d_{F}(k_I^2 k_F^2\lambda \alpha-k^2\chi\sin^2\theta_{\bf k})}{2\lambda \alpha(k^2\chi-k_I^2 k_F^2 \lambda \alpha)},
     \label{N_final}
 \end{align}
 is the demagnetization tensor, 
in which $\alpha \approx [1+(d_F+d_I)/\lambda]$, $k_{I(F)} = \omega \sqrt{\mu_0 \varepsilon_{I(F)}}$, and $\chi=d_Fk_I^2+d_Ik_F^2$.

The demagnetization fields (\ref{H_M}) affect the magnetization dynamics via the LLG equation
 \begin{align}
     {\partial{\mathbf M}_{1(2)}}/{\partial t}=-\gamma\mu_0{\mathbf M}_{1(2)}\times{\mathbf H}_{1(2)},
     \label{LLG_SFIFS}
 \end{align}
in which the magnetizations $\mathbf M_1$ and $\mathbf M_2$ of ${F_1}$ and ${F_2}$ layers are coupled via the non-diagonal elements of the magnetization tensor (\ref{N_final}). 
 The Meissner currents induced in the superconductors by the stray fields of magnons produce much stronger coupling compared to the bare dipolar interaction. When the superconductor layers are in the normal state, such that $\lambda \to \infty$ and $N \to 0$, the coupling is absent.
The eigenfrequencies of the coupled system are found from the linearized LLG equations (\ref{LLG_SFIFS})
\begin{align}
    \omega_{1,2}^2=&\frac{1}{2}\left[ \omega_{01}^2 + \omega_{02}^2  \pm\sqrt{(\omega_{01}^2 + \omega_{02}^2)^2-4 \omega_{01}^2 \omega_{02}^2(1-L^2)} \right],
    \label{modes_asymm}
\end{align}
where $\omega_{01(2)} =\mu_0 \gamma  \sqrt{(H_0 + M_{1(2)})(H_0 + N M_{1(2)})}$
and $L = N \sqrt{{M_1 M_2}/[{(H_0 + N M_1)(H_0 + N M_2)}}]$.

\begin{figure}[htp!]
    \centering
    \includegraphics[width=\linewidth]{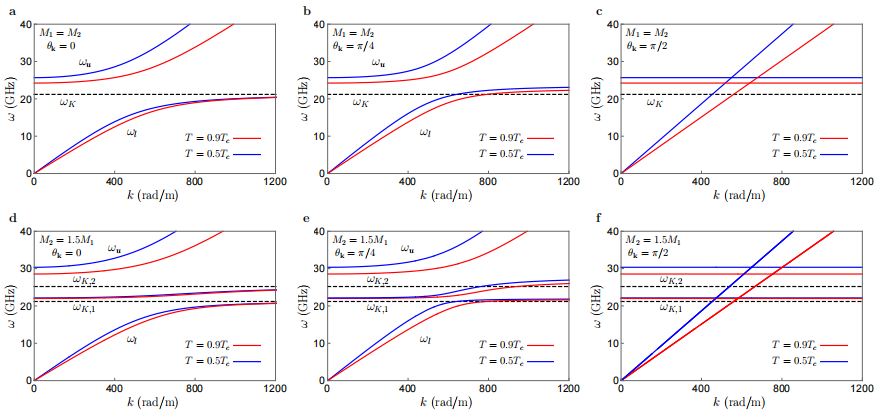}
    \caption{Frequency dispersion of the eigenmodes in the SC/F/I/F/SC structure in the symmetric configuration with $M_1=M_2$ [panels (a)-(c)] and in the asymmetric case with $M_2=1.5M_1$ [panels (d)-(f)]. Different columns correspond to different mode propagation directions $\theta_{\mathbf k}$. The blue and red curves in each panel correspond to different temperatures. \\
    \textit{Source:} The figure is taken from Ref.~\cite{Gordeeva2025}.}
    \label{fig:dispersion}
\end{figure}

The upper row of  Fig.~\ref{fig:dispersion} addresses the spectra of the SC/F/I/F/SC heterostructure with $M_1=M_2 = M$. 
When $T>T_c$, the two Kittel modes are degenerate with frequency $\omega_{K} =  \mu_0 \gamma\sqrt{H_0(H_0+M)}$, shown in Figs.~\ref{fig:dispersion} (a)-(c) by the dashed lines. At $T<T_c$, the Swihart mode strongly couples with one of the magnon modes, namely the acoustic mode, resulting in the same anticrossing as described in Sec.~\ref{magnon_polariton_theory}.
The spectrum of the resulting magnon-polaritons consists of the upper $\omega_u$ and lower $\omega_l$ branches. The second magnon mode, which is the optical mode, remains uncoupled with the Swihart
mode and, consequently, its frequency remains  $\omega_K$. This is because the magnonic stray fields in the optical mode with $M_1 = M_2$ are zero and, therefore, cannot induce Meissner currents in the superconductor layers. 
The new feature emerging in the SC/F/I/F/SC system compared to the SC/FI/SC structure discussed in Sec.~\ref{magnon_polariton_theory} is that the splitting between the acoustic and optic magnonic modes is giant, which is a manifestation of the ultra-strong magnon-magnon interaction.

The magnon-magnon interaction can be best addressed without the contamination of the magnon-photon coupling by analyzing the spectra at $k \gg k_K$, where $k_K \sim \omega_K \sqrt{\mu_0 \varepsilon_{FI}}$ is the wave vector of the photon at the Kittel frequency.  When $k \gg k_K$, the demagnetization factor in Eq.~(\ref{N_final}) becomes
\begin{align}
N = \frac{d_F \sin^2 \theta_{\bf k}}{2 (d_F + d_I + \lambda)}.
\label{eq:N_large_k}
\end{align}
The magnon frequencies when $k \gg k_K$ are expressed by $\omega_1 = \omega_K$ and $\omega_2=  \mu_0 \gamma\sqrt{(H_0 + M)(H_0 + N M)}$, where $N$ is expressed by Eq.~(\ref{eq:N_large_k}). 
The magnon frequency splitting 
$\Delta \omega_{m-m}(k \gg k_K) = \omega_2-\omega_1$ is large and anisotropic. The anisotropy is {\it opposite} to that of the magnon-photon interaction: the maximal magnon-magnon interaction is reached at $\theta_{\bf k} = \pi/2$, and vanishes at $\theta_{\bf k}=0$, as seen in the upper row of Fig.~\ref{fig:dispersion}. This maximal splitting at $\theta_{\bf k} = \pi/2$, as in Fig.~\ref{fig:dispersion}(c), is of the order of the bare frequency itself, and is much larger than the bare dipolar interaction in the F/I/F trilayer with the same parameters.

The spectra with $M_1 \neq M_2$ are presented in the bottom row of Fig.~\ref{fig:dispersion}. When $T>T_c$, there are again two uncoupled Kittel modes $\omega_{K,1(2)}=\mu_0 \gamma\sqrt{H_0(H_0+M_{1(2)})}$,  shown in Figs.~\ref{fig:dispersion}(d)-(f) by the dashed lines. 
When $T < T_c$, the interaction between the magnon modes and between the magnon and Swihart modes via the Meissner supercurrents is switched on. Crucially,  both magnon modes, acoustic and optical, are coupled to the photon because the magnon stray fields are not fully canceled even for the optical mode. 
As a result, both modes are coupled to the photon with coupling strength depending on the propagation direction $k$, as addressed in Figs.~\ref{fig:dispersion}(d)-(e). 
The only exception is $\theta_{\bf k} =\pi/2$, where the magnon-photon coupling vanishes.  The magnon-magnon coupling when $k \gg k_K$ is maximal at $\theta_{\bf k} =\pi/2$ and vanishes at $\theta = 0$. There is a giant frequency shift of several GHz for both magnonic modes at $\theta_{\bf k}=\pi/2$ when $k \gg k_K$.

\subsection{Plasmonic modes in SC$|$FM$|$SC Josephson junction}

We have discussed above the electromagnetic coupling that induces the Swihart mode in the SC$|$I$|$SC junction and the magnon-photon polariton in the SC$|$FM$|$SC junction. Neither of these discussions takes the Josephson effect into account. A supercurrent across the middle layer may arise when it is metallic or a thin insulator. Combining the Josephson current into Maxwell's equation, the collective modes are altered~\cite{Josephson_book,41_Eremin,1_Efetov,36_Bobkov}.

\begin{figure}[htp!]
    \centering
    \includegraphics[width=0.95\linewidth]{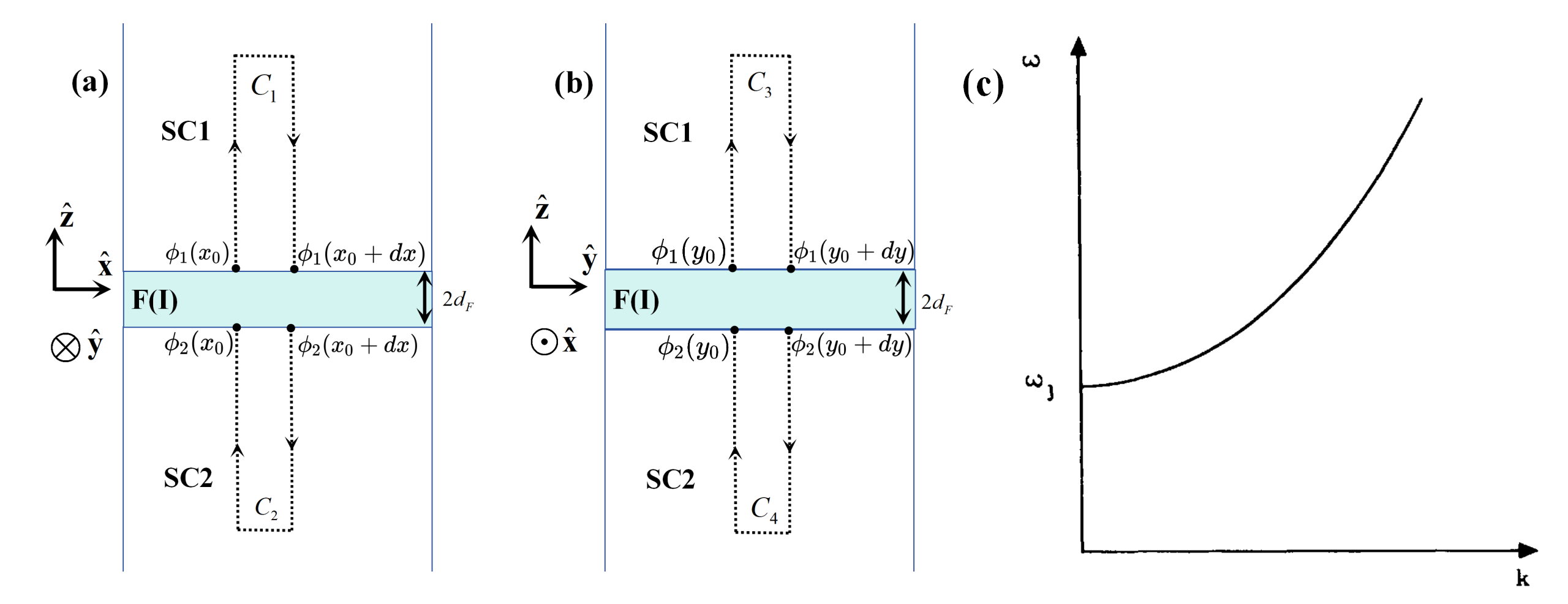}
    \caption{(a) and (b) illustrate the integral contours $\{C_1, C_2, C_3, C_4\}$ to derive the effect of the in-plane magnetic field on the SC$|$I$|$SC or SC$|$FM$|$SC Josephson junction. The thickness of the non-superconductor region is $2d_F$. (c) The dispersion of the collective modes in the SC$|$I$|$SC Josephson junction with a weak phase fluctuation $\varphi\propto e^{i{\bf k}\cdot {\bf r}-i\omega t}\ll 1$. The Swihart mode is gapped to the Josephson plasma frequency $\omega_J$. \\
 \textit{Source:} Figure (c) is taken from Ref.~\cite{Josephson_book}.}
    \label{magnetic_josephson}
\end{figure}

\textbf{Gapped Swihart mode by Josephson effect}.---We first address the role of the in-plane magnetic field ${\bf H}$ in the Josephson junction.  Figure~\ref{magnetic_josephson} illustrates a thin insulator (or a ferromagnet) with thickness  $2d_F$ along the normal $\hat{\bf z}$-direction, sandwiched by two thick superconductors labeled by ``SC1" and ``SC2".  The magnetic field is uniform across the thin insulator layer and decays inside the superconductors ``1" and ``2" with the decay length $\lambda_{\rm 1}$ and $\lambda_{\rm 2}$. 
The superconducting-phase gradients inside the two superconductors 
    \begin{align}
    \nabla \phi_{\rm 1} =\dfrac{2 \pi}{\Phi_0}\left(\mu_0 \lambda_1^2 {\bf J}_s+{\bf A}\right),\quad
    \nabla \phi_{\rm 2} =\dfrac{2 \pi}{\Phi_0}\left(\mu_0 \lambda_2^2 {\bf J}_s+{\bf A}\right),
    \label{SC_phase_with_A}
    \end{align}
are governed by both the supercurrent ${\bf J}_s$ and vector potential ${\bf A}$~\cite{Josephson_book,Josephson_book1},
where $\Phi_0=h/(2e)$ is the magnetic flux quantum.
To find the tunability of the phase difference $\varphi=\phi_{\rm 1}-\phi_{\rm 2}$ by the magnetic field ${\bf H}$, we construct four paths $\{C_1, C_2, C_3, C_4\}$ in the two superconductors.
As shown in Fig.~\ref{magnetic_josephson}(a), the paths $C_1$ and $C_2$ are located in the $x$-$z$ plane, which start from the point $\phi(x_0)$ and $\phi(x_0+dx)$, and end at $\phi(x_0+dx)$ and  $\phi(x_0)$.
The paths $C_3$ and $C_4$ in Fig.~\ref{magnetic_josephson}(b) are located in the $y$-$z$ plane, which start from the point $\phi(y_0)$ and $\phi(y_0+dy)$ and end at $\phi(y_0+dy)$ and $\phi(y_0)$.
Integrating Eq.~\eqref{SC_phase_with_A} along the paths $C_{ 1}$ and $C_{ 2}$ in Fig.~\ref{magnetic_josephson}(a) yields
\begin{align}
\phi_{1}(x_0+dx)-\phi_{1}(x_0) =\dfrac{2 \pi}{\Phi_0} \int_{C_1} \left(\mu_0 \lambda_1^2 {\bf J}_s+{\bf A}\right) \cdot d{\bf l},\nonumber\\
\phi_{2}(x_0)-\phi_{2}(x_0+dx) =\dfrac{2 \pi}{\Phi_0} \int_{C_2} \left(\mu_0 \lambda_2^2 {\bf J}_s+{\bf A}\right) \cdot d{\bf l}.
\label{inte_path}
\end{align}
When the thickness of the superconductor films is much larger than their London's penetration depth $\lambda_{1(2)}$, it is allowed to extend the contours deep inside the superconductors where the supercurrent ${\bf J}_s$ vanishes; thereby, the bulk supercurrents \({\bf J}_s\) are either perpendicular to the contour of integration or negligibly small. Taking the summation of the two equations of Eq.~\eqref{inte_path} leads to
\begin{align}
    \varphi(x_0+dx)-\varphi(x_0)=\dfrac{2 \pi}{\Phi_0} \left(\int_{C_1} {\bf A}\cdot d{\bf l} +\int_{C_2} {\bf A}\cdot d{\bf l}\right).
\end{align}

For the SC$|$I$|$SC Josephson junction, neglecting the barrier thickness $2d_F$, $C_1+C_2$ becomes a closed contour $C$, such that a phase difference
 \begin{align}
    \varphi(x_0+dx)-\varphi(x_0)=&\dfrac{2 \pi}{\Phi_0}\oint_C {\bf A}\cdot d{\bf l}=\dfrac{2 \pi}{\Phi_0}\oint_C {\bf B}\cdot {\bf n} d{s}=\dfrac{2 \pi}{\Phi_0} \mu_0 ( \lambda_1+\lambda_2 +2d_F)H_ydx
    \label{phase_difference_x}
\end{align}
is induced by the in-plane magnetic field.
When $dx\rightarrow0$, Eq.~\eqref{phase_difference_x} leads to 
     \begin{align}
	\partial_x \varphi=\dfrac{2 \pi}{\Phi_0} \mu_0 (\lambda_1+\lambda_2 +2d_F)H_y.
        \label{phi_x}
\end{align}
Similarly, integrating Eq.~\eqref{SC_phase_with_A} along the paths $C_{3}$ and $C_{4}$ in Fig.~\ref{magnetic_josephson}(b), we arrive at  
 \begin{align}
    \varphi(y_0+dy)-\varphi(y_0)=\dfrac{2 \pi}{\Phi_0}\oint {\bf A}\cdot d{\bf l}=\dfrac{2 \pi}{\Phi_0}\oint {\bf B}\cdot {\bf n} d{s}=-\dfrac{2 \pi}{\Phi_0} \mu_0 ( \lambda_1+\lambda_2 +2d_F)H_xdy,
        \label{phase_difference_y}
\end{align}
which when $dy\rightarrow 0$ leads to 
  \begin{align}
      \partial_y \varphi=-\dfrac{2 \pi}{\Phi_0} \mu_0 (\lambda_1+\lambda_2 +2d_F)H_x.
    \label{phi_y}
  \end{align}

 Equations~\eqref{phi_x} and~\eqref{phi_y} demonstrate that the in-plane magnetic fields induce the phase difference $\varphi$, which may result in the Josephson current ${\bf J}_J=j_c \sin{\varphi} \hat{\bf z}$ flowing along the $\hat{\bf z}$-direction across the insulator layer. 
On the other hand, the phase difference $\varphi$ satisfies the Josephson relation  $2eV=\hbar\partial_t\varphi$. The out-of-plane electric field $E_z$ is uniform in the thin insulator layer and vanishes inside the superconductor~\cite{52_Persistent_nodal_magnon-photon_polariton_in_ferromagnetic_heterostructures}, so $V=2d_F E_z$. 
  Inside the insulator layer, Maxwell's equations
\begin{align}
(\nabla\times {\bf H})_z=J_z+\varepsilon_0\varepsilon_r\partial_t E_z
\rightarrow\partial_x H_y-\partial_y H_x=j_c \sin{\varphi} +\varepsilon_0\varepsilon_r/(2d_F) \partial_t V.
\label{jz}
\end{align}
Substituting Eqs.~(\ref{phi_x}) and (\ref{phi_y}) into (\ref{jz}), we obtain the differential equation for the phase difference $\varphi$
\begin{align}
(\partial_x^2+\partial_y^2)\varphi -\frac{1}{\tilde{c}^2} \partial_t^2\varphi=\frac{1}{\Lambda^2} \sin{\varphi},
\label{charastic_phi}
\end{align}
where	$\tilde{c}=c \sqrt{d_F/[\varepsilon_r (\lambda +d_F)]}$ and by assuming $\lambda_1=\lambda_2=\lambda_L$, $\Lambda=\sqrt{\Phi_0/[j_c 4 \pi \mu_0 (\lambda_L +d_F)]}$. 
The collective photon modes can be found by solving such a nonlinear differential equation. In the absence of the Josephson current $j_c=0$, the transverse mode with phase fluctuation $\varphi\propto e^{i {\bf k}\cdot {\bf r}-i\omega t}$ oscillates with the frequency $\omega=\tilde{c} k$, which recovers the Swihart-mode frequency~\cite{Swihart_mode}. When $j_c\neq 0$, assuming $\varphi\ll 1$ with $\sin{\varphi}\sim \varphi$, it is found that $	\omega=\sqrt{\omega_J^2+\tilde{c}^2 k^2}$ with $\omega_J=\tilde{c}/\Lambda$, which is gapped by the Josephson current as illustrated in Fig.~\ref{magnetic_josephson}(c).

\textbf{Modulated ultrastrong magnon-photon coupling}.---In Sec.~\ref {magnon_polariton_theory}, we illustrate that the magnon mode couples with the Swihart photon mode in the SC$|$FM$|$SC heterostructure and results in the anticrossing spectrum; however, the effect of the Josephson effect is disregarded with a small Josephson current.
In the SC$|$FM$|$SC Josephson junction, the precession of a magnetization emits a magnetic field into the neighboring superconductors, which may induce a phase difference $\varphi$ across the two superconductors, which, in turn, generates the Josephson current $J_J=j_c \sin{\varphi}$ that acts back on the magnetization. 
Here, we review that the Josephson current modulates the magnon-photon coupling.

Including the magnetization $\bf M$ that is uniform across the thin ferromagnetic film along $\hat{\bf z}$, Eqs.~\eqref{phase_difference_x} and~\eqref{phase_difference_y} become 
\begin{align}
	\varphi(x_0+dx)-\varphi(x_0)=&\dfrac{2 \pi}{\Phi_0}\oint {\bf B}\cdot {\bf n} d{s}
	=\dfrac{2 \pi}{\Phi_0} \mu_0 \left[( \lambda_1+\lambda_2 +2d_F)H_y+2d_F M_y\right]dx,\nonumber\\
	\varphi(y_0+dy)-\varphi(y_0)=&\dfrac{2 \pi}{\Phi_0}\oint {\bf B}\cdot {\bf n} d{s}=-\dfrac{2 \pi}{\Phi_0} \mu_0 \left[( \lambda_1+\lambda_2 +2d_F)H_x+2d_F M_x\right]dy.
\label{phase_difference_sfs}
\end{align}
By taking $dx\rightarrow 0$, $dy\rightarrow0$, and London's penetration length $\lambda_1=\lambda_2=\lambda_L$, Eq.~\eqref{phase_difference_sfs} becomes~\cite{josephson1,Josephson_book,Josephson_book1}
\begin{align}
	a(\hat{\bf z}\times \nabla\varphi)={\bf H}_{\perp}+\tilde{d}_F{\bf M}_\perp,
	\label{h_m_phi}
\end{align}
where $a=\Phi_0/[4\pi \mu_0 (\lambda_L+d_F)]$, $\tilde{d}_F=d_F/(d_F+\lambda_L)$, ${\bf H}_\perp=H_x\hat{\bf x}+H_y\hat{\bf y}$, and ${\bf M}_\perp=M_x\hat{\bf x}+M_y\hat{\bf y}$. Both the magnetic field and magnetization induce the phase difference between superconductors.  Equation~\eqref{h_m_phi} is coupled with Maxwell's and LLG equations, resulting in the hybrid collective modes.

\begin{figure}[htp!]
    \centering
    \includegraphics[width=0.98\linewidth]{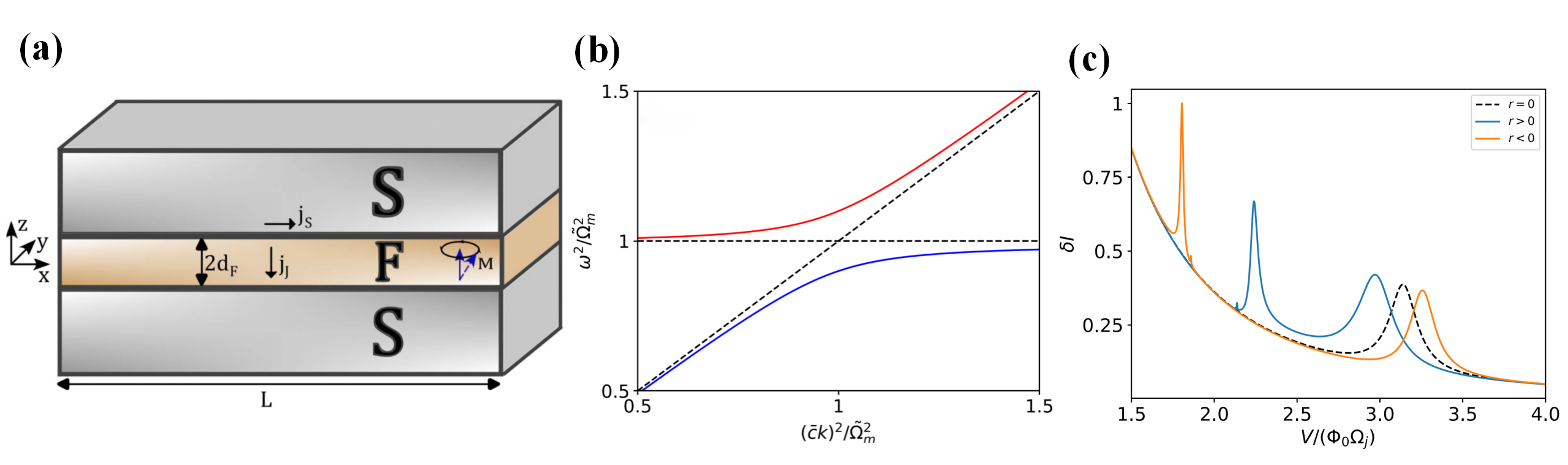}
    \caption{Josephson effect on the collective modes of the SC$|$FM$|$SC heterostructure. (a) In the SC$|$FM$|$SC multilayer, the magnetization $M_s$ is along the $\hat{\bf z}$ direction due to the out-of-plane anisotropy. The thickness of the ferromagnet is $2d_F$. $j_s$ and $j_J$ are the supercurrent and Josephson current, respectively. (b) Dispersion of the collective modes for $K>1$. The black dotted lines refer to the non-interacting dispersions. (d) Normalized corrections $\delta I$ in the $I$-$V$ curve. \\ 
      \textit{Source:} The figures are taken from Ref.~\cite{41_Eremin}.}
    \label{Josephson}
\end{figure}

In 2009, Volkov and Efetov considered the effect of Josephson current in the collective modes of the SC$|$I$|$FM$|$SC junction~\cite{1_Efetov}. Later in 2024, Derendorf \textit{et al}. solved  
the collective modes in the SC$|$FM$|$SC heterostructures in the existence of the Josephson current~\cite{41_Eremin}. They consider the ferromagnetic film with thickness $2d_F$ sandwiched by two thick adjacent superconductors, as shown in Fig.~\ref{Josephson}(a). The ferromagnet has an out-of-plane anisotropy field ${\bf H}_{\rm an}=KM_s \hat{\bf z}$, which biases the equilibrium magnetization ${\bf M}_s$ along the surface normal $\hat{\bf z}$-direction. 
The out-of-plane magnetization then leads to a diamagnetic field ${\bf H}_{\rm dem} =-M_s\hat{\bf z}$, resulting in the magnetic induction  $B_z=\mu_0({H}_{\rm dem}+M_s)=0$ in the ferromagnetic layer. 
Inside the ferromagnet, the dynamics of the magnetization is governed by the LLG equation $\partial_t {\bf M}=-\mu_0 \gamma ({\bf M}\times {\bf H}_{\rm eff})+(\alpha_G/M_s) {\bf M}\times \partial_t{\bf M}$, where the effective magnetic field ${\bf H}_{\rm eff}={\bf H}_{\rm dem}+{\bf H}_{\rm an}+{\bf H}_\perp$. In ${\bf M}=M_s \hat {\bf z}+ {\bf M}_{\perp}(t)$, the transverse magnetization ${\bf M}_{\perp}=\{M_x,M_y,0\}\ll M_s$ obeys
\begin{align}
	\dfrac{d{\bf M}_\perp}{dt}=-\Omega_m \hat{\bf z}\times({\bf H}_\perp+\kappa {\bf M}_\perp)+\dfrac{\alpha_G}{M_s} \hat{\bf z} \times \dfrac{d{\bf M}_\perp}{dt},
	\label{llg_perp}
\end{align}
where $\Omega_m=\mu_0 \gamma M_s$ and $\kappa=1-K$. Equation~\eqref{llg_perp} is coupled with the superconductor phase difference $\varphi$ [Eq.~\eqref{h_m_phi}] and Maxwell's equation Eq.~\eqref{jz}, resulting in the modulated magnon-photon mode. 
Combining Eqs.~\eqref{h_m_phi},~\eqref{llg_perp} and~\eqref{jz} and neglecting the Gilbert damping, the dynamics of  $\varphi$ reads
\begin{align}
	\dfrac{1}{\tilde{c}^2}
	\partial_t^2 \varphi-\nabla_\perp^2 \varphi +\dfrac{1}{\Lambda}\sin{\varphi} =- \dfrac{\tilde{d}_F R \tilde{\Omega}_m^2}{{\cal D}_m} \nabla_\perp^2 \varphi,
\end{align}
 where $\tilde{\Omega}_m=\Omega_m(\lambda_L/(\lambda_L+d_F)-K)$, $R=\Omega_m/\tilde{\Omega}_m$, and ${\cal D}_m=\omega^2-\tilde{\Omega}_m^2$. For a small phase difference $\varphi\propto e^{i {\bf k}\cdot {\bf r}-i\omega t}$, the characteristic equation reads
\begin{align}
	[\omega^2 -(\tilde{c}^2k^2+\omega_J^2)](\omega^2-\tilde{\Omega}^2_m)=-\tilde{d_F}R \tilde{\Omega}_m^2\tilde{c}^2k^2,
	\label{charastic_eq_jj}
\end{align}
which determines the spectra of the collective modes.

To show the effect of the Josephson current, we compare the case without the Josephson current ${\bf J}_J=0$; such that $\omega_J=0$. In this case, Eq.~\eqref{charastic_eq_jj} becomes $ 
(\omega^2 -\tilde{c}^2k^2)(\omega^2-\tilde{\Omega}^2_m)=-\tilde{d_F}R \tilde{\Omega}_m^2\tilde{c}^2k^2$, which is analogous to the case with the static in-plane magnetization as discussed in Sec.~\ref{magnon_polariton_theory}, i.e., Eq.~\eqref{Qum_dispersion_SC}.
Both equations show the coupling between the Swihart mode $\tilde{c}k$ and the (modified) magnon mode $\tilde{\Omega}_m$; their coupling is given by the right term in both equations.  Now the effect of the Josephson current is apparent: according to Eq.~\eqref{charastic_eq_jj}, the Josephson current first gaps the Swihart mode as $\sqrt{\tilde{c}^2k^2+\omega_J^2}$, as shown in Fig.~\ref{magnetic_josephson}(c), and then couples with the magnon mode $\tilde{\Omega}_m$, resulting in the anticrossing spectra as shown in Fig.~\ref{Josephson}(b).

Derendorf \textit{et al}.~\cite{41_Eremin} show that such collective modes could be revealed by the Fiske experiment~\cite{fiske}, i.e., by measuring the \(I\)-\(V\) characteristics in the presence of a magnetic field and voltage.
The correction to the current $\delta I_{\rm dc }$ as a function of $V$ is shown in Fig.~\ref{Josephson}(c) for different $r=\tilde{d_F} R$. 
When the modes in a ferromagnet
and a superconductor decouple ($r =0$), the correction of current $\delta I_{\rm dc}$ shows only one resonance, as shown in Fig.~\ref{Josephson}(c) by the dotted curve, which reflects the response of the Swihart mode. By contrast, when $r<0$ ($K>1$), for example, the resonance position shifts to a lower voltage. In addition, the author finds additional peaks at the lower voltages $V$, as shown by the blue curve in Fig.~\ref{Josephson}(c), which are caused by the FMR in the ferromagnet. Their positions and amplitudes depend on \(r\) and the damping strength.

\section{Exchange interaction between magnons and Cooper pairs}
\label{Exchange_interaction_between_magnons_and_Cooper_pairs}

\subsection{General theory}

\begin{figure}[tbh]
	\begin{center}
		\includegraphics[width=0.95\linewidth]{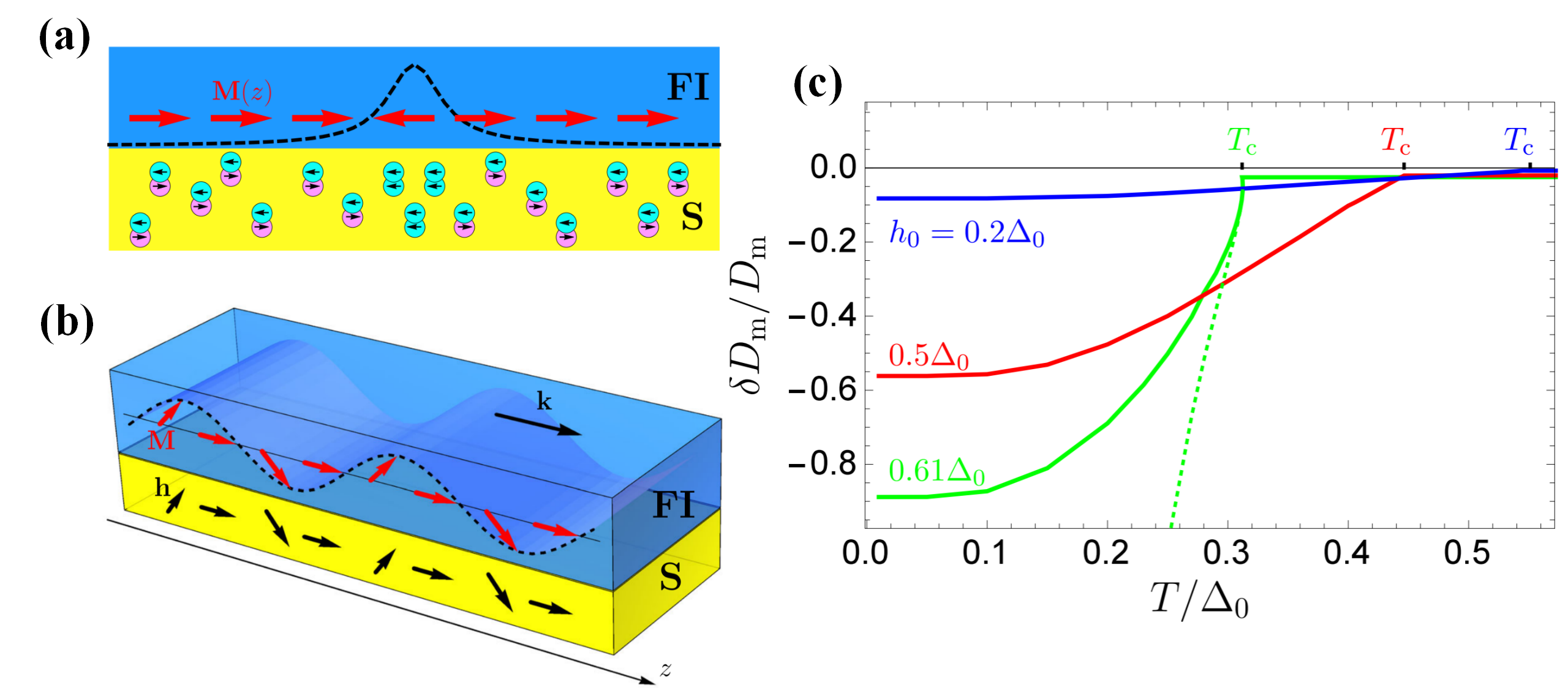}
		\caption{ (a) Sketch of the magnon-cooparon quasiparticle. The magnon is associated with the spatially varying magnetization (the dashed curve). Small black arrows in the superconductor illustrate the spins of electrons of the Cooper pairs. The localized spin-flip induces a surrounding cloud of spinful triplet Cooper pairs in S.  The magnetic moments point opposite to spins on account of the negative gyromagnetic ratio. (b) Noncollinear magnetization profile (the red arrows) is associated with a spin wave with wavenumber $k$, propagating in FI. It induces an analogous spatially varying exchange field by the black arrows in S, resulting in a spinful triplet condensate. (c) Renormalized spin stiffness $\delta D_{\mathrm{m}} / D_{\mathrm{m}}$ as a function of temperature $T$ for different static exchange field $h_0$ induced in S. $\Delta_0$ is the superconducting gap when $T = h_0 = 0$. The dashed curve plots the analytic result [Eq.~\eqref{eq:deltadm}] obtained in the limit $T \to T_{\mathrm{c}}$.\\
        \textit{Sources}: The figures are adapted from Ref.~\cite{Bobkova2022}.} 
        \label{Fig1}
	\end{center}
\end{figure}

In previous chapters, we focused on the non-local electromagnetic interaction between magnons in magnetic materials and Meissner currents in attached superconductors.  
 However, the interface exchange interaction also couples the electrons in the superconductor (S) with the magnetization in a magnet (F)~\cite{PhysRevLett.127.207001,PhysRevLett.127.207001,Bobkova2022,Bobkov2023,29_Magnon_influence_DOS_Bobkova,44_Giant_oscillatory_Gilbert_damping,12_Robinson}. 
 The exchange field produced by the uncompensated electron spins polarizes the quasiparticles in the superconductors and energetically favors triplet Cooper pairs with finite magnetic moments. In turn, spin-polarized quasiparticles and Cooper pairs renormalize the magnon dispersion. 
The quasiparticle contribution dominates at small magnon wave numbers $k$ and, therefore, mainly renormalizes the FMR $\omega(k=0)$~\cite{Bobkov2023}. However, to create the quasiparticle spin, the magnon frequency should be of the order of the superconducting gap, which corresponds to typical frequencies of the order of $10^{2}$-$10^{3}$~GHz. 
 Since such a frequency range is typical for antiferromagnets, it is reasonable to expect the effect of the quasiparticle renormalization in superconductor/antiferromagnet heterostructures~\cite{Bobkov2023}. However, it was also predicted that magnons with nonzero wave numbers universally induce a cloud of spinful triplet Cooper pairs around them in an adjacent conventional superconductor, even at GHz frequencies typical for ferromagnets. 
Here, we focus on the last effect. The resulting
composite quasiparticle, termed magnon-cooparon due to its similarity to the polaron quasiparticle, consists of a spin flip in the magnet screened
by a cloud of the spinful superfluid condensate [Fig.~\ref{Fig1}(a)] \cite{Bobkova2022}. Thus, it possesses a large effective mass and a reduced spin. This generation of spinful Cooper pairs in a conventional superconductor is only caused by the noncollinear magnetization profile of a spin wave with finite wavevector [Fig.~\ref{Fig1}(b)] and does not appear when considering FMR of the uniform magnon mode.

The most suitable heterostructure for implementing magnon-cooparons and studying them is a thin-film S/F bilayer. The sketch of the system is depicted in Fig.~\ref{Fig1}(b), in which a ferromagnetic insulator FI (e.g., YIG) is interfaced with a conventional spin-singlet s-wave superconductor S (e.g., Nb). The terminology ``thin-film bilayer" implies that the thicknesses $d_{\mathrm{FI}}$ and $d_{\mathrm{S}}$ ($\ll \xi_{\mathrm{S}}$, the superconducting coherence length) are considered such that the physical properties vary only in the in-plane direction. 
This condition is crucial not only because it simplifies calculations, but its key significance is that in this case, the region near the S/FI interface affected by the proximity effect, i.e., the appearance of triplet superconducting correlations, extends to the entire superconductor in the case of the S/FI bilayer or even to the bilayer as the whole in the case of a bilayer with a ferromagnetic metal. In its ground state, the FI is assumed to be magnetized along the $\hat{\bf z}$-direction. This magnetization induces an effective exchange magnetic field in the superconductor, which accounts for the interface exchange interaction between the conduction electrons of S and the magnetization of F. For the case of a thin-film superconductor in contact with a ferromagnetic insulator, it has the form $\mathbf{h} = J M_{\mathrm{s}} \mathbf{m}/ 2 \gamma d_{\mathrm{S}}$, where $M_{\mathrm{s}}$ the FI saturation magnetization, $J$ is the interfacial exchange interaction constant, and $\mathbf{m} = \mathbf{M}/M_s$ is the unit vector along the magnetization. 
For the S/FI heterostructures, the induced effective exchange field in the superconductor was observed experimentally by measurements of the spin-split DOS \cite{Bergeret2018}.
At the same time, for bilayers with ferromagnetic metals (S/FM), the well-pronounced homogeneous spin-split DOS was not reported. Strong suppression of superconductivity was typically demonstrated due to the leakage of Cooper pairs into the ferromagnetic region, which results in further destruction. The physical reason for the absence of well-pronounced spin-splitting of the density of states (DOS) in S/FM bilayers is not entirely apparent at present. Still, this fact forces us to make a preliminary conclusion that bilayers with a ferromagnetic \textit{insulator} are a preferable choice for generating triplet correlations in a superconductor.

On the other hand, in S/FM bilayers, triplet Cooper pairs can be induced in the ferromagnetic region due to the singlet-triplet conversion~\cite{Buzdin,Bergeret}. 
In this case, a cloud of triplet Cooper pairs, screening the magnon, could be generated directly in the ferromagnetic metal. It is expected that the qualitative physics of renormalization is similar.

To describe the renormalization of magnon properties by the cloud of triplet Cooper pairs, we combine the LLG equation to find the magnon spectra of the ferromagnet and the nonequilibrium Keldysh quasiclassical theory of superconductivity to describe triplet superconducting correlations induced by the ferromagnet in the superconductor as a result of the exchange proximity effect~\cite{Bobkova2022}. 
We assume the existence of a spin wave with wave vector $k \hat{\bf z}$ in the FI [Fig.~\ref{Fig1}(b)] such that the magnetization unit vector $\mathbf{m}(\mathbf{r},t) = \mathbf{m}_0 + \delta \mathbf{m}(\mathbf{r},t)$ consists of the equilibrium part $\mathbf{m}_0 = \hat{\bf z}$ and the excitation part $\delta \mathbf{m}(\mathbf{r},t) = \delta m \left[ \cos(kz + \omega t) \hat{\bf x} + \sin(kz + \omega t) \hat{\bf y} \right] \exp(- \kappa t)$, where $\kappa$ is the excitation decay rate. 
While we consider an excitation with wave vector along $\hat{\bf z}$, our analysis is general and valid for any in-plane wave vector. 
The magnetization dynamics is described within the LLG framework as 
\begin{align}\label{eq:llg}
\dot{\mathbf{m}} =  - \mu_0\gamma \left( \mathbf{m} \times \mathbf{H}_{\mathrm{eff}} \right) + \alpha \left(\mathbf{m} \times \dot{\mathbf{m}} \right) + \tilde{J} \left(\mathbf{m} \times \mathbf{s}\right),
\end{align}
where $\alpha$ is the Gilbert damping parameter, $\mathbf{H}_{\mathrm{eff}}$ is the effective magnetic field in the FI, and $\tilde{J} \equiv J/d_{\mathrm{FI}}$. The last term on the right-hand side of Eq.~\eqref{eq:llg} accounts for the spin torque exerted on the magnetization by the spin density $\mathbf{s}$ of electrons, which is induced in the superconductor by magnons~\cite{Ralph2008}.

The induced spin density in the superconductor is calculated using the quasiclassical Green's function. It is assumed that the superconductor is in the dirty limit. The corresponding quasiclassical theory is described in Sec.~\ref{Usadel_conventional}. 
To account for the spin effects induced in the S layer by the proximity effect with the FI layer, one should consider the spin space. The quasiclassical Green's function $\check g$ in the S layer is an $8 \times 8$ matrix in the direct product of Keldysh, spin, and particle-hole spaces. 
It obeys the Usadel equation, which takes the form 
\begin{eqnarray}
\mathrm i D \mathbf{\nabla} \bigl( \check g \otimes \mathbf{\nabla} \check g \bigr) = \bigl[ \varepsilon \hat \tau_z - (\mathbf h \cdot \hat {\pmb{\sigma}}) \hat \tau_z + \Delta \mathrm i \hat \tau_y, \check g \bigr]_\otimes,
\label{usadel_1}
\end{eqnarray}
where $D = v_F^2 \tau/3$ is the diffusion constant in the S layer. 
In comparison to Eq.~(\ref{Usadel}), Eq.~(\ref{usadel_1}) contains an additional term $\mathbf h \cdot \hat {\pmb{\sigma}}$, where $\hat {\pmb \sigma} = (\hat \sigma_x,  \hat \sigma_y, \hat \sigma_z)^T$ is a vector of the Pauli matrices in the spin space.
Physically, this term takes into account the exchange field $\mathbf h$ induced in S by the proximity to FI via the interface exchange coupling. 
Supposing we only consider the harmonic excitations, the $\otimes$-product is reduced to $ A(\varepsilon,t) \otimes B(\varepsilon, t) = A (\varepsilon-\Omega_B/2)  B(\varepsilon+\Omega_A/2,t)\exp[i (\Omega_A + \Omega_B) t]$, where  $A[B](\varepsilon, t) =  A[B] (\varepsilon) \exp[i \Omega_{A[B]} t]$. $\Delta $ is the superconducting order parameter of the S layer. 
The Green's function in the Keldysh space takes the form
\begin{eqnarray}
	\check g = 
	\left(
	\begin{array}{cc}
	\breve g^{\mathrm R} & \breve g^{\mathrm K} \\
	0 & \breve g^{\mathrm A}
	\end{array}
	\right),
	\label{keldysh_structure}
	\end{eqnarray}
where $\breve g^{{\mathrm R}({\mathrm A})}$ is the retarded (advanced) component of the Green's function and $\breve g^{\mathrm K}$ is the Keldysh component. The Keldysh component is expressed via the retarded, advanced Green's function, and the distribution function $\breve \varphi$ by $\breve g^{\mathrm K} = \breve g^{\mathrm R} \otimes \breve \varphi - \breve \varphi \otimes \breve g^{\mathrm A}$~\cite{Serene1983}.

The exchange field  $\mathbf{h}(\mathbf r,t) = h_0 \hat{\bf z}+\delta h \cos\left(\mathbf{k} \cdot \mathbf{r} + \omega t\right) \hat{\bf x} + \delta h \sin\left(\mathbf{k} \cdot\mathbf{r} + \omega t\right) \hat{\bf y}$
follows the time and spatial profiles of the magnetization of a circularly polarized magnon, where the magnon decay rate $\kappa$ is disregarded. Then $\delta \mathbf{h} \cdot {\pmb  \sigma} = \delta h {\mathrm e}^{-\mathrm i(\mathbf{k} \cdot \mathbf{r} + \omega t)\hat \sigma_z}\hat \sigma_x$.
The quasiclassical Green's function is to be found in the form $\check g = \check g_0 + \delta \check g$, where $\check g_0$ is the Green's function in the absence of the magnon and \begin{eqnarray}
\delta \check g = \hat U \otimes \left( \delta \hat g_{\mathrm mx} \hat \sigma_x \hat \tau_z + \delta \hat f_{\mathrm mx} \hat \sigma_x \mathrm i\hat \tau_y  \right) \otimes \hat U^{\dagger}
\label{transform}
\end{eqnarray}
 is the first-order correction with respect to $\delta \mathbf{h}$~\cite{Bobkova2022}, where $\hat U =  {\mathrm e}^{-\mathrm i(\mathbf{k} \cdot \mathbf{r} + \omega t)\hat \sigma_z/2}$. The correction to the anomalous Green's function $\hat f_{\mathrm mx}$ does not enter the expression of spin polarization. The correction to the normal Green's function 
\begin{eqnarray}
\delta \hat g_{\mathrm mx} = \frac{\delta h \bigl[ \hat f_{0,+}\hat \Lambda_{\mathrm{od}}^z - \hat g_{0,-}\hat \Lambda_{\mathrm d}^0 \bigr]}{2 \bigl[ \hat \Lambda_{\mathrm d}^0 \hat \Lambda_{\mathrm d}^z - \hat \Lambda_{\mathrm {od}}^0\hat \Lambda_{\mathrm {od}}^0 \bigr]}
\label{delta_g_sol2}
\end{eqnarray}
is time-independent. 
In Eq.~(\ref{delta_g_sol2}),
\begin{subequations}
\begin{align}
\hat g_{0,\pm} &= \hat g_{0,\uparrow}\left(\varepsilon+\frac{\omega}{2}\right) \pm \hat g_{0,\downarrow}\left(\varepsilon-\frac{\omega}{2}\right), 
\label{delta_g_def1} \\
\hat f_{0,\pm} &= \hat f_{0,\uparrow}\left(\varepsilon+\frac{\omega}{2}\right) \pm \hat f_{0,\downarrow}\left(\varepsilon-\frac{\omega}{2}\right), 
\label{delta_g_def2}\\
\breve \Lambda_{\mathrm{d}} &= \hat \Lambda_{\mathrm{d}}^0 + \hat \Lambda_{\mathrm{d}}^z \hat \sigma_z = \varepsilon + \frac{\mathrm i D k^2}{4}\hat g_{0,+} + \bigl( \frac{\omega}{2}-h_0 + \frac{\mathrm i D k^2}{4}\hat g_{0,-} \bigr)\hat \sigma_z,
\label{delta_g_def5} \\
\breve \Lambda_{\mathrm{od}} &= \hat \Lambda_{\mathrm{od}}^0 + \hat \Lambda_{\mathrm{od}}^z \hat \sigma_z = \Delta + \frac{\mathrm i D k^2}{4}\hat f_{0,+} + \frac{\mathrm i D k^2}{4}\hat f_{0,-} \hat \sigma_z,
\label{delta_g_def6}
\end{align}
\end{subequations}
where $\hat g_{0,\uparrow(\downarrow)}$ represent the bulk Green's functions for the superconductor in the exchange field $h_0$: 
\begin{eqnarray}
g_{0,\uparrow(\downarrow)}^{\mathrm R} = \frac{|\varepsilon \mp h_0|}{\sqrt{(\varepsilon + \mathrm i \delta \mp h_0)^2-\Delta^2}}, \label{delta_g_def3} \\
f_{0,\uparrow(\downarrow)}^{\mathrm R} = \frac{\Delta {\rm sgn}(\varepsilon \mp h_0)}{\sqrt{(\varepsilon + \mathrm i \delta \mp h_0)^2-\Delta^2}},
\label{delta_g_def4}
\end{eqnarray}
and $g(f)_{0,\uparrow(\downarrow)}^{\mathrm A} = -g(f)_{0,\uparrow(\downarrow)}^{{\mathrm R}*}$, $g(f)_{0,\uparrow(\downarrow)}^{\mathrm K} = [g(f)_{0,\uparrow(\downarrow)}^{\mathrm R} - g(f)_{0,\uparrow(\downarrow)}^{\mathrm A}] \tanh(\varepsilon/2T)$.

The distribution function also acquires a correction due to the magnon: $\breve \varphi = \breve \varphi_0 + \delta \check \varphi$. Here the equilibrium distribution function $ \varphi_0 = \tanh (\varepsilon/2T)$ and the correction $\delta \breve \varphi = \hat U \otimes\delta \breve \varphi_m \otimes \hat U^\dagger$. From the Keldysh part of the Usadel equation (\ref{usadel_1}), the first order correction to the distribution function $\delta \check \varphi_{\mathrm m} $ obeys
\begin{eqnarray}
\mathrm i Dk^2 (\delta \breve \varphi_{\mathrm m} - \breve g_{{\mathrm m}0}^{\mathrm R} \delta \breve \varphi_{\mathrm m} \breve g_{{\mathrm m}0}^{\mathrm A}) + \breve g_{{\mathrm m}0}^{\mathrm R} [\breve K, \delta \breve \varphi_{\mathrm m}] - [\breve K, \delta \breve \varphi_{\mathrm m}] \breve g_{{\mathrm m}0}^{\mathrm A} +
	\breve g_{{\mathrm m}0}^{\mathrm R} [\breve \varphi_{\mathrm m}, \delta h \hat \sigma_x \hat \tau_z] - [\breve \varphi_{\mathrm m}, \delta h \hat \sigma_x \hat \tau_z] \breve g_{{\mathrm m}0}^{\mathrm A} = 0, 
\label{distribution_correction}
\end{eqnarray}
where 
\begin{subequations}
\begin{align}
\breve K &= (\varepsilon+(\omega/2-h_0)\hat \sigma_z )\hat \tau_z +  \Delta \mathrm i \hat \tau_y,
\label{distribution_def_1}\\
\breve g_{{\mathrm m}0}^{\mathrm R,\mathrm A} &= \hat U^\dagger \otimes \breve g_0^{\mathrm R,\mathrm A} \otimes \hat U = (1/2)[(g_{0,+}^{\mathrm R,\mathrm A}+g_{0,-}^{\mathrm R,\mathrm A}
\hat \sigma_z)\hat \tau_z + (f_{0,+}^{\mathrm R,\mathrm A}+f_{0,-}^{\mathrm R,\mathrm A}\hat \sigma_z)\mathrm i\hat \tau_y].
\label{distribution_def_2}
\end{align}
\end{subequations}
The structure of Eq.~(\ref{distribution_correction}) dictates that 
\begin{eqnarray}
\delta \check \varphi_{\mathrm m} = 
\left(
\begin{array}{cc}
0 & \delta \varphi_{\mathrm m}^\uparrow \\
\delta \varphi_{\mathrm m}^\downarrow & 0
\end{array}
\right)\hat \tau_0.
\label{distribution_correction_2}
\end{eqnarray}
Substituting Eq.~(\ref{distribution_correction_2}) into Eq.~(\ref{distribution_correction}) we obtain 
\begin{eqnarray}
\delta \varphi_{\mathrm m}^\sigma = -\delta h \varphi_{{\mathrm {m-}}} \frac{2h_{0\omega}G_{-,\sigma}+\mathrm i \sigma D k^2 (g_{{\mathrm m}\sigma}^{\mathrm R} - g_{{\mathrm m} \bar \sigma}^{\mathrm A})}{4h_{0\omega}^2 G_{-,\sigma} + 4 h_{0\omega} {\mathrm i} \sigma D k^2 (g_{{\mathrm m}\sigma}^{\mathrm R} - g_{{\mathrm m} \bar \sigma}^{\mathrm A}) - (Dk^2)^2 G_{+,\sigma}},~~~~
\label{distribution_correction_3}
\end{eqnarray}
where we introduce the spin subband index $\sigma = \uparrow(\downarrow)$ in the subscripts/superscripts and $\sigma = \pm 1$ for spin-up (down) subbands, respectively. $\bar \sigma = -\sigma$, $h_{0\omega} = \omega/2-h_0$, ${g(f)}_{{\mathrm m}\sigma}^{{\mathrm R},{\mathrm A}} = g(f)_{0,\sigma}^{{\mathrm R},{\mathrm A}}(\varepsilon + \sigma \omega/2)$, and $G_{\pm,\sigma} = 1-g_{{\mathrm m}\sigma}^{\mathrm R} g_{{\mathrm m} \bar \sigma}^{\mathrm A} \pm f_{{\mathrm m}\sigma}^{\mathrm R} f_{{\mathrm m} \bar \sigma}^{\mathrm A}$.

The electron spin polarization in the superconductor 
\begin{align}
\mathbf{s} = -\frac{N_{\mathrm F}}{16} \int d\varepsilon {\rm Tr}_4 \Bigl( \hat {\sigma} \hat \tau_z \breve g^{\mathrm K} \Bigr)= s_0 \mathbf{m}_0 + \delta s_\parallel \delta \mathbf{m} + \delta s_\perp (\delta \mathbf{m} \times \mathbf{m}_0),
\label{spin2}
\end{align}
where $N_{\mathrm{F}}$ is the normal state density of states at the Fermi surface in S, 
\begin{eqnarray}
s_0  = -\frac{N_{\mathrm F}}{4} \int \limits_{-\infty}^\infty d \varepsilon \tanh \left(\frac{\varepsilon}{2T}\right) {\rm Re}\bigl( g_{0,\uparrow}^{\mathrm R} - g_{0,\downarrow}^{\mathrm R} \bigr)
\label{spin_0}
\end{eqnarray} 
is the equilibrium electron spin polarization in the superconductor, corresponding to the situation in the absence of magnons, and 
\begin{subequations}
\begin{align}
\delta s_\parallel  &= -\frac{N_{\mathrm F}h_0}{8\delta h} \int \limits_{-\infty}^\infty d \varepsilon \Bigl( 2 \varphi_{\mathrm m+} {\rm Re} (\delta g_{\mathrm mx}^{\mathrm R}) +  \sum \limits_\sigma (g_{\mathrm m\sigma}^{\mathrm R} - g_{\mathrm m \bar \sigma}^{\mathrm A}) \delta \varphi_{\mathrm m}^\sigma  \Bigr),
\label{spin_parallel}\\
\delta s_\perp  &= \frac{N_{\mathrm F}h_0}{8\delta h} \int \limits_{-\infty}^\infty d \varepsilon \Bigl( 2 \varphi_{{\mathrm m}-} {\rm Im} (\delta g_{{\mathrm m}x}^{\mathrm R}) +{\mathrm i}  \sum \limits_\sigma \sigma (g_{{\mathrm m}\sigma}^{\mathrm R} - g_{{\mathrm m} \bar \sigma}^{\mathrm A}) \delta \varphi_{\mathrm m}^\sigma\Bigr),
\label{spin_perpendicular}
\end{align}
\end{subequations}
describe the dynamic corrections to the spin polarization due to the magnon. Substituting $\mathbf{s}$ and $\mathbf{m}$ in Eq.~\eqref{eq:llg} and linearizing this equation with respect to $\delta \mathbf m$ we can obtain the eigenfrequencies of the composite magnetic excitations consisting of the magnon $\delta \mathbf m$ in FI and the dynamic corrections to the spin polarization in S.

\subsection{Magnon-cooparons}

The spin density $\mathbf{s}$ induced in the superconductor may renormalize both the excitation frequency and its lifetime.
 The renormalized eigenfrequency of the composite magnon mode
\begin{align}
 \omega = D_{\mathrm{m}} k^2 + \gamma K + \tilde{J} \left( \delta s_{\parallel} - s_0 \right), \label{eq:omega}
 \end{align}
 where $D_{\mathrm{m}}$ is the spin-wave stiffness and $K$ parameterizes a uniaxial anisotropy of the FI.
The magnon decay rate is renormalized according to 
 \begin{align}
\kappa = \alpha \omega - \tilde{J} \delta s_{\perp}. \label{eq:kappa}
\end{align}
The additional contribution to the magnon decay rate originates from the quasiparticle spin-flip processes.
 In the adiabatic limit $\hbar \omega \ll T$, one neglects $\hbar \omega $ with respect to the superconducting energies, such that $\delta s_\parallel - s_0 \propto Dk^2$ only renormalizes the magnon stiffness. The stiffness correction in the limit $T \to T_{\mathrm c}$~\cite{Bobkova2022}
\begin{align}\label{eq:deltadm}
\delta D_{\mathrm{m}} & = - \frac{\pi N_{\mathrm{F}} \gamma d_{\mathrm{S}} D \Delta^2}{16 T_{\mathrm{c}} d_{\mathrm{FI}} M_{\mathrm{s}}} \left( \frac{1}{x}\tanh x - \frac{1}{\cosh^2x} \right),
\end{align}
where $x=h_0/(2T_{\mathrm{c}})$ and $T_{\mathrm{c}}$ is the superconducting critical temperature taking into account the effect of the static exchange field $h_0$.

Numerically calculated $\delta D_{\mathrm{m}}$ is plotted in Fig.~\ref{Fig1}(c) versus the temperature. The renormalization of the stiffness depends strongly on the temperature when $T<T_c$. 
It shows the direct role of the superconducting condensate. It suggests that the temperature can control the effective mass $m_{\mathrm{eff}} = 1/(2\tilde{D}_{\mathrm{m}}) = 1/(2 D_{\mathrm{m}} + 2 \delta D_{\mathrm{m}})$ of the composite quasiparticle. Since $\delta D_{\mathrm{m}} < 0$ [Eq.~\eqref{eq:deltadm}], the effective mass of the composite quasiparticle is enhanced as compared to that of a magnon. The significant enhancement of the effective mass, shown in Fig.~\ref{Fig1}(c) with decreasing temperature, can be measured using, for example, the Brillouin light scattering technique~\cite{Demokritov2001} employed regularly in measuring magnon group velocities~\cite{Nembach2015}. It should also manifest in the typical non-local magnonic spin transport experiment~\cite{Cornelissen2015,Goennenwein2015}. Since the magnon spin conductivity scales as $\sim 1/\sqrt{D_{\mathrm{m}}}$~\cite{Cornelissen2016}, it is expected to be modified due to the formation of composite magnetic excitation.  With a rather large exchange field $h_0$, the renormalized stiffness $\tilde{D}_{\mathrm{m}}$ may become negative, implying that the assumed uniformly ordered magnetization is no longer the ground state~\cite{Bergeret2000} (the genuine ground state may correspond to a magnetic texture).

The adiabatic approximation $\hbar \omega \ll k_BT$ is relevant to SC/FI with typical ferromagnetic insulators, such as YIG, at not very low temperatures because the magnonic frequencies in such materials are of the order of GHz or tens of GHz. Then the condition $\hbar \omega \ll k_BT$ works well at typical superconducting temperatures of $1$-$10$~K. Under this condition, quasiparticles do not play an important role because their excitation by the magnon-induced spin flip is difficult due to the presence of the superconducting gap. In this limit, $\delta s_{\perp} \to 0$, such that the decay rate [Eq.~\eqref{eq:kappa}] of the composite magnetic excitation is practically not renormalized with respect to the bare magnon. On the other hand, in the adiabatic approximation, the frequency of the $k = 0$ mode is also not renormalized because
\begin{eqnarray}
\hbar (\omega_{\mathbf{k} = 0} - \gamma K) =  \frac{J}{d_{\mathrm {FI}}} \bigl(\delta s_\parallel(\omega,\mathbf{k}=0)-\delta s_\parallel(\omega=0,\mathbf{k}=0)\bigr) \to 0.
\label{renorm_zero_momentum}
\end{eqnarray}
Going beyond the adiabatic approximation $\hbar \omega \ll k_BT$, the magnon-induced quasiparticle spin-flip processes become essential and, therefore, a nonzero renormalization of the $k = 0$ mode frequency and the decay rate appear. But this regime is more appropriate for superconductor/antiferromagnet heterostructures~\cite{Bobkov2023}.

The cloud of spinful Cooper pairs that increases the effective mass of the composite magnetic excitation also screens its spin. A magnon spin current $\mathbf j_{\mathrm{m}} \mathbf{e}_z$ in FI is accompanied by a superfluid spin current $\hat j_{\mathrm S}$ in the superconductor. The spin current carried by the triplet pairs, induced in the superconductor by the magnon, can be calculated via the quasiclassical Green's function as
\begin{eqnarray}
\hat j_{\mathrm S} = \frac{\hbar N_{\mathrm F} D}{16}{\rm Tr}_4 \int d \varepsilon \hat { \sigma} \bigl( \check g {\mathbf {\nabla}} \check g \bigr)^{\mathrm K}.
\label{spin_current_1}
\end{eqnarray}
The first-order contribution with respect to $\delta {\bf h}$ to the spin current is zero after the time averaging. The DC spin current is of the second order with respect to $\delta {\bf h}$. It has the only non-zero component $\mathbf j_{S}$ carrying the $z$-component of spin along the $\mathbf k$ direction. In the adiabatic approximation and when $T \to T_c$,
\begin{eqnarray}
\mathbf{j}_{\mathrm S} = 2 \pi k_BT_{\mathrm c} \hbar N_{\mathrm F} D\mathbf{k} \sum \limits_{\omega_n>0} \frac{\Delta^2 \delta h^2 \omega_n^2}{(\omega_n^2 + h_0^2)^2(\omega_n + Dk^2/2)^2}.
\label{spin_current_5}
\end{eqnarray}
The spin current in the FI carried by the magnons can be expressed as 
\begin{eqnarray}
\mathbf{j}_{\mathrm m} = -\mathbf{k}  \Bigl( \frac{\delta h}{h_0} \Bigr)^2 \frac{\tilde D_{\mathrm m} M_{\mathrm s}}{\hbar \gamma}.
\label{estimates_1}
\end{eqnarray}
The total spin current equals the sum of both contributions, which when $\mathbf k = k \mathbf e_z$ reads
\begin{align}
\label{eq:magcoopspin}
\mathbf{j}_{\mathrm{m}} + \mathbf{j}_{\mathrm{S}} = S v_{k} n_{k} \mathbf{e}_z \equiv \left(1 + j_{\mathrm{S}}/j_{\mathrm{m}} \right) \hbar v_{k} n_{k} \mathbf{e}_z,
\end{align}
where $v_k = 2 \tilde{D}_{\mathrm{m}} k$ is the group velocity of the composite magnetic excitation, $n_k$ is the number of excitations, and $S$ becomes its net spin evaluated via 
\begin{align}\label{eq:spincurrratio}
\frac{j_{\mathrm{S}}}{j_{\mathrm{m}}} & = - \frac{8  \hbar^2 N_{\mathrm{F}} D \gamma}{ \tilde{D}_{\mathrm{m}} M_{\mathrm{s}}}~ \sum_{\omega_n > 0} \frac{ \pi k_BT_{\mathrm{c}} \Delta^2 h_0^2 ~ \omega_n^2}{\left( \omega_n^2 + h_0^2 \right)^2 \left(2 \omega_n + D k^2 \right)^2}.
\end{align}
Since $j_{\mathrm{S}}/j_{\mathrm{m}} < 0$, the net spin of the composite magnetic excitation is reduced from $\hbar$ because of the screening effect.
Taking YIG as the FI~\cite{Xiao2010} and Nb as the SC with $D_{\mathrm m} = 5 \cdot 10^{-40}~{\rm J} \cdot {\rm m}^2$, $\gamma = 1.76 \cdot 10^{11}~{\rm T}^{-1} \cdot {\rm s}^{-1}$, $M_{\mathrm s} = 1.4 \cdot 10^5~{\rm A \cdot m^{-1}}$, $h_0 = 0.61 \Delta_0$, $d_{\mathrm{FI}} = d_{\mathrm{S}}$, and $T = 0.9 T_{\mathrm{c}}$, the net spin of the excitation [Eq.~\eqref{eq:magcoopspin}] was evaluated as $0.4 \hbar$~\cite{Bobkova2022}, which is considerably smaller than $\hbar$ of the bare magnon. 

This composite quasiparticle shares some similarities with the polaron~\cite{Froehlich1954}, which is formed when the phonon cloud screens an electron. The quasiparticle under consideration is a magnon spin screened by a superconducting condensate of triplet Cooper pairs. Thus, the predicted composite magnetic quasiparticle was termed ``magnon-cooparon"~\cite{Bobkova2022}.

\subsection{non-local excitation of the magnon spin current via a superconductor}

\begin{figure}[htp!]
	\begin{center}
		\includegraphics[width=130mm]{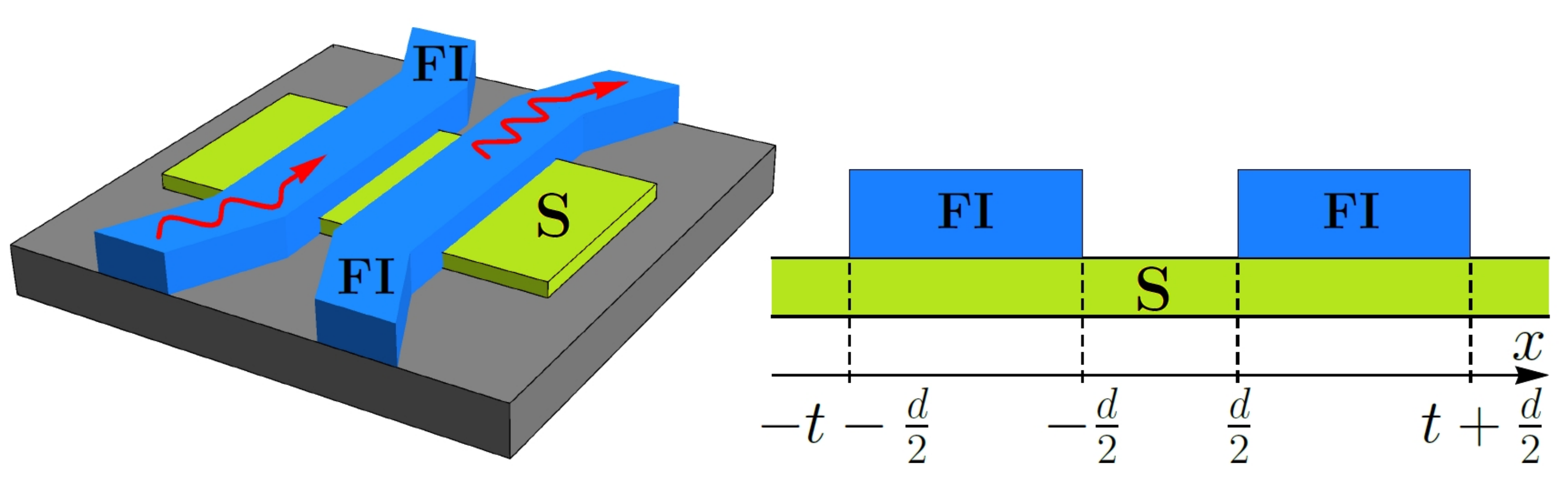}
		\caption{Sketch of the magnonic directional coupler based on magnon-cooparons. Top view (left) and side view (right). A spin wave propagating through one FI wire is controllably transferred to the second FI wire.\\
        The figure is adopted from Ref.~\cite{Bobkova2022}.}
        \label{Fig3}
	\end{center}
\end{figure}

In addition to the increased effective mass and screened spin, magnon-cooparons have a third important property, i.e., the spatial non-locality,  as a consequence of a cloud of Cooper pairs surrounding the magnon.  The Cooper pairs cloud extends over a correlation length $\sim \xi_{\mathrm{S}}$. This property enables non-local energy transfer from spin waves in one FI wire to another. One of the possible setups where the effect can manifest is the so-called magnon directional coupler, which consists of two FI wires deposited on a conventional superconductor within $\sim \xi_{\mathrm{S}}$ of each other, as shown in Fig.~\ref{Fig3}. Non-superconducting magnon directional couplers have been actively studied in the literature~\cite{Sadovnikov2015,Wang2018}. The physical principle of these devices is based on the dipolar interaction via the stray fields of magnons propagating in the waveguides. Magnon directional couplers are proposed to be key building blocks in wave-based logic and computing~\cite{Sadovnikov2016,Wang2020}. For the magnon-cooparon-based directional couplers, the underlying physical mechanism is not related to the dipolar interaction; instead, it is an interaction of a triplet Cooper pair cloud generated by one of the waveguides with the magnon in the other waveguide. It offers stronger coupling strengths, smaller footprint, additional control (e.g., via temperature), and universality (e.g., for antiferromagnets~\cite{Kamra2018,Bobkov2023}) as compared to the dipole-interaction-based designs considered previously~\cite{Sadovnikov2015,Wang2018}.  

In the setup depicted in Fig.~\ref{Fig3}, the width $t$ of each FI is smaller than the typical magnon wavelength, such that the inhomogeneities of the magnetization distribution in the FIs along the $\hat{\bf x}$-direction are disregarded. It is assumed that $t \gg \xi_{\mathrm S}$, in which case each of the FIs is semi-infinite from the point of view of the superconductor. One can solve the linearized Usadel equation~\cite{Bobkova2022} in the superconductor with the following spatial profile of exchange field
\begin{align}
\mathbf{h}(x)&= h_0(x)\mathbf{e}_z + \delta \mathbf{h}(x), \nonumber \\
h_0(x) &= h_0 \left[\Theta\left(-\frac{d}{2}-x\right) + \Theta\left(x-\frac{d}{2}\right)\right], \nonumber \\
\delta \mathbf{h}(x) &= \delta \mathbf{h}_{\mathrm l} \Theta\left(-\frac{d}{2}-x\right) + \delta \mathbf{h}_{\mathrm r} \Theta\left(x-\frac{d}{2}\right),
\nonumber\\
\delta \mathbf{h}_{{\mathrm l},{\mathrm r}} \hat {\pmb \sigma} &= \delta h_{{\mathrm l},{\mathrm r}} {\mathrm e}^{-{\mathrm i}(\mathbf{k} \cdot \mathbf{r} + \omega t)\hat \sigma_z} \hat \sigma_x .
\label{h_profile2}
\end{align}
Here, $h_0$ is the static exchange field in the superconductors below both FI wires, which is assumed to be the same for simplicity. However, distinct dynamic exchange fields $\delta \mathbf h_{\mathrm{l,r}}$ exist in superconductors below each FI wire since they are proportional to the respective spin-wave amplitudes $\delta m_{\mathrm{l,r}}$ of the two FIs. Solving the Usadel equation \eqref{usadel_1} under this exchange field profile, one can obtain the spin density in the superconductor $\mathbf{s}(x) = \mathbf{s}_{\mathrm{l}}(x) + \mathbf{s}_{\mathrm{r}}(x)$ with
\begin{align}\label{eq:snl}
\mathbf{s}_{\mathrm{l,r}}(x) & = s_0(x) \mathbf e_z + s_{\mathrm{loc}}(x) \frac{\delta \mathbf{h}_{\mathrm{l,r}}}{h_0} + s_{\mathrm{nl}}(x) \frac{\delta \mathbf{h}_{\mathrm{r,l}}}{h_0},
\end{align}
where $s_0(x)$ is the spin polarization induced by the equilibrium FI magnetization, $s_{{\mathrm {loc}}}(x)$ is the polarization induced in the left (right) covered superconducting region by the magnon travelling in the left (right) FI, and $s_{{\mathrm {nl}}}(x)$ is the non-local part of the polarization induced by the magnon travelling in the right (left) wire via the non-local triplet correlations. The non-local contribution $s_{\mathrm{nl}}(x)$ characterizes the spin density generated in the superconductor below the left FI by the right one, and vice versa.

Since the FI magnetization is homogeneous along the $\hat{\bf x}$-direction in each FI, only the averaged polarization $\overline{\bf s_{{\mathrm {l,r}}}}$ over the FI width enters the LLG equation.  
It is found~\cite{Bobkova2022}
\begin{align}
\overline{s_{{\mathrm {nl}}}}& = -N_{\mathrm F} \int d\varepsilon \tanh \frac{\varepsilon}{2T} {\rm Re}\Bigl[ \frac{h_0 \Delta \varepsilon {\mathrm e}^{-\lambda_k d}}{\lambda_k t [(\varepsilon + {\mathrm i} \delta)^2 - h_0^2]} \Bigl\{  \frac{\Delta \varepsilon}{2[(\varepsilon+{\mathrm i}\delta)^2-h_0^2](\varepsilon+{\mathrm i}Dk^2/2)} \nonumber \\
&+\frac{1}{{\mathrm i}D}\Bigl( \frac{C_{+,{\mathrm l}}}{\lambda_k(\lambda_k+\lambda_+)}+\frac{C_{-,{\mathrm l}}}{\lambda_k(\lambda_k+\lambda_-)} \Bigr)\Bigr\} \Bigr],
\label{s_non-local}
\end{align}
where $
C_{\pm {\mathrm l}} = C_{\pm {\mathrm r}} = \frac{1}{2}\left(\frac{\Delta}{\varepsilon \pm h_0}-\frac{\Delta_0}{\varepsilon} \right) \frac{({\lambda_\pm}/{\lambda})(1-\cosh \lambda d)-\sinh \lambda d}{(1+({\lambda_\pm^2}/{\lambda^2}))\sinh \lambda d + 2 ({\lambda_\pm}/{\lambda})\cosh \lambda d}$, 
$\lambda = \sqrt{-2{\mathrm i}\varepsilon/D}$, $\lambda_{\pm} = \sqrt{-2{\mathrm i}(\varepsilon \pm h_0)/D}$, and $\lambda_k = \sqrt{-2{\mathrm i}(\varepsilon+{\mathrm i}Dk^2/2)/D}$. The induced non-local spin density leads to a field-like spin torque in the LLG equations for both FIs
\begin{eqnarray}
\frac{\partial \mathbf{m}_{{\mathrm {l,r}}}}{\partial t} = - \gamma \mathbf{m}_{{\mathrm {l,r}}} \times \mathbf{H}_
{{\mathrm {eff}}} + \alpha \mathbf{m}_{{\mathrm {l,r}}} \times \frac{\partial \mathbf{m}_{{\mathrm {l,r}}}}{\partial t} + \frac{J}{d_{{\mathrm {FI}}}}\overline{s_{{\mathrm {nl}}}} \mathbf{m}_{{\mathrm {l,r}}} \times \mathbf{m}_{{\mathrm {r,l}}}.
\label{LLG_coupled}
\end{eqnarray}
The last term, representing the non-local torque, couples both magnetizations. Linearizing Eqs.~(\ref{LLG_coupled}) with respect to the magnon contribution $\delta \mathbf{m}_{{\mathrm {l,r}}}$ and solving them for the magnon dispersion leads to
\begin{eqnarray}
\omega_\pm (k) = \omega_0 + \tilde D_{\mathrm m} k^2 \mp \frac{J}{d_{{\mathrm {FI}}}}\overline{s_{{\mathrm {nl}}}},
\label{dispersion}
\end{eqnarray}
where $\omega_0 +\tilde D_{\mathrm m} k^2 $ is the magnon dispersion in a separate FI/SC bilayer with the renormalized stiffness, and the last term accounts for the hybridization of magnon modes due to the interaction via the superconductor. The resulted frequency splitting $\Delta \omega = (2J/d_{\mathrm {FI}})\overline{s_{\mathrm {nl}}}$ can be estimated as $\Delta \omega \sim 10^9 \exp(-d/\xi_{\mathrm S}) $~Hz for the material and geometric parameters used above and $t/\xi_{\mathrm S} \sim 10$.

Magnons with a specific frequency introduced into the coupled area possess different wave vectors $k_\pm \approx k_0 \pm \Delta k$ due to the frequency splitting $\Delta \omega$, where $\Delta k = k_0\Delta \omega/\tilde D_{\mathrm m} k_0^2$. A spin wave injected with frequency $\omega$ into the left FI transfers its energy via the spinful Cooper pair condensate to the right FI after traveling a so-called~\cite{Wang2020} coupling length 
\begin{align}\label{eq:L}
L & = \frac{2 \pi}{k_+ - k_-} = \frac{\pi \tilde{D}_{\mathrm{m}} (k_+ + k_-)}{\tilde{J} \bar{s}_{\mathrm{nl}}}.
\end{align} 
With the material parameters above, $h_0 = 0.61 \Delta_0$, $T = 0.9 T_{\mathrm{c}}$, $t = 10 \xi_{\mathrm{S}}$, $d = \xi_{\mathrm{S}}$, and $(k_+ + k_-)/2 = 10^7~\mathrm{m}^{-1}$ the coupling length $L$ [Eq.~\eqref{eq:L}] of the magnon-cooparon based directional coupler is evaluated as $\sim 100$ nm. This is an order of magnitude smaller than the coupling length provided by dipolar-interaction-based designs~\cite{Sadovnikov2015,Wang2018}. A smaller $L$ is desirable because it allows energy transfer and, consequently, implementation of logic operations in smaller devices.

\section{Superconducting vs. normal metals}
\label{Superconducting2normal_metals}

\subsection{Formalism conversion from superconductors to normal metals}
\label{transmutation_relation}

The electromagnetic proximity effect between a superconductor and a ferromagnet can be extended to that between a normal metal and a ferromagnet by replacing the penetration depth of the electromagnetic field in both cases. 
As addressed in Table~\ref{Extension}, the electric field behaves differently at the surface of superconductors and normal metals.
Inside the superconductors, the electric field drives the supercurrent ${\bf J}_s$ and decays exponentially with London's penetration depth $\lambda_L$. By contrast,  the electric field drives the normal current ${\bf J}_n$ inside the good normal metals, which oscillates and decays exponentially, governed by a complex penetration depth $\delta$.  Here,  $\delta=(1+i)/\sqrt{2\omega \mu_0 \sigma_c}$, in which $\sigma_c$ is the conductivity of normal metals.
Indeed, comparing Maxwell's equations inside the superconductor and the normal metal, the only difference lies in the penetration depth $\lambda_L$ and $\delta$; meanwhile, the boundary conditions of the electromagnetic fields remain unchanged, i.e., Table~\ref{boundary} applies to both superconductors and normal metals. 

\begin{table}[htp!]
\label{conversion_relation}
\centering
\caption{Comparison between superconductors and normal metals.}
\begin{tabular}{c|c|c}
\hline
~&\begin{minipage}[c]{.15\textwidth}	\vspace*{10pt} {Superconductor}\vspace*{10pt} \end{minipage} & {Normal metal} \\ \hline
\begin{minipage}[c]{.15\textwidth}
\centering	Electric field behavior
\end{minipage}
 &\begin{minipage}[m]{.32\textwidth}
	\centering \vspace*{5pt} \includegraphics[width=1\linewidth]{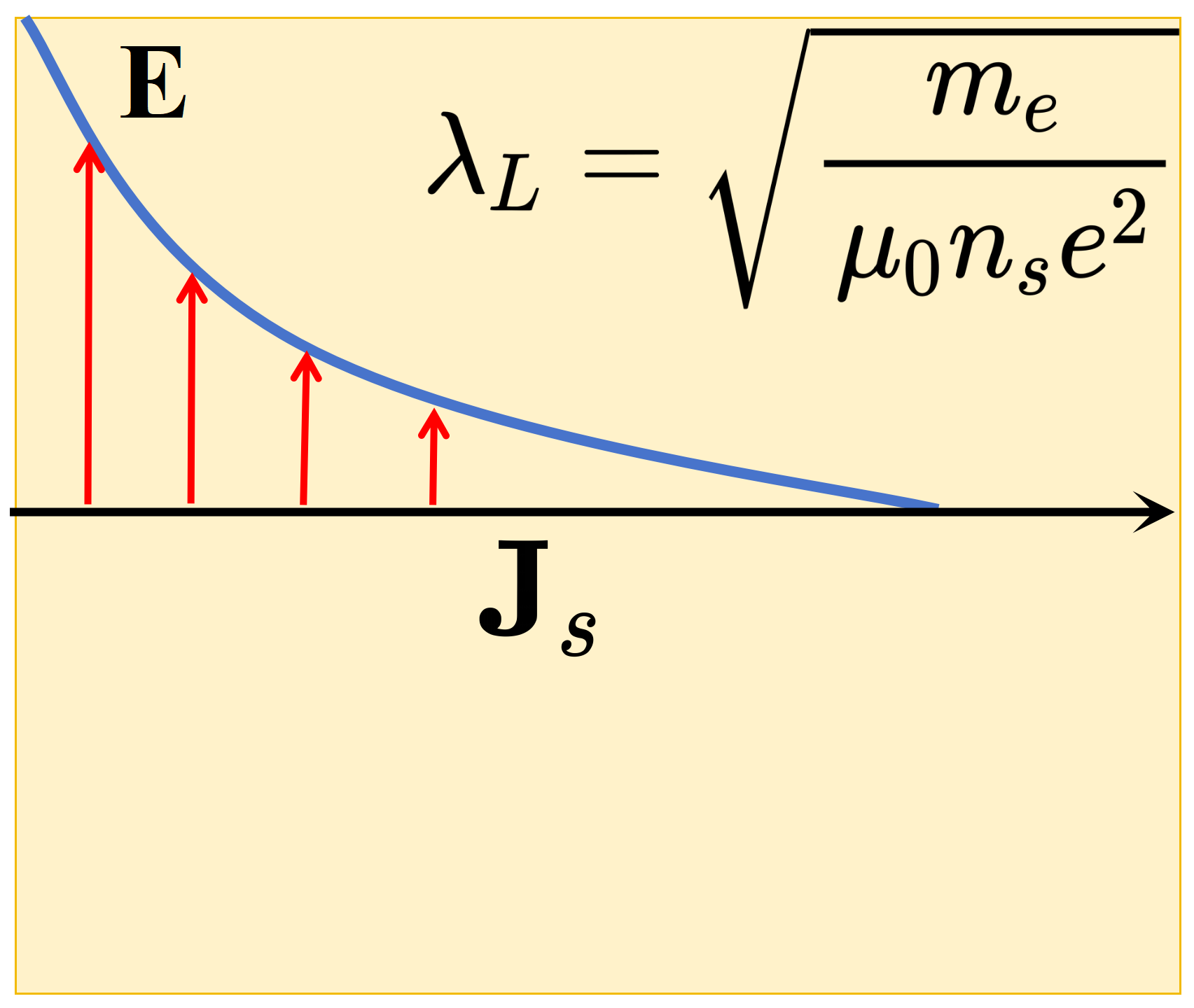} 
\end{minipage}	                & 
\begin{minipage}[m]{.32\textwidth}
	\centering \vspace*{5pt} 
	\includegraphics[width=1\linewidth]{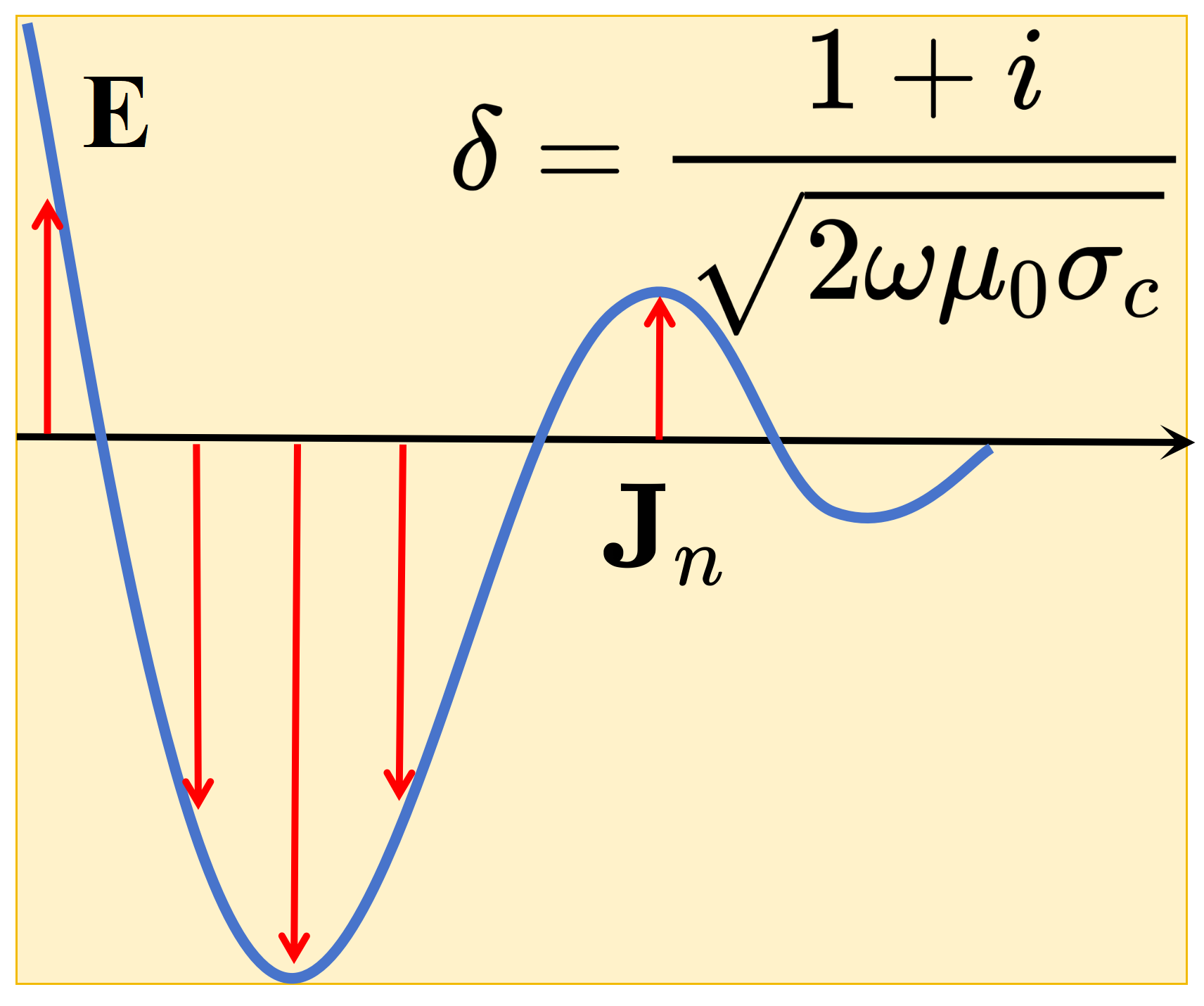}  
\end{minipage}	                   \\ \hline
\begin{minipage}[c]{.15\textwidth}
	\centering	Maxwell's equations
\end{minipage}&	\begin{minipage}[m]{.37\textwidth}
$$\begin{aligned}
&\nabla\cdot {\bf B}=0&&\nabla\times {\bf E}=-\dfrac{\partial{\bf  B}}{\partial t} \\ 
&\nabla\cdot {\bf E}=0 &&\nabla\times {\bf H}=\frac{i}{\omega\mu_0\lambda^2_L} {\bf E}+\varepsilon_0 \dfrac{\partial{\bf  E}}{\partial t}
\end{aligned}$$
\vspace*{0.4pt}  
\end{minipage} 
    &	\begin{minipage}[m]{.37\textwidth}
    $$\begin{aligned}
    &\nabla\cdot {\bf B}=0&&\nabla\times {\bf E}=-\dfrac{\partial{\bf  B}}{\partial t} \\ 
    &\nabla\cdot {\bf E}=0 &&\nabla\times {\bf H}=\frac{i}{\omega\mu_0\delta^2} {\bf E}+\varepsilon_0 \dfrac{\partial{\bf  E}}{\partial t}
    \end{aligned}$$\vspace*{0.4pt}  
    \end{minipage} 
\\
\hline
\begin{minipage}[c]{.15\textwidth}\centering	\vspace*{10pt} {Effect on magnons}\vspace*{10pt} \end{minipage}  &Chiral frequency shift& Chiral damping \\ \hline
\end{tabular}
\label{Extension}
\end{table}

Thereby, we are allowed to conveniently extend the electromagnetic proximity effect between a ferromagnet and a superconductor to that between a ferromagnet and a normal metal by replacing $\lambda_L$ with $\delta$.
Different from London's penetration depth $\lambda_L$, $\delta=(1+i)/\sqrt{2\omega \mu_0 \sigma_c}$ is complex, which brings an additional imaginary component to the magnon dispersion, leading to additional damping but preserving the chirality (Sec.~\ref{chiral_waves}). This extra damping originates from the dissipative Joule heat in normal metals caused by the normal current driven by ${\bf E}$, while the supercurrent in superconductors is non-dissipative. 
As a result, the additional damping brings new phenomena in the normal metal-superconductor heterostructures, such as the chiral damping~\cite{50_Chiral-damping-enhanced_magnon_transmission,45_Imaging_spin_wave_damping_underneath_metals},  non-Hermitian skin effect in the normal metal$|$ferromagnet bilayer~  (Sec.~\ref{Chiral_damping}), unidirectionally enhanced potential barrier penetration~\cite{50_Chiral-damping-enhanced_magnon_transmission}, magnon transistor~\cite{49_Chirality_enables_thermal_magnon_transistors,15_Liu,53_van_Wees}  (Sec.~\ref{Chiral_damping_for_magnon_transistors}), and the nodal magnon-photon polariton in the normal metal$|$ferromagnet$|$normal metal heterostructure~\cite{52_Persistent_nodal_magnon-photon_polariton_in_ferromagnetic_heterostructures} (Sec.~\ref{Nodal_magnon_photon_polaritons}). 
Different from the adjacent exchange interaction, the spin pumping via the dipolar interaction from the normal metal to ferromagnet shows a chiral locking between the injected magnon spin current and electron spin-accumulation directions~\cite{Chiral_Injection_of_Magnons,31_Reversal_enabling_magnon_memory_Grundler,38_Magnetization_Reversal_Grundler} (Sec.~\ref{Near_field_radiative_spin_transfer_to_magnons}).

\subsection{Chiral damping}
\label{Chiral_damping}

\begin{figure}[htp!]
 \centering
\includegraphics[width=1.0\textwidth,trim=0cm 0cm 0cm 0cm]{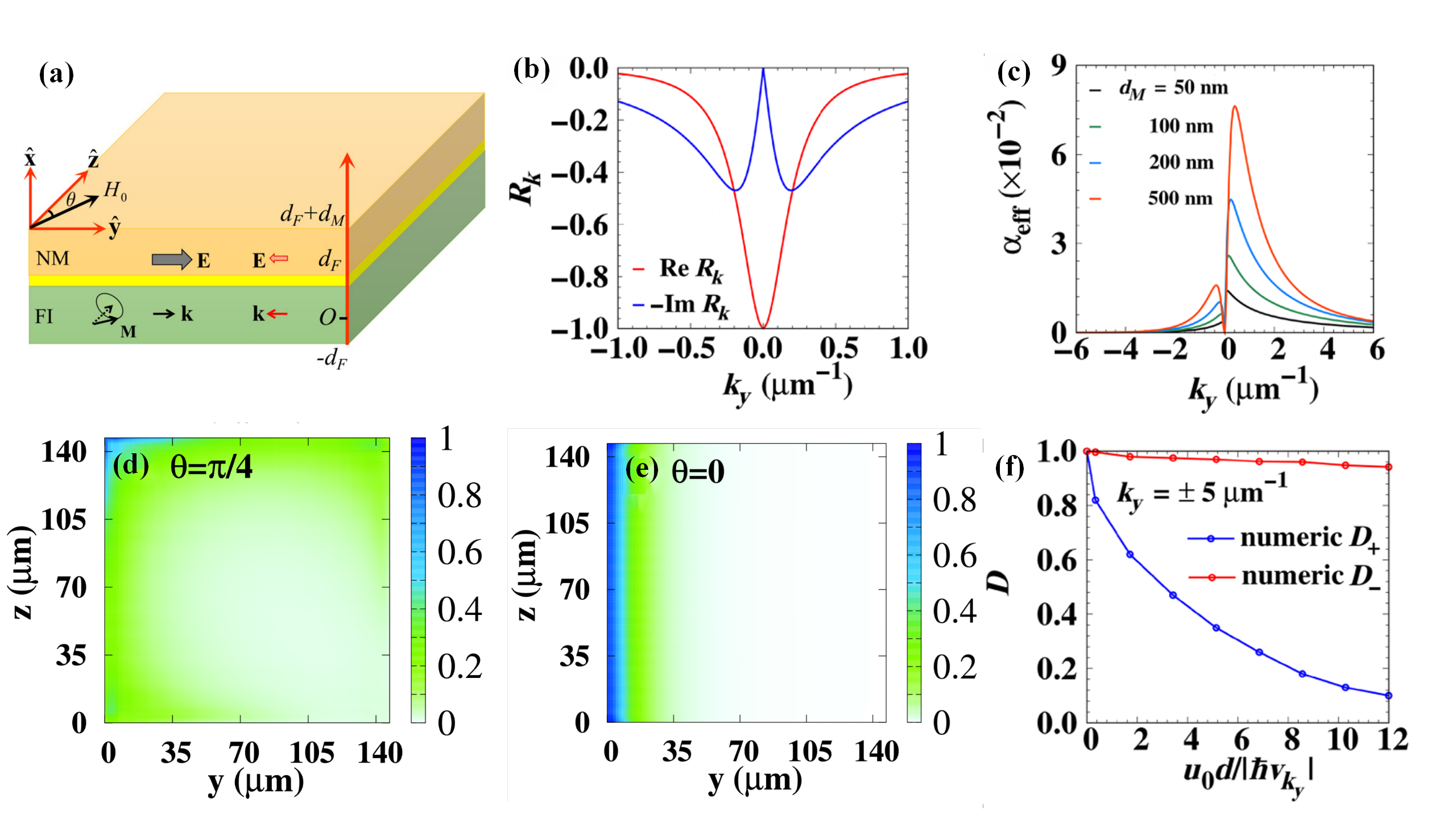}
    \caption{Magnon non-Hermitian skin effect and unidirectional potential barrier penetration caused by the chiral damping. 
    (a) Schematic of the NM$|$FI heterostructure with an in-plane magnetic field applied at an angle $\theta$ with respect to the $\hat{\bf z}$-direction. (b) Real and imaginary components of the reflection coefficient ${\cal R}_k$ of the electric field as a function of $k_y$. 
    (c) illustrates the strong $k_y$ dependence of the damping coefficient $\alpha_{\rm eff}$ caused by the normal metal (Cu) with different thicknesses $d_M = \{50, 100, 200, 500\}$~nm.
    (d) and (e)  Edge or corner aggregations of magnon eigenstates for typical magnetic configurations with $\theta \in\{\pi/4,0\}$. (f) Calculated transmissions $D_{\pm}$ of magnon wave packets across the potential barrier with wave vector $k_y=\pm 5~\mu{\rm m}^{-1}$ as a function of the barrier height $u_0$.\\
    \textit{Source:} These figures are taken from Ref.~\cite{50_Chiral-damping-enhanced_magnon_transmission}.}
    \label{skin}
\end{figure}

When a normal metal is placed near the spin waves, the eddy current driven by the dipolar electric field emitted from the spin waves may cause additional damping.  Recently, it has been shown that this damping is chiral by the chirality of the dipolar field emitted by the spin waves: the damping is larger for the spin waves propagating in one direction, while it is smaller for the spin waves propagating in the opposite direction~\cite{50_Chiral-damping-enhanced_magnon_transmission}.
Figure~\ref{skin}(a) illustrates a ferromagnetic insulator (FI)$|$normal metal (NM) heterostructure with the FI thickness of $2d_F$ and the NM thickness of $d_M$. An in-plane external magnetic field $H_0$ orients the magnetization by an angle \(\theta\) with respect to the $\hat{\bf z}$-direction.  A thin insulator spacer indicated by the yellow region between the NM and FI suppresses the exchange interaction and spin pumping. 
We focus on the Damon-Eshbach configuration with $\theta=0$ for  ${\bf H}_0=H_0\hat{\bf z}$, i.e.,  ${\bf k}\perp {\bf M}$ with $k_z=0$,  to demonstrate the chiral damping in this device. 

As addressed in Sec.~\ref{chiral_electric_fields}, only the electric field $E_z$ exists when radiated by the spin waves with $k_z=0$ of an isolated ferromagnetic film.
Such an electric field exhibits a special chirality, i.e., the electric field exists only above the film when the spin waves propagate in the positive $\hat{\bf y}$-direction, but becomes below the film when the spin-wave propagation direction is reversed.
For the NM$|$FI heterostructure, the electric field above the film emitted from the FI is then reflected by the normal metal. 
Here, the NM and FI divide the whole space into four different regions, i.e., the vacuum regions outside the heterostructure with $x>d_F+d_M$ and $x<-d_F$, the NM region with $d_F<x<d_F+d_M$, and the FI region with $-d_F<x<d_F$. Maxwell's equation in different regions reads
\begin{align}
    &\text{in vacuum:}~~~~~~~~~~~~~~~~~~~~~~~~~~~\nabla^2{\bf E}+k_0^2{\bf E}=0,\nonumber\\
    &\text{in normal metal:}~~~~~~~~~~~~~~~~~\nabla^2{\bf E}+k_m^2{\bf E}=0,\nonumber\\
    &\text{in ferromagnet insulator:}~~~~\nabla^2{\bf E}+k_0^2{\bf E}=-i\omega\mu_0\nabla\times{\bf M},
\end{align}
where $k_0=\omega\sqrt{\mu_0\varepsilon_0}$ and $k_m=\sqrt{\omega^2\mu_0\varepsilon_0-1/\delta^2}$.
Since only $E_z$ exists, we assume the solution $E_z={\cal E}(x) e^{i k_y y-i\omega t}$, where the  amplitude ${\cal E}(x)$ in different regions read
\begin{align}
      &\text{in vacuum}~(x>d_F+d_M):~~~~~~~~~~~~~~~~~~~~~~~~~~~~{\cal E}(x)={ A}e^{i {\cal A}_k x},\nonumber\\
    &\text{in normal metal}~(d_F<x<d_F+d_M):~~~~~~~~~~{\cal E}(x)={ B}_1e^{i {\cal B}_k x}+{ B}_2e^{-i {\cal B}_k x},\nonumber\\
    &\text{in ferromagnet insulator}~(-d_F<x<d_F):~~~~{\cal E}(x)={ C}_1e^{i {\cal A}_k x}+{ C}_2e^{-i {\cal A}_k x}+\dfrac{i a_{ k} \omega\mu_0 M_y}{{\cal A}_k^2}k_y,\nonumber\\
    &\text{in vacuum}~(x<-d_F):~~~~~~~~~~~~~~~~~~~~~~~~~~~~~~~~~~~~~~{\cal E}(x)={ D}e^{-i {\cal A}_k x},
\end{align}
where $a_{ k}=iM_x/M_y$ is the ellipticity of spin waves, ${\cal A}_k=\sqrt{k_0^2-k_y^2}$, and ${\cal B}_k=\sqrt{k_0^2-k_y^2-1/\delta^2}$. 
The coefficients $\{A,B_1,B_2,C_1,C_2,D\}$ are solved by matching the boundary conditions at the interfaces in Table~\ref{boundary} for the vacuum, superconductors, and normal metals. 
Inside the FI, the electric field 
\begin{align}
       E_z({\bf r},t)&=E_z^{(0)}({\bf r},t)+{\cal R}_kE_z^{(0)}({\bf r},t)|_{x=d_F}e^{-i{\cal A}_k(x-d_F)}
       \label{Ez_NM}
\end{align}
is a superposition of the electric field of an isolated FI 
\begin{align}
     E^{(0)}_z(\mathbf{r})&=\frac{ \omega \mu_0 M_ye^{i k_y y}}{2{\cal A}_k} 
\begin{cases}
(\beta_{ k}-1)\left(e^{i{\cal A}_k\left(x+d_F\right)}-e^{i{\cal A}_k\left(x-d_F\right)}\right), &x>d_F,\\
 -2\beta_{ k}+(\beta_{ k}-1)e^{i{\cal A}_k\left(x+d_F\right)}+(\beta_{k}+1)e^{-i{\cal A}_k\left(x-d_F\right)}, & |x| \leqslant d_F,   \\
(\beta_{ k}+1)\left(e^{-i{\cal A}_k\left(x-d_F\right)}-e^{-i{\cal A}_k\left(x+d_F\right)}\right), &x<-d_F, 
\end{cases} 
\label{E_z_single}
\end{align}
where $\beta_{ k}=-i a_{ k} k_y/{\cal A}_k$, and that reflected by the normal metal with the reflection coefficient
\begin{align}
    {\cal R}_k=\frac{\left({\cal A}_k^2-{\cal B}_k^2\right) e^{i {\cal B}_k d_M}-\left({\cal A}_k^2-{\cal B}_k^2\right) e^{-i {\cal B}_k d_M}}{({\cal A}_k-{\cal B}_k)^2 e^{i {\cal B}_k d_M}-({\cal A}_k+{\cal B}_k)^2 e^{-i {\cal B}_k d_M}}.
    \label{reflection_con_NM}
\end{align}
Equation~\eqref{Ez_NM} has the same form as Eq.~\eqref{electric_field_h},
which is a superposition of the electric field emitted by an isolated ferromagnetic insulator and its reflection by the superconductor via the reflection coefficient ${\cal R}_k$.  For normal metals, the complex penetration depth $\delta$ in ${{\cal B}_k}=\sqrt{k_0^2-k_y^2-1/\delta^2}$ leads to the complex reflection coefficient ${\cal R}_k=\Re({\cal R}_k)+i \Im({\cal R}_k)$. 
Figure~\ref{skin}(b) plots both the real and imaginary components of ${\cal R}_k$ as a function of $k_y$. Compared to the reflection coefficient in the SC$|$FI case [Fig.~\ref{enhancement}(b)], the real component of \({\cal R}_k\) exhibits analogous behavior, i.e., $\Re({\cal R}_k)\rightarrow0$ for large $|k_y|$ and $\Re({\cal R}_k)\rightarrow-1$ when $k_y=0$ due to the total reflection. By contrast, the imaginary component of \({\cal R}_k\) renders the reflected electric field in-phase with $M_y$, i.e., $i\Im ({\cal R}_k) E_z^{(0)}({\bf r},t)|_{x=d_F}\propto M_y$. 
This part of the electric field generates the magnetic field that is out-of-phase of $M_y$ according to Maxwell's equation $i\omega\mu_0({\bf H}+{\bf M})=\nabla \times {\bf E}$. The magnetic field then provides a damping-like torque to the magnetization dynamics, resulting in an additional damping effect. 
Note for the thick metal layer with large $d_M$, $e^{i{\cal B}_kd_M}\sim e^{-\sqrt{1/k^2+1/\delta^2}d_M}\rightarrow0$, ${\cal R}_k\approx ({\cal A}_k-{\cal B}_k)/({\cal A}_k+{\cal B}_k)$ recovers to the case with thick superconductors (as illustrated in Sec.~\ref{Chiral_gating_enhanced_magnon_transport}) by replacing $\delta$ with $\lambda_L$. 
This is consistent with the expected conversion relation in Table~\ref{conversion_relation}. 
On the other hand, according to Fig.~\ref{enhancement}(b),  ${\cal R}_k$ is an even function of $k_y$, which exhibits no obvious chirality. The chirality arises from the electric field $E_z^{(0)}({\bf r},t)|_{x=d_F}$. According to Eq.~\eqref{E_z_single}, $E_z^{(0)}({\bf r},t)|_{x=d_F}\propto( \beta_{ k}-1)\sim (-a_k k_y/|k_y|-k_y)$. With circular polarization $a_k\sim 1$ of spin waves, this electric field becomes much stronger for $k_y>0$ than for $k_y<0$, which shows a clear chirality.

The electric field inside the FI generates the magnetic field that modulates the dispersion, damping, and ellipticity of spin waves.
In the thin ferromagnetic insulator, the magnetic field decays with the penetration length $1/|k_y|\gg 2d_F$, allowing one to take the spatial average for  ${{\bf H}_d}$ over the film thickness, i.e., 
 \begin{align}
    H_{d,y}=&\dfrac{1}{2d_F}\int_{-d_F}^{d_F}\left(-\frac{1}{i\omega \mu_0}{\partial_x E_z}-M_y\right)dx=\zeta_y(k_y)M_y,\nonumber\\
    H_{d,x}=&\dfrac{1}{2d_F}\int_{-d_F}^{d_F}\left(\frac{k_y}{\omega \mu_0}E_z-M_x\right)dx=\zeta_x(k_y)M_x.
\end{align} 
By substitution into the LL equation, the  eigenfrequencies
and the ellipticities of spin waves are governed by
\begin{subequations}
\begin{align}
\omega({k_y})&=\mu_0 \gamma \sqrt{(\tilde{H}_0-\zeta_x({{k_y}})M_s)(\tilde{H}_0-\zeta_y({{k_y}})M_s)},\label{dispersion_eq}\\
a_{{k_y}}&=\sqrt{(\tilde{H}_0-\zeta_y({{k_y}})M_s)/(\tilde{H}_0-\zeta_x({{k_y}})M_s)},
\end{align}
\end{subequations}
where $\tilde{H}_0=H_0+\alpha_{\rm ex}k_y^2M_s$.  $\mathrm{Im}\, \omega({\bf k})\ne 0$ because of the Joule heating due to the eddy currents in the cap layer. Figure~\ref{skin}(c) plots the strong wave-vector dependence in the damping coefficient $\alpha_{\rm eff}=\Im(\omega({k_y}))/\Re(\omega({k_y}))$. The spin wave propagating in the positive $\hat{\bf y}$-direction suffers from a larger damping coefficient than those propagating in the opposite direction, resulting in a chiral damping.

\textbf{Non-Hermitian skin effect}.---The edges exist in any sample. The chiral damping implies that magnons propagate in the positive direction without damping but decay quickly when propagating in the opposite direction. Those reflected at the right boundary of the sample accumulate, which is a non-Hermitian skin effect~\cite{skin1,skin2,skin3,skin4,skin5,skin6,skin7}. 
The Hamiltonian formalism can well describe this phenomenon. By the construction of a two-dimensional square lattice model with
$\hat{m}_i=(1/\sqrt{N})\sum\limits_{i} \hat{m}_{\bf k} e^{i{\bf k}\cdot{\bf r}_i}$, where $i$ labels the lattice sites and $N$ is the site number, the Hamiltonian in the real space
\begin{align}
    \hat{H}_0=\sum\limits_{ij} t_{ji} \hat{m}_j^{\dagger}\hat{m}_i,
    \label{Ham}
\end{align} where $t_{ji}=({1}/{N})\sum\limits_{\bf k}\hbar 
\omega_{\bf k}e^{i{\bf k}\cdot({\bf r}_j-{\bf r}_i)}$ is a hopping amplitude between possibly distant sites \textit{i} and \textit{j}, and the summation is performed over the first Brillouin zone. Here $\omega_{\bf k}$ includes the chiral damping ${\rm Im}(\omega_{\bf k})\ne {\rm Im}(\omega_{-{\bf k}})$. When the frequencies   $\omega_{\bf k}$  are complex, the Hamiltonian (\ref{Ham}) is non-Hermitian and the hopping amplitude is non-reciprocal, \textit{i.e.},  
$t_{ji}\ne t_{ij}^*$. 
With a course-grained lattice constant of $a_y=a_z\sim O(100)$~nm, the first Brillouin zone is a square with the range $k_y\in[-\pi/a_y,\pi/a_y]$ and $k_z\in[-\pi/a_z,\pi/a_z]$.  This range is much larger than the range of wave vectors \(\bf k\) of the magnon of low frequencies.

Figure~\ref{skin}(d) and (e) are plots of the averaged spatial distributions $W({\bf r})=(1/N_m) \sum_{l=1}^{N_m}|\phi_l({\bf r})|^2$ of $N_m$ lowest-frequency eigenstates $\phi_l({\bf r})$ for $k_y \in [-3,3]~(\mu \mathrm{m})^{-1}$  and $k_z \in [-3,3]~(\mu \mathrm{m})^{-1}$~\cite{50_Chiral-damping-enhanced_magnon_transmission}. When the static magnetic field aligns with the sample boundary $z$-axis, i.e.,  $\theta=0$ in Fig.~\ref{skin}(e), the magnons tend to accumulate at the left edge. In the noncollinear configuration with $\theta = \pi/4$ [Fig.~\ref{skin}(d)], the maxima shifts to the upper-left corner.

\textbf{Anomalous barrier transmission}.---Natural and artificial potential barriers are useful instruments in magnonics
via confining and controlling spin-wave propagation. They may guide magnon transport~\cite{24_SF_shift_van_der_sar,Magnon_Blocking_Effect}, act as a magnonic logic gate~\cite{spin_wave_logic_gates}, induce magnon entanglement~\cite{YUAN20221,Magnonic_Hong_Ou_Mandel_effect}, and help to detect exotic magnon properties~\cite{Magnon_Dirac_materials,Spin_Wave_Doppler5,Yuan2023,Harms2022}. 
Chiral damping has a beneficial effect on spin transport, favoring penetration across the potential barrier in one direction. 
It is known from electron tunneling that the transmission across the potential barrier is symmetric, even when the barrier is asymmetric. It implies that the transmission across the potential barrier is the same whether the electrons propagate from left to right or from right to left. However, in the presence of the chiral damping, the transmission across the potential barrier becomes unidirectional: the spin waves propagating in one direction are totally reflected; in contrast, they penetrate across the potential barrier without any reflection when propagating in the opposite direction~\cite{50_Chiral-damping-enhanced_magnon_transmission}.

This phenomenon is quite universal. By using a delta function scattering potential and assuming the chiral damping of the waves, it can be demonstrated that the transmission is generally unidirectional. A one-dimensional scattering problem with a single short-range scattering potential $V \delta \left(  x\right)$ illustrates the general physics. The  dimensionless Schr\"{o}dinger or wave equation for the scattering states $\psi(x)$ with the kinetic energy $E=k^{2}>0$ reads 
\begin{align}
    \left(  -\partial^{2}+V\delta\left(  x\right)  -k^{2}\right)  \psi\left(
x\right)  =0.
\end{align}
We assume a chiral damping that affects only the right-moving waves. We use a Gilbert-type viscous damping proportional to the velocity, with a damping coefficient $\eta$.  The solutions to this problem are linear combinations
of plane waves. States \textit{coming from the right} read
\begin{equation}
\psi_{R}^{\mathrm{(c)}}=\left\{
\begin{array}
[c]{c}%
e^{-ikx}+r^{\mathrm{(c)}}e^{ikx}e^{-\eta kx}\\
t^{\mathrm{(c)}}e^{-ikx}%
\end{array}
\right.  \text{\, for  \,\,}%
\begin{array}
[c]{c}%
x>0\\
x<0
\end{array},
\end{equation}
where $r/t$ is the reflection/transmission coefficients that solve the scattering problem. The transmission and reflection probabilities 
\begin{subequations}
    \begin{align}
        | t^{\mathrm{(c)}}| ^{2} & =[\left(  \eta k\right)  ^{2}+4k^{2}]/[\left(  V+\eta k\right)  ^{2}+4k^{2}],\\
        | r^{\mathrm{(c)}}|^{2}&={V^{2}}/[\left(  V+\eta k\right)^{2}+4k^{2}].
    \end{align}
\end{subequations}
When the damping is sufficiently large $(\eta k\gg\left\vert V\right\vert )$, the transmission probability becomes unity and reflection vanishes,
irrespective of the scattering potential, i.e., which in essence is the anomalous transmission reported below for a magnetic device.

Figure~\ref{skin}(f) demonstrates a nearly unidirectional transmission of the wave packet through the potential barrier with height $u_0$ and finite width for the DE
configuration ${\bf k}\perp {\bf M}_s$; it is transparent for spin waves impinging from the left with a transmission $D_-\rightarrow 0$ at large $u_0$, but opaque for those from the right $D_-\rightarrow 1$.

\subsection{Chiral damping for magnon transistors}
\label{Chiral_damping_for_magnon_transistors}

The chiral gating effect~\cite{50_Chiral-damping-enhanced_magnon_transmission,Chiral_Damping_of_Magnons} raises an issue of whether it implies the Maxwell demon. A Maxwell demon can control the motion of molecules across two reservoirs. We assume a specific situation in which molecules in the left reservoir can move to the right reservoir, but molecules moving in the opposite direction are not allowed, making an analogy to the unidirectional penetration addressed above. 
At thermal equilibrium, the molecules move randomly to the left and right. Therefore, the demon moves more molecules to the right reservoir, causing it to become hotter. By the chiral gating, the thermal noise in the left reservoir can propagate to the right one, but the opposite propagation is forbidden, implying a Maxwell demon. However, the second Law of thermodynamics cannot be broken, so this paradox cannot happen because the ``gate" by normal metals, superconductors, or ferromagnets is passive; chirality diverts the thermal flow across the gate~\cite{49_Chirality_enables_thermal_magnon_transistors}.

\subsubsection{Theory}
\label{magnon_transistors_theory}

\begin{figure}[htp!]
 \centering
\includegraphics[width=0.8\textwidth,trim=1.2cm 0cm 0cm 0.1cm]{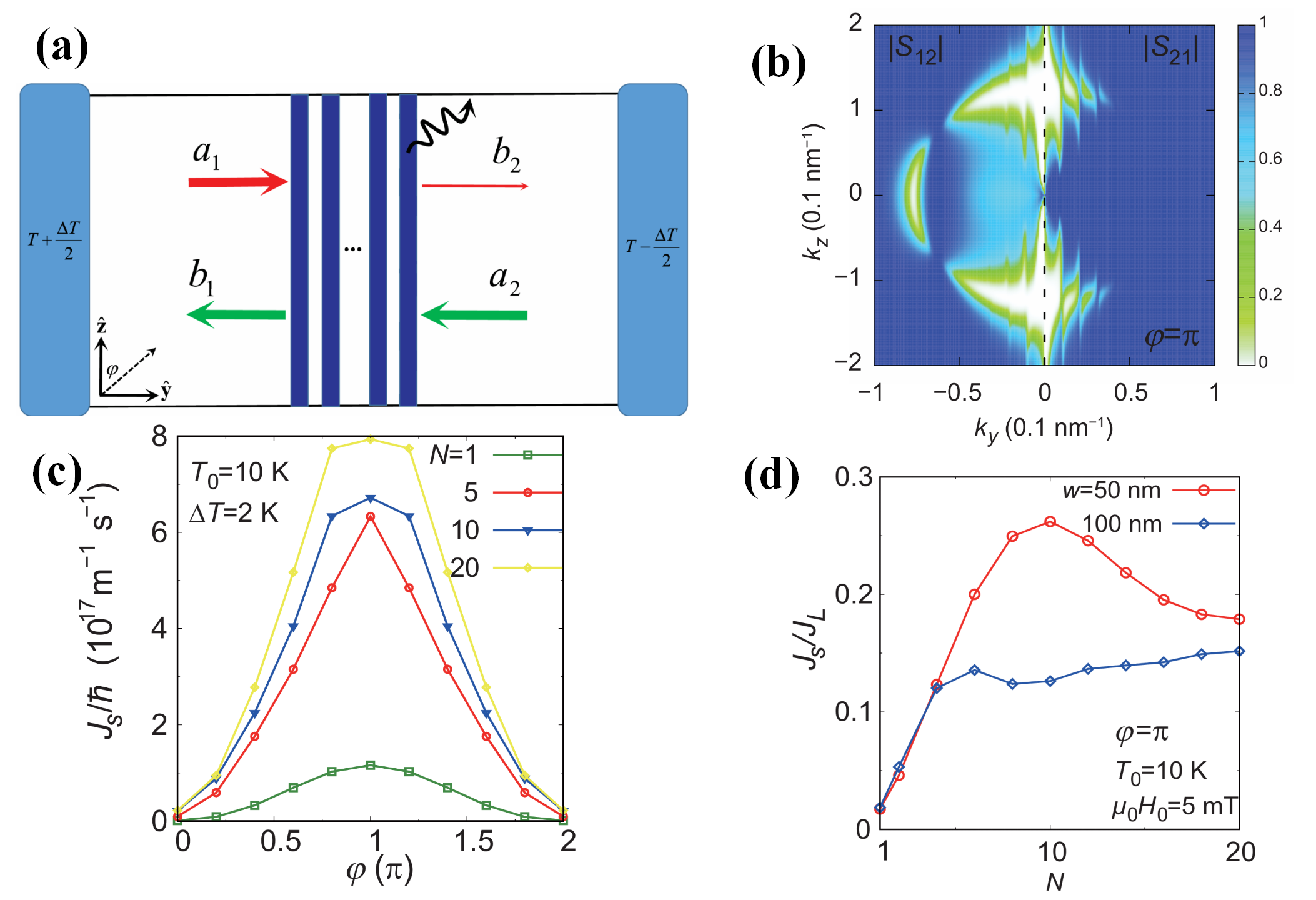}
\caption{Magnon transistor caused by chiral damping. (a) The exchange of magnons between a magnetic film that carries a heat current and a proximity magnetic nanowires (the blue-striped region) can be chiral, i.e., depending on the direction of the propagation of magnons in the film. $a_{1,2}$ are the input amplitudes of (spin) waves and $b_{1,2}$ are their output. (b) shows nearly total reflections when the film magnetization is antiparallel to that of the wires. (c) and (d) Spin current $J_s$ channeled off into the gate from the source to drain as a function of magnetization direction $\varphi$ [(c)], and the number of magnetic wires $N$ [(d)].\\
\textit{Source:} These figures are taken from Ref.~\cite{49_Chirality_enables_thermal_magnon_transistors}
\label{magnon_transistor}}
\end{figure}

We consider the thermal transport in a system composed of an array of magnetic nanowires loaded on a magnetic film~\cite{49_Chirality_enables_thermal_magnon_transistors}, biased by a magnetic field with an angle $\varphi$ to the wire $\hat{\bf z}$-direction, as illustrated in Fig.~\ref{magnon_transistor}(a). The magnetization of the wire is pinned along the $\hat{\bf z}$-direction, while the film magnetization is along the field direction. Magnons in the film propagate with wave vector $\kappa\hat{\bf y}+k_z\hat{\bf z}$ and in the wires propagate along the wire $\hat{\bf z}$-direction. When the magnetization between the wire and film is parallel, the interaction is not chiral, and the transmission of the spin waves across the gate is almost symmetric. Chirality appears when the magnetization between the wire and film is antiparallel. 

To show the non-unitarity of the scattering matrix, we expand the Hamiltonian into the magnon eigenstates $|\kappa,k_z\rangle$ of the magnetic film and $|l,k_z\rangle$ of the $l$-th magnetic wire as 
\begin{subequations}
     \begin{align}
\hat{H}_0/\hbar&=\sum_{\kappa}\sum_{k_z}\omega_{\kappa}(k_z)\left|\kappa,k_z\right\rangle \left\langle \kappa,k_z\right|+\sum_l\sum_{k_z} \Omega_{l}(k_z)|l,k_z\rangle\langle l,k_z|,\\
     \hat{H}_{\rm int}/\hbar&=\sum_l\sum_{\kappa}\sum_{k_z}g_{\kappa,l}(k_z)|l,k_z\rangle\langle \kappa,k_z|+\sum_l\sum_{\kappa}\sum_{k_z}g^*_{\kappa,l}(k_z)|\kappa,k_z\rangle\langle l,k_z|,
     \label{interaction_Hamiltonian}
 \end{align}
 \end{subequations}
where $\omega_{\kappa}(k_z)$ is the dispersion relation of the magnon in the film, $\Omega_l(k_z)$ is the frequency of the $l$-th wire, and $g_{\kappa,l}(k_z)$ denotes the coupling coefficient between magnons in the film and the $l$-th wire. 
According to quantum mechanics, the scattering states $|\psi_s,k_z\rangle$ are related to the initial states of the film $|\kappa,k_z\rangle$ by the $\hat{T}$ matrix, as addressed by the Lippmann-Schwinger equation~\cite{Mahan}:
 \begin{align}
|\psi_s,k_z\rangle=|\kappa,k_z\rangle+\frac{1}{\hbar\omega_{\kappa}(k_z)-\hat{H}_0+i0_+}\hat{H}_{\rm int}|\psi_s,k_z\rangle=\hat{T}|\kappa,k_z\rangle.
 \end{align}
The elements of the $\hat{T}$-matrix obey
\begin{subequations}
\begin{align}
    T_{l\kappa}(k_z)&=\langle l,k_z|\hat{T}|\kappa,k_z\rangle=\langle l,k_z|\psi_s,k_z\rangle=\frac{1}{\omega_{\kappa}(k_z)-\Omega_l(k_z)+i0_+}\sum_{\kappa'}g_{\kappa',l}(k_z)T_{\kappa'\kappa}(k_z),\label{Tlk}\\
    T_{\kappa'\kappa}(k_z)&=\langle\kappa',k_z|\hat{T}|\kappa,k_z\rangle=\delta_{\kappa'\kappa}+\frac{1}{\omega_{\kappa}(k_z)-\omega_{\kappa'}(k_z)+i0_+}\sum_{l'}g_{\kappa',l'}^*(k_z)T_{l'\kappa}(k_z),\label{Tkk}
\end{align}
\end{subequations}
combining which yields 
\begin{align}
    T_{l{\kappa}}(k_z)=\sum_{l'}{\cal G}^r_{k_z}(\omega_{\kappa,k_z})\big|_{ll'}g_{\kappa,l'}(k_z),\quad
     T_{{\kappa}'{\kappa}}(k_z)=\delta_{{\kappa}'{\kappa}}+\frac{{\cal M}_{{\kappa}'}^{\dagger}(k_z)
 {\cal G}^r_{k_z}(\omega_{\kappa,k_z})
 {\cal M}_{\kappa}(k_z)}{\omega_{\kappa}(k_z)-\omega_{{\kappa}'}(k_z)+i0_{+}}.
 \end{align}
 Here, ${\cal M}_{\kappa}(k_z)\equiv \begin{pmatrix}g_{\kappa,1}(k_z),\cdots,g_{\kappa,N}(k_z)
 \end{pmatrix}^T$ and ${\cal G}^r_{k_z}(\omega)=\left(\omega-{\cal H}_{k_z}(\omega)\right)^{-1}$ stands for the (retarded) Green's function of magnons in the nanowire for the non-Hermitian Hamiltonian 
 \begin{equation}
 {\cal H}_{k_z}(\omega)|_{ll'}=\Omega({k_z})(1-i\alpha_G)\delta_{ll'}+\Sigma_{k_z}(\omega)|_{ll'},
 \label{non_Hermitian_Hamiltonian}
 \end{equation}
 in which $\alpha_G$ is the Gilbert damping coefficient of magnons in the wires and 
\begin{align}
    \Sigma_{k_z}\left(\omega\rightarrow \omega_\kappa({k_z})\right)|_{ll'}=\sum_{\kappa'} \frac{g_{\kappa',l}(k_z)g^*_{\kappa',l'}(k_z)}{\omega_{\kappa}(k_z)-\omega_{\kappa'}(k_z)+i0_+}=-i \frac{L_y}{v_{\kappa}(k_z)}\begin{cases}\left|g_{-\kappa}(k_z)\right|^2 e^{-i \kappa\left(y_l -y_{l^{\prime}}\right)} & \left(l<l^{\prime}\right) \\ \left(|g_{\kappa}(k_z)|^2+|g_{-\kappa}(k_z)|^2\right)/2 & \left(l=l^{\prime}\right) \\ |g_{\kappa}(k_z)|^2 e^{i \kappa\left(y_l -y_{l^{\prime}}\right)}& \left(l>l^{\prime}\right)\end{cases}
    \label{self_energy}
\end{align}
denotes the elements of the self-energy matrix,
where $g_{\kappa,l}(k_z)=g_{\kappa}(k_z) e^{ik_y y_l}$,  $L_y$ is the length of the film, and $v_{\kappa}(k_z)=\partial \omega_{\kappa}(k_z)/\partial \kappa$ is the group velocity of the film magnons in the transport $\hat{\bf y}$-direction. With the non-Hermitian Hamiltonian \eqref{non_Hermitian_Hamiltonian}, the eigenstates $\psi_{\zeta}$ and $\phi_{\zeta}$ obey ${\cal H}_{k_z}(\omega_{\kappa,k_z})\psi_{\zeta}=\nu_{\zeta}\psi_{\zeta}$ and ${\cal H}^{\dagger}_{k_z}(\omega_{\kappa,k_z})\phi_{\zeta}=\nu^{\dagger}_{\zeta}\phi_{\zeta}$, in which the eigenvalue $\nu_{\zeta}$ is a complex number. 
 The retarded Green's function of the nanowire magnons  
  ${\cal G}^r_{k_z}(\omega_{\kappa,k_z})=\sum_{\zeta}{\psi_{\zeta}\phi_{\zeta}^{\dagger}}/({\omega_{\kappa}(k_z)-\nu_{\zeta}})$.

 For the initial state with $\kappa>0$, the scattering state far away from the scattering region 
 \begin{subequations}
     \begin{align}
     \langle y|\psi_s,k_z\rangle|_{y\rightarrow {+\infty}} &= \sum_{\kappa'}\langle y|\kappa'\rangle \delta_{\kappa \kappa'}+\frac{L_y}{2\pi}\int_{-\infty}^{\infty} d\kappa'
     \langle y|\kappa'\rangle \frac{{\cal M}_{{\kappa}'}^{\dagger}(k_z)
 {\cal G}^r_{k_z}(\omega_{\kappa,k_z})
 {\cal M}_{\kappa}(k_z)}{\omega_{\kappa}(k_z)-\omega_{{\kappa}'}(k_z)+i0_{+}}\nonumber\\
 &=\left(1-\frac{i L_y}{|v_{\kappa}(k_z)|}
 {\cal M}_{k}^{\dagger}(k_z)
 {\cal G}^r_{k_z}(\omega_{\kappa,k_z})
 {\cal M}_{k}(k_z)\right) \langle y|\kappa\rangle ,\\
\langle y|\psi_s,k_z\rangle|_{y\rightarrow {-\infty}} &= \frac{L_y}{2\pi}\int_{-\infty}^{\infty} d\kappa'
     \langle y|\kappa'\rangle \frac{{\cal M}_{{\kappa}'}^{\dagger}(k_z)
 {\cal G}^r_{k_z}(\omega_{\kappa,k_z})
 {\cal M}_{\kappa}(k_z)}{\omega_{\kappa}(k_z)-\omega_{{\kappa}'}(k_z)+i0_{+}}\nonumber\\
 &=\left(-\frac{i L_y}{|v_{\kappa}(k_z)|}
 {\cal M}_{-k}^{\dagger}
 {\cal G}^r_{k_z}(\omega_{\kappa,k_z}){\cal M}_{k}(k_z)\right)\langle y|-\kappa\rangle.
  \end{align}
\end{subequations}
Accordingly, the transmission and reflection amplitudes of magnons in the film are the elements of the 2\(\times\)2 scattering matrix $S$ that with $\kappa>0$ read
\begin{subequations}
 \begin{align}
 &S_{21}(\kappa,k_z)=1-\frac{iL_y}{|v_{\kappa}(k_z)|}
 {\cal M}_{k}^{\dagger}(k_z)
 {\cal G}^r_{k_z}(\omega_{\kappa,k_z})
 {\cal M}_{k}(k_z),\label{magnon_transmission}\\
 &S_{11}(\kappa,k_z)=-\frac{iL_y}{|v_{\kappa}(k_z)|}
 {\cal M}_{-k}^{\dagger}
 {\cal G}^r_{k_z}(\omega_{\kappa,k_z}){\cal M}_{k}(k_z).
 \end{align}
 \end{subequations}
 Figure~\ref{magnon_transistor}(b) illustrates that the non-reciprocity of the scattering matrix with respect to $k_y$ is pronounced in the
antiparallel $\varphi=\pi$ configuration.

We prove that the scattering matrix is unitary with $|S_{21}(\kappa,k_z)|^2+|S_{11}(\kappa,k_z)|^2=1$ only when the damping of the wire magnons $\alpha_G\rightarrow 0$. 
With the self-energy \eqref{self_energy}, 
${\Sigma}_{k_z}^{\dagger}-\Sigma_{k_z}=(i{L_y}/{|v_{\kappa}(k_z)|})\left({\cal M}_{-\kappa}(k_z){\cal M}_{-\kappa}^{\dagger}(k_z)+{\cal M}_{\kappa}(k_z){\cal M}_{\kappa}^{\dagger}(k_z)\right)$,
so when $\alpha_G\rightarrow 0$, 
$-{\cal H}_{k_z}+{\cal H}_{k_z}^{\dagger}=(i{L_y}/{|v_{\kappa}(k_z)|})\left({\cal M}_{-\kappa}(k_z){\cal M}_{-\kappa}^{\dagger}(k_z)+{\cal M}_{\kappa}(k_z){\cal M}_{\kappa}^{\dagger}(k_z)\right)$,
such that 
$({\cal G}_{k_z}^{a})^{-1}({\cal G}_{k_z}^a-{\cal G}_{k_z}^r)({\cal G}_{k_z}^r)^{-1}=({\cal G}_{k_z}^r)^{-1}-({\cal G}_{k_z}^{a})^{-1}=(i{L_y}/{|v_{\kappa}(k_z)|})\left({\cal M}_{-\kappa}(k_z){\cal M}_{-\kappa}^{\dagger}(k_z)+{\cal M}_{\kappa}(k_z){\cal M}_{\kappa}^{\dagger}(k_z)\right)$.
Thereby, 
\begin{align}
    {\cal G}_{k_z}^a-{\cal G}_{k_z}^r=i\frac{L_y}{|v_{\kappa}(k_z)|}{\cal G}_{k_z}^a\left({\cal M}_{-\kappa}(k_z){\cal M}_{-\kappa}^{\dagger}(k_z)+{\cal M}_{\kappa}(k_z){\cal M}_{\kappa}^{\dagger}(k_z)\right){\cal G}_{k_z}^r.
    \label{relation_1}
\end{align}
Accordingly, the scattering matrix is unitary $|S_{11}(\kappa,k_z)|^2+|S_{21}(\kappa,k_z)|^2=1$ with
\begin{subequations}
\begin{align}
\nonumber
    |S_{21}(\kappa,k_z)|^2&=1+\frac{iL_y}{|v_{\kappa}(k_z)|}{\cal M}_{\kappa}^{\dagger}(k_z)\left({\cal G}_{k_z}^a(\omega_{\kappa,k_z})-{\cal G}_{k_z}^r(\omega_{\kappa,k_z})\right){\cal M}_{\kappa}(k_z)\nonumber\\
    &+\frac{L_y^2}{|v_{\kappa}(k_z)|^2}{\cal M}_{\kappa}^{\dagger}(k_z){\cal G}_{k_z}^r(\omega_{\kappa,k_z}){\cal M}_{\kappa}(k_z){\cal M}_{\kappa}^{\dagger}(k_z){\cal G}_{k_z}^a(\omega_{\kappa,k_z}){\cal M}_{\kappa}(k_z),\\
|S_{11}(\kappa,k_z)|^2&=\frac{L_y^2}{|v_{\kappa}(k_z)|^2}{\cal M}_{-\kappa}^{\dagger}(k_z){\cal G}_{k_z}^r(\omega_{\kappa,k_z}){\cal M}_{\kappa}(k_z){\cal M}_{\kappa}^{\dagger}(k_z){\cal G}_{k_z}^a(\omega_{\kappa,k_z}){\cal M}_{-\kappa}(k_z).
\end{align}
\end{subequations}

\textbf{Chirality as the switch for spin injection}.---Yu \textit{et al.} consider the spin injection from the magnetic film that is biased by left and right reservoirs with temperature $T_L$ and $T_R$ [Fig.~\ref{magnon_transistor}(a)] to the wires and find that the chirality acts as the switch for such a spin injection~\cite{49_Chirality_enables_thermal_magnon_transistors}.

The spin injection rates from the film magnon $\hat{m}_{\kappa}(k_z)$ to the wire magnon $\hat{\beta}_{l,k_z}$ by the interaction (\ref{interaction_Hamiltonian}) read
\begin{align}
\nonumber
\left\langle\frac{d\hat{\rho}_{\kappa,k_z}}{dt}\right\rangle&=\frac{i}{\hbar}\left\langle \left[\hat{H}_{\rm int},\hat{m}^{\dagger}_{\kappa}(k_z)\hat{m}_{\kappa}(k_z)\right]\right\rangle=-\sum_lg_{\kappa,l}(k_z)G^<_{l,\kappa}(k_z;t,t)+\sum_lg_{\kappa,l}^*(k_z)G_{\kappa,l}^<(k_z;t,t)\\
&=2{\rm Re}\left(\sum_lg_{\kappa,l}^*(k_z)G^<_{\kappa,l}(k_z;t,t)\right),
\label{rate_of_change}
\end{align}
where the ``lesser" Green functions
$G^<_{l,\kappa}(k_z;t,t')=-i\langle\hat{\beta}_{l,k_z}^{\dagger}(t')\hat{m}_{\kappa}(k_z,t)\rangle$ and 
    $G^<_{\kappa,l}(k_z;t,t')=-i\langle\hat{m}^{\dagger}_{\kappa}(k_z,t')\hat{\beta}_{l,k_z}(t)\rangle$
account for both current directions.
The associated time-ordered Green function 
\[
G^t_{\kappa,l}(k_z;t,t')=-i\theta(t'-t)\left\langle \hat{m}^{\dagger}_{\kappa}(k_z,t')\hat{\beta}_{l,k_z}(t)\right\rangle
    -i\theta(t-t')\left\langle \hat{\beta}_{l,k_z}(t)\hat{m}^{\dagger}_{\kappa}(k_z,t')\right\rangle
    \]
obeys the equation of motion 
\begin{align}
    \left(-i\frac{\partial}{\partial t'}-\omega_{\kappa}(k_z)\right)G^t_{\kappa,l}(k_z;t,t')=\sum_{l'}g_{\kappa,l'}(k_z){\cal G}_{k_z,l'l}^t(t,t').
\end{align}
Here, ${\cal G}_{k_z,ll'}^t(t,t')=-i\theta(t'-t)\left\langle \hat{\beta}^{\dagger}_{l',k_z}(t')\hat{\beta}_{l,k_z}(t)\right\rangle-i\theta(t-t')\left\langle \hat{\beta}_{l,k_z}(t)\hat{\beta}^{\dagger}_{l',k_z}(t')\right\rangle$ is the time-ordered propagator of wire magnons. In the frequency domain, 
\begin{align}
G_{\kappa,l}^t(k_z,\omega)=\sum_{l'}g_{\kappa,l'}(k_z){\cal G}_{k_z,ll'}^t(\omega){G}_{\kappa,t}^{(0)}(k_z,\omega),
    \label{time_ordered}
\end{align}
where $G_{\kappa,t}^{(0)}(k_z,\omega)$ is the time-ordered Green function of the free magnons in films. Accordingly, by Langreth's theorem~\cite{Haug}, we return to the lesser Green function 
\begin{align}
G_{\kappa,l}^<(k_z,\omega)=\sum_{l'}g_{\kappa,l'}(k_z)\left({\cal G}^r_{k_z,ll'}(\omega){G}^{(0)}_{\kappa,<}(k_z,\omega)+{\cal G}_{k_z,ll'}^<(\omega){G}_{\kappa,a}^{(0)}(k_z,\omega)\right).
\end{align}
By substitution of this into Eq.~(\ref{rate_of_change}), 
\begin{align}
\nonumber
    \left\langle \frac{d\hat{\rho}_{\kappa,k_z}}{dt}\right\rangle&=\int\frac{d\omega}{\pi}{\rm Re}\Big(\sum_{ll'}g^*_{\kappa,l}(k_z){\cal G}_{k_z,ll'}^r(\omega)g_{\kappa,l'}(k_z)G_{\kappa,<}^{(0)}(k_z,\omega)+\sum_{ll'}g^*_{\kappa,l}(k_z){\cal G}^<_{k_z,ll'}(\omega)g_{\kappa,l'}(k_z){G}_{\kappa,a}^{(0)}(k_z,\omega)\Big),
\end{align}
where at temperature $T$, $G^{(0)}_{\kappa,<}(k_z,\omega)=2\pi i f(\omega,T)\delta\left(\omega-\omega_{\kappa}(k_z)\right)$ and $G^{(0)}_{\kappa,a}(k_z,\omega)=1/(\omega-\omega_{\kappa}(k_z)-i0^+)$
are the propagators for the free film magnons. For the wires
\begin{align}
    {\cal G}^<_{k_z}(\omega)=-f(\omega,T_0)\left[{\cal G}_{k_z}^r(\omega)-{\cal G}_{k_z}^{a}(\omega)\right]=-f(\omega,T_0)\left[{\cal G}_{k_z}^r(\omega)-{\cal G}_{k_z}^{r\dagger}(\omega)\right].
\end{align}
In terms of ${\cal M}_{\kappa}(k_z)=\begin{pmatrix}g_{\kappa,1}(k_z),\cdots,g_{\kappa,N}(k_z)
 \end{pmatrix}^T$, we arrive at the injection rates
\begin{align}
\nonumber
\left\langle {d\hat{\rho}_{\kappa}(k_z)}/{dt}\right\rangle&={\rm Re}\big[i{\cal M}^{\dagger}_{\kappa}(k_z){\cal G}_{k_z}^r(\omega_{\kappa,k_z}){\cal M}_{\kappa}(k_z)\big]\left[f(\omega_{\kappa,k_z},T)-f(\omega_{\kappa,k_z},T_0)\right]\nonumber\\
    &+{\rm Re}\big[i{\cal M}^{\dagger}_{\kappa}(k_z){\cal G}_{k_z}^r(\omega_{\kappa,k_z}){\cal M}_{\kappa}(k_z)\big]f(\omega_{\kappa,k_z},T)\nonumber\\
    &+{\rm Re}\big[i{\cal M}^{\dagger}_{\kappa}(k_z){\cal G}_{k_z}^{r\dagger}(\omega_{\kappa,k_z}){\cal M}_{\kappa}(k_z)\big]f(\omega_{\kappa,k_z},T_0).
    \label{rate_of_change_2}
\end{align}

The terms associated with the distribution $f(T)$ account for the spin injection from films to wires, which is balanced by the inverse processes proportional to $f(T_0)$.
When $T=T_0$,
$\left\langle {d\hat{\rho}_{\kappa}(k_z)}/{dt}\right\rangle={\rm Re}[i{\cal M}^{\dagger}_{\kappa}(k_z)({\cal G}_{k_z}^{r}(\omega_{\kappa,k_z})+{\cal G}_{k_z}^{r\dagger}(\omega_{\kappa,k_z}))
{\cal M}_{\kappa}(k_z)]f(\omega_{\kappa,k_z},T_0)=0$, so the injection vanishes as required by the principle of detailed balance, as do equilibrium magnon currents, despite the chirality of the interaction.

The spin-injection rates of magnons with $\kappa\hat{\bf y}+k_z\hat{\bf z}$ to the wire are related to the magnon transmission Eq.~(\ref{magnon_transmission}).   Here we distinguish the wave numbers $\kappa_L$ and $\kappa_R$ in the left ``L" and right ``R" reservoirs. The magnons in the film are weakly perturbed, such that  $\delta f_{\kappa}(k_z)=f(\omega_{\kappa,k_z},T)-f(\omega_{\kappa,k_z},T_0)$ is small.
Under the temperature bias $\Delta T=T_L-T_R$, 
\begin{align}
    \delta f_{\kappa}(k_z)=\frac{\partial f(\omega_{\kappa,k_z},T)}{\partial T_0}\left\{
    \begin{array}{c}
     \Delta T/2,~~~~\kappa\in L\\
     -\Delta T/2,~~~\kappa\in R
    \end{array}\right..
\end{align}
When $\Delta T>0$, the wires absorb the magnons with $\kappa_L>0$ as ``particles" and magnons with $\kappa_R<0$ as ``holes", so both the left and right reservoirs inject spin current into the wires.
The spin-injection rate from different reservoirs relates to the magnon transmission \eqref{magnon_transmission} as
 $\left\langle d\hat{\rho}_{\kappa,k_z}/dt\right\rangle={\delta f_{\kappa}(k_z)}{\Gamma_{\kappa}(k_z)}$,
\textit{viz.}, the wires dissipate magnon energy at the rates 
\[
    \Gamma_{\kappa}(k_z)=\frac{|v_{\kappa}(k_z)|}{L_y}\left\{
    \begin{array}{c}
     1-{\rm Re}S_{21}(\kappa,k_z),~~~\kappa_{L,R}>0\\
     1-{\rm Re}S_{12}(\kappa,k_z),~~~\kappa_{L,R}<0
    \end{array}\right.,
\]
which vanish when the transmission is high but become significant $\Gamma_{\kappa}(k_z)\rightarrow |v_{\kappa}(k_z)|/L_y$ otherwise, noting the real part of the transmission \eqref{magnon_transmission}
\begin{align}
    {\rm Re}S_{21}=1+\frac{iL_y}{2|v_{\kappa}(k_z)|}{\cal M}_{\kappa}^{\dagger}(k_z)\left({\cal G}^a_{k_z}(\omega_{\kappa,k_z})-{\cal G}^r_{k_z}(\omega_{\kappa,k_z})\right){\cal M}_{\kappa}(k_z).
\end{align}
Via Eq.~(\ref{relation_1}), $(1-{\rm Re}S_{21})+(1-{\rm Re}S_{12})=({1}/{2}){\rm Tr}\left[i(\Sigma_{k_z}^a-\Sigma_{k_z}^r){\cal G}_{k_z}^ai(\Sigma_{k_z}^a-\Sigma_{k_z}^r){\cal G}_{k_z}^r\right]$ is consistent with Landauer-B\"uttiker formula~\cite{Datta_1995}. In the chiral limits, when $S_{12}=1$, ${\rm Re}S_{21}\rightarrow -1$ with the wire dissipation $\alpha_G\rightarrow 0$ experiences a $\pi$-phase shift~\cite{magnon_trap_Yu,surface_magnetoelastic,Wang2021} (when $S_{21}=1$, ${\rm Re}S_{12}\rightarrow -1$), the injection rate $\Gamma_{\kappa}(k_z)$ is enhanced by a factor 2 because the source injects magnons into the wires while the drain contributes magnon ``holes".

Exchange magnons carry an angular momentum $\hbar$. The spin current density flowing from the film into the gate
\begin{align}
J_s &= \frac{\hbar}{L_z} \sum_{\kappa}\sum_{k_z}{\delta f_{\kappa}(k_z)}{\Gamma_{\kappa}(k_z)} \nonumber \\
    &=\frac{\hbar}{L_yL_z}\frac{\Delta T}{2}\sum_{\kappa>0,k_z}\frac{\partial f(\omega_{\kappa,k_z})}{\partial T_0}|v_{\kappa,k_z}|\Big({\rm Re}S_{12}(-\kappa,k_z)-{\rm Re}S_{21}(\kappa,k_z)\Big).
    \label{currents}
\end{align}
The right- and left-moving magnons in the film are distributed according to the temperature of their source. Without chirality ${\rm Re}\,S_{12}(-|\kappa|,k_z)={\rm Re}\,S_{21}(|\kappa|,k_z)$, the in- and out-going magnon currents cancel to $J_{s}=0$. However, a finite chirality, Eq.~(\ref{currents}), leads to net spin flow to wires and heating \textit{vs}. cooling of the wire, depending on the coupling parameter that can be tuned by the in-plane magnetic field and the sign of the applied temperature bias. The transistor can divert maximally $1/2$ of the spin current into the nanowires when the chiral damping is complete, such that all magnons from one direction are fully dissipated, while those flowing in the opposite direction are not affected at all,  i.e., $[{\rm Re}S_{12}(\kappa,k_z)-{\rm Re}S_{21}(-\kappa,k_z)]=1$ for all $k_z$. Figure~\ref{magnon_transistor}(c) and (d) address the injected spin current $J_s$ channeled off into the gate from the source to drain as a function of magnetization direction $\varphi$ [(c)] and the number of magnetic wires $N$ [(d)].

\subsubsection{Experiments}

The diode effect of magnon thermal transport can be achieved by exploiting the Dzyaloshinskii-Moriya~\cite{Lan_diode,Szulc_diode,14_Gambardella} or dipolar~\cite{Fripp_gate,53_van_Wees,15_Liu,31_Reversal_enabling_magnon_memory_Grundler,38_Magnetization_Reversal_Grundler} interactions. 
Using a ferroelectric insulator, the nonvolatile field effect transistor for thermal magnons may be realized via a coupling between magnons and ferrons (an introduction of ferron refers to Sec.~\ref{chiral_electric_fields})~\cite{30_Ding}. The non-reciprocal thermal transport and Hanle effect of magnons are observed in antiferromagnetic insulator hematite~\cite{32_Althammer}.

\begin{figure}[htp!]
    \centering
    \includegraphics[width=1.01\linewidth]{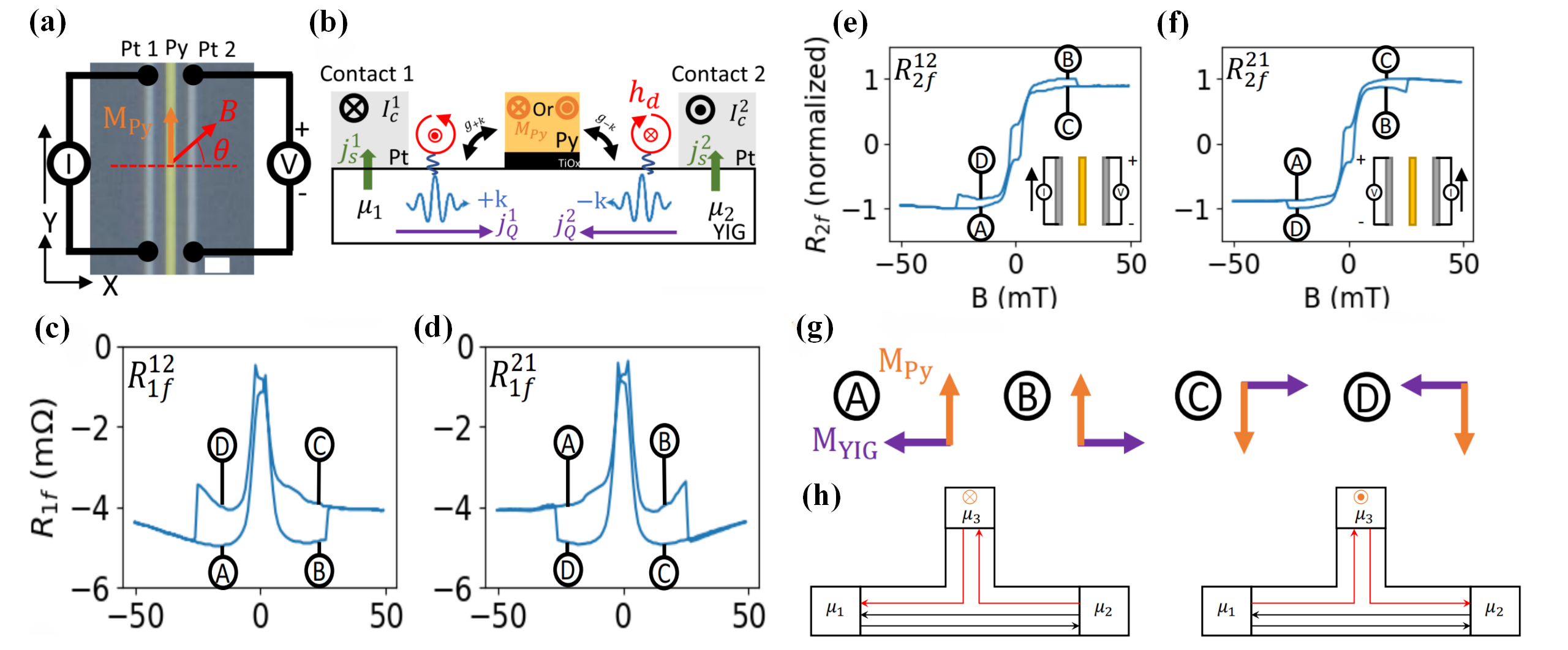}
    \caption{(a) Experimental configuration with the Pt (gray) and Py (yellow) wires fabricated on the YIG film. The magnetic field is applied parallel to the film at an angle $\theta$ with respect to the transport $\hat{\bf x}$-direction; the biased current is applied on one Pt strip, and the other Pt measures the voltage signal.  
    (b) Lateral view of the device geometry. The electric current applied to the $j$-th Pt strip creates the magnon accumulation with the chemical potential $\mu_j$. The YIG magnons couple with the Py magnons via the dipolar interaction with different coupling strengths for the magnon propagating along $\pm\hat {\bf x}$-direction.
    (c)-(f) Measured first and second harmonic signals. (g) Different magnetic configurations A-D for the Py and YIG magnetizations. 
    (h) Model of three-terminal magnon transport.\\
     \textit{Source:} The figures are taken from Ref.~\cite{53_van_Wees}.}
    \label{Non-Reciprocal_Trans}
\end{figure}

Recently, Cosset-Ch\'eneau \textit{et al.} measured the non-reciprocal thermal transport of magnons in the magnetic insulator films by a non-local transport setup~\cite{53_van_Wees}. 
As shown in Fig.~\ref{Non-Reciprocal_Trans}(a),
two Pt strips labeled as ``Pt~1" and ``Pt~2" are fabricated on top of the YIG film with a distance of around 3~$\mu$m. 
A long Py strip is sputtered between the two Pt strips, with the TiO$_x$ insertion between the YIG and Py.
Such an insertion suppresses the possible exchange coupling between YIG and Py but retains their long-range dipolar interaction. 
The AC electric current is applied to one Pt strip as the injector of spin angular momentum by the spin-Hall effect that creates magnon accumulation or depletion in the ferromagnetic insulator, and the other Pt strip acts as the detector to measure the voltage signal due to the spin transport~\cite{Cornelissen}. The biased electric current, due to the Joule heating, also creates a temperature gradient in the ferromagnetic insulator. This temperature gradient could also drive the thermal magnon flow from the injector to the detector.
Beneath the detector, the extra magnon accumulation results in the spin current being converted into a voltage signal by the inverse spin-Hall effect. The non-local response is defined as $R^{ij}_{1f}=V_{1f}^j/I^i_c$ and $R_{2f}^{ij}=V^j_{2f}/(I^i_c)^2$ (with $i,j=\{1,2\}$ and $i\neq j$), where $V^j_{1f}$ and \(V^j_{2f}\) are the first and second harmonics of the voltage signals measured by the detector ``$j$'', and $I^i_c\sim 1000$~$\mu$A is the bias current applied on the injector ``$i$''. Non-reciprocal transport is defined as a difference in the non-local signal when interchanging the magnon injector and detector without altering the magnetization configuration of the device.

The non-reciprocal thermal transport is measured by changing the magnitude of the magnetic field while fixing its direction $\theta=-4^{\circ}$ with respect to the $\hat{\bf x}$-direction. The field is first increased from a negative to a positive value and then decreased from a positive to a negative value. 
As shown in Fig.~\ref{Non-Reciprocal_Trans}(c)-(f), all the measured signals have abrupt jumps at ${\bf B}=\pm 25$~mT, which are attributed to the reversal of the magnetization direction of the Py wire.
On the other hand, the YIG magnetization direction is the same as the magnetic-field direction when the applied field is above 5~mT, leading to four possible magnetization configurations as shown in Fig.~\ref{Non-Reciprocal_Trans}(g). 
For the first harmonic non-local response as in Fig.~\ref{Non-Reciprocal_Trans}(c) and (d), $R^{12}_{1f}(S)\neq R^{21}_{1f}(S)$, where $S = \{A,B,C,D\}$ represents four different magnetization configurations in Fig.~\ref{Non-Reciprocal_Trans}(g).

When the magnetization of Py orients along the $\hat{\bf y}$-direction, as in Fig.~\ref{Non-Reciprocal_Trans}(g) for the A and B configurations, the measured non-local resistivity $|R^{12}_{1f}(A/B)|>|R^{21}_{1f}(A/B)|$. 
When reversing the magnetization direction of the gate, i.e., ${\bf M}_{\rm Py}$ is along the $-\hat{\bf y}$-direction, $|R^{12}_{1f}(C/D)|<|R^{21}_{1f}(C/D)|$. 
This difference shows the non-reciprocity of the electrically generated magnon transport according to the definition, and indicates that the non-reciprocity relates to the direction of the gate ${\bf M}_{\rm Py}$. 
For the second harmonic (thermal) signals, a similar effect was observed as shown in Fig.~\ref{Non-Reciprocal_Trans}(e) and (f).
Similar to the case of the first harmonic signals, $R^{12}_{2f}(S)\neq R^{21}_{2f}(S)$, showing the  non-reciprocity in the thermal-transport signals.
In particular, for a Py magnetization along the $\hat{\bf y}$-direction, the non-local signal generated by a temperature gradient in the $+\hat{\bf x}$ direction is larger than that by a temperature gradient in the $-\hat{\bf x}$-direction, i.e., $|R^{12}_{2f}(A/B)|>|R^{21}_{2f}(A/B)|$. $ |R^{12}_{2f}(C/D)|< |R^{21}_{2f}(C/D)|$ when the Py magnetization is reversed to the $-\hat{\bf y}$ direction. These results demonstrate non-reciprocity in the transport of electrically and thermally generated magnons, which is controlled by the direction of the gate Py magnetization.

The mechanism behind this is the chiral coupling between the wire and gate magnons governed by the long-range dipolar stray field~\cite{Magnon_Magnon_Coupling,Tao_chiral_pumping,49_Chirality_enables_thermal_magnon_transistors,15_Liu,Kruglyak_Chiral_magnonic_resonators} (Sec.~\ref{chiral_magnetic_fields}),  
which is addressed by Yu \textit{et al.}~\cite{Tao_chiral_pumping}.
As shown in Fig.~\ref{Non-Reciprocal_Trans}(b), for the experimental configuration with ${\bf M}_{\rm YIG}\parallel {\bf H}_0$ at $\theta=-4^\circ$, the $\hat{\bf y}$-component of the static magnetization ${\bf M}_{\rm YIG}\cdot {\hat{\bf y}}=M_{\rm YIG} \sin\theta\ll M_{\rm YIG}$ is negligibly small, such that the static magnetization is almost parallel to the transport $\hat{\bf x}$-direction and normal to the wire magnetization.
In such a configuration, the dipolar field emitted by the spin waves propagating along the $\hat{\bf x}$-axis exists both above and below the YIG film, which, however, holds opposite handedness for different magnon propagation directions, i.e., a spin-momentum locking effect. 
The dipolar field above the film emitted by the spin waves propagating along the $\hat{\bf x}$-axis reads~\cite{Tao_chiral_pumping}
\begin{align}
\begin{pmatrix}
h_z(\mathbf{r}) \\
h_x(\mathbf{r})
\end{pmatrix}
=\frac{1}{2}e^{i(kx - \omega t)}
\begin{pmatrix}
1 \\
-i{\rm sgn}(k)
\end{pmatrix}
e^{-|k||z|} (1-e^{-|k|s})M_z,
\end{align}
where $s$ is the thickness of the YIG film and $k$ is the wave vector along the $\hat{\bf x}$-axis. 
As illustrated in Fig.~\ref{Non-Reciprocal_Trans}(b), for the magnon with the positive $k$, the stray field above the film holds right-handed polarization with the polarization vector along the $-\hat{\bf y}$-direction, i.e., $h_z=ih_x$; by contrast, for the negative $k$, the polarization vector of the stray field is along the $\hat{\bf y}$-direction ($h_z=-ih_x$). 
Meanwhile, the polarization of the stray field below the film is reversed. 
Such a polarized magnetic field is then coupled with the transverse magnetization of the Py strip when their handedness matches. 
For the static magnetization ${\bf M}_{\rm Py}$ along the $+\hat{\bf y}$-direction, the dynamics of the magnetization is right-handed, i.e., $\tilde{M}_z=-i\tilde{M}_x$ with $\tilde{\bf M}$ being the transverse magnetization of the Py strip. 
Thereby, the magnons in the Py strip are more likely to couple with the spin waves with negative $ k$ due to the match of the handedness, hence leading to a coupling constant \(g_k\) depending on the magnon propagation direction. 
While reversing the ${\bf M}_{\rm Py}$ direction, the coupling is reversed. 
Yu \textit{et al.}~\cite{Tao_chiral_pumping} investigated the coupling strength $g_k$ for such a dipolar coupling. For the B configuration in Fig.~\ref{Non-Reciprocal_Trans}(g) with ${\bf M}_{\rm YIG}\cdot {\hat{\bf x}}>0$ and ${\bf M}_{\rm Py}\cdot {\hat{\bf y}}>0$, when the Py strip has the same thickness and width $d$, the coupling strength ~\cite{15_Liu,magnon_trap_Yu2,49_Chirality_enables_thermal_magnon_transistors,Tao_chiral_pumping} 
\begin{align}
 g_k=-\dfrac{1}{2\sqrt{ds}} \mu_0\gamma \hbar\sqrt{M_{\rm YIG}M_{\rm Py}} \dfrac{1}{k|k|^2}\sin(\dfrac{k d}{2})(1-e^{-|k|d})(1-e^{-|k|s})(|k|-k)
\end{align}
depends on the propagation direction. 
Cosset-Ch\'eneau \textit{et al.} explained the measurements by a three-terminal magnon transport configuration as addressed in Sec.~\ref{magnon_transistors_theory}~\cite{53_van_Wees}, where the Py strip acts as the third terminal connected to the ground with $\mu_3=0$ as shown in Fig.~\ref{Non-Reciprocal_Trans}(h). The magnon transport from the source to the drain can be described as an unbalanced transmission of magnons flowing in opposite directions, i.e, the individual magnons can couple differently with the third terminal as a function of their propagation direction and the Py magnetization direction, as shown by the red arrows in Fig.~\ref{Non-Reciprocal_Trans}(h).
For a larger coupling strength, the individual magnon is more likely to relax in the Py wire, resulting in larger damping and lower transmission signals. The chirality of $g_k$ agrees with both electrically-generated magnons and thermally-generated magnons signals, i.e., when $g_k<g_{-k}$, $|R^{12}_{1f}(B)|>|R^{21}_{1f}(B)|$ and $|R^{12}_{2f}(B)|>|R^{21}_{2f}(B)|$.
According to Yu \textit{et al.}~\cite{49_Chirality_enables_thermal_magnon_transistors}, the chirality also enables the injection of magnons from YIG to the Py strip (Sec.~\ref{magnon_transistors_theory}), which may drive the magnetization dynamics.

\begin{figure}[htp!]
    \centering
    \includegraphics[width=\linewidth]{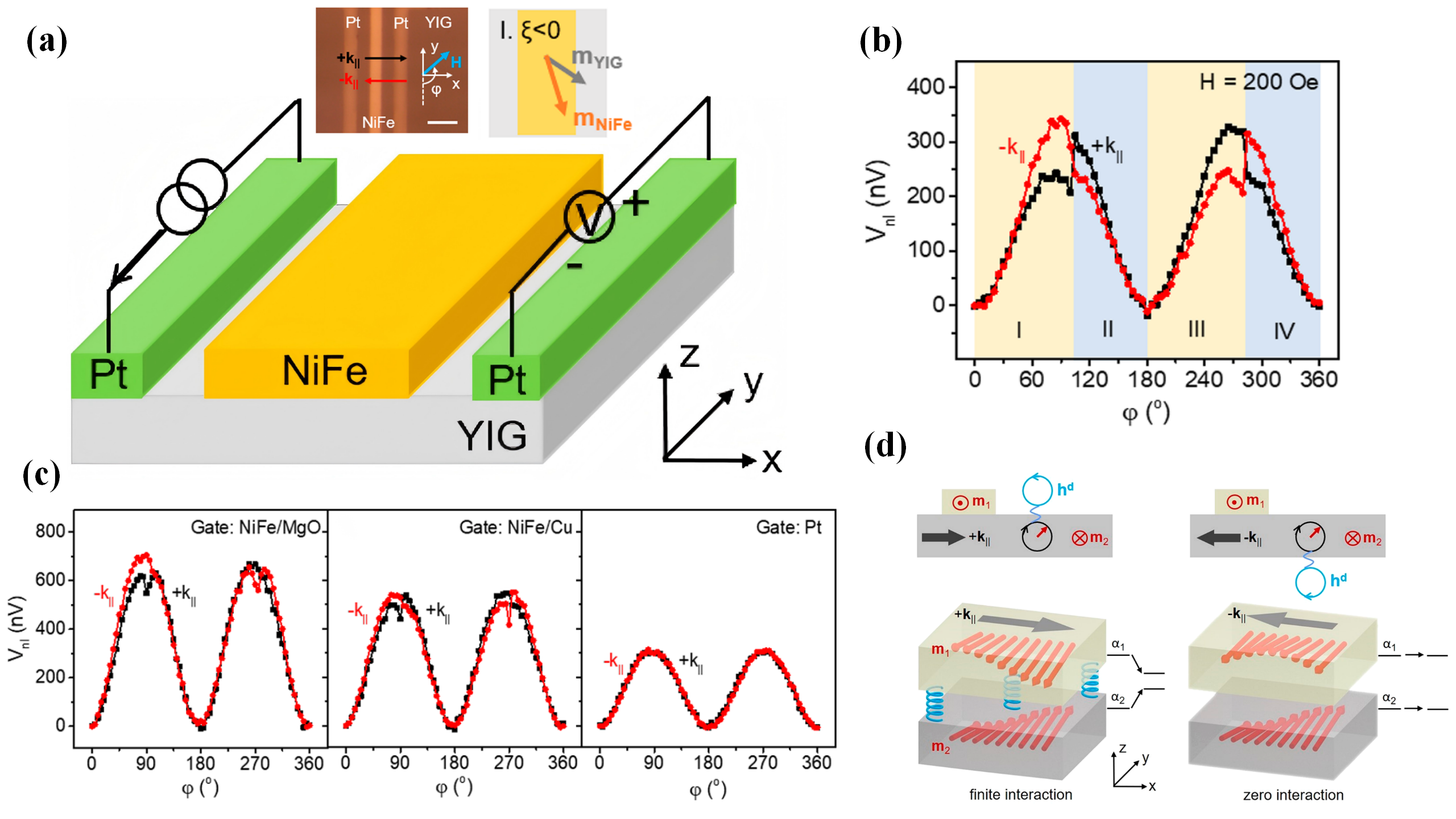}
    \caption{(a) Experimental configuration with a NiFe long strip as a gate on top of the YIG film. The magnetic field is applied parallel to the film; the magnetization directions of YIG and NiFe have an angle, as the NiFe film exhibits an uniaxial anisotropy with an easy axis along the $\hat{\bf y}$-direction. (b) Angular dependence of $V_{\rm nl}$ for magnon flowing in opposite directions when $H=200$~Oe. (c) Experimental results with controlled samples, where i) MgO and Cu are inserted between NiFe and YIG, and ii) Pt replaces NiFe. (d) Mechanism of non-reciprocal magnon transport.  \\
    \textit{Source:} The figures are taken from Ref.~\cite{15_Liu}.}
    \label{Non-Reciprocal_Trans_Liu}
\end{figure}

Han \textit{et al.} used the permalloy strip to gate the transport of incoherent magnons and found that the magnon flow was non-reciprocal~\cite{15_Liu} when the transport is normal to the magnetization. The authors also attributed such non-reciprocity to the chirality of the stray field~\cite{Magnon_Magnon_Coupling,Tao_chiral_pumping,49_Chirality_enables_thermal_magnon_transistors,15_Liu,Kruglyak_Chiral_magnonic_resonators}.
The experimental configuration is shown in Fig.~\ref{Non-Reciprocal_Trans_Liu}(a). 
A pair of Pt electrodes is fabricated to inject and detect the magnon flow. A NiFe strip is deposited on the YIG thin film between the two Pt electrodes and is electrically isolated.
The external magnetic field ${\bf H}$ is applied parallel to the film plane with the angle $\varphi$  with respect to the $-\hat{\bf y}$-direction, which biases the YIG magnetization ${\bf M}_{\rm YIG}$ parallel to the magnetic field, i.e.,  ${\bf H}\parallel{\bf M}_{\rm YIG}$.
On the other hand, the NiFe has an easy axis along the $\hat{\bf y}$-direction due to the long strip shape; thus, the magnetization of NiFe and YIG does not generally align in the same direction. 
When changing the angle $\varphi$, the magnetization direction ${\bf M}_{\rm NiFe}\cdot\hat{\bf y}$ flips at some specific angle since ${\bf  M}_{\rm NiFe}$ tends to align along its easy axis. During the measurement, a low-frequency ($<$50 Hz) alternating current is applied to the Pt injector to excite broadband magnons via the spin Hall effect. 
The nonequilibrium magnons diffuse toward the Pt detector and are converted to a charge current via the inverse spin Hall effect, which is detected as the first harmonic voltage.
Figure~\ref{Non-Reciprocal_Trans_Liu}(b) plots the angular dependence of the magnon transport under the influence of the NiFe and the bias magnetic field $H=200$~Oe within the film plane. $\pm k_{\parallel}$ denotes the right-/left-going magnon flow, which is experimentally realized by swapping the role of injection/detection Pt electrodes. Non-reciprocity emerges in magnon transport, i.e., under the same field conditions, the transport signals change when the direction of the magnon transport is reversed. This non-reciprocity is most significant when $\varphi\sim 90^\circ$ and $\varphi\sim 270^\circ$, where a sudden jump occurs caused by the flipping of the $y$-component of ${\bf M}_{\rm NiFe}$. 

To verify that the long-range dipolar interaction causes such an effect, Han \textit{et al.} investigated several controlled samples, as shown in Fig.~\ref{Non-Reciprocal_Trans_Liu}(c). Firstly, they fabricated the sample by inserting a spacer layer of 3~nm MgO or 8~nm Cu between NiFe and YIG. Both spacers can eliminate the interlayer exchange interaction. In this configuration, a weaker but finite signal was observed, which is consistent with the result in Fig.~\ref{Non-Reciprocal_Trans_Liu}(c), indicating that the interlayer exchange coupling is not responsible for the non-reciprocity. Besides this, it is speculated that the interfacial DMI can lead to the non-reciprocal magnon transmission~\cite{DMI_Casiraghi}, so they also organized another controlled sample by replacing the center NiFe strip with nonmagnetic Pt. They find the voltage difference between $+k_\parallel$ and $-k_\parallel$ is too small to identify compared to the NiFe/YIG device, which shows the DMI is probably not the origin of the non-reciprocity as in Fig.~\ref{Non-Reciprocal_Trans_Liu}
(c). Han \textit{et al.}~\cite{15_Liu} attributed such an effect to the long-range magnetic dipolar interaction, as shown in Fig.~\ref{Non-Reciprocal_Trans_Liu}(d). For the single magnetic layer with ${\bf k}\perp {\bf m}$, the dynamic dipolar field generated by magnons is spatially chiral as we addressed in Sec.~\ref{chiral_magnetic_fields} [Fig.~\ref{gating_01}(a)]. Consequently, in a bilayer with antiparallel magnetizations, the magnon modes of the two layers exhibit finite interaction for $+k_\parallel$ but no interaction for $-k_\parallel$.

\subsection{Nodal magnon-photon polaritons }
\label{Nodal_magnon_photon_polaritons}

Exceptional point---special spectral singularities at which two or more eigenvalues, and their corresponding eigenvectors, coalesce and become identical---was proposed and measured in the ferromagnet$|$normal metal$|$ferromagnet heterostructures~\cite{EP1,Ep2,Ep3}. Achieving an exceptional point in these setups requires fine-tuning of parameters, which is often challenging in heterostructures. The electromagnetic proximity effect in normal metal|ferromagnet|normal metal renders a persistent realization of the exceptional point or a nodal magnon-photon polariton.

\textbf{Dissipative coupling}.---The magnetization is quantized by combining the LLG equation with the normalization relation~\cite{Walkermode,spinwaveExcitation} from Eq.~\eqref{MP polariton_SC} as [similar to Eq.~\eqref{magnetization_operator_SC}]
\begin{align}
    \hat{M}_x&=-i\int\frac{d{\bf k}}{2\pi} \sqrt{\frac{\hbar M_s\omega_N}{4d_F\mu_0(H_0+M_s)}}\left(e^{i{\bf k}\cdot{\pmb \rho}}\hat{m}_{\bf k}-e^{-i{\bf k}\cdot{\pmb \rho}}\hat{m}_{\bf k}^{\dagger}\right),\nonumber\\
    \hat{M}_y&=\int\frac{d{\bf k}}{2\pi} \sqrt{\frac{\hbar\mu_0\gamma^2(H_0+M_s)M_s}{4d_F\omega_N}}\left(e^{i{\bf k}\cdot{\pmb \rho}}\hat{m}_{\bf k}+e^{-i{\bf k}\cdot{\pmb \rho}}\hat{m}_{\bf k}^{\dagger}\right),
    \label{magnetization_NM}
\end{align}
where  $\hat{m}_{\bf k}^{\dagger}$ and $\hat{m}_{\bf k}$ are the creation and annihilation operators of magnon. According to Sec.~\ref{transmutation_relation}, replacing London's penetration depth $\lambda$ in Eq.~\eqref{frequency_SC} by the penetration depth of the normal metal $\delta=(1+i)/\sqrt{2\omega\mu_0\sigma_{c}}$ leads to the eigenfrequency $\omega_N=\mu_0\gamma\sqrt{(H_0+M_s)(H_0+M_sd_F/(d_F+\delta)}$ of magnon mode in the NM$|$FI$|$NM heterostructure~\cite{Kittel_mode}. $\omega_N$ is no longer a purely real quantity due to the Ohmic dissipation, which is reflected in $\delta$. The operators $\hat{M}_{x,y}$ are not  Hermitian in this ansatz, distinguishing this approach from the conventional quantization scheme.

\begin{figure}[htp!]
    \centering
    \includegraphics[width=1\linewidth]{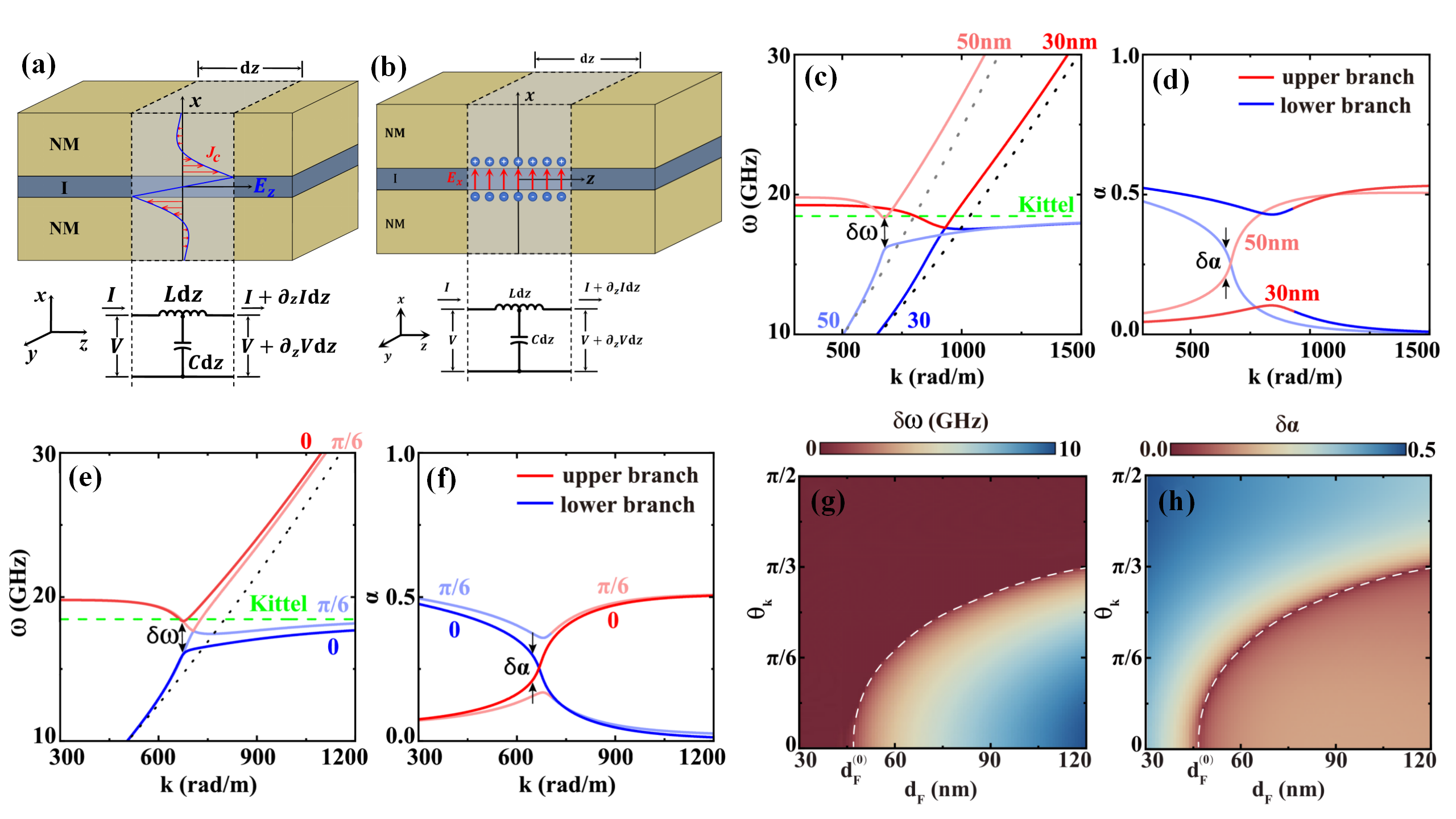}
    \caption{(a) and (b): Current and charge distribution in NM$|$I$|$NM heterostructures with equivalent circuit models for in-plane (a) and out-of-plane (b) electric components.  (c) Dispersion relation of nodal magnon-photon polariton as a function of wavevector ${\bf k}$ with upper and lower branches separated by a frequency gap $\delta\omega$ at different ferromagnetic insulator layer thicknesses $d_F=\{30, 50\}$ nm. The green dashed line represents the bare magnon mode, while the black dotted curve shows the photon mode. (d) presents the corresponding damping $\alpha$ of the coupled modes. (g) and (h) depict phase diagrams of the frequency gap $\delta\omega$ and damping gap $\delta\alpha$ as functions of ferromagnetic layer thickness $d_F$ and propagation direction $\theta_{\bf k}$. The white dashed curve marks the emergence of exceptional points.\\
     \textit{Source:} Figures (a), (c)-(h) are taken from Ref.~\cite{52_Persistent_nodal_magnon-photon_polariton_in_ferromagnetic_heterostructures}.}
    \label{nodal MP polariton}
\end{figure}

The photon modes are found in an NM$|$non-magnetic insulator(I)$|$NM heterostructure, as in Fig.~\ref{nodal MP polariton}(a) and (b). The thickness of the non-magnetic insulator is set to $2d_{F}$, and the dielectric constant is $\varepsilon_{\rm I}$, identical to those of the ferromagnetic insulator layer. Maxwell's equations governing the electric fields are given by~\cite{Jackson}
\begin{align}
    &\text{in I}: ~~~\nabla^2 \mathbf{E}({\bf r}, t)+k_0^2 \mathbf{E}({\bf r}, t)=0,\nonumber\\
    &\text{in NMs}: ~~~\nabla^2 \mathbf{E}({\bf r}, t)+k_n^2 \mathbf{E}({\bf r}, t)=0.
    \label{field_NM}
\end{align}
The propagation is isotropic in the $y$-$z$ plane. In a plane-wave solution ${\bf E}=\tilde{\bf E}(x)e^{ikz-i\omega t}$, the amplitudes 
\begin{align}
    &\text{In I},~~\tilde{\bf E}(x)=\pmb{ \cal E}_0e^{i{\cal A}_k x}+\pmb{ \cal E}'_0e^{-i{\cal A}_k x},\nonumber\\
    &\text{In NMs}~(x>d_{F}~\text{or}~ x<-d_F),~~\tilde{\bf E}(x)=\pmb{ \cal E}_{1}e^{\pm i{\cal B}_k x},  \nonumber
\end{align}
in which ${\cal A}_k=\sqrt{k_0^2-k^2}$ and ${\cal B}_k=\sqrt{k_n^2-k^2}$. The boundary condition with the continuity of $H_y$ and $E_z$ at the interfaces leads to the secular equation for $\{{\cal E}_{0x},{\cal E}'_{0x}\}$ and 
$i(R_{k}^2+1)\sin(2{\cal A}_kd_{F})=2R_{k}\cos(2{\cal A}_kd_{F})$.
Since the wavelength $2\pi/|{\bf k}|$ is much larger than the thickness $d_{F}$, $\sin(2{\cal A}_kd_{F})\approx2{\cal A}_kd_{F}$ and $\cos(2{\cal A}_kd_{F})\approx1$. When the frequency  $\omega\lesssim 10$~GHz, $\omega\mu_0\sigma_c\gg\omega^2\mu_0\varepsilon_0$, leading to ${\cal B}_k\approx \sqrt{i\omega\mu_0\sigma_c}=i/\delta$ and $ R_k={({\cal A}_k^2+k^2){\cal B}_k}/{{\cal A}_k({\cal B}_k^2+k^2)}\approx-i{k_0^2}\delta/{{\cal A}_k}\ll1$.
Accordingly, the secular equation for $\{{\cal E}_{0x},{\cal E}'_{0x}\}$ simplifies to $ k^2=\Omega_n^2\mu_0\varepsilon_{\rm I}\left(1+{\delta}/{d_{F}}\right)$, so the eigenfrequency of photon modes
\begin{align}
\Omega_n=\sqrt{d_{F}/(d_{F}+\delta)\mu_0\varepsilon_{\rm I}}|{\bf k}|
\label{photon_dispersion}
\end{align}
is not purely real since $\delta$ contains an imaginary component. 

The electromagnetic field corresponds to the transverse magnetic (TM) modes~\cite{Swihart_mode}, characterized by $E_y=H_x=H_z=0$ with only three non-zero components $\{E_z,E_x,H_y\}$. 
The amplitudes ${\cal E}_{0x}'=(1-{R}_k)/(1+{R}_k)e^{2i{\cal A}_kd_F}{\cal E}_{0x}\approx{\cal E}_{0x}$. In the insulators $-d_F<x<d_F$, 
\begin{align}
    \tilde{E}_{x,k}(x)&={\cal E}_{0x}e^{i{\cal A}_k x}+{\cal E}_{0x}'e^{-i{\cal A}_k x}\approx 2{\cal E}_{0x},\nonumber\\
    \tilde{E}_{z,k}(x)&=\frac{{\cal A}_k}{k}\left({\cal E}_{0x}e^{i{\cal A}_k x}-{\cal E}_{0x}'e^{-i{\cal A}_k x}\right)\approx i2\frac{{\cal A}_k^2}{k}{\cal E}_{0x}x,\nonumber\\
    \tilde{H}_{y,k}(x)&=\frac{{\omega \varepsilon_{\rm I} }}{k}\left({\cal E}_{0x}e^{i{\cal A}_k x}+{\cal E}_{0x}'e^{-i{\cal A}_k x}\right)\approx2\frac{{\omega \varepsilon_{\rm I} }}{k}{\cal E}_{0x}.
    \label{EM_in_FI}
\end{align}
$H_y\equiv H_{0y}$ and $E_x\equiv E_{0x}$ are then uniform across the thin insulator. Out of the insulator with ``$+$" and ``$-$" for the regions $x>d_F$ and $x<-d_F$, 
\begin{align}
    \tilde{E}_{x,k}(x)&=\frac{\varepsilon_{\rm I}}{\varepsilon_0+i\sigma_c/\omega}\tilde{E}_{x,k}(\pm d_F^{\mp})e^{\pm i{\cal B}_k (x\mp d_F)},\nonumber\\  \tilde{E}_{z,k}(x)&=\tilde{E}_{z,k}(\pm d_F)e^{\pm i{\cal B}_k (x\mp d_F)},\nonumber\\
    \tilde{H}_{y,k}(x)&=\tilde{H}_{y,k}(\pm d_F)e^{\pm i{\cal B}_k (x\mp d_F)}.
    \label{EM_in_NM}
\end{align}

The electric field in the system is oriented either in-plane, $E_z\hat{\bf z}$ [Fig.~\ref{nodal MP polariton}(a)], or out-of-plane, $E_x\hat{\bf x}$ [Fig.~\ref{nodal MP polariton}(b)]. The in-plane electric field $E_z\hat{\bf z}$ remains continuous across the interfaces, allowing penetration into the NM layers. As shown in Fig.~\ref{nodal MP polariton}(a), this field induces eddy currents in the NMs, which exhibit an oscillatory decay within the skin depth $|\delta|$ away from the NM$|$I interfaces. The out-of-plane electric field $E_x\hat{\bf x}$ induces charge accumulation at the top and bottom surfaces of the NM layers, creating a voltage across the insulator. Due to this charge buildup at the interfaces, $E_x$ vanishes within the NM layers and only exists within the insulator. Figure~\ref{nodal MP polariton}(b) shows the discontinuity of the $E_x$ across the interfaces, leading to the accumulation of bound charges.

Using photon dispersion \eqref{photon_dispersion} and field distributions \eqref{EM_in_FI} and \eqref{EM_in_NM}, the photon magnetic fields is quantized using LC circuits~\cite{52_Persistent_nodal_magnon-photon_polariton_in_ferromagnetic_heterostructures}, as shown in Fig.~\ref{nodal MP polariton}(a) and (b). To represent the system as an LC circuit, we define the voltage $V$ across the heterostructure and the total current $I$ per unit length in the $\hat{\bf y}$-direction. The voltage is determined by the out-of-plane $x$-component of the electric field, given by $V=\int{\rm d}x E_x=2E_{0x}d_F$. The in-plane component drives the eddy current. The total current $I=H_{0y}$ follows Amp\`ere's circuital Law, calculated using a loop that crosses the insulator in the $x$-$y$ plane.  This loop has one unit width along the $\hat{\bf y}$-direction and extends infinitely in the $\hat{\bf x}$-direction.

The dynamics of voltage and current are not independent of each other. Integrating Maxwell's equation $\nabla \times \mathbf{E} = -\partial\mathbf{B}/\partial t$ yields $  \int^{+\infty}_{-\infty}{\rm d}x\partial_z E_x=-\mu_0\int^{+\infty}_{-\infty}{\rm d} x \dot{H}_y$ and $2\partial_z E_{0x}d_F=-2\mu_0\dot{H}_{0y} d_F(1+\delta/d_F)$, where $\int_{-\infty}^{-d_F} {\rm d} x H_y=\int_{d_F}^{+\infty} {\rm d}x H_{y}=H_{0y}\delta$. Substituting $E_{0x}$ and $H_{0y}$ with $V$ and $I$ leads to $\partial_z V=-2\mu_0d_F(1+{\delta}/{d_F})\dot{I}$. Comparing this with the circuit equation $\partial_z V = -L \dot{I}$, the effective inductance $L=2\mu_0d_F(1+\delta/d_F)$. Similarly, substituting $E_{0x}$ and $H_{0y}$ with $V$ and $I$ into Maxwell's equation $\partial_z H_{0y}=-\varepsilon_{\rm I} \dot{E}_{0x}$ yields $\partial_z I=-\varepsilon_{\rm I}/(2d_F) \dot{V}$. Comparing this with the circuit equation $\partial_z I=-C \dot{V}$, the effective capacitance $C=\varepsilon_{\rm I}/(2d_F)$.

Using these effective inductance and capacitance, the Hamiltonian of the NM$|$I$|$NM heterostructure can be expressed as  $\hat{H}=\int {\rm d}z~(1/2)( L\hat{I}^2+C \hat{V}^2)$. In the wave-vector space, with ${\bf k}=k\hat{\bf z}$, 
\begin{equation}
    \hat{H}=\int {\rm d}k~\frac{1}{2}\left(L \hat{I}_k \hat{I}_{-k}+\frac{CL^2}{k^2}\dot{\hat{I}}_{k}\dot{\hat{I}}_{-k}\right),
    \label{quantization_2}
\end{equation}
where $\hat{I}_k(t)=\int {\rm d}z/\sqrt{2\pi} \hat{I}(z,t) e^{-ikz}$.

We define the canonical conjugate variables: $\hat{q}_k = \hat{I}_k$ represents the generalized coordinate and $\hat{\pi}_k = CL^2\dot{\hat{I}}_k/k^2 $ is the canonical momentum. The effective Hamiltonian Eq.~\eqref{quantization_2} then becomes $\hat{H}=\int {\rm d}k\left(m_k\Omega_n^2 \hat{q}_k \hat{q}_{-k}+\hat{\pi}_k \hat{\pi}_{-k}/m_k\right)/2$, where $m_k=CL^2/k^2$. The annihilation and creation operators of photon modes 
\begin{align}
    \hat{p}_{k}=\frac{1}{\sqrt{2m_k\hbar\Omega_n}}(m_k\Omega_n \hat{q}_k + i \hat{\pi}_{-k}),~~~~~~~~ \hat{p}_{k}^\dagger=\frac{1}{\sqrt{2m_k\hbar\Omega_n}}(m_k\Omega_n \hat{q}_{-k} + i \hat{\pi}_{k}),
\end{align}
obey the bosonic commutation relation $[\hat{p}_k, \hat{p}_{k'}^\dagger] = \delta(k-k')$. Inversely, $\hat{q}_k=\sqrt{{\hbar}/{2m_k\Omega_n}}(\hat{p}_k+\hat{p}_{-k}^\dagger)$ and $\hat{\pi}_k=i\sqrt{{m_k\hbar\Omega_n}/{2}}\left(\hat{p}_{k}^\dagger-\hat{p}_{-k}\right)$ due to the dissipation in photon dispersion $\Omega_n$.

The magnetic field $\hat{\bf H}=\hat{H}_y\hat{\bf y}+\hat{H}_z\hat{\bf z}$ with 
\begin{align}
     \hat{H}_y&=\int \frac{d{\bf k}}{2\pi}~ \sqrt{\frac{\hbar}{2m_k\Omega_n}}\cos\theta_{\bf k}\left( e^{i{\bf k}\cdot{\pmb \rho}}\hat{p}_{\bf k}+e^{-i{\bf k}\cdot{\pmb \rho}}\hat{p}_{\bf k}^\dagger\right),\nonumber\\ 
     \hat{H}_z&=-\int \frac{d{\bf k}}{2\pi}~ \sqrt{\frac{\hbar}{2m_k\Omega_n}}\sin\theta_{\bf k} \left(e^{i{\bf k}\cdot{\pmb \rho}}\hat{p}_{\bf k}+e^{-i{\bf k}\cdot{\pmb \rho}}\hat{p}_{\bf k}^\dagger\right).
     \label{magnetic_field_NM}
\end{align}
Using the magnetization operator \eqref{magnetization_NM} and magnetic-field operator \eqref{magnetic_field_NM}, the magnon and photon couple via  
\[
\hat{H}_{\rm int}=-\mu_0\int d{\bf r}\hat{\bf M}\cdot\hat{\bf H}=\int d{\bf k}~g_{\bf k}\left(\hat{m}_{\bf k}\hat{p}_{\bf k}^{\dagger}+\hat{m}_{\bf k}^{\dagger}\hat{p}_{\bf k}-\hat{m}_{\bf k}^{\dagger}\hat{p}_{\bf -k}^{\dagger}-\hat{m}_{\bf k}\hat{p}_{\bf -k}\right),
\]
where the coupling constant $g_{\bf k}/\hbar=\cos\theta_{\bf k} \frac{\hbar\gamma}{2} \sqrt{\frac{\mu_0(H_0+M_s)M_s}{2m_k\Omega_n\omega_Nd_F}}$ strongly depends on the propagation direction of the collective modes. The total Hamiltonian of the system 
\begin{align}
    \hat{H}_{\rm tot}&=\int d{\bf k} \left(\hbar\omega_N\left(\hat{m}_{\bf k}^{\dagger}\hat{m}_{\bf k}+\frac{1}{2}\right)+\hbar\Omega_n\left(\hat{p}_{\bf k}^{\dagger}\hat{p}_{\bf k}+\frac{1}{2}\right)+g_{\bf k}\left(\hat{m}_{\bf k}\hat{p}_{\bf k}^{\dagger}+\hat{m}_{\bf k}^{\dagger}\hat{p}_{\bf k}-\hat{m}_{\bf k}^{\dagger}\hat{p}_{\bf -k}^{\dagger}-\hat{m}_{\bf k}\hat{p}_{\bf -k}\right)\right)
\label{Hamiltonian_NM}
\end{align}
is non-Hermitian, as $g_{\bf k}\ne g_{\bf k}^*$, which goes beyond perturbation theory without invoking the rotating-wave approximation. Using the Bogoliubov transformation leading to the characteristic equation  
\begin{equation}
    \omega^4-\omega^2\left(\omega_N^2+\Omega_n^2\right)+\omega_N^2\Omega_n^2-4\omega_N\Omega_n{g_{\bf k}^2}/{\hbar^2}=0.
    \label{Qum_dispersion_NM}
\end{equation}
Accordingly, the dispersion of the two collective modes 
\begin{equation}
\omega_{u(l)}^2=\frac{1}{2}\left(\Omega_n^2+\omega_N^2\pm\sqrt{\left(\Omega_n^2-\omega_N^2\right)^2+16\Omega_n\omega_N\frac{g_{\bf k}^2}{\hbar^2}}\right).
\label{dispersion_NM}
\end{equation}

\textbf{Nodal magnon-photon polariton}.---YIG is an excellent example to illustrate the collective modes in the NM$|$FI$|$NM heterostructure, with parameters including a saturation magnetization $\mu_0M_s=0.17$ T, intrinsic Gilbert damping $\alpha_G=10^{-4}$, and a dielectric constant $\varepsilon_{\rm FI}=8\varepsilon_0$~\cite{45_Imaging_spin_wave_damping_underneath_metals,YIG_parameters2}. The YIG layer is sandwiched between two thick copper films, each with a conductivity $\sigma_{c}=5.96\times 10^{-7}$~$\Omega^{-1} \cdot {\rm m}^{-1}$. An in-plane bias magnetic field $\mu_0H_0=50$ mT is applied along the $\hat{\bf z}$-direction.

The coupled magnon and photon modes exhibit persistent non-Hermitian topological phenomena in the wave-vector space, driven by eddy-current-induced anisotropic damping $\alpha$. For modes propagating parallel to the magnetization ($\theta_{\bf k}=0$), we plot the real and imaginary parts of the eigenfrequencies obtained from Eq.~\eqref{Qum_dispersion_NM}, as shown in Fig.~\ref{nodal MP polariton}(c) and (d), respectively. The frequency gap $\delta\omega$ and the damping gap $\delta\alpha$ are highly sensitive to the thickness of the ferromagnetic insulator layer, $d_F$. For $d_F=50$ nm, the frequency gap remains open, while for $d_F=30$ nm, it closes, leading to gapless modes at specific wavevectors. Notably, these gaps can collapse simultaneously at a critical thickness, indicating the presence of exceptional points in the wave vector space.

Exceptional points can also arise for wave propagation in various directions due to the anisotropic nature of the magnon-photon coupling $g_{\bf k}$. For example, when spin waves propagate at an angle $\theta_{\bf k}=\pi/6$, the upper and lower branches intersect, as shown in Fig.~\ref{nodal MP polariton}(e). This directional dependence enables exceptional points to persist at different wave vectors, even when specific parameters, such as $d_F$, remain fixed. Figure~\ref{nodal MP polariton}(g) and (h) demonstrate that exceptional points persist in the wave vector space for ferromagnetic insulator thicknesses exceeding a critical value $d_F^{(0)}$, with exceptional lines indicating points of simultaneous frequency and damping closure. By adjusting parameters $\{d_F,H_0,\theta_{\bf k}\}$, experimental discovery of these exceptional points becomes feasible, which could facilitate synchronization of oscillators due to the unique coalescence of eigenfrequencies and dissipation rates at the exceptional points.

\subsection{Near-field radiative spin transfer}
\label{Near_field_radiative_spin_transfer_to_magnons}

\subsubsection{Experiments}
Recent experiments~\cite{31_Reversal_enabling_magnon_memory_Grundler,inject_exp} accumulate evidence for spin transfer mediated by the dipolar interaction.

\begin{figure}[htp!]
    \centering
    \includegraphics[width=0.92\linewidth]{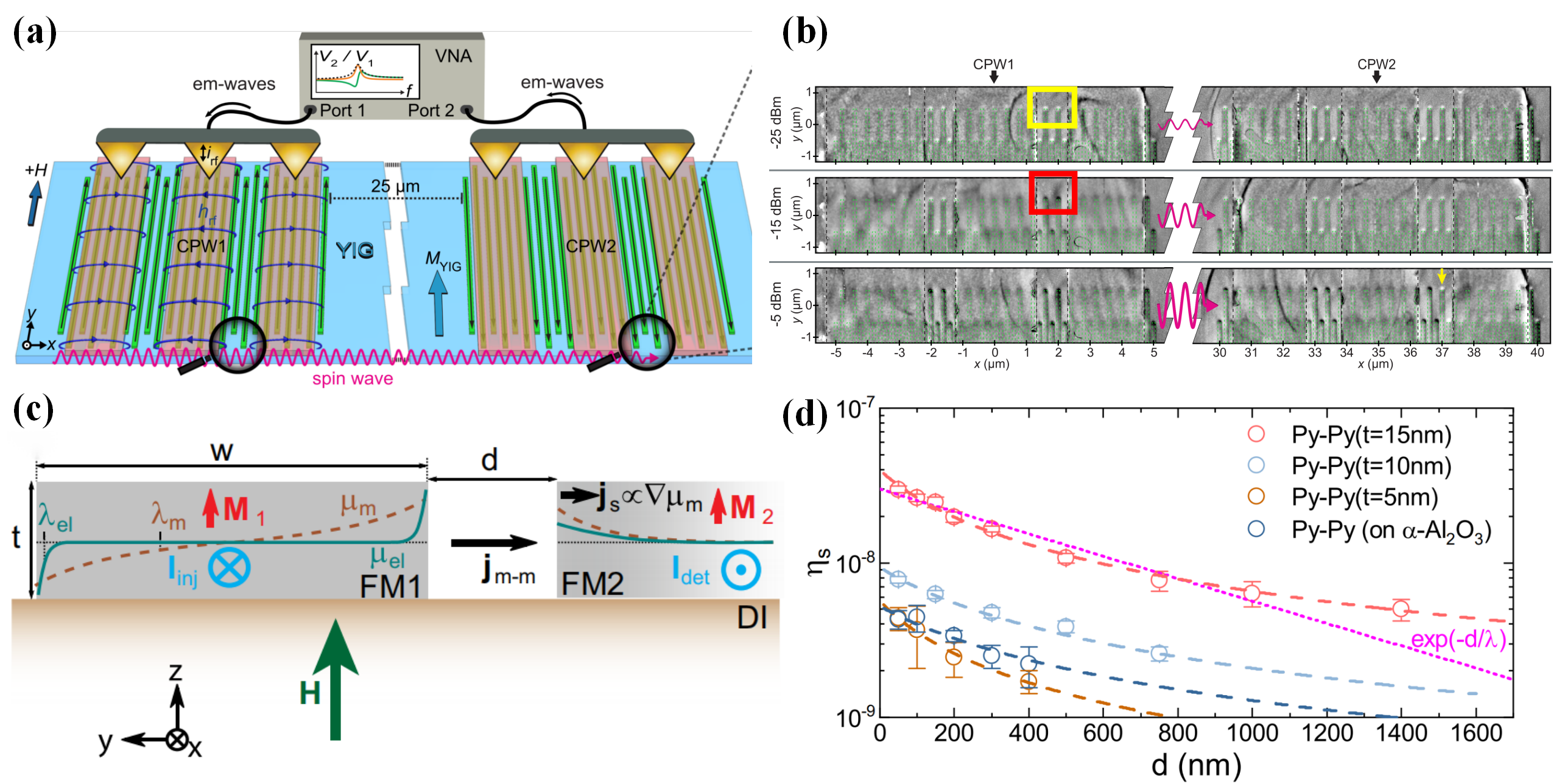}
    \caption{(a) Experimental configuration with two  Py stripes on a YIG film. (b) Magnetic force microscopy measurements were performed on the sample excited by different powers $P_{\rm irr}=\{-25, -15,-5\}$~dBm. (c) Spin transport between two separated ferromagnet strips on a diamagnetic insulator. (d) Spin transfer efficiency $\eta_s$ as a function of distance $d$ with four different samples by varying  substrate and Py thickness $t$.\\
    Sourse: Figures (a) and (b) are taken from Ref.~\cite{31_Reversal_enabling_magnon_memory_Grundler}; (c) and (d) are taken from Ref.~\cite{inject_exp}.}
    \label{injection_experiment}
\end{figure}

Baumgaertl and Grundler reported the switch of magnetization of magnetic nanowires by spin waves propagating in a thin YIG film~\cite{31_Reversal_enabling_magnon_memory_Grundler}.
In the configuration Fig.~\ref{injection_experiment}(a), two Py nanostripe arrays with a 25~$\mu$m spacing are sandwiched by the coplanar waveguides (CPWs) and YIG film. 
The magnetization ${\bf M}_{\rm Py}$ of Py strips is along the $\pm \hat{\bf y}$-direction due to the shape anisotropy. 
The emitter (left) CPW that emits linearly polarized magnetic field ${\bf h}_{\rm rf}$ inside the YIG film and generates the spin wave propagates in the $\pm \hat{\bf x}$-direction with same amplititute is connected to Port 1 of the VNA, as shown in Fig.~\ref{injection_experiment}(a). 
The detector (right) CPW is connected to Port 2. As in Fig.~\ref{injection_experiment}(b), with the white/black end (encircled by a yellow/red box) of the strips indicating the magnetization of Py along $\mp\hat{\bf y}$-direction, all the nanostrip magnetizations are initially set to the $-\hat{\bf y}$-direction. 
When increasing the excitation powers $P_{\rm irr}$, more Py magnetizations on the right-hand side of CPW1 are reversed than the Py strips on the left-hand side.

The experiments suggest the dipolar interaction is responsible. Since the frequency of the excitation $\sim3$~GHz is far less than the FMR of Py strips $\sim10$~GHz, no magnons are excited in Py. The spin waves then inject spin current carried by electrons in the Py strip, which provides a torque that switches the magnetization. 
Although both exchange~\cite{RevModPhys.77.1375} and dipolar interactions~\cite{Yu_Spin_Pumping,cai_spin_radiation,doi:10.1126/science.aav8076} inject spin current, only the dipolar interaction shows chirality as in Fig.~\ref{gating_01}(a). Such chirality is consistent with the experimental observation that more Py magnetizations on the right-hand side of CPW 1 are reversed~\cite{31_Reversal_enabling_magnon_memory_Grundler}. A more direct evidence is later provided by Mucchietto \textit{et al.}~\cite{38_Magnetization_Reversal_Grundler}, who observed a similar magnetization switch inside the Py strip above the YIG film when an insulator spacer was inserted between them to suppress the exchange interaction.

Schlitz \textit{et al.} observed the angular momentum flow between separated ferromagnets deposited on top of a diamagnetic insulator substrate~\cite{inject_exp}. As in Fig.~\ref{injection_experiment}(c), with the magnetization biased by a magnetic field along the surface normal $\hat{\bf z}$, a current $I_{\rm inj}\parallel \hat{\bf x}$ in one ferromagnetic strip ``FM1" generates a voltage in the other ferromagnetic strip ``FM2".
The dipolar coupling~\cite{DZYAPKO201383,10.1063/1.4936207} and the phonon-mediated spin transport through the substrate~\cite{PhysRevLett.121.027202,Holanda2018,10.1063/5.0035690,PhysRevB.101.060407,PhysRevLett.124.117201,PhysRevB.101.104402,PhysRevB.103.144430} could transport the (spin) angular momentum from FM1 to FM2, but the efficiency by phonon may be smaller~\cite{surface_acoustic_wave,surface_magnetoelastic,surface_acoustic_wave2,PhysRevB.107.L100410}.
The spin transfer efficiency $\eta_s(d)=I_{\rm det}/I_{\rm inj}$ as a function of distance $d$ between two ferromagnet strips is shown in Fig.~\ref{injection_experiment}(d). Exponential decay $\eta_s\sim e^{-d/\lambda_{ph}}$ by the phonon mechanism is different from the power law decay $\sim 1/d^2$ by dipolar interaction. The fitting by exponential and power-law decay suggests the dipolar interaction may be responsible.

\subsubsection{Theory}

The key to the electromagnetic proximity effect is the near-field dipolar interaction between electrons in conductors and ferromagnets. The magnetic nanostructures, when excited, generate the stray field that can carry photon spin~\cite{PhysRevApplied.22.034042}, which pumps the electron spin currents in conductors~\cite{Yu_Spin_Pumping,cai_spin_radiation,achilli2025opticalspinpumpingsilicon,zuev2025superconductingphotocurrentsinducedstructured}. The inverse process is the pumping of magnon spin current by the stray field emitted by the spin-non-equilibrium population of electrons. Zhou \textit{et al.} showed that the dipolar field emitted by the spin accumulation inside the metals can transfer the angular momentum to the magnon spin current via the long-range dipolar interaction, where the injected magnon spin current and electron spin-accumulation direction exhibit a chiral locking relation~\cite{Chiral_Injection_of_Magnons}.

\begin{figure}[htp!]
    \centering
    \includegraphics[width=0.8\linewidth]{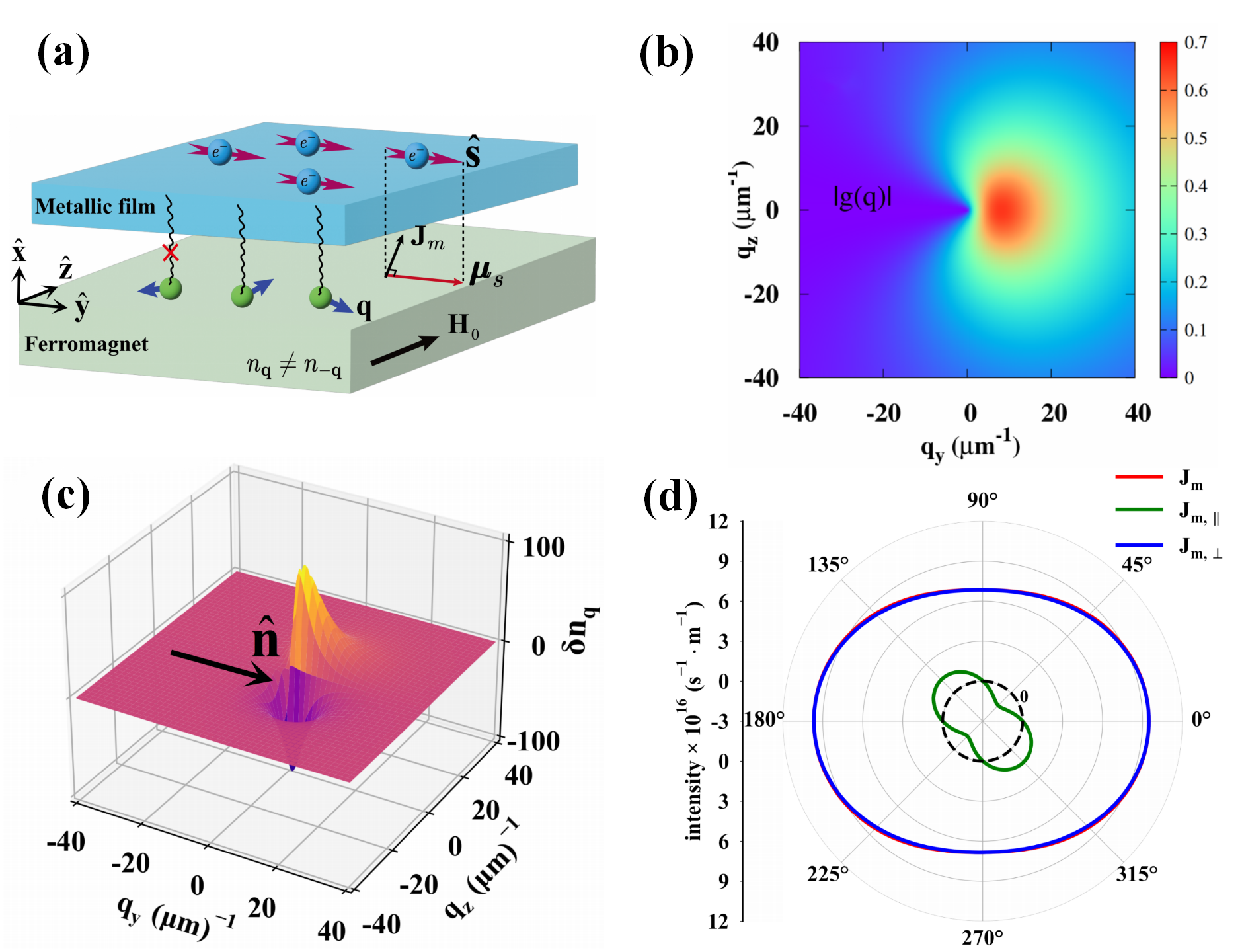}
    \caption{(a) Chiral injection of magnons with $n_{\bf q}\ne n_{-{\bf q}}$ by nearby spin accumulation ${\pmb \mu}_s$ in metals that generates the magnon current ${\bf J}_m\perp {\pmb \mu}_s$. (b) Chirality of interlayer dipolar coupling strength $|g({\bf q})|$. (c) Distribution of injected magnons $\delta n_{\bf q}$ in the Brillouin zone when biased by spin-accumulation $\parallel \hat{\bf y}$. (d) The magnon flow ${\bf J}_m$, ${\bf J}_{m,\parallel}$, and ${\bf J}_{m,\perp}$ as a function of spin-accumulation directions $\phi$ with respect to the saturation magnetization direction.\\
    \textit{Source:} These figures are taken from Ref.~\cite{Chiral_Injection_of_Magnons}.}
    \label{Chiral_Injection_of_Magnons}
\end{figure}

As shown in a bilayer geometry in Fig.~\ref{Chiral_Injection_of_Magnons}(a),  a heavy metal that allows a spin accumulation ${\pmb \mu}_s$ of electrons biases a thin ferromagnet with thickness $d$. An external magnetic field ${\bf H}_0$ biases the saturation magnetization to $\hat{\bf z}$. A thin insulator spacer, such as Al$_2$O$_3$, is inserted between them to suppress the adjacent exchange interaction. 
The electron spin \(\hat{\bf s}({\bf r})\) in the metallic film interacts with the stray magnetic field \(\hat{\bf h}_d({\bf r})\) of magnons via the dipolar coupling
\begin{align}
\hat{H}_{\rm int}&=\mu_0\gamma_e\hbar\int_0^s d{\bf r}\hat{\bf s}({\bf r},t)\cdot \hat{{\bf h}}_d({\bf r},t)
=\hbar\sum_{{\bf k}{\bf q}}\left({\bf g}({\bf q})\cdot {\pmb \sigma} \right)_{\delta\xi}\hat{f}_{{\bf k}+{\bf q},\delta}^{\dagger}\hat{f}_{{\bf k},\xi}\hat{\alpha}_{\bf q}+{\rm H.c.},
\label{dipolar}
\end{align}
where $\hat{\alpha}_{{\bf q}}$ is the magnon  operator with wave vector ${\bf q}$ and $ \hat{f}_{{\bf k},\xi}$ is the electron operator with momentum $\bf k$ and spin $\xi$.
It describes the spin non-conserved scattering process with momentum conservation, governed by the scattering amplitude $\left({\bf g}({\bf q})\cdot {\pmb \sigma} \right)_{\delta\xi}$ with 
\begin{align}
g_{x}({\bf q})&\rightarrow -i\sqrt{1/(4{\cal A}d)}\eta F(|{\bf q}|)(1-e^{-|{\bf q}|d})(1+\cos\theta_{\bf q}),\nonumber\\
g_{y}({\bf q})&\rightarrow -i\cos\theta_{\bf q}g_x({\bf q}),\quad g_{z}({\bf q})\rightarrow -i\sin\theta_{\bf q}g_x({\bf q}),
\label{coupling_constants_g}
\end{align}
where ${\cal A}$ is the area of sample, $\theta_{\bf q}$ is the angle between $\bf q$ and $\hat{\bf y}$, $\eta\equiv -({\mu_0\gamma_e}/{4})\sqrt{{2M_s}{\gamma\hbar}}$, and $F(|{\bf q}|)={1}/{(|{\bf q}|s)}\left(1-e^{-|{\bf q}|s}\right)$. ${g}_x({\bf q})$ and ${g}_y({\bf q})$ induce the spin-flip scattering for electrons, while $g_z({\bf q})$ causes spin-conserved scattering. By the spatial chirality addressed in  Fig.~\ref{gating_01}(a), ${\bf g}({\bf q})$ mainly exists when $q_y>0$, but vanishes when $q_y<0$, exhibiting the chirality as in Fig.~\ref{Chiral_Injection_of_Magnons}(b).

The chirality induces the unbalanced magnon distribution $n_{\bf q}\neq n_{\bf -q}$ that generates a net magnon current ${\bf J}_m=(1/{\cal A}) \sum_{\bf q} {\bf v}_{\bf q} n_{\bf q}$, where the magnon group velocity ${\bf v}_{\bf q}= \partial \Omega_{\bf q}/(\hbar\partial{\bf q})$ in terms of the magnon energy $\Omega_{\bf q}$. The exchange interaction induces no magnon current in this geometry. The injection to the magnon population $n_{\bf q}$ is governed by 
\begin{align}
    \frac{\partial n_{\bf q}}{\partial t}\Big|_s
    =& {2\pi}{\hbar}\sum_{{\bf k}}\delta(\varepsilon_{{\bf q}+{\bf k}}-\varepsilon_{{\bf k}}-\Omega_{\bf q})\times\Big\{{\rm Tr}\left[\rho_{{\bf q}+{\bf k}}({\bf g}({\bf q})\cdot{\pmb \sigma})(1-\rho_{{\bf k}})({\bf g}({\bf q})\cdot{\pmb \sigma})^{\dagger}\right](1+n_{\bf q})\nonumber\\
-&{\rm Tr}\left[(1-\rho_{{\bf q}+{\bf k}})\left({\bf g}({\bf q})\cdot{\pmb \sigma}\right)\rho_{{\bf k}}\left({\bf g}({\bf q}\right)\cdot{\pmb \sigma})^{\dagger}\right]n_{\bf q}\Big\},
\end{align}
where $\rho_{{\bf k}}=\dfrac{{\cal F}_{\bf k,\uparrow}+{\cal F}_{\bf k,\downarrow}}{2}{\cal \pmb I}_{2\times 2}+\dfrac{{\cal F}_{\bf k,\uparrow}-{\cal F}_{\bf k,\downarrow}}{2}{\pmb \sigma\cdot \hat{\bf n}}$
is the density matrix of electrons~\cite{Wu_review,PhysRevA.88.043634}, in which 
 ${\cal{\pmb I}}_{2\times 2}$ is the $2 \times 2$ identity matrix and ${\cal F}_{\bf k,\zeta}=(e^{\beta {\cal E}_{\bf k,\zeta}}+1)^{-1} $ is the Fermi-Dirac distribution with ${\cal E}_{\bf k,\uparrow}=\varepsilon_{\bf k}-\mu_{\uparrow}$ and ${\cal E}_{\bf k,\downarrow}=\varepsilon_{\bf k}-\mu_{\downarrow}$. With different chemical potentials $\mu_\uparrow=\varepsilon_{F}+\mu_s/2$ and $\mu_\downarrow=\varepsilon_{F} -\mu_s/2$, the spin accumulation ${\pmb \mu}_s\parallel \hat{\bf n}$.
The injected magnon relaxes to the equilibrium distribution $N_{\bf q}$ within the relaxation time $\tau_{\bf q}$, with which the injected magnons
\begin{align}
    \delta n_{\bf q}=n_{\bf q}-N_{\bf q}\approx -8 \pi\hbar N_{\bf q}I_{\bf q}\tau_{\bf q}\mu_s g_x^2({\bf q}) \cos{(\theta_{\bf q}+\phi)},
\end{align}
with the condition that $N_{\bf q}\gg 1$ and $\mu_s\gg \Omega_{\bf q}$,
where $I_{\bf q}=-\sum_{\bf k}\delta(\varepsilon_{\bf k+q}-\varepsilon_{\bf k}-\Omega_{\bf q})\delta(\varepsilon_{\bf k}-\varepsilon_{F})$ and $\phi$ is the angle of the spin-polarization direction $\hat{\bf n}$ with respect to the saturation magnetization \(\hat{\bf z}\)-direction.

The magnon accumulation or absorption is governed by $g_x^2({\bf q}) \cos(\theta_{\bf q}+\phi)$.
$g_x({\bf q})$ mainly exists when $q_y>0$ [Fig.~\ref{Chiral_Injection_of_Magnons}(b)]; $\cos{(\theta_{\bf q}+\phi)}$ determines whether magnons are excited or absorbed. 
Note $I_{\bf q}<0$ and $g_x^2({\bf q})<0$ when $\cos{(\theta_{\bf q}+\phi)}<0$, the magnons are excited; while they are absorbed by the metal when $\cos{(\theta_{\bf q}+\phi)}>0$. As an example, when the spin-polarization is along $\hat{\bf y}$ ($\phi=\pi/2$), the magnons are excited for $0<\theta_{\bf q}<\pi/2$ since $\cos{(\theta_{\bf q}+\phi)}<0$ and absorbed for $-\pi/2<\theta_{\bf q}<0$ with $\cos{(\theta_{\bf q}+\phi)}>0$, as shown in Fig.~\ref{Chiral_Injection_of_Magnons}(c). This implies a magnon flux ${\bf J}_m$ along $\hat{\bf z}$, which is perpendicular to $\hat{\bf n}$. Indeed, the chiral locking relation ${\bf J}_m \perp \hat{\bf n}$ is a universal feature in the near-field spin transfer: with the parallel and normal components of ${\bf J}_m$
\begin{align}
    {\bf J}_{m,\parallel}\equiv \hat{\bf n}\cdot {\bf J}_{m}\propto\sin(2\phi),&& {\bf J}_{m,\perp}\equiv \hat{\bf n}\times {\bf J}_{m}\propto-(6+\cos(2\phi)),
\end{align}
\({\bf J}_{m,\parallel}/{\bf J}_{m,\perp}\ll1\), i.e., the parallel components is negligible small.  Figure~\ref{Chiral_Injection_of_Magnons}(d) plots ${\bf J}_{m,\perp}$, ${\bf J}_{m,\parallel}$, and ${\bf J}_{m}$ as a function of spin-accumulation directions $\phi$. The chiral locking is exact \({\bf J}_{m,\parallel}/{\bf J}_{m,\perp}=0\) when $\phi=\{ 0,\pi/2,\pi,3\pi/2\}$.

\section{Summary and outlook}
\label{summary}

In conclusion, we have reviewed the proximity effect mediated by the evanescent stray electromagnetic field between magnetic/ferroelectric excitations and electrons in superconductors or normal metals. The proximity of metallic gates affects a magnet through diamagnetic supercurrents or eddy currents induced by the magnetic stray fields emitted by spin waves. We demonstrate that the evanescent field between the magnet and conductor is generally chiral in that the locking of the wave vector, spin, and surface normal directions is universally described by the right-hand rule, which can be interpreted as non-relativistic ``geometrical spin-orbit interaction"~\cite{Yu_chirality,skin7,PhysRevApplied.22.034042}. Examples include the Damon-Eshbach surface spin waves~\cite{DEmode,DE2}, stray fields emitted by spin waves in thin magnetic films~\cite{Tao_chiral_pumping,Tao_chiral_excitation}, electric field emitted by surface
plasmon polariton~\cite{SPP1,SPP2,SPP3,SPP4,SPP5}, surface acoustic waves~\cite{surface_acoustic_wave,surface_magnetoelastic,surface_acoustic_wave2}, microwaves at the boundaries of metallic waveguides~\cite{Tao_Magnon_Accumulation}, the magnetic field of microwave striplines~\cite{Yu_chirality}, and surface ferrons~\cite{ferron,bauer_ferron_trans,ferroelectric,Capacitors,Point_Contacts,ferron_transport_exp,Coherent_exp,Diffuse_exp,fe_polarition,bulkmode_ferron2,surface_ferron,bulkmode_ferron,ferron_landau,ferron_landau1,LiNbO3,LiNbO3_g,LKT1,LKT2,LKT3,LKT4}. The interaction between magnons and adjacent conventional superconductor results in new collective modes or quasiparticles such as the magnon-Meissner collective modes by the electromagnetic proximity effect~\cite{46_Efficient_Gating,21_Fraerman,48_Giant_enhancement_of_magnon_transport,42_Centala,5_magnon_fluxon_Nat_Chumak,19_magnon_fluxonics_Chumak,34_Berakdar,39_Cherenkov_radiation_Buzdin,Golovchanskiy_gating3_exp,20_shift_Silaev,47_Gating_ferromagnetic_resonance,52_Persistent_nodal_magnon-photon_polariton_in_ferromagnetic_heterostructures} and magnon cooparon~\cite{PhysRevLett.127.207001,Bobkova2022,Bobkov2023} by the exchange proximity effect~\cite{PhysRevLett.127.207001,Bobkova2022,Bobkov2023,Bergeret2018,Buzdin,Bergeret,Nembach2015,Demokritov2001,Cornelissen2015,Goennenwein2015,Kamra2018}. In the Josephson SC$|$FM$|$SC heterostructure, it also induces the Josephson plasmonic modes~\cite{41_Eremin,1_Efetov,36_Bobkov}; when the temperature is larger than the superconducting transition temperature, the superconductor becomes the normal metal, the Josephson plasmonic modes coalescence to the nodal magnon-photon polariton without any splitting gap at specific wave vectors~\cite{52_Persistent_nodal_magnon-photon_polariton_in_ferromagnetic_heterostructures}.

\begin{figure}[htp]
    \centering
    \includegraphics[width=1\linewidth]{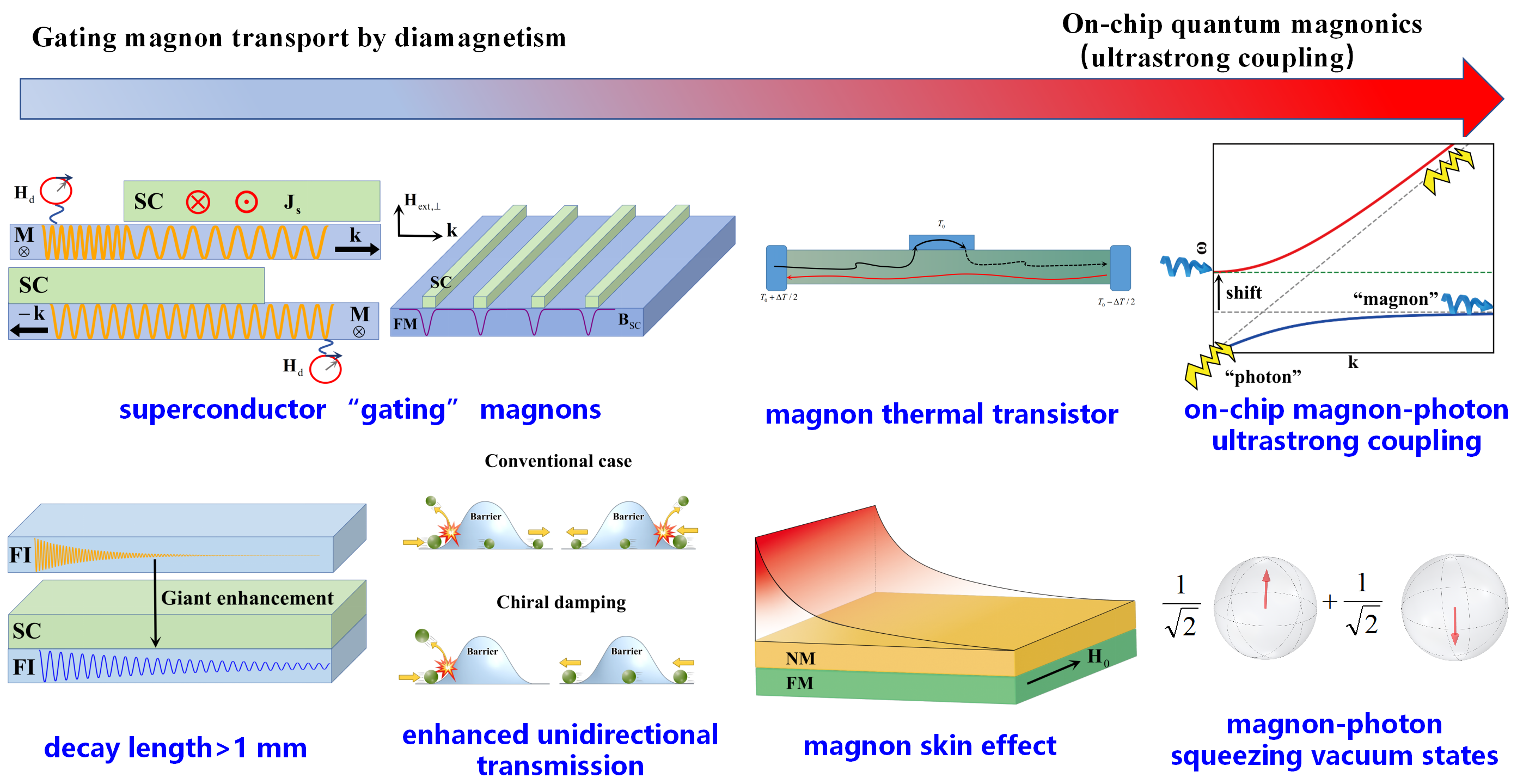}
    \caption{Typical effects by electromagnetic proximity effect, including superconductor ``gating'' effect, magnon thermal transistor, on-chip magnon-photon ultrastrong coupling, giant enhancement of magnon transport, enhanced unidirectional transmission, magnon non-Hermitian skin effect, and magnon-photon squeezing vacuum states.}
    \label{summary_functionalities}
\end{figure}

Metallic gates modulate and control the magnetodipolar interaction that causes chirality of the magnetization and energy transport by magnetic/ferroelectric excitations at interfaces and in thin films. 
Metallic and superconducting contacts thereby generate new functionalities in magnonic circuits and devices, as summarized in Fig.~\ref{summary_functionalities}.  In the SC$|$FM bilayer, the superconductor locally shifts the magnon frequency and therefore acts as non-dissipative repulsive potential barriers for propagating spin waves~\cite{46_Efficient_Gating,Golovchanskiy_gating2,Golovchanskiy_gating4,21_Fraerman,37_klos,55_Burmistrov,42_Centala,48_Giant_enhancement_of_magnon_transport,5_magnon_fluxon_Nat_Chumak,19_magnon_fluxonics_Chumak,34_Berakdar,39_Cherenkov_radiation_Buzdin,Golovchanskiy_gating1_exp,Golovchanskiy_gating3_exp,8_Pt_S_F_vortex_Blamire,13_Passamani,54_L.Tao,Seshadri_gating,Mruczkiewicz_gating1,24_SF_shift_van_der_sar,2_FS_for_Magnonic_application_Ustinov,41_Eremin}. When sandwiched between two superconductors, the superconductor reflects the emitted stray magnetic field back and forth that shifts the FMR giantly~\cite{4_sun,7_SFS_shift_Blamire,10_SFS_shift_Ustinov,25_SFS_shift_Stolyarov,16_Giant_demagnetization_effects_Buzdin,20_shift_Silaev,47_Gating_ferromagnetic_resonance,bai_shift,shift_yangguang,23_magnon_photon_coupling_Silaev} and induces the ultrastrong coupling between magnons and Swihart photons~\cite{18_photon-to-magnon_coupling_Ustinov,17_On-Chip_Photon-To-Magnon_Coupling_Ustinov,35_magnon_photon_with_SC_resonator_Affronte,40_magnon-photon_bilayer_Affronte,23_magnon_photon_coupling_Silaev,52_Persistent_nodal_magnon-photon_polariton_in_ferromagnetic_heterostructures,20_shift_Silaev,Gordeeva2025,26_AFM_Resonances_in_SFS_Stolyarov}; both can be exploited in the future quantum applications generating squeezed quantum states and transferring entanglement
between different types of quantum systems~\cite{20_shift_Silaev,23_magnon_photon_coupling_Silaev,Hopfield_model}. 
The eddy currents in normal metals are dissipated according to Ohm’s Law and thereby cause an additional damping channel that increases the Gilbert damping constant on top of its intrinsic value. This effect can be distinguished from that of spin-pumping by a thin insulating barrier that suppresses any exchange interaction. Moreover, the chirality of the stray field translates into non-local chiral damping~\cite{50_Chiral-damping-enhanced_magnon_transmission,45_Imaging_spin_wave_damping_underneath_metals,Chiral_Damping_of_Magnons}, i.e. affects only spin waves that obey the right-hand rule. The induced eddy currents also generate Oersted magnetic fields that blue shift the magnon frequencies. The spin wave propagation and transport with chiral damping is a non-Hermitian chiral Hamiltonian problem. Its numerical solution for finite systems leads to a chiral spin wave skin effect in which magnon accumulates at one edge~\cite{50_Chiral-damping-enhanced_magnon_transmission}. Three-terminal thin film magnon ``transistors” may consist of thin films of magnetic insulators gated by magnetic metals. Here, the angle between the thin film and gate magnetization controls the heat and spin currents between the source and drain, leading to a switchable spin wave Peltier effect~\cite{49_Chirality_enables_thermal_magnon_transistors,53_van_Wees,15_Liu}. The two-terminal non-Hermitian problem can be mapped onto a three-terminal Hermitian one by modelling the dissipative gate by a thermodynamic reservoir. In the case of ballistic spin waves, the problem is equivalent to the integer quantum Hall effect in large quantum dots. The Landauer-B\"uttiker scattering theory of transport reveals that up to half of the source-drain currents may be diverted into the gate~\cite{49_Chirality_enables_thermal_magnon_transistors}. Surprisingly, chiral damping also affects the scattering of spin waves at semi-transparent potential barriers~\cite{50_Chiral-damping-enhanced_magnon_transmission}. Spin waves with the fitting chirality may profit from a sufficiently strong damping by an anomalous transmission probability that can be strongly enhanced from the intrinsic one up to unity. The dipolar interaction between electrons and magnons also extends the conventional spin pumping to the near-field spin pumping~\cite{31_Reversal_enabling_magnon_memory_Grundler,38_Magnetization_Reversal_Grundler,inject_exp,Chiral_Injection_of_Magnons}, governed by a universal chiral locking of magnon flow and electron spin accumulation directions.

The electromagnetic proximity effect may be useful for future spin-wave computing applications by enabling easy reprogramming and heat management of magnonic circuits. Currently, intriguing effects for magnons in ferromagnets modulated by conventional superconductors are being extensively explored in the linear regime. The backaction of the magnons on the superconductor order is disregarded at large. We envision that the modulation effects are strong for magnons in antiferromagnets~\cite{RevModPhys.90.015005} and ferronic excitations in ferroelectric and ionic crystals. Choices of superconductors may be extended to high-temperature $d$-wave superconductors and triplet superconductors, which exhibit richer charge and spin dynamics. A superconductor can divert the thermal magnon flow, acting as a useful transistor, and enhance the nonlinear response of magnetization dynamics by creating a cavity.

\addcontentsline{toc}{section}{Declaration of competing interest}
\section*{Declaration of competing interest}
The authors declare no competing financial interests that could have appeared to influence the work reported in this paper.

\addcontentsline{toc}{section}{Acknowledgments}
\section*{Acknowledgments}
This work is financially supported by the National Key Research and Development Program of China under Grant No.~2023YFA1406600 and the National Natural Science Foundation of China under Grant No.~12374109. The JSPS KAKENHI Grants No.~22H04965 and JP24H02231 support G.B. financially. I.V.B. acknowledges the financial support by the Russian Science Foundation via the project No.~23-72-30004. We gratefully acknowledge our many valuable discussions with Sergio M. Rezende, Teono van der Sar, Mihail Silaev, Eugene Kamenetskii, Lihui Bai, Guang Yang, Junxue Li, Ke Xia, Ji Zou, Ping Tang, Mehrdad Elyasi, Hanchen Wang, Xiyin Ye, Zhiping Xue, Ping Li, Zhuolun Qiu, and Chengyuan Cai.

\addcontentsline{toc}{section}{References}

\normalem
\bibliography{refs}
\bibliographystyle{apsrev4-2}

\end{document}